\documentclass[letterpaper,11pt]{article}
\usepackage[utf8]{inputenc}

\pdfoutput=1
\usepackage{jheppub}
\usepackage{preambles_Saebyeok}
\usepackage{preambles_Nathan}

\title{{Miura operators as R-matrices
from M-brane intersections}}
\author[a]{Nathan Haouzi}
\author[b,c]{and Saebyeok Jeong}

\affiliation[a]{School of Natural Sciences, Institute for Advanced Study, \\ Einstein Drive, Princeton, NJ 08540 USA}
\affiliation[b]{Department of Theoretical Physics, CERN, \\ 1211 Geneva 23, Switzerland}
\affiliation[c]{Center for Geometry and Physics, Institute for Basic Science, \\ Pohang 37673, South Korea}

\emailAdd{nathanh@ias.edu}
\emailAdd{saebyeok.jeong@gmail.com}

\preprint{CERN-TH-2024-094}

\vspace{2cm}
\abstract{We propose that Miura operators are R-matrices of certain infinite-dimensional quantum algebras. We test our proposal by realizing Miura operators of $q$-deformed $W$- and $Y$-algebras in terms of R-matrices of the quantum toroidal algebra of $\fgl(1)$. 
Physically, the representations of this toroidal algebra arise from the algebra of local operators on M2-branes and M5-branes, in M-theory subject to an $\O$-background. 
We associate an R-matrix to each M2-M5 brane crossing, by studying its description as a gauge-invariant intersection of a topological line defect and a holomorphic surface defect in 5-dimensional non-commutative Chern-Simons theory. The Miura transformation is engineered using multiple M2-M5 intersections, relying crucially on the properties of the underlying R-matrices. We thereby identify each R-matrix with a Miura operator.
In a dual type IIB frame, the components of the Miura transformation are shown to coincide with the half-index of a 3d supersymmetric gauge theory on a Hanany-Witten system of D3-NS5 branes. As a further application, we demonstrate that $qq$-characters can be algebraically constructed from the Miura transformation.}

\begin{document}
\maketitle
\section{Introduction}

In two-dimensional conformal field theory, the Miura transformation\footnote{In the present work, the Miura transformation always refers to the \textit{quantum} Miura transformation, unless specified otherwise.} is a procedure used to construct the holomorphic generators of the theory in terms of free bosons. The most celebrated example is the construction of the $\mathcal{W}_N$-algebra, a higher spin generalization of the Virasoro 
vertex algebra consisting of $N$ generators (we add an additional free boson to the usual $W$-algebra of $\fsl(N)$, and call it the $W$-algebra of $\fgl(N)$). In this case, the Miura transformation states that the generating currents $U_k(z)$ of the $\cW_N$-algebra are constructed by
\begin{equation}\label{Miuraintro}
(\ve_3 \partial_z - \ve_1 \ve_2 J_N (z) )(\ve_3 \partial_z - \ve_1 \ve_2 J_{N-1} (z) ) \cdots (\ve_3 \partial_z - \ve_1 \ve_2 J_1 (z) )=\sum_{k=0}^N U_k(z) \left(\ve_3 \partial_z\right)^{N-k},
\end{equation}
where $U_0 (z) = 1$. Here, $(\ve_1,\ve_2,\ve_3) \in \BC^{ 3}$, constrained by $\ve_1+\ve_2+\ve_3= 0$, parametrize the central charge by $c = N\left(1+ (N^2-1)\left( \frac{\ve_2}{\ve_1} + \frac{\ve_1}{\ve_2} + 2 \right)  \right)$.
%via $c=(N-1)(1+\alpha_0^2\, N \, (N+1))$.
The $J_i (z)$ is the current of the $\widehat{\fgl}(1)$ vertex algebra with the operator product expansion (OPE) given by $J_i (z) J_j (w) \sim -\frac{1}{\ve_1 \ve_2} \frac{1}{(z-w)^2} \d_{i,j}$, whose mode algebra is the free boson algebra. Each factor $\ve_3 \partial_z - \ve_1 \ve_2 J_i (z) $ on the left-hand side of \eqref{Miuraintro} is called a \emph{Miura operator}. 

The $\mathcal{W}_N$-algebra arises as the vertex algebra of local operators of a 6-dimensional $\EN=(2,0)$ superconformal field theory subject to the $\O$-background. This is the 6-dimensional origin of the celebrated correspondence between $\mathcal{W}_N$-algebra conformal blocks and  partition functions of 4-dimensional $\EN=2$ supersymmetric $U(N)$ gauge theories \cite{Alday:2009aq, Wyllard:2009hg,Yagi:2012xa,Beem:2014kka}.

In a trivial background, the 6d theory is the low energy effective field theory on $N$ M5-branes in M-theory. This prompted Costello to define a twisted version of M-theory on the $\O$-background \cite{Costello:2016nkh}, where $N$ parallel M5-branes are reinterpreted as a surface defect in 5-dimensional non-commutative $\fgl(1)$ Chern-Simons theory.
 The $\CalW_N$-algebra is then viewed as the vertex algebra of local operators on the surface defect whose mode algebra acts on the Hilbert space of this surface defect theory, which is identified with the equivariant cohomology of the moduli space of $U(N)$ instantons on $\BR^4$. The configuration can be augmented to incorporate M5-branes oriented in different directions, extending the relevant vertex algebra to the $Y$-algebra \cite{Gaiotto:2017euk,Gaiotto:2020dsq}.

In this \textit{twisted} M-theory setting, it was proposed that the Miura operator for the $\CalW_N$-algebra should be realized as the transverse intersection of an M2-brane and an M5-brane \cite{Gaiotto:2020dsq}. It was observed that this intersection point supports a fermionic zero-mode, with the classical expectation value given by the distance between the two branes in the transverse holomorphic plane. Quantum corrections are then suggested to uplift this expectation value to $\ve_3 \p_z - \ve_1 \ve_2 J(z)$, the Miura operator for the $\CalW_N$-algebra. This twisted M-theory approach was further extended to include \textit{pseudo-differential} Miura operators for more general $Y$-algebras \cite{Prochazka:2018tlo,Gaiotto:2020dsq}, incorporating non-transverse intersections of the M2- and M5-branes that partially overlap in their worldvolume.

Meanwhile, a definition of a multiplicative uplift of the $Y$-algebra was recently proposed in \cite{Harada:2021xnm}, in a purely algebraic framework. In particular, it was shown that Miura operators for a $q$-deformed $Y$-algebra could be defined in terms of certain $q$-difference operators, replacing the previous differential operators.
It is tempting to guess that these Miura operators should likewise have a physical interpretation as brane intersections in twisted M-theory. One of the main goals of this work is to make this intuition precise, and prove that the algebraic Miura operators of \cite{Harada:2021xnm} are exactly supported on M2-M5 intersections in the multiplicative uplift of the twisted M-theory background considered in \cite{Gaiotto:2020dsq} (up to a slight redefinition we will motivate).

We will start by reviewing the twisted M-theory and 5-dimensional non-commutative Chern-Simons theory \cite{Costello:2016nkh}. Then, we will introduce the central idea of the present work: that the M2-M5 intersection gives rise to an \emph{R-matrix}.

\subsection{Twisted M-theory and 5-dimensional Chern-Simons theory}
Let us consider the M-theory defined on the 11-dimensional worldvolume $\BR^7 \times \mathcal{C}$, where $\mathcal{C}$ is a complex two-dimensional manifold to be chosen. Three main cases are $\mathcal{C} = \BC \times \BC$; $\mathcal{C} = \BC \times \BC ^\times$; and $\mathcal{C} = \BC^\times \times \BC^\times$. While we will mostly focus on the last case throughout the work, it is helpful to treat them all at once here to illustrate their differences.

With the flat metric on the 11-dimensional worldvolume, there is a nilpotent supercharge $\mathcal{Q}_{\ve_1=\ve_2=0}$, under which $\BR^7$ is topological and $\mathcal{C}$ is holomorphic in its cohomology. Now, we further introduce the 3-form background $C^{(3)} = (\ve_1 V_1 ^\flat + \ve_2 V_2 ^\flat) \wedge \bar{\o} $, where $V_i = \p_{\varphi_i} - \p_{\varphi_3}$, $i=1,2$, are the vector fields generating the rotations on the three $\BR^2$-planes ($\varphi_i$ is the angular coordinate for the $i$-th $\BR^2$-plane) and $V_i ^\flat$ are the dual 1-forms. Also, $\bar{\o}$ is a $(0,2)$-form on $\mathcal{C}$ given by the complex conjugate of the holomorphic symplectic $(2,0)$-form $\o$,
\begin{align}
\bar{\o} = \begin{cases}
    d\bar{x} \wedge d\bar{z} \qquad\qquad &\mathcal{C} = \BC_x \times \BC_z \\
    d\bar{x} \wedge \frac{d\bar{Z}}{\bar{Z}} &\mathcal{C} = \BC_x \times \BC^\times _Z \\
     \frac{d\bar{X}}{\bar{X}} \wedge \frac{d\bar{Z}}{\bar{Z}} & \mathcal{C} = \BC^\times _X  \times \BC^\times _Z
\end{cases},
\end{align}
where we indicated the holomorphic coordinates on the complex planes by subscripts. At this point, let us also write the topological part of the worldvolume as $\BR^2 _{\ve_1} \times \BR^2 _{\ve_2}\times \BR^2_{\ve_3} \times \BR_t$, where we used the subscripts to invoke the background parameters associated to the rotations on the respective $\BR^2$-planes $(\ve_3 = -\ve_1-\ve_2)$ and $t$ is the coordinate on the remaining topological line $\BR$. In the presence of the 3-form background, the preserved supercharge gets deformed to $\mathcal{Q}_{\ve_1,\ve_2}$, which squares to the isometry
\begin{align}
    \mathcal{Q}_{\ve_1,\ve_2} ^2 = \ve_1 V_1 + \ve_2 V_2.
\end{align}
The M-theory subject to the $\mathcal{Q}_{\ve_1,\ve_2}$-cohomology is called the twisted M-theory \cite{Costello:2016nkh}.\\

Since the supersymmetry squares to the rotations of the $\BR^2$-planes, the twisted M-theory localizes to an effective 5-dimensional theory on the fixed locus $\{0\} \times \BR_t \times \mathcal{C}$. The localization procedure can be understood in the following way. We may continuously deform the flat $\BR^2 _{\ve_2} \times \BR^2 _{\ve_3}$ to the single-centered Taub-NUT space. Reducing the M-theory along the circle fiber of the Taub-NUT space, we arrive at the IIA theory with an emergent D6-brane supported on $\BR^2 _{\ve_1} \times \BR_t \times \mathcal{C}$ at the location of the center. The effective field theory on the worldvolume of the D6-brane is the 7-dimensional supersymmetric $U(1)$ gauge theory. The role of the 3-form background $C^{(3)}$ is two-fold. First, it descends to a B-field which is a mixture of the back-reaction of the emergent D6-brane and the constant B-field background. The back-reaction of the D6-brane can be subtracted by taking its asymptotics at infinity, leaving $B = \ve_2 \bar{\o}$. Because of this constant B-field, the usual wedge product of the differential forms in the 7d effective gauge theory is deformed to the non-commutative Moyal product defined by
\begin{align}
    f\star_{\ve_2} g = \sum_{n=0} ^\infty \frac{\ve_2 ^n}{2^n n!} \epsilon_{i_1 j_1} \cdots \epsilon_{i_n j_n} \left(\frac{\p}{\p z_{i_1}} \cdots \frac{\p}{\p z_{i_n}} 
 f\right)\wedge\left(\frac{\p}{\p z_{j_1}} \cdots \frac{\p}{\p z_{j_n}} g \right),
\end{align}
where $\epsilon_{ij}$ is the anti-symmetric symbol and $z_1 = x$ or $\log X$ and $z_2 = z$ or $\log Z$ depending on whether the corresponding complex plane is $\BC$ or $\BC^\times$. Second, the 3-form also descends to a RR 3-form background which, after subtracting the one sourced by the D6-brane in the presence of the constant B-field, implements the $\O$-background for the 7d effective gauge theory associated with the rotation on $\BR^2 _{\ve_1}$. 

In total, the 7-dimensional gauge theory is localized into the 5-dimensional non-commutative $\fgl(1)$ Chern-Simons theory on $\BR_t \times \mathcal{C}$, whose action is given by
\begin{align}
    S =  \frac{1}{\ve_1}\int_{\BR_t \times \mathcal{C}} \o \wedge \left( A \star_{\ve_2} dA + \frac{2}{3} A \star_{\ve_2} A \star_{\ve_2} A \right),
\end{align}
where $A$ is the complex-valued gauge field with three components
\begin{align}
    A= A_t dt + A_{\bar{z}_1} d\bar{z}_1 + A_{\bar{z}_2} d\bar{z}_2.
\end{align}
Here, $\bar{z}_{1,2}$ are the anti-holomorphic coordinates on $\mathcal{C}$.

\begin{table}[h!]
    \centering
    \begin{tabular}{  c||c|c|c|c|c|c|c|c|c|c|c  } 
          \text{M-branes} & 0 & 1 & 2 & 3 & 4 & 5 & 6 & 7 & 8 & 9 &10 \\ \hline\hline 
          $\text{M2}_{1,0,0}$ &  \rm{x} & \rm{x}  &  &   &  &   & \rm{x}  & &  &   &  \\ 
           $\text{M2}_{0,1,0}$ &  &  & \rm{x}  & \rm{x}   &  &   & \rm{x}  & &  &   & \\
           $\text{M2}_{0,0,1}$ &  &  &  &   & \rm{x}  & \rm{x}  & \rm{x}  & &  &   &
          \\ \hline  $\text{M5}_{1,0,0} ^{\mathbf{(1,0)}}$ &  & & \rm{x}& \rm{x} & x & x & & \rm{x} & \rm{x} &  & \\   
            $\text{M5}_{0,1,0} ^{\mathbf{(1,0)}}$ & \rm{x} & \rm{x}& &  & x & x & & \rm{x} & \rm{x} &  & \\
            $\text{M5}_{0,0,1} ^{\mathbf{(1,0)}}$ & \rm{x} & \rm{x}& \rm{x}& \rm{x} &  & & & \rm{x} & \rm{x} &  & \\ \hline
          $\text{M5}_{0,0,1} ^{\mathbf{(0,1)}}$ & \rm{x} & \rm{x}& \rm{x}& \rm{x} &  & & &  
 &  & \rm{x}  & \rm{x} \\ \hline
           $\text{M5}_{0,0,1} ^{\mathbf{(p,q)}}$ & \rm{x} & \rm{x}& \rm{x}& \rm{x} &  & & &  \multicolumn{4}{c}{$C^{\mathbf{(p,q)}}$} 
    \end{tabular} 
    \caption{M-brane configuration in twisted M-theory on $\BR^2_{\ve_1} \times \BR^2 _{\ve_2} \times \BR^2 _{\ve_3} \times \BR_t \times \BC^\times _X \times \BC^\times _Z$. The M5-brane $\text{M5}_{c} ^{\mathbf{(p,q)}}$ is supported on $\BR^2 _{\ve_{c+1}} \times \BR^2 _{\ve_{c-1}} \times C^{\mathbf{(p,q)}}$ where $C^{\mathbf{(p,q)}} = \{X^{\mathbf{q}} Z^{-\mathbf{p}} = \text{const} \} \subset \BC^\times _X\times \BC^\times _Z$.}
     \label{table:twmbrane}
\end{table}

\begin{table}[h!]
    \centering
    \begin{tabular}{c|c|c|c|c|c}
        $\BR^2 _{\ve_1}$ & $\BR^2 _{\ve_2}$ & $\BR^2 _{\ve_3}$ & $\mathbb{R} _t $  & $\BC ^\times _X $ & $\BC^\times _Z $  \\ \hline
        $x^0,x^1$ & $x^2,x^3$ & $x^4, x^5$ & $x^6$  & $x^7,x^8$ & $ x^9, x^{10}$ 
    \end{tabular}
    \caption{Spacetime of the twisted M-theory}
    \label{table:spacetime}
\end{table}

The M-branes in the twisted M-theory engineer defects in the 5d Chern-Simons theory. We may introduce M2-branes, yielding topological line defects; and M5-branes, producing holomorphic surface defects (see table \ref{table:twmbrane} and \ref{table:spacetime}, for our main case of $\mathcal{C} = \BC_X ^\times \times \BC_Z ^\times$).
\begin{itemize}
    \item The M2-branes can be supported on $\BR^2 _{\ve_c} \times \BR_t$, $c \in\{1,2,3\}$, yielding topological line defects of the 5d Chern-Simons theory on the line $\BR_t$. In the most general case, the line defect descends from $(l,m,n)$ M2-branes supported on $\BR^2 _{\ve_1} \times \BR_t$, $\BR^2 _{\ve_2} \times \BR_t$, and $\BR^2 _{\ve_3} \times \BR_t$, respectively. We call this line defect $\mathcal{L}_{l,m,n}$. The local operators on this line defect form a non-commutative associative algebra defined by their OPEs, which we denote by $\text{M2}_{l,m,n}$. 

    \item The M5-branes can be supported on $\BR^2 _{\ve_{c+1}}\times \BR^2 _{\ve_{c-1}} \times C$, $c \in \{1,2,3\}$, where $C \subset \mathcal{C}$ is a holomorphic curve. They induce holomorphic surface defects of the 5d Chern-Simons theory lying on the curve $C$. In the most general case, the surface defect descends from $(L,M,N)$ M5-branes supported on $\BR^2 _{\ve_{2}}\times \BR^2 _{\ve_{3}} \times C$, $\BR^2 _{\ve_{1}}\times \BR^2 _{\ve_{3}} \times C$, and $\BR^2 _{\ve_{1}}\times \BR^2 _{\ve_{2}} \times C$, respectively. We call this surface defect $\mathcal{S}_{L,M,N} ^C$. The local operators on this surface defect form a chiral algebra defined by their OPEs. We denote its mode algebra by $\text{M5}_{L,M,N} ^C$.\footnote{In this work, we restrict our attention to the mode algebra of local operators on the holomorphic surface defect. In our multiplicative setting, where $\CalC=\BC^\times \times \BC^\times$, the chiral algebra on the surface defect is anticipated to be a quantum vertex algebra. Consequently, the associated surjective map $\r_{\text{M5}_{L,M,N} ^C }$ is expected to extend to a quantum vertex algebra homomorphism. See footnote \ref{fn:qva}.}
\end{itemize}
The coupling of the defect is built by applying a descent procedure to a combination of local operators on the defect and local operators of the 5d Chern-Simons theory at the locus of the defect (modes of the ghost field).\footnote{In fact, the descent procedure for the coupling of the surface defect requires more care, since we need to take account of the back-reaction of the M5-branes sourcing a singularity of the fields. See \cite{Costello:2016nkh}.} The BRST-invariance of this combination requires the algebra of local operators on the defect to be a representation of a certain universal associative algebra, which we generically denote by $\EA_{\ve_1,\ve_2}$. Namely, there exist surjective algebra homomorphisms,
\begin{align} \label{eq:algsurhom}
\begin{split}
    &\r_{\text{M2}_{l,m,n}}: \EA_{\ve_1,\ve_2} \twoheadrightarrow \text{M2}_{l,m,n}, \qquad\quad \r_{\text{M5}_{L,M,N} ^C}: \EA_{\ve_1,\ve_2} \twoheadrightarrow \text{M5}_{L,M,N} ^C,
\end{split}
\end{align}
for the topological line defect from the M2-branes and the holomorphic surface defect from the M5-branes, respectively.\\

The same kind of defects may lie on top of each other, defining a new defect by a fusion. The coupling of the fused defect should be built by the couplings of the individual defects before the fusion operation. Since the coupling is encoded in the representation \eqref{eq:algsurhom} of the universal associative algebra carried by the defect, the fusion operation implies that there must be an algebra homomorphism,
\begin{align} \label{eq:univcop}
    \D : \EA_{\ve_1,\ve_2} \to \EA_{\ve_1,\ve_2} \, \widehat{\otimes} \, \EA_{\ve_1,\ve_2},
\end{align}
from which the the representation assigned to the fused defect would be reconstructed by composing $\Delta$ with the respective representations assigned to the individual defects before the fusion. The completed tensor product was used so that the image may be given as a sum of infinitely many terms. This algebra homomorphism, called the coproduct, equips the universal associative algebra $\EA_{\ve_1,\ve_2}$ with a coalgebra structure (namely, $\EA_{\ve_1,\ve_2}$ is a bialgebra).\footnote{The coassociativity $(\Delta \otimes \text{id})\Delta = (\text{id} \otimes \Delta)\Delta$ follows from the independence of the ordering of two consecutive fusion operations. The existence of the counit $\epsilon : \EA_{\ve_1,\ve_2} \to \BC$, satisfying $(\text{id} \otimes \epsilon)\Delta = (\epsilon \otimes \text{id}) \Delta = \text{id}$, is guaranteed by the existence of the trivial defect.} See section \ref{eq:mbrep} for more details about the fusion operations of the line defects and the surface defects in our main case $\CalC = \BC^\times _X \times \BC^\times _Z$.\\

\begin{table}[h!]
    \centering
\begin{tabular}{c||c|c|c}
 & $\BC \times \BC$ & $\BC 
\times  \BC^\times$ & $\BC ^\times \times \BC^\times $ \\ \hline\hline
 Universal associative algebra & $Y_1 (\widehat{\fgl}(1))$ & $Y (\widehat{\fgl}(1))$ & $\qta$ \\ \hline
 Parallel M2-branes & $\mathbf{Sr\ddot{H}}$ & $\mathbf{St\ddot{H}}$ & $\sh$ \\ \multirow{2}{*}{Non-parallel M2-branes} & Generalized & Generalized  & Generalized \\  & rat. Calogero & trig. Calogero & Macdonald \\  \hline
   Parallel M5-branes & $\mathcal{W}^{\leq 0}$ & $\mathcal{W}$ or $\hbar\mathcal{W}^{\leq 0}$ & $q\mathcal{W}$ \\
   Non-parallel M5-branes & $Y^{\leq 0}$ & $Y$ or $\hbar Y^{\leq 0}$ & $qY$
\end{tabular} 
\caption{Universal associative algebra for 5d $\fgl(1)$ Chern-Simons theory on $\BR_t \times \mathcal{C}$, where $\mathcal{C} = \BC\times \BC$; $\mathcal{C} = \BC\times \BC^\times$; or $\mathcal{C} = \BC^\times \times \BC^\times $, and their representations formed by the algebra of local operators on topological line defects and holomorphic surface defects from M2- and M5-branes. Here, $\mathbf{S\ddot{H}}$ is the spherical double affine Hecke algebra (spherical DAHA; see section \ref{sssec:parallelM2}), while $\mathbf{Sr\ddot{H}}$ and $\mathbf{St\ddot{H}}$ are its rational and trigonometric degeneration, respectively. Also, $qY$ (resp. $\hbar Y$) stands for the $q$-deformed (resp. $\hbar$-deformed) $Y$-algebra. We will discuss the $\hbar$-deformed $Y$-algebra in a separate work. We mainly focus on the case of $\BC^\times \times \BC^\times$ (the last column) in the present work.}\label{table:twmcases}
\end{table}

Since the 5d Chern-Simons theory depends on the choice of the complex manifold $\mathcal{C}$, the universal associative algebra also depends on it (see table \ref{table:twmcases}). Previous studies have primarily focused on the cases where $\mathcal{C} = \BC \times \BC$ and $\mathcal{C} = \BC \times \BC^\times$ \cite{Costello:2016nkh,Costello:2017fbo,Gaiotto:2019wcc,Gaiotto:2020vqj,Gaiotto:2020dsq}. For these cases, it was shown that the universal associative algebra is the 1-shifted affine Yangian of $\fgl(1)$ ($Y_1 (\widehat{\fgl}(1))$) and the affine Yangian of $\fgl(1)$ ($Y(\widehat{\fgl}(1))$, respectively. Correspondingly, the algebra of local operators on the line defect from M2-branes is given by the (generalized) rational (resp. trigonometric) Calogero representation of $Y_1 (\widehat{\fgl}(1))$ (resp. $Y (\widehat{\fgl}(1))$). The algebra of local operators on the surface defect from M5-branes was shown to be the $Y$-algebra \cite{Gaiotto:2017euk}.

In this work, we examine the case where $\mathcal{C} = \BC^\times \times \BC^\times$. We propose that the universal associative algebra in this setting is the quantum toroidal algebra of $\fgl(1)$, denoted by $\qta$. We claim that the algebra of local operators $\text{M2}_{l,m,n}$ on the line defect, originating from M2-branes, is given by the spherical double affine Hecke algebra (spherical DAHA) and its generalization (see section \ref{subsec:m2}). Furthermore, we assert that the mode algebra $\text{M5}_{L,M,N}^C $ of local operators on the surface defect, originating from M5-branes, is the $q$-deformed $Y$-algebra (see section \ref{subsec:m5}).\footnote{A qualitative argument for the appearance of $q$-deformed algebras is as follows: after replacing $\BC \times \BC^\times$ with $\BC^\times \times \BC^\times$, there is now a circle transverse to all $\text{M5}^{\mathbf{(1,0)}}$-branes: the radius of this circle is a finite scale which breaks the $\EN=(2,0)$ superconformal symmetry of the 6-dimensional theory on the M5-branes; the resulting theory is known as the $\EN=(2,0)$ little string theory \cite{Seiberg:1997zk,Losev:1997hx}. The conformal symmetry of the chiral algebra is likewise broken by the radius scale: when all M5-branes are parallel to each other, the conformal $\mathcal{W}$-algebra was shown to deform to a $q\mathcal{W}$-algebra \cite{Aganagic:2015cta,Haouzi:2017vec}. In our work, we will encounter an arbitrary configuration of $\text{M5}^{\mathbf{(1,0)}}$-branes, with different orientations, and the associated chiral $Y$-algebra is therefore expected to deform to the $qY$-algebra.}

\subsection{M2-M5 intersections as R-matrices and Miura operators} \label{subsec:intrmat}
In the present work, we establish the Miura transformation for the algebra of local operators on the surface defect, called the $q$-deformed $W$- and $Y$- algebras, by the fusion of the intersections between a single M2-brane and multiple M5-branes. The most crucial point is to realize the M2-M5 intersection as an R-matrix of the universal associative algebra and to utilize its properties. Since the idea penetrates the central theme of the present work, we will elucidate how the M2-M5 intersection is associated with an R-matrix here.

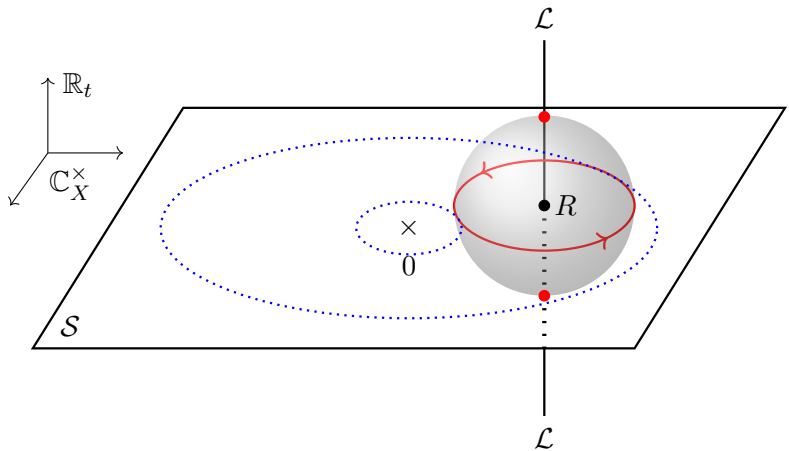
\begin{figure}[h!]\centering
\begin{tikzpicture}[arrowmark/.style 2 args={decoration={markings,mark=at position #1 with \arrow{#2}}}]

\draw[line width=0.3mm] (-2,1.6) -- (6,1.6) -- (4,-1.6) -- (-4,-1.6) 
      -- cycle;

\draw[line width=0.3mm, loosely dotted, ultra thick] (2.8,0.3) -- (2.8,-1.6);
\draw[line width=0.3mm] (2.8,-1.6) -- (2.8,-2.5);

\draw[line width=0.3mm] (2.8,2.5) -- (2.8,0.3);

\draw[postaction={decorate},
    arrowmark={0.4}{>},arrowmark={0.9}{>},red,line width=0.3mm] (2.8,0.3) ellipse (1.2cm and 0.6cm);
    
\shade[ball color = gray!40, opacity = 0.4] (2.8,0.3) circle (1.2cm);

\draw[blue,line width=0.3mm, dotted] (1,0) ellipse (0.7cm and 0.35cm);

\draw[blue,line width=0.3mm, dotted] (1,0) ellipse (3.3cm and 1.2cm);

\filldraw[black] (2.8,0.3) circle (2pt) node[anchor=west]{$R$};

\filldraw[red] (2.8,1.48) circle (2pt) node[anchor=west]{};

\filldraw[red] (2.8,-0.9) circle (2pt) node[anchor=west]{};

 \node at (1,0) {$\times$};
 \node at (1,-0.5) {$0$};

   \node at (-3.5,-1.3) {$\mathcal{S}$};

    \node at (2.8,2.8) {$\mathcal{L} $};
     \node at (2.8,-2.8) {$\mathcal{L} $};

\draw[->] (-3.8,1) -- (-3.8,2);
\draw[->] (-3.8,1) -- (-2.8,1);
\draw[->] (-3.8,1) -- (-4.3,0.3);

\node at (-3.5,0.6) {$\BC^\times _X$};
\node at (-3.4,1.9) {$\BR_t$};
    
\end{tikzpicture} \caption{M2-M5 intersection as an R-matrix. Classically, the gauge variation of the local operator $R$ (black dot) at the intersection receives linear contributions from the line defect (red dots) and the surface defect (red line). The contribution from the surface defect is decomposed into a linear combination of two: one from inward and the other from outward (blue dotted lines).} \label{fig:m2m5intsec} 
\end{figure}

Consider the configuration 
where M2-branes and M5-branes are present simultaneously. To develop the argument in a universal manner, we denote a generic line defect engineered by the M2-branes simply as $\mathcal{L}$, and the algebra of local operators on the line defect by $\text{M2}$. Similarly, we denote a generic surface defect engineered by the M5-branes as $\mathcal{S}$, and the mode algebra of local operators on the surface defect by $\text{M5}$. The corresponding surjective algebra homomorphisms \eqref{eq:algsurhom} from the universal associative algebra $\EA_{\ve_1,\ve_2}$ are simply called $\r_{\CalL}$ and $\r_{\CalS}$. The holomorphic support of the M5-branes are assumed to be identical, say $\BC^\times _X$, so that they define a single surface defect in the 5d Chern-Simons theory. The intersection between the line defect $\CalL$ and the surface defect $\CalS$ can then be visualized as shown in Figure \ref{fig:m2m5intsec}.

The intersection supports a space of local operators. A local operator $R$ at the intersection can be built from the local operators on the line defect and the surface defect, so that it is an element of the completed tensor product
\begin{align}
    R\in \text{M2} \,\widehat{\otimes} \,\text{M5},
\end{align}
subject to the gauge-invariance condition. Equivalently, we may view the local operator $R$ as acting on the space $\CalV \otimes \CalF$, where $\CalV$ is the Hilbert space of the line defect worldline theory attached to the boundary at the negative infinity of $\BR_t$ and $\CalF$ is the Hilbert space of the surface defect theory attached to the boundary at the origin of $\BC^\times _X$. The result of this action is another state in the same space attached to the positive infinity of $\BR_t$ and the infinity of $\BC^\times _X$, respectively. Thus, we recognize that $R \in \text{End}(\CalV)\, \widehat{\otimes}\, \text{End}(\CalF) = \text{M2}\, \widehat{\otimes}\, \text{M5}$. \\

The gauge variation of the local operator $R$ gets contributions from both the line defect and the surface defect, which have to be canceled with each other. At the classical limit $\ve_1 = 0$, there are only linear contributions from the two defects. These contributions can be visualized by drawing a small sphere surrounding the intersection point, picking up the gauge transformations whenever the sphere intersects the defects.

Let $g \in \EA_{\ve_1=0,\ve_2}$ be the element paired with a fixed mode of the ghost field. The line defect contribute $\r_{\CalL} (g)$ from both above and below, ordered in the increasing $\BR_t$-direction. The surface defect contributes by a contour integral of the generating current of the chiral algebra along the small circle around the intersection point. The contour is decomposed into the linear combination of two encircling the origin, radially ordered together with the intersection point itself. Thus, the contour integral produces two modes $\r_{\CalS} (g)$ as well, acting from the inward and the outward in the radial ordering. In total, the classical gauge-invariance condition reads
\begin{align}
    R  \left(\r_{\CalL} (g) \otimes \text{id} + \text{id} \otimes \r_{\CalS}  (g) \right) =   \left(\r_{\CalL} (g) \otimes \text{id} + \text{id} \otimes \r_{\CalS}  (g) \right)  R,
\end{align}
for any $g \in  \EA _{\ve_1=0,\ve_2}$.

Turning on $\ve_1 \neq 0$, the gauge-invariance condition gets non-linear quantum corrections. We postulate that the quantum corrections respect the algebra structure, uplifting the gauge-invariance condition to\footnote{In the case of $\mathcal{C} = \BC\times \BC$ or $\mathcal{C} = \BC \times \BC^\times$, the support of the holomorphic surface defect $\CalS$ can be chosen to be $\BC$ instead of $\BC^\times$. In this case, we may consider the local operator $R$ acting on the vacuum $\vert \varnothing \rangle \in \mathcal{F}$ of the Hilbert space of the surface defect theory; namely, $R(1\otimes \vert \varnothing \rangle)$. This can be thought of as the effect of capping $\BC^\times$ to $\BC$. The gauge-invariance condition then reads
\begin{align}
        R (1\otimes \vert\varnothing \rangle )   \r_{\CalL} (g)  =  \Delta^{\text{op}}_{ \text{M2},\text{M5} } (g)   R (1\otimes \vert\varnothing \rangle ). \nonumber
\end{align}
Up to the exchange of the two factors, this recovers the gauge-invariance condition assigned to the M2-M5 intersection in \cite{Gaiotto:2020dsq}. We suggest that the gauge-invariance condition should be extended to \eqref{eq:ginvr}, without 
having $R$ act on the vacuum in particular, when the support of the holomorphic surface defect is $\BC^\times$.
}

\begin{align}\boxed{ \label{eq:ginvr}
        R \, \Delta_{\text{M2} , \text{M5}} (g)  =  \Delta^{\text{op}}_{\text{M2}, \text{M5}}  (g)  R, \qquad \text{for any } g\in \EA_{\ve_1,\ve_2}.}
\end{align}
Here, we defined the \textit{mixed} coproduct by composing the coproduct \eqref{eq:univcop} of the universal associative algebra $\EA_{\ve_1,\ve_2}$ and the representations \eqref{eq:algsurhom} assigned to the M2- and M5-branes,
\begin{align}
\begin{split}
  &  \D_{\text{M2},\text{M5} } := (\r_{\CalL} \otimes \r_{\CalS} )  \Delta \quad : \quad \EA_{\ve_1,\ve_2} \to \text{M2} \,\widehat{\otimes}\, \text{M5},
\end{split}
\end{align}
and $\Delta^{\text{op}} = \s \circ \D $ is the opposite coproduct obtained by composing the exchange operator $\s(g_1 \otimes g_2) = g_2 \otimes g_1$.\\

\begin{figure}[h!]\centering
\resizebox{.4\textwidth}{!}{
\begin{tikzpicture}
\draw[line width=0.3mm] (-1,1.5) -- (5,1.5) -- (3,-0.5) -- (-3,-0.5) 
      -- cycle;

\draw[line width=0.3mm] (-1,-0.1) -- (5,-0.1) -- (3,-2.1) -- (-3,-2.1) 
      -- cycle;

\draw[line width=0.3mm, loosely dotted, ultra thick] (2.5,0.5) -- (2.5,-0.5);
\draw[line width=0.3mm] (2.5,-0.5) -- (2.5,-1.1);
\draw[line width=0.3mm, loosely dotted, ultra thick] (2.5,-1.1) -- (2.5,-2.1);
\draw[line width=0.3mm] (2.5,-2.1) -- (2.5,-3);

\draw[line width=0.3mm] (2.5,2.5) -- (2.5,0.5);

\filldraw[black] (2.5,0.5) circle (2pt) node[anchor=west]{$R_{\mathcal{L}, \mathcal{S}'}$};

\filldraw[black] (2.5,-1.1) circle (2pt) node[anchor=west]{$R_{\mathcal{L},\mathcal{S}}$};

 \node at (1,0.5) {$\times$};
 \node at (1,0.1) {$0$};

  \node at (1,-1.1) {$\times$};
 \node at (1,-1.5) {$0$};

   \node at (-3.4,-0.4) {$\mathcal{S}'$};
      \node at (-3.4,-2) {$\mathcal{S}$};
  \node at (2.5,-3.3) {$\mathcal{L}$};
    \node at (2.5,2.8) {$\mathcal{L}$};

\draw[->] (-3,2) -- (-3,3);
\draw[->] (-3,2) -- (-2,2);
\draw[->] (-3,2) -- (-3.5,1.4);

\node at (-2.7,1.6) {$\BC^\times _X$};
\node at (-2.6,2.9) {$\BR_t$};
    
\end{tikzpicture}} \hspace{4mm} \raisebox{2.5\height}{\begin{tikzpicture}
    \draw[->] (-0.5,0) -- (0.5,0);
    \node at (0,0.5) {fusion};
\end{tikzpicture}} \raisebox{0.13\height}{\resizebox{.4\textwidth}{!}{
\begin{tikzpicture}
\draw[line width=0.3mm] (-1,1.5) -- (5,1.5) -- (3,-0.5) -- (-3,-0.5) 
      -- cycle;

\draw[line width=0.3mm, loosely dotted, ultra thick] (2.5,0.5) -- (2.5,-0.5);
\draw[line width=0.3mm] (2.5,-0.5) -- (2.5,-1.5);

\draw[line width=0.3mm] (2.5,2.5) -- (2.5,0.5);

\filldraw[black] (2.5,0.5) circle (2pt) node[anchor=west]{$R_{\mathcal{L},\mathcal{S}\circ \mathcal{S}'}$};

 \node at (1,0.5) {$\times$};
 \node at (1,0.1) {$0$};

   \node at (-3.6,-0.4) {$\mathcal{S}\circ \mathcal{S}'$};
 
  \node at (2.5,-1.9) {$\mathcal{L} $};
    \node at (2.5,2.8) {$\mathcal{L} $};  
\end{tikzpicture}}}
\caption{Fusion of intersections between one line defect and two surface defects} \label{fig:fusionint1} 
\end{figure}
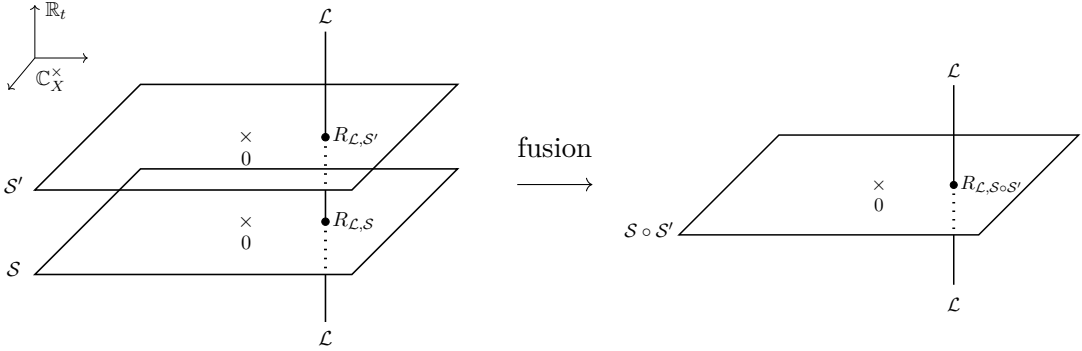

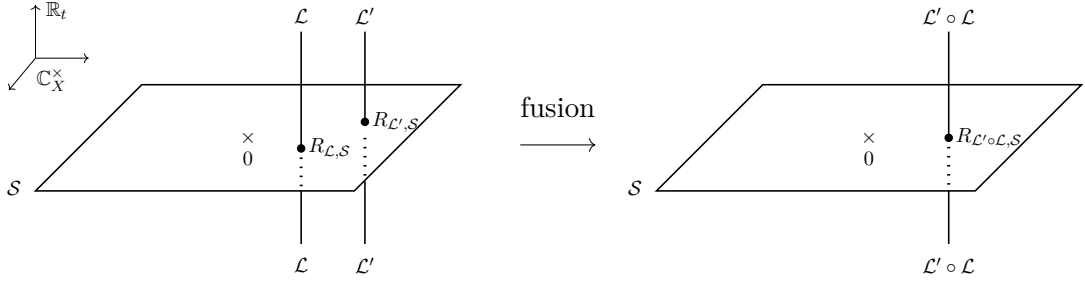
\begin{figure}[h!]\centering
\resizebox{.4\textwidth}{!}{
\begin{tikzpicture}
\draw[line width=0.3mm] (-1,1.5) -- (5,1.5) -- (3,-0.5) -- (-3,-0.5) 
      -- cycle;

\draw[line width=0.3mm] (2,2.5) -- (2,0.3);
\draw[line width=0.3mm, loosely dotted, ultra thick] (2,0.3) -- (2,-0.5);
\draw[line width=0.3mm] (2,-0.5) -- (2,-1.5);

\draw[line width=0.3mm] (3.2,2.5) -- (3.2,0.8);
\draw[line width=0.3mm, loosely dotted, ultra thick] (3.2,0.5) -- (3.2,-0.32);
\draw[line width=0.3mm] (3.2,-0.32) -- (3.2,-1.5);

\filldraw[black] (2,0.3) circle (2pt) node[anchor=west]{$R_{ \mathcal{L},\mathcal{S}}$};

\filldraw[black] (3.2,0.8) circle (2pt) node[anchor=west]{$R_{\mathcal{L}',\mathcal{S}}$};

 \node at (1,0.5) {$\times$};
 \node at (1,0.1) {$0$};

   \node at (-3.4,-0.45) {$\mathcal{S}$};

  \node at (2,-1.9) {$\mathcal{L} $};
    \node at (2,2.8) {$\mathcal{L} $};  
 
  \node at (3.2,-1.9) {$\mathcal{L}' $};
    \node at (3.2,2.8) {$\mathcal{L}' $};  

\draw[->] (-3,2) -- (-3,3);
\draw[->] (-3,2) -- (-2,2);
\draw[->] (-3,2) -- (-3.5,1.4);

\node at (-2.7,1.6) {$\BC^\times _X$};
\node at (-2.6,2.9) {$\BR_t$};
    
\end{tikzpicture}} \hspace{4mm} \raisebox{2\height}{\begin{tikzpicture}
    \draw[->] (-0.5,0) -- (0.5,0);
    \node at (0,0.5) {fusion};
\end{tikzpicture}} \resizebox{.4\textwidth}{!}{
\begin{tikzpicture}
\draw[line width=0.3mm] (-1,1.5) -- (5,1.5) -- (3,-0.5) -- (-3,-0.5) 
      -- cycle;

\draw[line width=0.3mm, loosely dotted, ultra thick] (2.5,0.5) -- (2.5,-0.5);
\draw[line width=0.3mm] (2.5,-0.5) -- (2.5,-1.5);

\draw[line width=0.3mm] (2.5,2.5) -- (2.5,0.5);

\filldraw[black] (2.5,0.5) circle (2pt) node[anchor=west]{$R_{\mathcal{L}'\circ \mathcal{L},\mathcal{S}}$};

 \node at (1,0.5) {$\times$};
 \node at (1,0.1) {$0$};

   \node at (-3.4,-0.45) {$\mathcal{S}$};
 
  \node at (2.5,-1.9) {$\mathcal{L}'\circ \mathcal{L}$};
    \node at (2.5,2.8) {$\mathcal{L}' \circ\mathcal{L}$};  
\end{tikzpicture}}
\caption{Fusion of intersections between two line defects and one surface defect} \label{fig:fusionint2} 
\end{figure}

Further, let us consider the next-to-simplest configurations involving two intersections: $(a)$ one line defect $\mathcal{L}$ passing through two surface defects $\mathcal{S}$ and $\mathcal{S}'$ (see Figure \ref{fig:fusionint1}); $(b)$ two line defects $\mathcal{L}$ and $\mathcal{L}'$ passing through one surface defect $\mathcal{S}$ (see Figure \ref{fig:fusionint2}).

In the first case, the two surface defects $\mathcal{S}$ and $\mathcal{S}'$, located at different points in $\BR_t$, can approach each other and fuse into a new surface defect $\mathcal{S} \circ \mathcal{S}'$. Accordingly, the intersections between the line defect and the two surface defects undergo a fusion, implying that the local operator $R_{\mathcal{L}, \mathcal{S}\circ \mathcal{S}'}$ at the intersection between the line defect and the fused surface defect should simply be the product of the local operators $R_{\mathcal{L},\mathcal{S}}$ and $R_{\mathcal{L}, \mathcal{S}'}$ assigned to the individual intersections before the fusion. Namely, we have
\begin{align} \label{eq:rfuse1}
    R_{\mathcal{L},\mathcal{S}\circ \mathcal{S}'} = R_{\mathcal{L}, \mathcal{S}'} R_{\mathcal{L},\mathcal{S}},
\end{align}
where the product in the right hand side is taken as elements in the algebra $\text{M2}$ of local operators on $\mathcal{L}$.

Similarly, in the second case the two line defects $\CalL$ and $\CalL'$, located at different points in $\BC^\times _X$, can approach each other to be fused into the a new line defect $\CalL' \circ \CalL$. The intersections also fuse correspondingly, implying that the local operator $R_{\CalL' \circ \CalL , \CalS}$ supported at the fused intersection is the simple product of the local operators $R_{\CalL,\CalS}$ and $R_{\CalL', \CalS}$ attached at the individual intersections before the fusion:
\begin{align} \label{eq:rfuse2}
    R_{\mathcal{L}'\circ \mathcal{L},\mathcal{S} } = R_{\mathcal{L}',\mathcal{S} } R_{\mathcal{L},\mathcal{S} },
\end{align}
where the product is taken as elements in the mode algebra $\text{M5}$ of the local operators on the surface defect $\CalS$.

All in all, we discover that the local operator supported at the intersection should obey the constraint \eqref{eq:ginvr} and behave as \eqref{eq:rfuse1} and \eqref{eq:rfuse2} under the fusion. We claim that these are precisely the manifestation of the R-matrix properties. This is most apparent when there exists the universal R-matrix $\CalR \in \EA_{\ve_1,\ve_2}\, \widehat{\otimes}\, \EA_{\ve_1,\ve_2}$ of the universal associative algebra, whose defining relations are
\begin{subequations} 
\begin{align}
    &\CalR \Delta (g) = \Delta^{\text{op}} (g) \CalR,\qquad \text{for any } g\in \EA_{\ve_1,\ve_2} , \label{eq:r1} \\
        & (\text{id} \otimes \Delta) \CalR = \CalR_{13} \CalR_{12} \label{eq:r2} \\
    &(\Delta \otimes \text{id} ) \CalR = \CalR_{13} \CalR_{23}. \label{eq:r3}
\end{align}
\end{subequations}
It is now straightforward that \eqref{eq:ginvr} is \eqref{eq:r1} composed with $\r_{\CalL} \otimes \r_{\CalS}$; \eqref{eq:rfuse1} is \eqref{eq:r2} composed with $\r_{\CalL} \otimes \r_{\CalS} \otimes \r_{\CalS'}$; and \eqref{eq:rfuse2} is \eqref{eq:r3} composed with $\r_{\CalL'} \otimes \r_{\CalL} \otimes \r_{\CalS}$. In our case of $\CalC = \BC^\times _X \times \BC^\times _Z$, we suggested that the universal associative algebra is the quantum toroidal algebra of $\fgl(1)$, $\EA_{\ve_1,\ve_2} = \qta$. Following from its quantum double construction, the quantum toroidal algebra of $\fgl(1)$ does possess the universal R-matrix (see section \ref{subsubsec:univr} for more detail). Let us remark that the suggested relation between the R-matrix and the M2-M5 intersection can be directly verified within the perturbation theory of the 5d non-commutative Chern-Simons theory \cite{Ishtiaque:2024orn} (see also \cite{Ishtiaque:2018str,Oh:2020hph,Oh:2021wes,Ashwinkumar:2024vys} for other works on the perturbative study of the 5d CS theory).

As we emphasized earlier, the R-matrix properties of the M2-M5 intersections established here are key to achieving the Miura transformation for the algebra $\text{M5}_{L,M,N} ^C$ of local operators on the most general surface defect $\CalS_{L,M,N} ^C$. In particular, the axiom \eqref{eq:r2} implies that the R-matrix for the intersection between the line defect $\CalL_{0,0,1}$ and the surface defect $\CalS_{L,M,N} ^{C}$ is factorized into the product of $L+M+N$ basic R-matrices, yielding
\begin{align}
\boxed{
R_{\text{M2}_{0,0,1} , \text{M5}_{L,M,N} ^{C }} = R_{L+M+N} ^{(c_{L+M+N})} \cdots R_2 ^{(c_2)} R_1 ^{(c_1)} = \sum_{m=0} ^\infty (-1)^m T_m (X_1) q_3 ^{-m D_{X_1}}
}
\end{align}

\noindent
where in the second equality we expanded the R-matrix as a series in the difference operator $q_3 ^{-D_{X_1}}$, which is a generator of the algebra $\text{M2}_{0,0,1}$. In section \ref{sec:miura}, we will verify that this expansion of the R-matrix provides the Miura transformation for the algebra $\text{M5}_{L,M,N} ^{C}$. We thereby identify the basic R-matrix associated with the intersection between a single M2-brane and a single M5-brane as the Miura operator for the $q$-deformed $Y$-algebra. Moreover, we show that the change of ordering of any two basic R-matrices results in isomorphic $q$-deformed $Y$-algebras by a direct consequence of the universal Yang-Baxter equation satisfied by the universal R-matrix.

\subsection{Corner, $\mathbf{(p,q)}$-web, and gauge origami} \label{subsec:duality}
The twisted M-theory is connected to different IIB theories through string dualities. It is illuminating to examine how the intersecting M-branes are transformed via these dualities and study their implications.\\

\paragraph{Chiral algebra at the corner} The M-theory on a toric Calabi-Yau 3-fold is dual to the $\mathbf{(p,q)}$-web of fivebranes in the IIB theory, where the web diagram is given by the toric diagram \cite{Leung:1997tw}. Regarding $\BR^2 _{\ve_1}\times \BR^2 _{\ve_2} \times \BR^2 _{\ve_3}$ as the toric Calabi-Yau 3-fold for the duality, the twisted M-theory is dual to the $\mathbf{(p,q)}$-web composed of one NS5-brane, one D5-brane, and one $(1,1)$-fivebrane, all of which are supported on a semi-infinite line on a $\BR^2$-plane joining at a single point (see table \ref{table:iibcorner} and \ref{table:spacetimecorner}). The M5-branes are dualized to D3-branes, filling the faces between the fivebranes and ending on them. The holomorphic part of their support remains the same holomorphic curve $C \subset \mathcal{C}$.

In the case of $\mathcal{C} = \BC\times \BC^\times$, consider $(L,M,N)$ M5-branes whose supports are $\BR^2 _{\ve_2} \times \BR^2 _{\ve_3} \times \BC^\times$, $\BR^2 _{\ve_1} \times \BR^2 _{\ve_3} \times \BC^\times$, and $\BR^2 _{\ve_1} \times \BR^2 _{\ve_2} \times \BC^\times$, respectively. By the above duality with a proper choice of the $SL(2,\BZ)$ frame of the IIB theory, they are mapped to the $(L,M,N)$ D3-branes filling the face between the NS5-brane and the $(1,1)$-fivebrane; between the D5-brane and the $(1,1)$-fivebrane; and between the NS5-brane and the D5-brane, respectively, while sharing $\BC^\times$ in common. The effective field theory on each stack is the 4d GL-twisted $\EN=4$ gauge theory with the gauge group $U(L)$, $U(M)$, and $U(N)$, respectively, making interfaces across codimension-one boundaries assigned by the fivebranes. Since the path integral of the 4d GL-twisted gauge theory with a (deformed) Neumann boundary passes to that of the analytically continued 3d Chern-Simons theory \cite{Witten:2010cx,Witten:2010zr}, the configuration boils down to an interface between two 3d Chern-Simons theories with the gauge groups $U(N\vert L)$ and $U(M \vert L)$. The local operators on the two-dimensional interface supported on $\BC^\times$ form a chiral algebra $Y_{L,M,N}$ by their OPEs. This is how the $Y$-algebra was originally constructed as the vertex algebra at the corner \cite{Gaiotto:2017euk}. The M2-branes, which are dualized to 1-branes lying on top of the fivebranes, provide degenerate modules for this vertex algebra.

In the present work, instead, we mainly focus on the case of $\mathcal{C} = \BC^\times _X \times \BC^\times_Z$. The string duality to the IIB theory operates as previously described, except for the change in the worldvolume to $\mathcal{C} = \BC^\times _X \times \BC^\times _Z$. Let us re-examine the D3-branes supported on $\BC^\times _X$ in this setting. The effective 4d theory for the D3-D3 open string now contains infinite numbers of winding modes around the circle of $\BC^\times _Z \simeq \BR_z \times S^1$, resulting in its uplift to a 5d $\EN=2$ gauge theory compactified on a circle. We expect that the interfaces of these 5d $\EN=2$ gauge theories provides the multiplicative uplift of the corner vertex algebra. In this sense, it is appropriate to refer to the algebra $\text{M5}_{L,M,N} ^C$ of local operators on the holomorphic surface defect $\CalS_{L,M,N} ^C$ as the \textit{$q$-deformed} $Y$-algebra.\footnote{Even though we only restrict our attention to the mode algebra, we expect that the $q$-deformed $Y$-algebra $qY_{L,M,N}$ can be endowed with a \textit{quantum} vertex algebra structure, and the surjective map $\r_{\text{M5}_{L,M,N} ^C}: \qta\twoheadrightarrow qY_{L,M,N}$ would be uplifted to a quantum vertex algebra homomorphism. In this sense, $qY_{L,M,N}$ would be called the \textit{quantum} vertex algebra at the corner. See footnote \ref{fn:qva}.}\\

\begin{table}[h!]
    \centering
    \begin{tabular}{  c||c|c|c|c|c|c|c|c|c|c  } 
          \text{IIB branes} & 0 & 1 & 2 & 3 & 4 & 5 & 6 & 7 & 8 & 9  \\   \hline\hline
          Fivebranes & \multicolumn{2}{c|}{$\vdash$}  & &  &  & x & x & x & x & x \\ \hline
           1-branes &   \multicolumn{2}{c|}{$\vdash$}   &   &   &   & \rm{x}  & &  &   &
          \\ \hline  $\text{D3}_{c} ^{\mathbf{(1,0)}}$ &  $\vdash$ &  $\vdash$ &  & & & & \rm{x} & \rm{x} &  &  \\ \hline
          $\text{D3}_{c} ^{\mathbf{(0,1)}}$ &  $\vdash$ & $\vdash$&  &  & & &  
 &  & \rm{x}  & \rm{x} \\ \hline
           $\text{D3}_{c} ^{\mathbf{(p,q)}}$ &  $\vdash$ &  $\vdash$ & 
           &  & & &  \multicolumn{4}{c}{$C^{\mathbf{(p,q)}}$} 
    \end{tabular} 
    \caption{IIB brane configuration for the corner. The symbol $\vdash$ indicates semi-infinite filling.}
     \label{table:iibcorner}
\end{table}

\begin{table}[h!]
    \centering
    \begin{tabular}{c|c|c|c|c}
        $\BR^2 $ & $\BR^3$  & $\mathbb{R} _t $  & $\BC ^\times _X $ & $\BC^\times _Z $  \\ \hline
        $x^0,x^1$ & $x^2,x^3 ,x^4$  & $x^5$  & $x^6,x^7$ & $ x^8, x^{9}$ 
    \end{tabular}
    \caption{Spacetime of the IIB theory for the corner.}
    \label{table:spacetimecorner}
\end{table}

\paragraph{$\mathbf{(p,q)}$-web of fivebranes and 3d $\EN=2$ half-indices}
Note that the holomorphic part $\BC^\times _X \times \BC^\times _Z$ of the 11-dimensional worldvolume contains a two-torus. The twisted M-theory on that torus is the same theory as type IIB string theory on a circle, in the twisted background.\footnote{Note this type IIB frame was not available in the works \cite{Gaiotto:2017euk,Prochazka:2018tlo,Gaiotto:2020dsq}, since the M-theory background considered there is $\CalC= \mathbb{C} \times\mathbb{C}  $ or $\CalC= \mathbb{C} \times\mathbb{C} ^\times $.}

The M5-brane supported on $\BR^2 _{\ve_{c+1}} \times \BR^2 _{\ve_{c-1}} \times C^{\mathbf{(p,q)}}$ becomes the $\mathbf{(p,q)}$-fivebrane supported on $\BR^2 _{\ve_{c+1}} \times \BR^2 _{\ve_{c-1}} \times l^{\mathbf{(p,q)}} \times S^1$, where $l^{\mathbf{(p,q)}}=\{ \mathbf{q} x - \mathbf{p} z = \text{const} \}$ is a line in the two-dimensional plane $\BR_x \times \BR_z$ (see table \ref{table:2am} and \ref{table:stpq}). We choose the $SL(2,\BZ)$ frame of type IIB such that $\mathbf{(1,0)} = \text{NS5}$ and $\mathbf{(0,1)} = \text{D5}$. When the fivebranes share the same $\BR^2 _{\ve_{c+1}} \times \BR^2 _{\ve_{c-1}}$ in their worldvolume, they can make a non-transverse trivalent intersection if the $\mathbf{(p,q)}$-charge is conserved. These non-transverse intersections can be consecutively joined to form the $\mathbf{(p,q)}$-web of fivebranes. The M2-brane supported on $\BR^2 _{\ve_c} \times \BR_t$ is mapped to a D3-brane supported on $\BR^2 _{\ve_c} \times \BR_t \times S^1$. 

In \cite{Awata:2011ce}, it was observed that the web of $q$-boson representations of the quantum toroidal algebra of $\fgl(1)$, depicted based on the $\mathbf{(p,q)}$-web of fivebranes, gives a reconstruction of the partition function of the topological vertices and the 5d $\EN=1$ effective gauge theory realized on the fivebrane web. The contents in section \ref{eq:mbrep} and section \ref{sec:qqchar} clarify the physical meaning of this algebraic formulation: the vacuum expectation value of the \textit{web} of intersecting defects in the 5d non-commutative $\fgl(1)$ Chern-Simons theory on $\BR_t\times \BC^\times _X\times \BC^\times _Z$.\\

The M-brane configuration of the Miura transformation, namely, a single M2-brane passing through multiple M5-branes, is translated in type IIB to a single D3-brane passing through multiple NS5-branes with different orientations. This is precisely the IIB brane engineering of a 3-dimensional supersymmetric abelian quiver gauge theory \cite{Hanany:1996ie}. The amount of supersymmetry and the matter content depends on the  orientation of the various NS5-branes relative to each other as the D3-brane goes through them. when the D3-brane and an NS5-brane are supported on a common $\mathbb{R}^2_{\epsilon_c}$ plane, there is 3d $\EN=4$ bifundamental hypermultiplet matter from quantizing open strings at the intersection. When the D3-brane support is totally transverse to that of an NS5-brane, the matter content is that of a 1d $\EN=4$ hypermultiplet multiplet instead.\\

In this type IIB frame, we will show that the various components of the Miura transformation are equal to a certain \emph{half-index} of the 3d theory on the D3 brane.\footnote{It was first discovered in \cite{Aganagic:2013tta,Aganagic:2014oia} that the half-index of 3d supersymmetric gauge theories is sometimes equal to a deformed conformal blocks of a $q\mathcal{W}$-algebra, a statement closely related to the Alday-Gaiotto-Tachikawa-Wyllard correspondence \cite{Alday:2009aq,Wyllard:2009hg}. The result was later reinterpreted via the representation theory of the quantum toroidal algebra of $\fgl(1)$ \cite{Mironov:2016yue,Zenkevich:2018fzl,Zenkevich:2020ufs}. Our present paper directly falls under the scope of this research program.} For us, this half-index is a count of BPS states defined in the UV, for any Lagrangian 3d ${\EN} = 2$ theory \cite{Dimofte:2017tpi,Yoshida:2014ssa,Gadde:2013wq,Gadde:2013sca}: one replaces $\BR^2_{\ve_c}$ with a finite disk/hemisphere $D^2$, and imposes a specific set of 1/2-BPS $\EN=(0,2)$ boundary conditions at finite distance on $S^1(R)\times S^1_{D^2}=T^2$, which will flow to a superconformal point in the IR.\footnote{The index is sometimes called ``A-twisted'', but we do not perform the topological twist of any supercharge here.} 
%This half-index can be computed via equivariant localization, as a trace over the Hilbert space of states on $D^2$.
In geometry, the half-index we compute is a generating function of quasimaps $\mathbb{CP}^1/\{0\}\rightarrow X$ of all degrees in equivariant quantum K-theory, where $X$ is the Higgs branch of the 3d theory \cite{2009arXiv0908.4446C,2010arXiv1005.4125K,2011arXiv1106.3724C,Okounkov:2015spn}; by excising the origin of $\mathbb{CP}^1$, we are allowing for line operator insertions supported on $\{0\}\times S^1$ to contribute to the quasimap count (see for instance \cite{Aganagic:2017gsx,Aganagic:2017smx}); these are the 1-dimensional degrees of freedom supported at a totally transverse NS5-D3 brane intersection. In section \ref{sec:typeIIB}, we find:\\

\noindent\fbox{%
    \parbox{\textwidth}{%
       Each component of the Miura transformation is the half-index of a 3d supersymmetric abelian (quiver) gauge theory on $D^2\times S^1$, possibly coupled to 1d line defects.
    }%
}

\vspace{4mm}

The Miura transformation admits an integral representation, whose components are recovered from the index by imposing Neumann boundary conditions for the vector multiplets and a fixed polarization for the hypermultiplets (a specific holomorphic Lagrangian splitting of the $\EN=4$ hypermultiplets). The residue sum of this Miura transformation corresponds to giving Dirichlet boundary conditions to the vector multiplets and hypermultiplets.

The fact that we are able to recover the components of the Miura transformation straight from a type IIB computation reinforces the validity of the duality we propose here.\\

\begin{table}[h!]
    \centering
    \begin{tabular}{  c||c|c|c|c|c|c|c|c|c|c } 
         \text{IIB branes} & 0 & 1 & 2 & 3 & 4 & 5 & 6 & 7 & 8 & 9  \\ \hline\hline 
        $\text{D3}_{1,0,0}$ &  \rm{x} & \rm{x}  &  &   &    &   & \rm{x}  & &  &  \rm{x} \\ 
             $\text{D3}_{0,1,0}$ &  &  & \rm{x}  & \rm{x}   & &  & \rm{x}  & &  &  \rm{x} \\
            $\text{D3}_{0,0,1}$ &  &  &  &   & \rm{x}  & \rm{x}  & \rm{x}  & &  &  x  \\
          \hline  $\text{NS5}_{1,0,0} $ &  & & \rm{x}& \rm{x} & x & x & & \rm{x} &  & \rm{x} \\
          $\text{NS5}_{0,1,0} $ & \rm{x} & \rm{x}& &  & x  & x & & \rm{x} &  & \rm{x} \\  $\text{NS5}_{0,0,1} $ & \rm{x} & \rm{x}& \rm{x}& \rm{x} &  & & & \rm{x} &  & \rm{x}   \\ \hline
           $\text{D5}_{0,0,1} $ & \rm{x} & \rm{x}& \rm{x}& \rm{x} &  & & &  &  \rm{x} & \rm{x} \\ \hline
           $\mathbf{(p,q)}_{0,0,1} $ & \rm{x} & \rm{x}& \rm{x}& \rm{x} &  & & &   \multicolumn{2}{c|}{$l^{\mathbf{(p,q)}}$} & \rm{x}    
    \end{tabular} 
   \caption{IIB-brane configuration for the $\mathbf{(p,q)}$-web. Here, $\mathbf{(p,q)}$ indicates 
the $\mathbf{(p,q)}$-fivebrane, supported on the line $l^{\mathbf{(p,q)}} = \{ \mathbf{q} x - \mathbf{p} z = \text{const} \} \subset \BR_x \times \BR_z$. }
     \label{table:2am}
\end{table}

\begin{table}[h!]
    \centering
    \begin{tabular}{c|c|c|c|c|c|c}
        $\BR^2 _{\ve_1}$ & $\BR^2 _{\ve_2}$ & $\BR^2 _{\ve_3}$ & $\mathbb{R} _t $  & $\BR _x $ & $\BR _z $ & $S^1$  \\ \hline
        $x^0,x^1$ & $x^2,x^3$ & $x^4, x^5$ & $x^6$  & $x^7$ & $ x^8$ & $x^9$ 
    \end{tabular}
    \caption{Spacetime of the IIB theory for the $\mathbf{(p,q)}$-web. The $\mathbf{(p,q)}$-fivebrane web is depicted on the $\BR_x \times \BR_z$ plane.}
    \label{table:stpq}
\end{table}

\paragraph{Gauge origami and $qq$-characters}
It was suggested in \cite{Jeong:2023qdr} that the IIB theory for the gauge origami \cite{Nikita:I,Nikita:III} is also connected to the twisted M-theory by a string duality, up to certain unrefinement of the background. The gauge origami is the configuration of intersecting D3-branes in the IIB theory defined on $X \times \BC^\times _X$. The $\O$-background can be implemented with respect to the $U(1)^3 \subset SU(4)$ isometry of the Calabi-Yau 4-fold $X$, parametrized by the four $\O$-background parameters $(\ve_1,\ve_2,\ve_3,\ve_4)$ constrained by $\ve_1+\ve_2+\ve_3+\ve_4=0$. The D3-branes occupy real four-cycles in $X$ which are invariant under the isometry, located at certain points on the transverse holomorphic plane $\BC^\times _X$.

We set $X = \BR^2 _{\ve_1} \times \BR^2 _{\ve_2} \times (\BR^2 _{\ve_3} \times \BR^2 _{\ve_4})/\BZ_l$ and deform the $A_{l-1}$-singularity into the $l$-centered Taub-NUT space (see table \ref{table:oribrane} and \ref{table:orist}). The subscript of each $\BR^2$-plane denotes the $\O$-background parameter associated to its rotation. By unrefining the $\O$-background, the T-duality along the Taub-NUT circle can be performed in the simplest manner, yielding the IIA theory defined on $\BR^2 _{\ve_1} \times \BR^2 _{\ve_2} \times \BR^2 _{\ve_3} \times \BR_t \times  \BC^\times _X \times S^1$ ($\ve_1+\ve_2+\ve_3=0$) with $l$ NS5-branes supported on $\BR^2 _{\ve_1} \times \BR^2 _{\ve_2} \times \BC^\times _X$. We may further decompactify the circle $S^1$ to a line $\BR_z$, making one NS5-brane disappear to \textit{infinity} in $\BR_z$, and uplift to M-theory with the M-theory circle $S^1 _M$, which combines into another holomorphic plane $\BC^\times _Z$ \cite{Witten1997}. In this way, we arrive at the twisted M-theory on $\BR^2 _{\ve_1} \times \BR^2 _{\ve_2} \times \BR^2 _{\ve_3} \times \BR_t \times \BC^\times _X \times \BC^\times _Z$.

Let us examine how two different types of D3-branes in the IIB theory of the gauge origami are transformed under the duality. The first type is the ones supported on $\BR^2 _{\ve_1} \times \BR^2 _{\ve_2}$, each of which is assigned  a $\BZ_l$-charge. After the T-duality, they are mapped to the D4-branes stretched between the newly created NS5-branes in the $\BR_z$-direction. The number of D4-branes between each consecutive pair of NS5-branes is determined by their $\BZ_l$-charge assignment \cite{Douglas:1996sw}. This is precisely the IIA engineering of the 5-dimensional $\EN=1$ quiver gauge theory compactified on a circle, as the effective field theory obtained from quantizing the D4-D4 open strings. Both the D4-branes and the NS5-branes become M5-branes in the twisted M-theory, sharing $\BR^2 _{\ve_1} \times \BR^2_{\ve_2}$ in their worldvolume but wrapping different holomorphic plane, $\BC^\times _Z$ and $\BC^\times _X$, respectively. By their non-transverse intersections on $\BC^\times _X \times \BC^\times _Z$, they form a \textit{web} of M5-branes.

The second type is supported on $(\BR^2_{\ve_3} \times \BR^2_{\ve_4})/\BZ_l$, for which we also assign a $\BZ_l$-charge. Equivalently, the D3-brane can wrap different compact $\BP^1$'s after resolving the $A_{l-1}$ singularity into the $l$-centered Taub-NUT space, where the choice is encoded in the $\BZ_l$-charge $c\in \{0,1,\cdots, l-1\}$. Performing the T-duality, this D3-brane becomes a D2-brane supported on $\BR^2_{\ve_3} \times \BR_t$, sharing the $\BR_z$-coordinate with one of the NS5-branes ($l$ NS5s before the decompactification or $l-1$ NS5s after the decompactification), determined by the $\BZ_l$-charge. At last, under the uplift to the twisted M-theory, the D2-brane becomes an M2-brane supported on $\BR^2 _{\ve_3} \times \BR_t$. Note that the resulting M2-brane is totally transverse to the M5-branes that we obtained above in the 11-dimensional worldvolume.\\

Recall that the crossed D3-brane configuration in the gauge origami involves both types of D3-branes present simultaneously. In particular, due to their transverse intersection, the second type of D3-brane provides a codimension-4 defect lying along the circle in the 5-dimensional $\EN=1$ effective gauge theory from the first type, referred to as the \textit{$qq$-characters} \cite{Nikita:I}.\footnote{In more detail, by $qq$-character, we mean the gauge and string theoretical realization of the generating currents of the $q\mathcal{W}$-algebra (and by extension  those of the $qY$-algebra as well), originally constructed algebraically by Frenkel and Reshetikhin  \cite{Frenkel:1998} (see also \cite{Shiraishi:1995rp,Feigin:1995sf,Awata:1995zk,Bouwknegt:1998da}), and revisited recently in a wide variety of examples
\cite{Kimura:2015rgi,Kimura:2016dys,Kimura:2017hez,Kojima2019QuadraticRO,Kojima2021QuadraticRO2,2023arXiv231216856K}. The relevance of $qq$-characters to the representation theory of quantum toroidal algebras was studied in \cite{2015arXiv151208779B,Bourgine:2016vsq,2020arXiv200304234F,Harada:2021xnm}.} In the dual IIB theory for the $\mathbf{(p,q)}$-fivebrane web on the $\mathbb{R}_x\times\mathbb{R}_z$ plane (see table \ref{table:2am}), the crossed D3-brane is mapped to a D3-brane at a point on that plane.\footnote{It has been suggested that $qq$-characters could be realized via \emph{different} brane defects in a  $\mathbf{(p,q)}$-fivebrane web \cite{Kimura:2017auj,Zenkevich:2023cza}. Unfortunately, none of these proposals coincide with the original stringy crossed instanton definition in \cite{Nikita:I}, nor are they obviously dual to it. Our construction of the $qq$-characters from transverse M2-branes is directly dual to the original construction, as explained above.}  Physically, the $qq$-character is the BPS count of this brane system; equivalently, it is the Witten index of the supersymmetric quantum mechanics on instantonic D1-branes wrapping the circle, in the presence of the heavier branes \cite{NaveenNikita,Kim:2016qqs,Chang:2016iji,Assel:2018rcw,Agarwal:2018tso,Haouzi:2019jzk,Haouzi:2020bso}. Mathematically, this index is defined as the (properly regularized) integral over the crossed instanton moduli space of the generalized gauge theory, evaluated in equivariant K-theory.

We just analyzed that, in the twisted M-theory, the corresponding M-brane configuration is a single M2-brane placed upon a transverse web of M5-branes, enumerated by $c=1,2,\cdots, l-1$, which determines the M5-brane that it shares the $\BC^\times _Z$-coordinate with. As we discussed earlier, the M2-M5 intersection assigns an R-matrix of the quantum toroidal algebra of $\fgl(1)$ (i.e., the Miura operator by our result in section \ref{sec:miura}). In section \ref{sec:qqchar}, we show that:\\

\noindent\fbox{%
    \parbox{\textwidth}{%
The insertion of the R-matrix associated with the M2-M5 intersection on the M5-brane web of $q$-bosons exactly reproduces the $qq$-characters.
    }%
}
\\

\noindent
This explicitly confirms the duality we suggest here. 

Already back in the 90's  \cite{Frenkel:1998}, it was appreciated that $qq$-characters are traces of transfer matrices for certain TQ-integrable systems. Here, we are  showing that the generating function of all the $qq$-characters is a product of R-matrices of the quantum toroidal algebra, making the integrability even more manifest. See \cite{Kimura:2015rgi,Kimura:2017hez,Bourgine:2015szm,Jeong:2017pai,Jeong:2018qpc,Lee:2020hfu,Jeong:2021bbh,Jeong:2023qdr,Grekov:2023fek,Jeong:2024hwf,Jeong:2024onv} for studies of the quantum integrability using the $qq$-characters as gauge theory observables. See also \cite{Mironov:2016yue,Kim:2016qqs,Elliott:2018yqm,Frenkel:2020iqq,Jeong:2017mfh,Jeong:2019fgx,Jeong:2020uxz,Assel:2018rcw,Agarwal:2018tso,Haouzi:2019jzk,Haouzi:2020bso,Haouzi:2020yxy,Haouzi:2020zls,2022arXiv220307072L} for related studies on the $qq$-characters.

\begin{table}[h!]
    \centering
    \begin{tabular}{  c||c|c|c|c|c|c|c|c|c|c  } 
          \text{IIB branes} & 0 & 1 & 2 & 3 & 4 & 5 & 6 & 7 & 8 & 9  \\ \hline\hline 
           D3 & x & x  & x  & x  &   &   &   & &  &   
          \\ D3 &  & & &  & x & x & x& \rm{x} &  &   \\  
          \hline
            KK5 & \rm{x} & \rm{x}&  x&  x&  &  & &  & \rm{x} & x  
    \end{tabular} 
    \caption{IIB configuration for the gauge origami. The first row is the D3-branes of first type, while the second row is the D3-branes of second type. The Kaluza-Klein 5-monopoles simply indicate the spacetime is modified to the $l$-centered Taub-NUT space.}
     \label{table:oribrane}
\end{table}

\begin{table}[h!]
    \centering
    \begin{tabular}{c|c|c|c|c}
        $\BR^2 _{\ve_1}$ & $\BR^2 _{\ve_2}$ & $\BR^2 _{\ve_3}$ & $\mathbb{R}^2 _{\ve_4} $  & $\BC^\times _X $   \\ \hline
        $x^0,x^1$ & $x^2,x^3$ & $x^4, x^5$ & $x^6,x^7$  & $x^8 ,x^9$  
    \end{tabular}
    \caption{Spacetime of the IIB theory for the gauge origami}
    \label{table:orist}
\end{table}

\subsection{Outline}
In section \ref{eq:mbrep}, we identify the algebra of local operators on the line defects and the surface defects of 5d Chern-Simons theory from the M2-branes and the M5-branes, respectively, and realize them as representations of the quantum toroidal algebra of $\fgl(1)$. In section \ref{sec:miura}, we solve the R-matrix between the representations assigned to a single M2-brane and a single M5-brane. We construct the Miura transformation for the $q$-deformed $Y$-algebra using the R-matrix properties of the multiple M2-M5 intersections. We also obtain the screening charges from M2-branes placed between each consecutive M5-branes, and verify that the $q$-deformed $Y$-algebra is exactly characterized as the commutants of these screening charges. In section \ref{sec:typeIIB}, we show that the half-index of the 3d $\EN=2$ quiver $U(1)$ gauge theory with 1d defect, engineered as the effective field theory on the D3-brane passing through multiple NS5-branes, exactly matches with the components of the Miura transformations. In particular, the relevant boundary conditions are exactly identified. In section \ref{sec:qqchar}, we present different incarnations of the $q$-boson representations of the quantum toroidal algebra associated to the M5-branes supported on general holomorphic curves. It is shown that their non-transverse intersections lead to the intertwiners between those $q$-boson representations. We consider the configuration where a single M2-brane passes through a web of M5-branes, and verify that the insertion of the R-matrix associated to the M2-M5 intersection precisely yields the $qq$-characters of the 5d $\EN=1$ effective gauge theory realized on the web of M5-branes. We conclude with discussions in section \ref{sec:discussion}. The appendices contain generalities about the quantum toroidal algebra of $\fgl(1)$ and the spherical double affine Hecke algebra; computational details for solving the R-matrices and the screening charges; and details of the $q$-boson intertwiners of the quantum toroidal algebra of $\fgl(1)$.

\paragraph{Acknowledgment}
The authors thank Mina Aganagic, Sibasish Banerjee, Davide Gaiotto, Alba Grassi, Nafiz Ishtiaque, Ahsan Khan, Shota Komatsu, Norton Lee, Evgeny Mukhin, Nikita Nekrasov, Jihwan Oh, Miroslav Rap\v{c}\'{a}k, Edward Witten, Yegor Zenkevich, and Yehao Zhou for discussions and collaboration on related subjects. The authors are grateful to the organizers of Simons Physics Summer Workshop 2023, where the collaboration was initiated. The work of NH is supported by NSF Grant PHY-2207584 and the Sivian Fund at the Institute for Advanced Studies. The work of SJ is supported by the Institute for Basic Science under the project IBS-R003-Y3. This work was also partly supported by CERN and CKC fellowship.

\vspace{4mm}

\section{M-branes and representations of quantum toroidal algebra of $\fgl(1)$} \label{eq:mbrep}
The twisted M-theory subject to the $\O$-background reduces to the 5-dimensional non-commutative Chern-Simons theory on $\BR_t \times \BC^\times _X \times \BC^\times _Z$ \cite{Costello:2016nkh}. The M2-branes and the M5-branes descend to topological line defects and holomorphic surface defects, respectively. The BRST-invariance of the coupling of these defects require the algebra of local operators on the defects to be representations of certain universal associative algebra. In this section, we suggest this universal associative algebra is the quantum toroidal algebra of $\fgl(1)$, $\qta$, for our case. We also specify the representations of $\qta$ associated to the M2-branes and the M5-branes. 

\vspace{4mm}

\subsection{M2-branes and topological line defects} \label{subsec:m2}
The M2-branes in the twisted M-theory can be supported on $\BR^2 _{\ve_c} \times \BR_t$, where $c\in \{1,2,3\}$. Once reduced to the 5d $\fgl(1)$ Chern-Simons theory, they give rise to line defects lying along $\BR_t$, located at points on $\BC^\times _X \times \BC^\times _Z$. Let us focus on the case where all the M2-branes are located at a single point on $\BC^\times _X \times \BC^\times _Z$, giving a single line defect. In the most general case, there are $l$, $m$, and $n$ M2-branes wrapping $\BR^2 _{\ve_1} \times \BR_t$, $\BR^2 _{\ve_2} \times \BR_t$, and $\BR^2 _{\ve_3} \times \BR_t$, respectively. We call the resulting line defect $\mathcal{L}_{l,m,n}$. Let us define $q_c = e^{\ve_c} \in \BC^\times$, $c\in \{1,2,3\}$.

The local observables of the 5d $\fgl(1)$ Chern-Simons theory localized on the line $\BR_t \times \{\text{point} \}$ form an algebra $\text{Obs}_{q_1,q_2} ^{\BR_t}$, with the product defined by their OPEs. At the classical level ($q_1 = 1$), the algebra $\text{Obs}_{q_1=1,q_2} ^{\BR_t}$ has a simple description; the solutions to the equations of motion is trivial, and the only local observables are the functions of the modes of the ghost under the Laurent expansion around the locus of the line in $\BC^\times_X \times \BC^\times _Z$, for which the holomorphic coordinates $X$ and $Z$ are subject to the non-commutativity relation
\begin{align} \label{eq:noncom}
    X Z = q_2 Z X,
\end{align}
due to the B-field $B = \ve_2 \frac{d\bar{X}}{\bar{X}} \wedge \frac{d\bar{Z}}{\bar{Z}}$.

The Laurent expansion $c = \sum_{m,n \in \BZ} c_{m,n} Z^m X^n$ indicates that the gauge Lie algebra is generated by $W_{m,n} \equiv Z^m X^n$, satisfying the commutation relations $\left[W_{m,n},W_{m',n'} \right] = \left(q_2 ^{m' n} - q_2 ^{m n'} \right) W_{m+m', n+n'}$. This is precisely the quantum torus Lie algebra (see e.g. \cite{Nakatsu:2007dk}), which we denote by $\mathscr{O}_{q_2} (\BC^\times \times \BC^\times)$. Note that its universal enveloping algebra is the quantum torus algebra, $U(\mathscr{O}_{q_2} (\BC^\times \times \BC^\times)) = \BC\left\langle X^{\pm 1},Z^{\pm 1} \right\rangle/(XZ = q_2 ZX)$. Equipped with the BRST differential, the functions of the ghost thus form the Chevalley-Eilenberg cochain complex,
\begin{align} \label{eq:cecpx}
    \text{Obs}_{q_1 = 1,q_2} ^{\BR_t} = C^\bullet (\mathscr{O}_{q_2} (\BC^\times \times \BC^\times ) ).
\end{align}

A line defect of the 5d $\fgl(1)$ Chern-Simons theory can be defined by coupling a topological quantum mechanics, whose algebra of local observables is, say, $\EuScript{B}$. Explicitly, the line defect coupling is implemented by inserting the generalized Wilson line,
\begin{align}
    \text{Pexp} \sum_{m,n \in \BZ}\int_{\BR_t} dt\, P_{m,n} (A_t)_{m,n},
\end{align}
into the path integral, where $(A_t)_{m,n}$ are the modes of the gauge field under the Laurent expansion in $Z$ and $X$, and $P_{m,n} \in \EB$ are the local operators on the line defect. The BRST-invariance of the coupling is guaranteed if the coupling is the topological descent of a Maurer-Cartan element in $\text{Obs}_{q_1 ,q_2} ^{\BR_t} \otimes \EuScript{B}$. This is, by the Koszul duality, equivalent to requiring that there is a surjective algebra homomorphism
\begin{align}
    \EA_{q_1,q_2}:= \left(\text{Obs}_{q_1,q_2} ^{\BR_t} \right)^! \twoheadrightarrow \EuScript{B},
\end{align}
where $ \EuScript{A}_{q_1,q_2}= \left(\text{Obs}_{q_1,q_2 }  ^{\BR_t} \right)^!$ is the Koszul dual of $\text{Obs}_{q_1,q_2} ^{\BR_t}$. In this sense, the Koszul dual algebra $\EuScript{A}_{q_1,q_2}$ can be said to be the algebra of local observables on the \textit{universal} line defect. It is called the universal associative algebra in short.

In the classical limit $q_1 = 1$, the standard result tells that the Koszul dual of the operator algebra \eqref{eq:cecpx} is (see \cite{Paquette:2021cij} for a review)
\begin{align} \label{eq:kdclassical}
    \EuScript{A}_{q_1=1,q_2} = \left(\text{Obs}_{q_1 = 1,q_2}  ^{\BR_t} \right) ^! = U(\mathscr{O}_{q_2} (\BC^\times \times \BC^\times )).
\end{align}
Turning on $q_1 \neq  1$, the quantum corrections deform the operator algebra $\text{Obs}_{q_1,q_2} ^{\BR_t}$, and therefore its Koszul dual $\EuScript{A}_{q_1,q_2} = (\text{Obs}_{q_1,q_2} ^{\BR_t})^!$, in a non-trivial way. In particular, the Koszul dual would be a deformation of the quantum torus algebra \eqref{eq:kdclassical}, which we can formally denote by
\begin{align}
    \EuScript{A}_{q_1,q_2} = U_{q_1}(\mathscr{O}_{q_2} (\BC^\times \times \BC^\times )).
\end{align} 

The quantum corrections for $\text{Obs}_{q_1,q_2} ^{\BR_t}$ can in principle be determined by computing the Feynman diagrams of the 5d $\fgl(1)$ Chern-Simons theory on $\BR_t \times \BC^\times _X  \times \BC^\times_Z $ relevant to the operator products of the mode (Laurent) expansions of the ghost fields.\footnote{See \cite{Costello:2017fbo} for the explicit evaluation of the Feynman diagrams relevant to the quantum corrections to the operator algebra in the case of the 5d $\fgl(k)$ Chern-Simons theory on $\BR \times \BC^2$. In this case, the operator algebra is an $\ve_1$-deformation of $C^\bullet (\fgl(k)\otimes \mathscr{O}_{\ve_2} (\BC^2))$, and its Koszul dual, i.e., the universal associative algebra, is proven to be the 1-shifted affine Yangian of $\fgl(k)$, $U_{\ve_1} (\fgl(k) \otimes \mathscr{O}_{\ve_2} (\BC^2)) = Y_1 (\widehat{\fgl}(k))$.} Also, the defining relations of the universal associative algebra can in principle be achieved as conditions for the anomalous Feynman diagrams to cancel each other (see \cite{Costello:2017dso, Costello:2017fbo, Oh:2020hph}). We do not take such a direct approach of the perturbation theory in the present work.

Rather, we will make a natural suggestion for the universal associative algebra $\EA_{q_1,q_2}$, and assign appropriate representations to the topological line defects $\CalL_{l,m,n}$ created by the M2-branes. Our proposal is that the universal associative algebra $\EuScript{A}_{q_1,q_2}$ is precisely the quantum toroidal algebra of 
$\fgl(1)$, namely, $\EuScript{A}_{q_1,q_2} = \qta$.\footnote{Recall that $q_3 = q_1 ^{-1} q_2 ^{-1}$ so that we could have omitted the dependence on $q_3$ and write $U_{q_1,q_2} (\widehat{\widehat{\fgl}}(1))$. We write $q_3$ explicitly only to emphasize the triality of exchanging $q_1$, $q_2$, and $q_3$ with each other.} We review the definition and the relevant properties of the quantum toroidal algebra $\qta$ in appendix \ref{app:qta}. In the present work, we will provide evidences for our claim, leaving the exact proof of the statement through perturbative study of 5d non-commutative $\fgl(1)$ Chern-Simons theory on $\BR_t \times \BC^\times_X \times \BC^\times _Z$ to future work.

\subsubsection{Single M2-brane and quantum torus algebra} \label{subsubsec:singlem2}
Let us begin with a single M2-brane supported on $\BR^2 _{\ve_2} \times \BR_t$. In the IIB dual frame, the M2-brane passes to a fundamental string attached to the NS5-brane, ending at the junction of the fivebranes. In the worldvolume theory of the D5-brane, the fundamental string provides a boundary Wilson line of charge 1 \cite{Gaiotto:2019wcc}. The position of the Wilson line along the junction of the fivebranes, which we denote by $(X_1,Z_1) \in \BC _X ^\times \times \BC_Z ^\times $, are subject to the relation 
\begin{align}
X_1 Z_1 = q_2 Z_1 X_1 
\end{align}
by the non-commutativity \eqref{eq:noncom}. They can be thought of as the generators of the algebra $\text{M2}_{0,1,0}$ of local operators on this Wilson line, which is nothing but the quantum torus algebra. Namely, we have $ \text{M2}_{0,1,0} = U(\mathscr{O}_{q_2} (\BC^\times \times \BC^\times ))$.

The quantum torus algebra is a representation of the quantum toroidal algebra of $\fgl(1)$. Indeed, there is a surjective algebra homomorphism 
\begin{align}
    \r_{\text{M2}_{0,1,0}} : \qta  \twoheadrightarrow \text{M2}_{0,1,0}
\end{align}
given by
\begin{align}
\begin{split}
    &E(X) \mapsto \frac{1}{1-q_2}  \d\left( \frac{X_1}{X} \right) Z_1 ,\quad F(X) \mapsto \frac{1}{1-q_2 ^{-1}} Z_1 ^{-1} \d \left( \frac{X_1}{X} \right)    \\
    &K^\pm (X) \mapsto \frac{(1- (q_1 ^{-1} X_1/X)^{\pm 1}) (1- (q_3 ^{-1} X_1/X)^{\pm 1})  }{ (1- ( X_1/X)^{\pm 1}) (1- (q_2 X_1/X)^{\pm 1}) } \\
    &C  \mapsto 1.
\end{split}
\end{align}
Note $(C,C^\perp) \mapsto (1,1)$ in particular. The map $\r_{\text{M2}_{0,1,0}}$ can be written equivalently in terms of the modes as
\begin{align} \label{eq:vecrep}
\begin{split}
    &E_k \mapsto \frac{1}{1-q_2} X_1 ^k Z_1,\quad F_k \mapsto \frac{1}{1-q_2 ^{-1}} Z_1^{-1}  X_1 ^k,\qquad k \in \BZ, \\
    & H_{\pm r} \mapsto \frac{1}{1- q_2 ^{\mp r}} X_1 ^{\pm r} , \qquad r \in \BZ_{>0}, \\
    &C\mapsto 1,\quad C^\perp \mapsto 1.
\end{split}
\end{align}
In particular, the homomorphism in fact does not depend on $q_1$.  

For any given weight parameter $u \in \BC^\times $, there is a faithful representation of the quantum torus algebra $\text{M2}_{0,1,0}$ on the vector space $\CalV_2 (u) = \bigoplus_{i\in \BZ} \BC [u]_i ^{(2)}$, defined by
\begin{align} \label{eq:vectorr}
\begin{split}
    Z_1 [u]_i ^{(2)} = [u]_{i+1} ^{(2)}, \quad\quad X_1 [u]_i ^{(2)} = u q_2 ^i [u]_i ^{(2)}.
\end{split}
\end{align}
More concretely, the representation can be thought of as the $q_2$-difference operators
\begin{align}
    Z_1 = q_2 ^{-D_{X_1}}, \qquad X_1 = X_1
\end{align}
acting on the space of Laurent polynomials
\begin{align}
    \CalV_2 (u) = u ^{- D_{X_1} } \BC \left(\!\left(q_2 ^{- D_{X_1}}\right)\!\right) = \bigoplus_{i \in \BZ} \BC q_2 ^{-\left(\frac{\log u}{\log q_2} + i \right) D_{X_1}}.
\end{align}
Here, we used the notation $D_{X_1} = X_1 \p_{X_1}$ so that $q_2 ^{-D_{X_1}} = q_2 ^{-X_1 \p_{X_1}}$ is the $q_2 ^{-1}$-shift operator for $X_1$ (i.e., $q_2 ^{-D_{X_1}} f(X_1) = f(q_2 ^{-1} X_1) q_2 ^{-D_{X_1}} $ for any function $f(X_1)$). This is precisely what is called the \textit{vector} representation of the quantum toroidal algebra of $\fgl(1)$ (cf. \cite{Tsymbaliuk:2014fvq}), factored through the quantum torus algebra. Namely, what we obtained can be summarized as $\qta  \twoheadrightarrow \text{M2}_{0,1,0} \hookrightarrow \text{End}( \CalV_2 (u))$.\\

For later use, note that the grading operator $d$ (see section \ref{subsec:grading}) acts on the generators as
\begin{align}
    q^d X_1 = q X_1 q^d\qquad q^d Z_1 = Z_1 q^d,
\end{align}
for any $q\in \BC^\times$. It follows that the representation on $\CalV_2 $ can be extended to incorporate the grading operator by regarding it as the shift operator for the module parameter: $q^d [u]^{(2)}_i = [u q^{-1}]_i ^{(2)}$.\\

As stressed above, the homomorphism $\r_{\text{M2}_{0,1,0}}:\qta \to \text{M2}_{0,1,0}$ is not affected by tuning of the value of $q_1$. In this sense, the quantum torus algebra, as a representation of the quantum toroidal algebra, does not deform from $q_1=1$, and it is expected to still provide the M2-brane algebra with all the quantum corrections taken into account. Such a rigidity is indeed anticipated from the boundary Wilson line description for the single M2-brane.\\

Now, let us turn to the single M2-brane wrapping $\BR^2 _{\ve_1} \times \BR_t$ (resp. $\BR^2 _{\ve_3} \times \BR_t$). These M2-branes can simply be obtained by shuffling the $\BR^2$-planes in $\BR^2 _{\ve_1} \times \BR^2 _{\ve_2} \times \BR^2 _{\ve_3}$, which is equivalent to exchanging the parameters $q_1$, $q_2$, and $q_3$ with each other. Therefore, the algebra of local operators on this M2-brane is expected to be the quantum torus algebra $\text{M2}_{1,0,0} = U(\mathscr{O}_{q_1} (\BC^\times \times \BC^\times ))$ (resp. $\text{M2}_{0,0,1} = U(\mathscr{O}_{q_3} (\BC^\times \times \BC^\times ))$) with the non-commutativity of the generators controlled by $q_1$ (resp. $q_3$) instead of $q_2$.

Indeed, as apparent from its defining relations, the quantum toroidal algebra $\qta$ is invariant under the triality of permuting $(q_1,q_2,q_3)$. The algebra homomorphisms to the other two quantum torus algebras thus exist, obtained simply by replacing $q_2$ by $q_1$ or $q_3$ in the map $\r_{\text{M2}_{0,1,0}}$ \eqref{eq:vecrep}. We denote these algebra maps by $\r_{\text{M2}_{1,0,0}} : \qta \to \text{M2}_{1,0,0}$ and $\r_{\text{M2}_{0,0,1}} : \qta \to \text{M2}_{0,0,1}$, respectively. We can define faithful representations of these quantum torus algebras on the vector spaces $\CalV_c (u) = \bigoplus_{i\in \BZ} \BC [u]_i ^{(c)}$, $c\in \{1,2,3\}$, simply by permuting $(q_1,q_2,q_3)$ on \eqref{eq:vectorr}.

\subsubsection{Parallel M2-branes and spherical DAHA}
\label{sssec:parallelM2}
Next, we examine multiple M2-branes wrapping the same $\BR^2 _{\ve_c} \times \BR_t$, $c\in \{1,2,3\}$. We will explain that the two cases, $c=1$ and $2$, independently lead to the same conclusion that the algebra of local operators on the parallel M2-branes is given by the spherical subalgebra of the double affine Hecke algebra (spherical DAHA). Then, we will show that the fusion of the parallel M2-branes is compatible with the coproduct of the quantum toroidal algebra.\\

\paragraph{M2-branes on $\BR^2 _{\ve_2} \times \BR_t$ and boundary Wilson lines} 
As studied in the previous section, the $m$ M2-branes supported on $\BR^2 _{\ve_2} \times \BR_t$ reduce to the $m$ boundary Wilson lines on the worldvolume theory of the D5-brane. We denote the algebra of local operators on these M2-branes by $\text{M2}_{0,m,0}$. 

Let $(X_i,Z_i)_{i=1} ^m$ be the positions of the $m$ Wilson lines on the holomorphic surfaces $\BC^\times _X \times \BC^\times _Z$. Just as the case of single M2-brane, these holomorphic coordinates are the generators of the algebra of local operators, subject to the non-commutativity \eqref{eq:noncom}. In the classical limit $q_1 = 1$, there is no quantum correction coming from the interaction between these Wilson lines. The algebra of local operators is thus simply given by the symmetric product of the $m$ quantum torus algebras, $\text{M2}_{0,m,0} ^{q_1 = 1} =  U(\mathscr{O}_{q_2} \left(\BC^\times \times \BC^\times  \right))^{\otimes m} /S_m$, each of which is from individual Wilson line. As discussed earlier, the universal associative algebra at $q_1 = 1$ is the quantum torus algebra \eqref{eq:kdclassical}, and there is indeed a natural homomorphism
\begin{align}\label{eq:classmorphdaha}
    U( \mathscr{O}_{q_2} (\BC^\times \times \BC^\times )) \twoheadrightarrow \frac{ U(\mathscr{O}_{q_2} \left(\BC^\times \times \BC^\times  \right)) ^{\otimes m}}{S_m},
\end{align}
which encodes the coupling of the line defect in the classical limit $q_1 =1$.\\

Turning on $q_1 \neq 1$, we suggested that the universal associative algebra gets deformed to the quantum toroidal algebra of $\fgl(1)$, $\qta = U_{q_1}(\mathscr{O}_{q_2} (\BC^\times \times \BC^\times )) $. The algebra $\text{M2}_{0,m,0}$ of local operators on the line defect should also be a $q_1$-deformation of $\text{M2}_{0,m,0} ^{q_1=1} = U( \mathscr{O}_{q_2} (\BC^\times \times \BC^\times )) ^{\otimes m} /S_m$, and the coupling of the line defect would be encoded in a $q_1$-deformation of the surjective homomorphism \eqref{eq:classmorphdaha}, which we may write as
\begin{align}
    \r_{\text{M2}_{0,m,0}}:\qta  \twoheadrightarrow \text{M2}_{0,m,0}.
\end{align}

There is indeed a natural $q_1$-deformation of the symmetric product of quantum torus algebras: the spherical subalgebra of the double affine Hecke algebra (spherical DAHA) of $GL(m)$, which we denote by $\sh^{(m)} _{q_2,q_1} $. The spherical DAHA can be presented as the algebra generated by the difference operators,
\begin{align}
\begin{split}
    &P_{0,r} ^{(m)} = q_2 ^r \sum_{i=1} ^m X_i ^r,\qquad P_{0,-r} ^{(m)} =  \sum_{i=1} ^m X_i ^{-r},\qquad r \in \BZ_{>0} \\
    \begin{split}
    &P_{1,k} ^{(m)} = q_2 \sum_{i=1} ^m \left( \prod_{j\neq i} \frac{ q_1 X_i -  X_j }{X_i -X_j} \right) X_i ^k q_2 ^{-D_{X_i}} \\
    &P_{-1,k} ^{(m)} = \sum_{i=1} ^m \left( \prod_{j\neq i} \frac{ q_1 ^{-1} X_i -  X_j }{X_i -X_j} \right) q_2 ^{D_{X_i}} X_i ^k
    \end{split} ,\qquad k \in \BZ,
\end{split}
\end{align}
acting on the space of symmetric polynomials $\BC(q_1,q_2) [X_1 ^{\pm 1} ,\cdots, X_m ^{\pm 1}]^{S_m}$. We review the definition and the relevant properties of the spherical DAHA of $GL(m)$ in appendix \ref{app:sdaha}.

Correspondingly, the algebra homomorphism \eqref{eq:classmorphdaha} admits a $q_1$-deformation,
\begin{align} \label{eq:sdrep}
    \qta  \twoheadrightarrow \sh^{(m)} _{q_2,q_1},
\end{align}
given by
\begin{align} \label{eq:sdrepexp}
\begin{split}
\begin{split}
  &E_k \mapsto \frac{1}{q_2 (1-q_2)} P_{1,k} ^{(m)} = \frac{1}{1-q_2}  \sum_{i=1} ^m \left( \prod_{j\neq i} \frac{ q_1 X_i -   X_j}{ X_i- X_j} \right) X_i ^{k} q_2 ^{-D_{X_i}}, \\  
   &F_k \mapsto  \frac{1}{1-q_2 ^{-1}} P_{-1,k} ^{(m)} = \frac{1}{1-q_2 ^{-1}} \sum_{i=1} ^m \left( \prod_{j\neq i} \frac{ q_1 ^{-1} X_i -   X_j}{X_i - X_j} \right) q_2 ^{D_{X_i}} X_i ^k , \\
   \end{split} \qquad k \in \BZ, \\
   \begin{split}
    & H_{\pm r} \mapsto \pm \frac{1}{1-q_2 ^r} P_{0,\pm r }^{(m)} = \frac{1}{1- q_2 ^{\mp r}} \sum_{i=1} ^ m X_i ^{\pm r} , \qquad r \in \BZ_{>0}, \\
    &C\mapsto 1,\quad C^\perp \mapsto 1.
\end{split}
\end{split}
\end{align}
This is a surjective algebra homomorphism for each $m \in \BZ_{>0}$ \cite{Schiffmann_Vasserot_2011}. Note that in the classical limit $q_1 =1$, the spherical DAHA simplifies to the symmetric product $\sh^{(m)} _{q_2,1} =U\left(  \mathscr{O}_{q_2} ( \BC^\times \times \BC^\times ) \right) ^{\otimes m}/{S_m} $ and the algebra homomorphism reduces to \eqref{eq:classmorphdaha}, as promised (compare with \eqref{eq:vecrep}). At the same time, for $m=1$ we simply recover the algebra map \eqref{eq:vecrep} for the single M2-brane, $\r_{\text{M2}_{0,1,0}} : \qta \to \text{M2}_{0,1,0} $, which is rigid under the $q_1$-deformation. Therefore, we suggest that the algebra of local operators on the parallel $m$ M2-branes wrapping $\BR^2 _{\ve_2} \times \BR_t$ is given by the spherical DAHA, $\text{M2}_{0,m,0} = \sh^{(m)} _{q_2,q_1}$, and the algebra map \eqref{eq:sdrep} is nothing but the representation $\r_{\text{M2}_{0,m,0}} : \qta \to \text{M2}_{0,m,0}$ that determines the coupling of the associated line defect $\CalL_{0,m,0}$.\\

Recall that the quantum toroidal algebra $\qta$ is invariant under the triality of permuting $(q_1,q_2,q_3)$. As explicit from \eqref{eq:sdrepexp}, the spherical DAHA $\sh^{(m)} _{q_2,q_1}$ as a representation of $\qta$ partially breaks this triality. However, the symmetry of exchanging $q_1$ and $q_3 (= q_1^{-1} q_2 ^{-1})$ remains to be unbroken. Indeed, the relations imposed on the generators of the spherical DAHA $\sh^{(m)} _{q_2,q_1}$ are explicitly symmetric under the exchange of $q_1$ and $q_3$, while $q_2$ plays a distinct role. Moreover, at the level of its representation in $q_2$-difference operators \eqref{eq:sdrepexp}, it is straightforward to show that the conjugation by
\begin{align}
    g(X_1,X_2 \cdots, X_m) = \prod_{i=1} ^m X_i ^{\frac{(m-i)\log \frac{q_3}{q_1}}{\log q_2} } \prod_{1 \leq i<j \leq m} \frac{\left( q_3 ^{-1} \frac{X_j}{X_i} ; q_2  \right)_\infty}{\left( q_1^{-1} \frac{X_j}{X_i} ; q_2  \right)_\infty}
\end{align}
yields exactly the same $q_2$-difference operators except that $q_1$ is replaced by $q_3$, due to the relation $q_1 q_2 q_3 = 1$. Thus, the conjugation establishes the isomorphism $\text{M2}_{0,m,0 } = \sh^{(m)} _{q_2,q_1} \simeq \sh^{(m)} _{q_2 ,q_3}$. Note that such a breaking pattern of the triality is precisely what we would expect from the associated brane configuration in the twisted M-theory, since the support of the M2-branes distinguishes $q_2$ out of the triple $(q_1,q_2,q_3)$ but still there is a remaining symmetry of exchanging the two transverse topological planes $\BR^2 _{\ve_1}$ and $\BR^2 _{\ve_3}$.\\

\paragraph{M2-branes on $\BR^2 _{\ve_1} \times \BR_t$ and Hilbert scheme of points in $\BC^\times \times \BC^\times$} 
So far, we have investigated the algebra of local operators on the parallel M2-branes supported on $\BR^2 _{\ve_2} \times \BR_t$. To study parallel M2-branes, we can alternatively take a stack of $l$ M2-branes wrapping $\BR^2 _{\ve_1} \times \BR_t$. A crucial point of this consideration is the following. On one hand, such a change merely amounts to swapping the topological planes $\BR^2 _{\ve_1}$ and $\BR^2 _{\ve_2}$ in the twisted M-theory setup compared to the previous case, so that the algebra of local operators should be again given by the spherical DAHA of $GL(l)$, albeit with the roles of $q_1$ and $q_2$ exchanged: $\text{M2}_{l,0,0} = \sh^{(l)} _{q_1,q_2}(\simeq \sh^{(l)} _{q_1,q_3})$. On the other hand, the IIA reduction of the twisted M-theory seemingly breaks the triality, so that the emergence of the spherical DAHAs, with the $q_1$ and $q_2$ exchanged, from the two different parallel M2-branes is a non-trivial consistency check.\footnote{When the holomorphic surfaces are $\BC\times \BC$, the appearance of the spherical rational DAHA $\mathbf{Sr\ddot{H}}$ from two different parallel M2-branes, with the roles of $\ve_1$ and $\ve_2$ swapped, 
was discussed in \cite{Gaiotto:2019wcc}. What we discuss here is the multiplicative uplift of this manifestation of the triality, obtained by replacing the holomorphic surfaces $\BC \times \BC$ by $\BC^\times \times \BC^\times$.}

\begin{figure}[h!]
\centering
\begin{tikzpicture}[square/.style={%
            draw,
            minimum width=width("#1"),
            minimum height=width("#1")+2*\pgfshapeinnerysep,
            node contents={#1}}]
    \node at (-1,0) [minimum size=0.65cm,draw] (v) {$1$}; 
  \node at (1,0) [circle,draw] (l) {$l$};
  \draw[->] (v) -- (l);
   \draw [->] (l.south)arc(-160:160:1);
\end{tikzpicture}
\caption{Jordan quiver} \label{fig:jordan}
\end{figure}
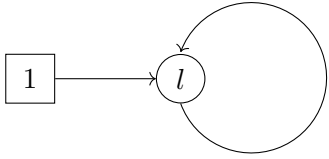

Reducing the twisted M-theory along the Taub-NUT circle, we arrive at a configuration of $l$ D2-brane wrapping $\BR^2 _{\ve_1} \times \BR_t$ lying on top of $1$ D6-brane wrapping $\BR^2 _{\ve_1} \times \BR_t \times \BC^\times _X \times \BC^\times _Z$. The effective 3-dimensional $\EN=4$ theory is obtained by quantizing the D2-D2 open string and the D2-D6 open string, where the former contains infinite numbers of winding modes around the two circles of $\BC^\times _X \times \BC^\times _Z$. The Higgs branch of this 3d $\EN=4$ theory is then given by the Hilbert scheme of $l$ points on $\BC^\times \times \BC^\times$, which can be constructed as the multiplicative quiver variety for the Jordan quiver (Figure \ref{fig:jordan}) \cite{Gorsky:1999rb,Kapustin:2000ek,AOblomkov2003DoubleAH,JORDAN2014420}.

The multiplicative quiver variety $\mathcal{M}_{q_2} ^{(l)}$ for the Jordan quiver is constructed as follows. Let us consider the space of quadruples $(A,B,I,J) \in \text{End}(\BC^l) \times \text{End}(\BC^l) \times \text{Hom}(\BC,\BC^l) \times \text{Hom}(\BC^l ,\BC)$ satisfying the relation,
\begin{align}
    AB - q_2 BA + IJ = 0\qquad (\text{equivalently,}\;\; \text{rk}(AB-q_2 BA) = 1).
\end{align}
There is a natural $GL(l)$-action on this space given by, 
\begin{align}
    (A,B,I,J) \mapsto (g A g^{-1}, gB g^{-1}, gI, Jg^{-1}),\qquad g \in GL(l).
\end{align}
It is known that the action is free for generic $q_2 \in \BC^\times$, so that the quotient is a smooth variety \cite{AOblomkov2003DoubleAH}. The multiplicative quiver variety $\mathcal{M}_{q_2} ^{(l)}$ for the Jordan quiver is given by this quotient. It is straightforward to see $\dim_\BC \mathcal{M}_{q_2} ^{(l)} = 2l$.

The multiplicative quiver variety $\mathcal{M}_{q_2} ^{(l)}$ is equipped with a Poisson structure \cite{Fock:1992xy, Chalykh:2017urw}, which furnishes a complete integrable structure associated to the classical trigonometric Ruijsenaars-Schneider model (the relativistic uplift of the trigonometric Calogero-Moser model) of $GL(l)$-type. The precise definition of the Poisson structure will not be very important to us, and we only present it here by giving the associated Darboux coordinates on an open dense subset $\left(\mathcal{M}_{q_2} ^{(l)} \right) ^\circ = \{A \text{ is invertible} \} \subset \mathcal{M}_{q_2} ^{(l)}$, following \cite{Chalykh:2017urw}. Let us define
\begin{align}
    (\BC^\times)^l _\text{reg} = \{(x_1,x_2,\cdots, x_l) \in (\BC^\times)^l \;\vert \; x_i \neq x_j,\; x_i \neq q_2 x_ j \,  \text{ for any } i,j=1,2,\cdots, l\}.
\end{align}
Then we may parameterize $\left(\mathcal{M}_{q_2} ^{(l)} \right) ^\circ $ by the map
\begin{align}
    \xi: \left((\BC^\times)^l  _\text{reg} \times (\BC^\times)^l \right)/S_l \longrightarrow\left(\mathcal{M}_{q_2} ^{(l)} \right) ^\circ ,
\end{align}
defined by $\xi\left((x_i)_{i=1}^ l, (p_i)_{i=1} ^l \right) =(A,B)$ with
\begin{align} \label{eq:coord}
\begin{split}
    &A = \text{diag} (x_1,x_2,\cdots, x_l ) \\
    &B=(B_{ij})_{i,j=1} ^l,\quad B_{ij}= p_j ^{-1} \frac{x_j (1-q_2)}{x_i - q_2 x_j} \prod_{k \neq j} \frac{q_2 x_j - x_k}{x_j - x_k}.
\end{split}
\end{align}
Here, $S_l$ acts by the simultaneous permutation of $(x_i)_{i=1} ^l$ and $(p_i)_{i=1} ^l$. When $\left((x_i)_{i=1} ^l, (p_i)_{i=1} ^l \right)$ gets permuted, the image $(A,B)$ is conjugated by the matrix of that permutation. Thus, the coordinates are well-defined on $\left(\mathcal{M}_{q_2} ^{(l)} \right) ^\circ$. This is a Darboux coordinate system on $\left(\mathcal{M}_{q_2} ^{(l)} \right) ^\circ$, in the sense that their Poisson brackets are given by
\begin{align} \label{eq:pb}
    &\{x_i,x_j\} = \{p_i,p_j\} =0, \quad \{p_i,x_j\} = p_i x_i \d_{i,j} ,\quad i,j=1,2,\cdots, l.
\end{align}

It is straightforward from \eqref{eq:coord} and \eqref{eq:pb} that the trace invariants $\text{Tr}\, A^k$, $k\in \BZ$, are mutually Poisson-commuting. It can also be shown that $\text{Tr}\, B^k$, $k\in \BZ$, are mutually Poisson-commuting. By taking suitable linear combinations of the products of $\text{Tr}\, B^{-k}$ with $k=1,2,\cdots, l$, we obtain
\begin{align} \label{eq:classrsh}
    &H_r = q_2^{-\frac{r(r-1)}{2}} \sum_{\substack{I \subset \{1,2,\cdots, l\} \\ \vert I \vert = r }} \left( \prod_{\substack{i \in I \\ j \notin I } } \frac{q_2 ^{-1} x_i - x_j}{x_i - x_j} \right) \prod_{i\in I} p_i , \qquad r=1,2,\cdots, l,
\end{align}
which in turn also Poisson-commute with each other, $\{H_r,H_s\}=0$, $r,s=1,2,\cdots, l$. They are precisely the classical Hamiltonians of the trigonometric Ruijsenaars-Schneider model of $GL(l)$-type.\\

In the classical limit $q_1 =1$, the effective field theory on the worldvolume of the M2-branes is the 3d $\EN=4$ theory subject to the Rozansky-Witten twist, which flows to the 3d topological sigma model with the target space being the Higgs branch $\mathcal{M}_{q_2} ^{(l)}$ \cite{Rozansky1996HyperKhlerGA}. The algebra $\text{M2}_{l,0,0} ^{q_1=1}$ of local operators of this theory is thus the commutative algebra of holomorphic functions on the space $\mathcal{M}_{q_2} ^{(l)}$. It was shown in \cite{AOblomkov2003DoubleAH} that this algebra is identical to $\sh^{(l)} _{1,q_2}$, namely, the $q_1 =1$ limit of the spherical DAHA $\sh^{(l)} _{q_1,q_2}$. Note that this limit is different from the other limit to the symmetric product of quantum torus algebras we observed earlier, due to the crucial exchange of the $q_1$ and $q_2$ parameters.

Turning on $q_1 \neq 1$, the Rozansky-Witten twisted 3d $\EN=4$ theory on $\BR^2 _{\ve_1} \times \BR_t$ gets $\O$-deformed, localized onto the line $\{0\} \times \BR_t$. The local operators then form a non-commutative associative algebra $\text{M2}_{l,0,0}$, which is a deformation of the algebra of functions on $\mathcal{M}_{q_2} ^{(l)}$ \cite{Yagi:2014toa}. The quantization takes place in the way that the canonical Poisson brackets \eqref{eq:pb} are uplifted to the canonical commutation relations,
\begin{align}
    [X_i,X_j] = \left[q_1 ^{D_{X_i}} , q_1 ^{D_{X_j}}\right] = 0,\qquad q_1 ^{D_{X_i}} X_j  = q_1 ^{\d_{i,j}}  X_j q_1 ^{D_{X_i}},\qquad i,j=1,2,\cdots l.
\end{align}
The so-obtained quantization of the multiplicative quiver variety $\mathcal{M}_{q_2} ^{(l)}$ precisely yields the spherical DAHA $\sh^{(l)} _{q_1,q_2}$ \cite{JORDAN2014420,Balagovic2016TheHI,Wen:2023}. The construction involves the quantum Hamiltonian reduction of the algebra of quantum differential operators on $GL(l)$. In particular, the classical Hamiltonians \eqref{eq:classrsh} uplift to the mutually commuting Macdonald operators,
\begin{align} \label{eq:qrsh}
    &\hat{H}_r = q_2^{-\frac{r(r-1)}{2}} \sum_{\substack{I \subset \{1,2,\cdots, l\} \\ \vert I \vert = r }} \left( \prod_{\substack{i \in I \\ j \notin I } } \frac{q_2 ^{-1} X_i - X_j}{X_i - X_j} \right) \prod_{i\in I} q_1 ^{D_{X_i}} , \qquad r=1,2,\cdots, l,
\end{align}
furnishing the quantum Hamiltonians of the trigonometric Ruijsenaars-Schneider model of $GL(l)$-type. Thus, we conclude that the algebra of local operators on the parallel M2-branes is given by the spherical DAHA, $\text{M2}_{l,0,0} = \sh_{q_1,q_2} ^{(l)}$.

By the trialiy of $\qta$, the surjective algebra homomorphism from the quantum toroidal algebra of $\fgl(1)$ to the spherical DAHA $\sh_{q_1,q_2} ^{(l)}$ is obtained simply by exchanging $q_1$ and $q_2$ in \eqref{eq:sdrepexp}. For completeness, we write out here the surjective algebra homomorphism
\begin{align}
    \r_{\text{M2}_{l,0,0}}: \qta \twoheadrightarrow \text{M2}_{l,0,0}
\end{align}
in terms of the modes as
\begin{align} 
\begin{split}
\begin{split}
  &E_k \mapsto \frac{1}{1-q_1}  \sum_{i=1} ^l \left( \prod_{j\neq i} \frac{ q_2 X_i -   X_j}{ X_i- X_j} \right) X_i ^{k} q_1 ^{-D_{X_i}}, \\  
   &F_k \mapsto \frac{1}{1-q_1 ^{-1}} \sum_{i=1} ^l \left( \prod_{j\neq i} \frac{ q_2 ^{-1} X_i -   X_j}{X_i - X_j} \right) q_1 ^{D_{X_i}} X_i ^k , \\
   \end{split} \qquad k \in \BZ, \\
   \begin{split}
    & H_{\pm r} \mapsto \frac{1}{1- q_1 ^{\mp r}} \sum_{i=1} ^ l X_i ^{\pm r} , \qquad r \in \BZ_{>0}, \\
    &C\mapsto 1,\quad C^\perp \mapsto 1.
\end{split}
\end{split}
\end{align}
In particular, the Macdonald operators \eqref{eq:qrsh} are achieved by taking suitable combinations of $F_0$ and the commutators of $\text{ad}^{i-1} _{F_0} F_{\pm 1}$ with $i=1,2,\cdots, l$, and picking up the images under the map $\r_{\text{M2}_{l,0,0}}$.\\

Recall that the Higgs branch of the effective 3d $\EN=4$ theory on the stack of $l$ D2-branes, wrapping $\BR^2 _{\ve_1} \times \BR_t$, placed upon $1$ D6-brane, wrapping $\BR^2 _{\ve_1} \times \BR_t \times \BC \times \BC$, is the Hilbert scheme of $l$ points on $\BC\times \BC$, which is given by the Nakajima quiver variety for the same Jordan quiver (Figure \ref{fig:jordan}). The algebra of local operators on the stack of D2-branes is the quantized algebra of holomorphic functions on this quiver variety, which turns out to be the spherical subalgebra of the rational double affine Hecke algebra (spherical rational DAHA, $\mathbf{Sr\ddot{H}}$) of $GL(l)$ \cite{Costello:2017fbo,Gaiotto:2020vqj}. This is precisely what was called the (rational) Calogero representation of the 1-shifted affine Yangian of $\fgl(1)$ ($Y_1 (\widehat{\fgl}(1))$) in \cite{Gaiotto:2020dsq}. What we have here is its multiplicative uplift $-$ the spherical DAHA of $GL(l)$ as a representation of the quantum toroidal algebra of $\fgl(1)$ $-$ established by replacing the complex manifold $\BC \times \BC$ by $\BC^\times \times \BC^\times$ (see table \ref{table:twmcases}).\footnote{The intermediate step, namely, the spherical trigonometric DAHA $\mathbf{St\ddot{H}}$ as a representation of the affine Yangian of $\fgl(1)$ ($Y (\widehat{\fgl}(1))$), can also be achieved as the quantization of the Hilbert scheme of points in $\BC \times \BC^\times$. See \cite{Finkelberg2010}, for instance. The associated twisted M-theory background is thus expected to have the complex manifold $\BC\times \BC^\times$ (see table \ref{table:twmcases}). See \cite{Kodera:2016faj,Gaiotto:2019wcc} for an approach from the 3d mirror dual side.} See also \cite{Gaiotto:2020vqj,Gukov:2022gei} for another incarnation of the spherical DAHA as the quantized Coulomb branch of the 4d $\EN=2^*$ theory compactified on a circle.\\

\paragraph{Fusion of parallel M2-branes and coproduct of quantum toroidal algebra}
Having identified the algebra of local operators on the parallel M2-branes with the spherical DAHA, let us examine how the statement is compatible with the fusion operation of the line defects. Due to the triality, we may restrict to the case of the M2-branes supported on $\BR^2 _{\ve_2} \times \BR_t$ without losing generality.

Let us first remind that the fusion operation for universal line defects implies the existence of the coproduct of the universal associative algebra \cite{Gaiotto:2020dsq}, since the fusion exhibits a representation of the universal associative algebra in the tensor product with itself.\footnote{Recall that the coassociativity, $(\Delta \otimes \text{id})\Delta = (\text{id} \otimes \Delta)\Delta$, follows from the fact that changing the ordering of two consecutive fusion operations results in the same line defect. We also remind that the counit $\epsilon: \qta \to \BC$ is provided by the trivial line defect, i.e., $\epsilon = \r_{\text{M2}_{0,0,0}}$, since the fusion with a trivial line defect is the identity operation: $(\text{id}\otimes \epsilon)\Delta = (\epsilon \otimes \text{id})\Delta = \text{id}$.\label{fn:counit}} In our case, the quantum toroidal algebra of $\fgl(1)$ is indeed equipped with the coproduct $\Delta: \qta \to \qta \, \widehat{\otimes} \, \qta$ given by \eqref{eq:cprod}.

\begin{figure}[h!]\centering
\begin{tikzpicture} 
 \draw [line width=0.3mm] (-2,2) -- (-2,-2);
 \draw [line width=0.3mm] (0,2) -- (0,-2); 
 \draw [line width=0.3mm] (4,2) -- (4,-2);
 \draw[->] (1.3,0) -- (2.7,0);
  \node at (2,0.4) {fusion};
 \node at (-1.35,1.7) {$\mathcal{L}_{0,m,0}$};
 \node at (0.7,1.7) {$\mathcal{L}_{0,m',0}$};
 \node at (4.95,1.7) {$\CalL_{0,m+m',0}$};
  \node at (-2,-2.8) {$\text{M2}_{0,m,0}$};
    \node at (0,-2.8) {$\text{M2}_{0,m',0}$};
      \node at (4,-2.8) {$\text{M2}_{0,m+m',0}$};
       \draw[left hook-stealth] (2.7,-2.8) -- (1.3,-2.8) ;
  \node at (2.05,-2.5) {$\iota_{m,m'}$};
        \node at (-1,-2.8) {$\widehat{\otimes}$};
\end{tikzpicture} \caption{Fusion of line defects and algebra embedding induced by the coproduct} \label{fig:m2fuse} 
\end{figure}

Then, we specialize to the line defect $\CalL_{0,m+m',0}$ produced by $m+m'$ parallel M2-branes ($m,m' \geq 0$). We may split them into two groups composed of $m$ and $m'$ M2-branes, yielding two line defects $\CalL_{0,m,0}$ and $\CalL_{0,m',0}$ lying on top of each other. The original line defect $\CalL_{0,m+m',0}$ should be recovered once the two line defects $\CalL_{0,m,0}$ and $\CalL_{0,m',0}$ are fused together (Figure \ref{fig:m2fuse}). Thus, the coupling of the composite line defect $\CalL_{0,m+m',0}$ is built from the couplings of the base line defects $\CalL_{0,m,0}$ and $\CalL_{0,m',0}$. This implies there is an embedding of the algebra $\iota_{m,m'}:\text{M2}_{0,m+m',0} \hookrightarrow \text{M2}_{0,m,0} \otimes \text{M2}_{0,m',0} $, induced from the coproduct $\Delta$ through the surjective algebra maps associated to these line defects. Namely, we have an isomorphism,
\begin{align} \label{eq:fuse}
    (\r_{\text{M2}_{0,m,0}} \otimes \r_{\text{M2}_{0,m',0}} )\D \simeq \r_{\text{M2}_{0,m+m',0}},
\end{align}
which can also be represented by the commutative diagram:
\begin{equation}
\begin{tikzcd}[row sep=huge]
\qta \arrow{r}{\Delta}\arrow[two heads]{d}{\r_{\text{M2}_{0,m+m',0}}} & \qta \, {\widehat{\otimes}} \, \qta \arrow[two heads]{d}{\r_{\text{M2}_{0,m,0}}  \otimes \, \r_{\text{M2}_{0,m',0}}}\\
 \sh^{(m+m')} _{q_2,q_1} \arrow[ur, phantom, "\scalebox{1.5}{$\circlearrowleft$}" description] \arrow[hookrightarrow]{r}{\iota_{m,m'}} &  \sh^{(m)} _{q_2,q_1}{\widehat{\otimes}} \, \sh^{(m')} _{q_2,q_1}
\end{tikzcd}
\end{equation}

It is straightforward to verify that the suggested relation \eqref{eq:fuse} holds. The left hand side can be computed using the explicit forms of the $q_2$-difference operators \eqref{eq:sdrepexp} for the 
Macdonald representation $\r_{\text{M2}_{0,m,0}}$ and the coproduct $\Delta$ \eqref{eq:cprod} of $\qta$. The resulting $q_2$-difference operators is not symmetric under the action of the extended permutation group $S_{m+m'}$, but they can be easily mapped to the symmetric ones by conjugating
\begin{align}
    \tilde{g} (\{X_i\}_{i=1} ^m, \{X_j \}_{j=m+1}^{m+m'} ) = \prod_{i=1} ^m X_i ^{m' \frac{\log q_1}{\log q_2}}  \prod_{j=m+1} ^{m+m'}  \frac{\left( q_1 ^{-1} \frac{X_j }{X_i} ; q_2  \right)_\infty}{\left(  \frac{X_j }{X_i} ; q_2  \right)_\infty}.
\end{align}
The outcome is precisely the $q_2$-difference operators \eqref{eq:sdrepexp} for $\r_{\text{M2}_{0,m+m',0}}$, i.e., the right hand side of \eqref{eq:fuse}. In other words, the conjugation by $\tilde{g}$ establishes the desired algebra embedding $\iota_{m,m'} : \text{M2}_{0,m+m',0} \hookrightarrow \text{M2}_{0,m,0} \otimes \text{M2}_{0,m',0}   $.

\subsubsection{Non-parallel M2-branes and generalized Macdonald representation}
Finally, let us examine the most generic case for the topological line defects in the 5d Chern-Simons theory created by non-parallel M2-branes. Namely, we put $l+m+n$ M2-branes, where $l$ of them wrap $\BR^2 _{\ve_1} \times \BR_t$; $m$ of them wrap $\BR^2 _{\ve_2} \times \BR_t$; and $n$ of them wrap $\BR^2 _{\ve_3} \times \BR_t$. We denote the algebra of local operators on this line defect by $\text{M2}_{l,m,n}$. Again by the Koszul duality, there must be a surjective homomorphism from the universal associative algebra
\begin{align} \label{eq:m2lmn}
    \r_{\text{M2}_{l,m,n}} : \qta  \twoheadrightarrow \text{M2}_{l,m,n},
\end{align}
which controls the coupling of the line defect $\CalL_{l,m,n}$.

The line defect $\CalL_{l,m,n}$ can be constructed by the consecutive fusion of three line defects $-$ $\CalL_{l,0,0}$, $\CalL_{0,m,0}$, and $\CalL_{0,0,n}$ $-$ each of which originates from the M2-branes wrapping three different topological planes. Algebraically, the fusion induces a reconstruction of the above algebra map \eqref{eq:m2lmn} as the composition of the coproducts of the universal associative algebra with the tensor product of three representations for these parallel M2-branes (surjective algebra maps to spherical DAHAs $\sh_{q_1,q_2} ^{(l)}$, $\sh_{q_2,q_3} ^{(m)}$, and $\sh_{q_3,q_1} ^{(n)}$). Namely,\footnote{Due to the coassociativity of $\qta$, $\Delta^2 := (\text{id} \otimes \Delta) \Delta =(\Delta \otimes \text{id}) \Delta$.}
\begin{align} \label{eq:m2lmncons}
    \r_{\text{M2}_{l,m,n}} = ( \r_{\text{M2}_{l,0,0}} \otimes \r_{\text{M2}_{0,m,0}} \otimes \r_{\text{M2}_{0,0,n}} ) \Delta^{ 2}.
\end{align}
The homomorphism obtained in this way can be presented in terms of the modes as
\begin{align} 
\begin{split}
\begin{split}
 & E_k \mapsto  \frac{1}{1-q_1} \sum_{i=1}^l \prod_{\substack{j=1 \\ j \neq i}} ^l \frac{q_2 X_i -  X_j}{X_i - X_j} \prod_{j=1} ^m \frac{q_1 X_i -  X_j '}{X_i - X_j'} \prod_{j=1} ^n \frac{q_1 X_i -  X_j ''}{X_i - X_j ''} X_i ^k q_1 ^{-D_{X_i}} \\
  &\qquad  + \frac{1}{1-q_2} \sum_{i=1}^m \prod_{j=1 } ^l \frac{q_2 X'_i -  X_j}{X'_i - X_j} \prod_{\substack{j=1 \\ j \neq i}} ^m \frac{q_3 X'_i -  X_j '}{X' _i - X_j'} \prod_{j=1} ^n \frac{q_2 X'_i -  X_j ''}{X' _i - X_j ''} (X_i ') ^k q_2 ^{-D_{X'_i}}  \\  
  &\qquad  + \frac{1}{1-q_3} \sum_{i=1}^n \prod_{j=1 } ^l \frac{q_3 X''_i -  X_j}{X''_i - X_j} \prod_{j=1 } ^m \frac{q_3 X''_i -  X_j '}{X'' _i - X_j'} \prod_{\substack{j=1\\ j\neq i}} ^n \frac{q_1 X''_i -  X_j ''}{X'' _i - X_j ''} (X_i '') ^k q_3 ^{-D_{X''_i}} \\
   &F_k \mapsto  \frac{1}{1-q_1 ^{-1}} \sum_{i=1}^l \prod_{\substack{j=1 \\ j \neq i}} ^l \frac{q_2 ^{-1} X_i -   X_j}{X_i - X_j} \prod_{j=1} ^m \frac{q_2 ^{-1} X_i -  X_j '}{q_1 q_2 ^{-1} X_i - X_j'} \prod_{j=1} ^n \frac{q_3 ^{-1} X_i - X_j ''}{q_1 q_3 ^{-1} X_i - X_j ''} q_1 ^{D_{X_i}} X_i ^k  \\
  &\qquad  + \frac{1}{1-q_2^{-1}} \sum_{i=1}^m \prod_{j=1 } ^l \frac{q_1 ^{-1} X'_i -  X_j}{q_1 ^{-1} q_2 X'_i - X_j} \prod_{\substack{j=1 \\ j \neq i}} ^m \frac{q_3 ^{-1} X'_i -  X_j '}{X' _i - X_j'} \prod_{j=1} ^n \frac{q_3^{-1} X'_i -   X_j ''}{q_2 q_3 ^{-1} X' _i - X_j ''}q_2 ^{D_{X'_i}} (X_i') ^k   \\  
  &\qquad  + \frac{1}{1-q_3 ^{-1}} \sum_{i=1}^n \prod_{j=1 } ^l \frac{q_1 ^{-1} X''_i - X_j}{q_1 ^{-1} q_3 X''_i - X_j} \prod_{j=1 } ^m \frac{q_2 ^{-1} X''_i -  X_j '}{q_2 ^{-1}q_3  X'' _i - X_j'} \prod_{\substack{j=1\\ j\neq i}} ^n \frac{q_1 ^{-1} X''_i - X_j ''}{X'' _i - X_j ''} q_3 ^{D_{X''_i}}  (X_i'') ^k ,  
   \end{split} \qquad k \in \BZ, \\
   \begin{split}
    & H_{\pm r} \mapsto \frac{1}{1- q_1 ^{\mp r}} \sum_{i=1} ^ l X_i ^{\pm r} + \frac{1}{1- q_2 ^{\mp r}} \sum_{i=1} ^ m (X_i ') ^{\pm r} + \frac{1}{1- q_3 ^{\mp r}} \sum_{i=1} ^ n (X_i '') ^{\pm r} , \qquad r \in \BZ_{>0}, \\
    &C\mapsto 1,\quad C^\perp \mapsto 1.
\end{split}
\end{split}
\end{align}
Here, we conjugated the right hand side of \eqref{eq:m2lmncons} by
\begin{align}
\begin{split}
    &\hat{g} (\{X_i\}_{i=1} ^l, \{X'_j \}_{j=1}^{m} 
, \{X''_k \}_{k=1}^{n} ) \\
&= \prod_{i=1} ^l X_i ^{m+n} \prod_{j=1} ^m (X_j ')^{n  } \prod_{i=1}^l \prod_{j=1} ^m  \left( 1- \frac{q_1 ^{-1} X_j ' }{X_i} \right) \prod_{j=1} ^m \prod_{k=1} ^n \left( 1- \frac{q_2 ^{-1} X_k '' }{X_j'} \right) \prod_{i=1} ^l \prod_{k=1}^n \left( 1- \frac{q_1 ^{-1} X_k '' }{X_i} \right)
\end{split}
\end{align}
to put the resulting $q$-difference operators in a symmetric fashion as above. Since this generalizes the Macdonald representation $\r_{\text{M2}_{0,0,n}}$, we call it the generalized Macdonald representation of the quantum toroidal algebra $\qta$.

What we established is the multiplicative uplift of the generalized (rational) Calogero representation of the 1-shifted affine Yangian of $\fgl(1)$ obtained in \cite{Gaiotto:2020dsq}, as a result of changing the complex manifold from $\BC\times \BC$ to $\BC^\times \times \BC^\times$. In this regard, we may refer to it as the generalized trigonometric Ruijsenaars-Schneider representation of the quantum toroidal algebra of $\fgl(1)$ as well.

\vspace{4mm}

\subsection{M5-branes and holomorphic surface defects} \label{subsec:m5}
The M5-branes in the twisted M-theory can be supported on $\BR^2 _{\ve_{c+1}} \times \BR^2 _{\ve_{c-1}} \times C^{\mathbf{(p,q)}}$, where $c \in \{1,2,3\}$ and $C^{\mathbf{(p,q)}} = \{ X^\bq Z^{-\bp} = \text{const} \} \subset \BC^\times _X \times \BC^\times _Z$ is a holomorphic curve. Passing to the 5d $\fgl(1)$ Chern-Simons theory, the M5-branes descend to a holomorphic surface defect supported on the curve $C^{(\bp,\bq)}$. Fixing the holomorphic part of the support as $C^{\mathbf{(p,q)}}$, the most generic configuration comprises $L$, $M$, and $N$ M5-branes wrapping $\BR^2 _{\ve_2} \times \BR^2 _{\ve_3} \times C^{\mathbf{(p,q)}}$, $\BR^2 _{\ve_1} \times \BR^2 _{\ve_3} \times C^{\mathbf{(p,q)}}$, and $\BR^2 _{\ve_1} \times \BR^2 _{\ve_2} \times C^{\mathbf{(p,q)}}$, respectively. We call the induced holomorphic surface defect $\mathcal{S}_{L,M,N} ^{\mathbf{(p,q)}}$.\footnote{We emphasize that the $c$-th subscript of the surface defect $\mathcal{S}_{L,M,N} ^{\mathbf{(p,q)}}$ indicates the number of M5-branes wrapping the $(c\pm 1)$-th topological planes, whereas the $c$-th subscript of the line defect $\mathcal{L}_{l,m,n}$ denotes the number of M2-branes wrapping the $c$-th topological plane.}

The local operators of the 5d $\fgl(1)$ Chern-Simons theory localized at the holomorphic curve $C^{\mathbf{(p,q)}}$ are the modes of the ghost field under the Laurent expansion along the transverse holomorphic direction. They form a chiral algebra, say $\text{Obs}_{q_1,q_2} ^{C^{\mathbf{(p,q)}}}$, by their OPEs. The BRST-invariant coupling of a surface defect, carrying a chiral algebra $\EuScript{V}$ of local operators, is expected to be classified by certain universal chiral algebra $\left(\text{Obs}_{q_1,q_2} ^{C^{\mathbf{(p,q)}}} \right)^!$ and surjective homomorphisms $\left(\text{Obs}_{q_1,q_2} ^{C^{\mathbf{(p,q)}}} \right)^! \twoheadrightarrow \EuScript{V}$ \cite{Costello:2020jbh}. We anticipate this universal chiral algebra is again given by the quantum toroidal algebra of $\fgl(1)$, enhanced by a proper chiral algebra structure.\footnote{More precisely, we expect that the quantum toroidal algebra of $\fgl(1)$ is equipped with a \textit{quantum} vertex algebra structure \cite{Frenkel:1996nz,Etingof2000QuantizationOL}, and that the surjective map $\r_{\text{M5}_{L,M,N} ^{\mathbf{(p,q)}} }: \qta \twoheadrightarrow \text{M5}_{L,M,N} ^{\mathbf{(p,q)}}$ is uplifted to a quantum vertex algebra homomorphism. For instance, the $q$-boson algebra for a single M5-brane and the $q$-deformed $W$-algebras for multiple parallel M5-branes are known to admit the quantum vertex algebra structure \cite{Frenkel:1996nz}. Nevertheless, the quantum vertex algebra structure for the whole quantum toroidal algebra of $\fgl(1)$ is not well understood, and we will restrict our discussion to the mode algebras in this work. See \cite{Chen:2024lsc} for a study of quantum vertex algebra structure for higher-rank quantum toroidal algebras.\label{fn:qva}} 

In principle, we can derive the OPEs of local operators in the following way. Let us take the example of $C^{\mathbf{(1,0)}} = \BC^\times _X$. We can explicitly write the surface defect coupling as
\begin{align}
    \sum_{m\in \BZ} \int_{\BC^\times _X} W^{(m)} (A_{\bar{X}})_m,
\end{align}
where $(A_{\bar{X}})_m$ are the modes of the gauge field under the Laurent expansion along the transverse plane $\BC^\times _Z$ and $W^{(m)}$ are local operators on the surface defect. Then, the OPEs of the local operators can be deduced as conditions for the anomalous Feynman diagrams to cancel each other (see \cite{Costello:2020jbh} for the case of surface defect in holomorphic Chern-Simons theory). We will not conduct such a perturbative analysis of the 5d Chern-Simons theory on $\BR_t \times \BC^\times_X \times \BC^\times _Z$ in the present work. Instead, we will restrict our attention to the mode algebra, referred to as $\text{M5}_{L,M,N} ^{\mathbf{(p,q)}}$, of the local operators on the surface defect $\mathcal{S}_{L,M,N} ^{\mathbf{(p,q)}}$. We will suggest natural candidates for them and the surjective algebra homomorphisms $\r_{\text{M5}_{L,M,N} ^{\mathbf{(p,q)}}}:\qta \twoheadrightarrow \text{M5}_{L,M,N} ^{\mathbf{(p,q)}}$.

\subsubsection{Single M5-brane and free $q$-boson} \label{subsubsec:singlem5}
In this section, we will only consider M5-branes whose holomorphic part of the support is $\BC^\times _X$ (namely, $C^{\mathbf{(1,0)}} = \BC^\times _X$). We will come back to M5-branes supported on general holomorphic curves $C^{\mathbf{(p,q)}}$ in section \ref{sec:qqchar}. Consider a single M5-brane wrapping $\BR^2 _{\ve_2} \times \BR^2 _{\ve_3} \times \BC^\times _X$. Deforming $\BR^2 _{\ve_2} \times \BR^2 _{\ve_3}$ to the one-centered Taub-NUT space and reducing the twisted M-theory along the Taub-NUT circle, the M5-brane becomes a D4-brane which intersects the emergent single D6-brane along the holomorphic surface $\BC^\times_X$. 

If the holomorphic plane transverse to the D4-brane were $\BC$, the D4-D4 open string would give rise to the 5d $\EN=2$ $U(1)$ gauge theory, coupled to the chiral fermions living on $\BC^\times _{X}$ arising from the D4-D6 open string \cite{Green:1996dd,Dijkgraaf:2007sw}. Due to the $\O$-background, the 5d gauge theory is localized to the 3d $U(1)$ Chern-Simons theory \cite{Yagi:2013fda,Lee:2013ida,Luo:2014sva}. Therefore, the algebra of local operators on the surface defect of the 5d Chern-Simons theory is given by the algebra of chiral fermions coupled to the 3d Chern-Simons theory, which is identified to be the free boson algebra $\widehat{\fgl}(1)$ \cite{Costello:2016nkh}.

In our case, however, the open strings may wind around the compact circle of the plane $\BC^\times _Z \simeq \BR_z \times S^1$ transverse to the D4-brane, producing infinite numbers of winding modes. The effective 5d gauge theory for the D4-D4 open string now contains infinite numbers of Kaluza-Klein modes, which can be uplifted to the 6d $\EN=(1,1)$ $U(1)$ gauge theory compactified on a circle, as a result. Due to the $\O$-background, the 6d gauge theory is localized into the 4d $\fgl(1)$ Chern-Simons theory on $\BR \times S^1 \times \BC^\times _X$ \cite{Costello:2018txb}. The chiral fermions on $\BC^\times_X$ couple to the 4d $\fgl(1)$ Chern-Simons theory, and the quantum corrections would uplift the algebra of local operators to a representation of the quantum affine algebra of $\fgl(1)$, namely the free $q$-boson algebra $U_{q} (\widehat{\fgl}(1))$.\footnote{We stress, however, that this claim regarding the quantum corrections to the algebra of local operators on the holomorphic surface defect in the 4d Chern-Simons theory is an independent statement that needs to be proven within its perturbation theory, just as was done for the surface defect in the holomorphic Chern-Simons theory in \cite{Costello:2020jbh}. To the best of our knowledge, such a perturbative analysis for the holomorphic surface defect has not yet been established in the 4d Chern-Simons theory, unlike the studies for topological line defects in \cite{Costello:2017dso} and for topological surface defects in \cite{Costello:2019tri}.} Since the representation has to be non-trivial, it should be the $q$-boson algebra itself. Therefore, we suggest that the algebra $\text{M5}_{0,1,0} ^{\mathbf{(1,0)}}$ of local operators on the surface defect $\mathcal{S}_{0,1,0} ^{\mathbf{(1,0)}}$ is the free $q$-boson algebra.\\

From the triality of the twisted M-theory, we expect that a single M5-brane on $\BR^2 _{\ve_{c+1}} \times \BR^2 _{\ve_{c-1}} \times \BC^\times _X$, $c\in \{1,2,3\}$, would also support a $q$-boson algebra $\text{M5}_c ^{\mathbf{(1,0)}}$ with the permutation on $(q_1,q_2,q_3)$. We used the short-hand notation $ (\d_{c,a})_{a=1} ^3 \equiv c $ for the subscript only for the cases with single M5-brane. The free $q$-boson algebra $\text{M5}_c ^{\mathbf{(1,0)}}$ is the algebra generated by the modes $a_r^{(c)}$, $r\in \BZ$, subject to the commutation relation
\begin{align}\label{eq:qbos}
    [a_r ^{(c)},a_s ^{(c)}] = - \d_{r+s,0} \frac{r}{\k_r} (q_c ^{r/2} - q_c ^{-r/2})^3.
\end{align}

Meanwhile, we suggested that the universal associative algebra for the 5d $\fgl(1)$ Chern-Simons theory on $\BR_t \times \BC^\times _X \times \BC^\times _Z$ is the quantum toroidal algebra $\qta$, from which there is a surjective algebra homomorphism to each $\text{M5}_{c} ^{\mathbf{(1,0)}}$. Indeed, it is straightforward to see that two of the defining relations \eqref{eq:eq:cartan2} and \eqref{eq:cartans} for $\qta$ are equivalent to the $q$-boson commutation relation satisfied by $H_r$,
\begin{align}
    [H_r,H_s] = \d_{r+s,0} \frac{r}{\k_r} (C^r - C^{-r}).
\end{align}
At $(C,C^\perp) = \left( q_c ^{1/2},1 \right)$ where $c\in \{1,2,3\}$, this vertex representation extends to the whole quantum toroidal algebra $\qta$. Namely, we have surjective algebra homomorphisms
\begin{align} \label{eq:singlem5}
    \r_{\text{M5}_c ^{\mathbf{(1,0)}}}: \qta \twoheadrightarrow \text{M5}_c ^{\mathbf{(1,0)}},
\end{align}
given by
\begin{align} \label{eq:fockrep}
\begin{split}
    &E(X) \mapsto -e^{a_0 ^{(c)} \frac{\log q_{c+1 } \log q_{c-1}}{\log q_c}} \frac{1-q_c }{\k_1} \eta_c (X),\quad F(X)\mapsto  e^{-a_0 ^{(c)} \frac{\log q_{c+1 } \log q_{c-1}}{\log q_c}}  \frac{1-q_c  ^{-1} }{\k_1} \xi_c (X), \\  
    &H_r \mapsto \frac{a_r ^{(c)}}{q_c ^{r/2} -q_c ^{-r/2}},\qquad (C,C^\perp) \mapsto \left( q_c ^{1/2},1 \right),
\end{split}
\end{align}
where we used the indices in cyclical manner so that $\{c-1,c,c+1\} = \{1,2,3\}$. Here, the vertex operators are given by
\begin{align}
\begin{split}
    &\eta_c (X) = \exp \left( -\sum_{r=1} ^\infty \frac{\k_r}{r} \frac{1}{( q_c ^{r/2} - q_c ^{-r/2} )^2 }a_{-r} ^{(c)} X^r \right) \exp\left( - \sum_{r=1} ^\infty \frac{\k_r}{r} \frac{q_c ^{-{r/2}}}{ (q_c ^{r/2} - q_c ^{-r/2} )^2 } a_r ^{(c)} X^{-r} \right) \\
    &\xi _c (X) =  \exp \left( \sum_{r=1} ^\infty \frac{\k_r}{r} \frac{ q_c ^{r/2} }{ (q_c ^{r/2} - q_c ^{-r/2 } )^2 }a_{-r} ^{(c)} X^r \right) \exp\left(  \sum_{r=1} ^\infty \frac{\k_r}{r} \frac{1}{ (q_c ^{r/2} - q_c ^{-r/2} )^2 } a_r ^{(c)} X^{-r} \right).
\end{split}
\end{align}
For later convenience, we also denote the image of $K^\pm (X)$ by vertex operators $\varphi^\pm _c (X)$,
\begin{align}
    K^\pm (X) \mapsto \varphi_c ^\pm (X) \equiv \exp \left(  \sum_{r=1} ^\infty \frac{\k_r}{r} \frac{a_{\pm r} ^{(c)}}{q_c ^{r/2} -q_c ^{-r/2}} X^{\mp r} \right).
\end{align}

We may represent the $q$-boson algebra \eqref{eq:qbos} on the Fock space $\CalF_c ^{\mathbf{(1,0)}} (v) = \bigoplus_{\{\l\}} \BC \vert v,\l \rangle^{(c)}$, where the basis elements are enumerated by a single Young diagram $\l = (\l_i)_{i=1} ^\infty $, $\l_i \geq \l_{i+1} \geq 0$. The whole Fock space can be generated by acting the creation operators $a_{-r} ^{(c)}$, $r\in \BZ_{>0}$ on the vacuum $\vert v, \varnothing  \rangle ^{(c)} \in \CalF_c ^{\mathbf{(1,0)}} (v)$ satisfying 
\begin{align}\label{eq:zeromode}
   a_{r} ^{(c)} \vert v,\varnothing  \rangle^{(c)} =0,\quad r>0 ,\qquad  a_0 ^{(c)} \vert v, \varnothing  \rangle^{(c)} = \frac{\log q_c}{\log q_{c+1} \log q_{c-1}} \log v  \vert v, \varnothing  \rangle^{(c)},
\end{align}
where $v \in \BC^\times$ is a module parameter. The seemingly unusual normalization of the zero-mode $a_0 ^{(c)}$ will be explained soon. This is a faithful representation of the $q$-boson algebra, $\text{M5}_{c} ^{\mathbf{(1,0)}}$. Thus, what we found for single M5-brane can be summarized into $\qta \twoheadrightarrow \text{M5}_c ^{\mathbf{(1,0)}} \hookrightarrow \text{End}(\CalF_c ^{\mathbf{(1,0)}} (v))$ where $c \in \{1,2,3\}$.\\

For later use, we stress that the grading operator $d^\perp$ plays the role of the conjugate variable to the zero-mode $a_0 ^{(c)}$. Indeed, from its definition \eqref{eq:grading} it follows that
\begin{align}
    [d^\perp , a_m ^{(c)} ]= \frac{\log q_c}{\log q_{c+1} \log q_{c-1}} \d_{m,0},\qquad m\in \BZ.
\end{align}
This implies that the Fock representation on $\mathcal{F}^{\mathbf{(1,0)}} _c $ can be extended to incorporate the grading operator $d^\perp$ by regarding it as the shift operator for the module parameter: $q^{d^\perp} \vert v, \l \rangle ^{(c)} = \vert v q^{-1}, \l \rangle ^{(c)}$ for any $q \in \BC^\times$.\\

Note that the representations \eqref{eq:singlem5} break the triality of $\qta$, in a way that the duality of exchanging $q_{c+1}$ and $q_{c-1}$ is still preserved. This is indeed to be expected from the corresponding M-brane configuration, since the support of the M5-brane distinguishes two parameters $q_{c\pm 1}$ out of the triple $(q_1,q_2,q_3)$.

\subsubsection{Parallel M5-branes and $q$-deformed $W$-algebra}
Next, let us consider multiple parallel M5-branes all of which are supported on $\BR^2 _{\ve_{c+1}} \times \BR^2 _{\ve_{c-1}} \times \BC^\times _X$. By triality, it is enough to consider one of the three cases $c \in \{1,2,3\}$, and let us fix $c=3$ and the number of M5-branes to be $N$. The resulting holomorphic surface defect $\mathcal{S}_{0,0,N} ^{\mathbf{(1,0)}}$ carries the algebra $\text{M5}_{0,0,N} ^{\mathbf{(1,0)}}$ of local operators.

Under the reduction along the circle of $\BC^\times _X \simeq \BR_x \times S^1$, the $N$ M5-branes become $N$ D4-branes supported on $\BR^2 _{\ve_1} \times \BR^2 _{\ve_2} \times \BR_x$. Due to the $\O$-background, the local operators form a non-commutative associative algebra $\text{M5}_{0,0,N} ^{\mathbf{(1,0)}}$ on the fixed locus $\{0\} \times \BR_x$ of the isometry. Since there are infinite numbers of winding modes along the transverse circle, the effective 5d $U(N)$ gauge theory for the D4-D4 open string contains infinite numbers of Kaluza-Klein modes. Its Hilbert space is the equivariant K-theory of the moduli space of $U(N)$ instantons, on which the algebra $\text{M5}_{0,0,N} ^{\mathbf{(1,0)}}$ of local operators can act. It is indeed known that the equivariant K-theory of the moduli space of $U(N)$ instantons admits an action of the quantum toroidal algebra $\qta$ \cite{2009arXiv0905.2555S}, which would factor through a surjective algebra homomorphism $\r_{\text{M5}_{0,0,N} ^{\mathbf{(1,0)}}} : \qta \twoheadrightarrow \text{M5}_{0,0,N} ^{\mathbf{(1,0)}}$.\\

\begin{figure}[h!]\centering
\begin{tikzpicture} 
\draw[line width=0.3mm] (-2,1) -- (1,1) -- (-1,-0.5) -- (-4,-0.5) 
      -- cycle;
\draw[line width=0.3mm] (-2,0) -- (1,0) -- (-1,-1.5) -- (-4,-1.5) 
      -- cycle;

\draw[line width=0.3mm] (6,0.5) -- (9,0.5) -- (7,-1) -- (4,-1) 
      -- cycle;
      
 \node at (-5,-0.5) {$\CalS^{\mathbf{(1,0)}}_{0,0,N}$};
 \node at (-5,-1.5) {$\CalS ^{\mathbf{(1,0)}} _{0,0,N'}$};
 \node at (4.5,0.5) {$\CalS ^{\mathbf{(1,0)}}_{0,0,N+N'}$};
  \node at (2.5,0) {fusion};
 
  \node at (-3,-2.8) {$\text{M5}^{\mathbf{(1,0)}} _{0,0,N}$};
    \node at (-1,-2.8) {$\text{M5} 
^{\mathbf{(1,0)}}  _{0,0,N'}$};
      \node at (6,-2.8) {$\text{M5} 
^{\mathbf{(1,0)}}  _{0,0,N+N'}$};

 \draw[->] (1.7,-0.4) -- (3.2,-0.4);
       \draw[left hook-stealth] (3.2,-2.8) -- (1.7,-2.8) ;
  \node at (2.45,-2.5) {$\d_{N,N'}$};
        \node at (-2,-2.8) {$\otimes$};
\end{tikzpicture} \caption{Fusion of surface defects and algebra embedding induced by the coproduct} \label{fig:m5fuse} 
\end{figure}
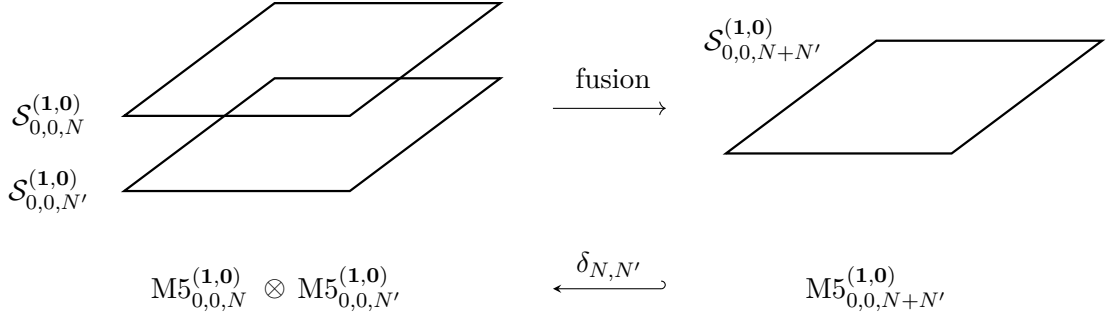

In section \ref{sec:miura}, we will show that this algebra $\text{M5}_{0,0,N} ^{\mathbf{(1,0)}}$ is precisely the $q$-deformed $W$-algebra of $\fgl(N)$ \cite{frenkel_reshetikhin_1995,Shiraishi:1995rp,Feigin:1995sf,Awata:1995zk}, which we denote by $q\mathcal{W}_N$. To verify this, it is crucial to study the fusion of the surface defects and its implications. First, note that the surface defect $\mathcal{S}_{0,0,N+N'} ^{\mathbf{(1,0)}}$ can be obtained by fusing two surface defects $\mathcal{S}_{0,0,N} ^{\mathbf{(1,0)}}$ and $\mathcal{S}_{0,0,N'} ^{\mathbf{(1,0)}}$ on top of each other (see Figure \ref{fig:m5fuse}). Just as in the case of the topological line defects, the fusion implies that the coupling of the fused surface defect can be built from the couplings of the individual surface defects before the fusion. Namely, there is an algebra embedding $\d_{N,N'} : \text{M5}_{0,0,N+N'} ^{\mathbf{(1,0)}} \hookrightarrow \text{M5}_{0,0,N} ^{\mathbf{(1,0)}} \otimes \text{M5}_{0,0,N'} ^{\mathbf{(1,0)}}$ induced by the coproduct of $\qta$, which can be represented by the following commutative diagram:
\begin{equation} \label{eq:commsur}
\begin{tikzcd}[row sep=huge]
\qta \arrow{r}{\Delta}\arrow[two heads]{d}{\r_{\text{M5}^{\mathbf{(1,0)}} _{0,0,N+N'}}} & \qta \, {\widehat{\otimes}} \, \qta \arrow[two heads]{d}{\r_{\text{M5}^{\mathbf{(1,0)}}_{0,0,N}}  \otimes \, \r_{\text{M5}^{\mathbf{(1,0)}}_{0,0,N'}}}\\
 q\mathcal{W}_{N+N'} \arrow[ur, phantom, "\scalebox{1.5}{$\circlearrowleft$}" description] \arrow[hookrightarrow]{r}{\d_{N,N'}} &  q\mathcal{W}_{N} \,{\otimes} \, q\mathcal{W}_{N'}
\end{tikzcd}
\end{equation}
Then, by consecutively applying this diagram, we achieve an embedding of the algebra
\begin{align}
    \text{M5}_{0,0,N} ^{\mathbf{(1,0)}} \hookrightarrow \text{M5}_{0,0,1} ^{\mathbf{(1,0)}} \otimes \cdots \otimes  \text{M5}_{0,0,1} ^{\mathbf{(1,0)}},
\end{align}
into the tensor product of free $q$-boson algebras. The fusion of surface defects thus provides the free field realization of the M5-brane algebra $\text{M5}_{0,0,N} ^{\mathbf{(1,0)}}$.\\

In section \ref{sec:miura}, we will show that this free field realization exactly matches with that of the $q$-deformed $W$-algebra, given in \cite{Feigin:1995sf,Awata:1995zk}. The proof is given in two steps. First, we establish the Miura transformation for the algebra $\text{M5}_{0,0,N} ^{\mathbf{(1,0)}}$ which explicitly realizes the commutative diagram \eqref{eq:commsur} by the product of newly defined Miura operators. Then, we prove that the algebra generated by such a Miura transformation is completely characterized as the commutants of screening charges in the tensor product of $q$-bosons, which precisely recovers the ones in \cite{Feigin:1995sf,Awata:1995zk}. 

We emphasize that our result therefore derives the Miura transformation for the $q$-deformed $W$-algebra, which can be regarded as an independent result within itself.

\subsubsection{Non-parallel M5-branes and $q$-deformed $Y$-algebra}
\label{ssec:nonpar}
We further study more general configurations of non-parallel M5-branes, which leads to an extension of the result for the parallel case. Here, by \textit{non-parallel}, we indicate that the topological part of the support may differ for different M5-branes. This should not be confused with the holomorphic part of the support, which is chosen to be $\BC^\times _X = C^{\mathbf{(1,0)}}$ for all the M5-branes. We will return to the investigation of M5-branes supported on generic holomorphic curves $C^{\mathbf{(p,q)}}$ in section \ref{sec:qqchar}.

Let us be given with $L+M+N$ M5-branes which share the common holomorphic part of the support $\BC^\times _X$. The topological part of the support is determined by the function $c:I \mapsto c_I \in \{1,2,3\}$, $I\in \{1,2,\cdots, L+M+N\}$, where the topological part of the support of the $I$-th M5-brane is assigned to be $\BR^2 _{\ve_{c_I+1}} \times \BR^2 _{\ve_{c_I-1}}$. Let us set $(\vert c^{-1} (1) \vert, \vert c^{-1} (2) \vert, \vert c^{-1} (3) \vert) = (L,M,N)$. We call the induced holomorphic surface defect in the 5d $\fgl(1)$ Chern-Simons theory $\mathcal{S}_{L,M,N} ^{\mathbf{(1,0)}}$, and the algebra of local operators on this surface defect is denoted by $\text{M5}_{L,M,N} ^{\mathbf{(1,0)}}$. 

Just as in the parallel case, the fusion operation of surface defects implies that the coproduct of the quantum toroidal algebra $\qta$ induces the algebra embedding
\begin{align}
    \text{M5}_{L,M,N} ^{\mathbf{(1,0)}} \hookrightarrow \text{M5}_{c_1} ^{\mathbf{(1,0)}} \otimes \text{M5}_{c_2} ^{\mathbf{(1,0)}} \otimes \cdots \otimes \text{M5}_{c_{L+M+N}} ^{\mathbf{(1,0)}}
\end{align}
into the tensor product of $q$-boson algebras. Thus, the algebra $\text{M5}_{L,M,N} ^{\mathbf{(1,0)}}$ admits a free field realization.\\

In section \ref{sec:miura}, we show that this free field realization is implemented by a Miura transformation, with newly defined Miura operators. We also show that the algebra generated by the Miura transformation can be completely characterized as the commutants of properly defined screening charges in the tensor product of $q$-bosons. Due to the equivalence that we prove, the latter may be employed as the definition of the algebra $\text{M5}_{L,M,N} ^{\mathbf{(1,0)}}$. In particular, we verify that changing the ordering of the $q$-boson algebras in the tensor product produces an isomorphic algebra, as a consequence of the universal Yang-Baxter equation for the universal R-matrix of $\qta$. Hence, the algebra of local operators on the fused surface defect $\mathcal{S}_{L,M,N} ^{\mathbf{(1,0)}}$ can be denoted as $\text{M5}_{L,M,N} ^{\mathbf{(1,0)}}$ without ambiguity.

The algebra $\text{M5}_{L,M,N} ^{\mathbf{(1,0)}}$ can be thought of as the multiplicative uplift of the $Y$-algebra studied in \cite{Gaiotto:2017euk}. As we will explain, the multiplicative uplift is achieved by replacing $\BC\times \BC^\times$ by $\BC^\times \times \BC^\times$ for the holomorphic part of the worldvolume of the twisted M-theory. In this sense, we call $\text{M5}_{L,M,N} ^{\mathbf{(1,0)}}$ the $q$-deformed $Y$-algebra as well, which may be denoted by $q{Y}_{L,M,N}$.

\vspace{4mm}

\section{M2-M5 intersections, R-matrices, and Miura operators} \label{sec:miura}
So far, we have introduced the algebra of local operators on M2- and M5-branes and surjective homomorphisms from the quantum toroidal algebra $\qta$. In this section, we study the intersections of M2-branes and M5-branes. Passing to the 5d Chern-Simons theory on $\BR_t \times \BC^\times _X \times \BC^\times _Z$, the M2-M5 intersections descend to the intersections between topological line defects and holomorphic surface defects. Note that the M2-brane wraps one, say $\BR^2 _{\ve_c}$, of the three topological planes, while an M5-brane wraps two, say $\BR^2 _{\ve_{c'+1}} \times \BR^2 _{\ve_{c'-1}}$, of them. Depending on the relative orientation, we distinguish two possibilities:
\begin{itemize}
    \item If $c = c'$, the two defects are completely transverse so that, when visualized in the 3-dimensional worldvolume that they span, the line defect can only pass through the surface defect. By slightly abusing the terminology, we call this configuration the \textit{transverse} M2-M5 intersection.

    \item If $c \neq c'$, the line defect can end on the surface defect at the \textit{non-transverse} M2-M5 intersection. Thus the line defect may terminate at the intersection, or may continue across the intersection. In the former case, the M2-brane lies on a semi-infinite line in $\BR_t$, stretched from the intersection point to either positive or negative infinity.
\end{itemize}
We will investigate these two cases separately.

In this section, we will restrict to the case where the holomorphic part of the support of M5-branes is $\BC^\times _X = C^{\mathbf{(1,0)}}$. We will come back to more generic M5-brane configurations in section \ref{sec:qqchar}.

\vspace{4mm}

\subsection{R-matrix for transverse M2-M5 intersection} \label{subsec:transr}
Let us begin with the cases where there are a single M2-brane and a single M5-brane. Due to the triality, we may fix the M2-brane to be supported on $\BR^2 _{\ve_3} \times \BR_t$. Then, the M5-brane wrapping $\BR^2 _{\ve_1}\times \BR^2 _{\ve_2} \times \BC^\times _X$ is totally transverse to the M2-brane. Even though they do not intersect at all in the 11-dimensional worldvolume, we shall call such a configuration a transverse M2-M5 intersection.

The transverse M2-M5 intersection supports a space of local operators. We can write a generic element of this space as
\begin{align} \label{eq:transrmat}
    R^{(3) } = R _{\text{M2}_{0,0,1}, \text{M5}_{0,0,1} ^{\mathbf{(1,0)}    }} \in \text{M2}_{0,0,1} \widehat{\otimes} \, \text{M5}_{0,0,1} ^{\mathbf{(1,0)}}.
\end{align}
Under the condition that the support of the M2-brane is $\BR^2 _{\ve_3} \times \BR_t$ and the holomorphic part of the support of the M5-brane is $\BC^\times _X = C^{\mathbf{(1,0)}}$, we use the abbreviation $R^{(3)}$ where the superscript indicates the topological part of the M5-brane support $\BR^2 _{\ve_1} \times \BR^2 _{\ve_2} $ (by $\{1,2\} = \{1,2,3\} \setminus\{3\}$). Passing to the 5d $\fgl(1)$ Chern-Simons theory, the transverse M2-M5 intersection should descend to a gauge-invariant intersection between a line defect and a surface defect. The local operators at the intersection are thereby constrained by the gauge invariance condition.\\

As we discussed in section \ref{subsec:intrmat}, the fully quantum-corrected gauge-invariance condition is expected to be given by
\begin{align}\label{eq:rmatm2m5}
    R^{(3)}  \Delta_{\text{M2}_{0,0,1},\text{M5} _{0,0,1} ^{\mathbf{(1,0)}}} (g) = \Delta_{\text{M2}_{0,0,1},\text{M5} _{0,0,1} ^{\mathbf{(1,0)}}} ^{\text{op}} (g)  R^{(3)},\quad \text{for any } g \in \qta,
\end{align}
where $\D_{\text{M2}_{0,0,1},\text{M5} _{0,0,1} ^{\mathbf{(1,0)}}}$ is the mixed coproduct, obtained by composing the coproduct with the respective representations,
\begin{align}
    \Delta_{\text{M2}_{0,0,1},\text{M5} _{0,0,1} ^{\mathbf{(1,0)}}} = (\r_{\text{M2}_{0,0,1}} \otimes \r_{\text{M5} _{0,0,1} ^{\mathbf{(1,0)}}}  ) \Delta \quad : \quad \qta \to \text{M2}_{0,0,1} \, \widehat{\otimes} \, \text{M5} _{0,0,1} ^{\mathbf{(1,0)}} ,
\end{align}
and $\Delta^{\text{op}} = \s \circ \Delta$ is the opposite coproduct.

\begin{figure}[h!]\centering
\begin{tikzpicture} 
\draw[line width=0.3mm] (-1,1.5) -- (5,1.5) -- (3,-1.5) -- (-3,-1.5) 
      -- cycle;
\draw[line width=0.3mm] (2.5,3) -- (2.5,0.5);
\draw[line width=0.3mm, loosely dotted, ultra thick] (2.5,0.5) -- (2.5,-1.5);
\draw[line width=0.3mm] (2.5,-1.5) -- (2.5,-3);

\filldraw[black] (2.5,0.5) circle (2pt) node[anchor=west]{$R^{(3)}$};
 \node at (1,0) {$\times$};
 \node at (1,-0.5) {$0$};
  \node at (-3.6,-1.3) {$\mathcal{S}_{0,0,1} ^{\mathbf{(1,0)}}$};
  \node at (2.5,3.3) {$\mathcal{L}_{0,0,1} $};

\draw[->] (-3,2) -- (-3,3);
\draw[->] (-3,2) -- (-2,2);
\draw[->] (-3,2) -- (-3.5,1.4);

\node at (-2.7,1.6) {$\BC^\times _X$};
\node at (-2.6,2.9) {$\BR_t$};
    
\end{tikzpicture} \caption{Transverse M2-M5 intersection} \label{fig:m2m5int} 
\end{figure}
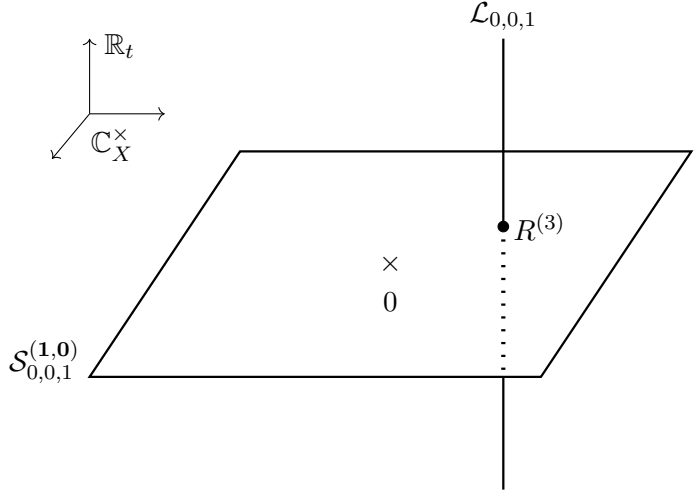

The constraint \eqref{eq:rmatm2m5} can be solved exactly. We show in appendix \ref{sec:solver} that it fully determines $R^{(3)}$ up to an overall scaling. The solution is obtained as the product:
\begin{align} \label{eq:transr}
R^{(3)} = \bar{R}^{(3)} K^{(3)},
\end{align}
with the Cartan part  given as
\begin{align} \label{eq:rcartan3}
\begin{split}
    K ^{(3)} = \left(q_3 ^{\frac{d}{2}} \otimes 1 \right)  \exp \left[ \sum_{r=1} ^\infty \frac{\k_r}{r} \frac{X_1 ^{-r}}{1-q_3 ^r} \otimes \frac{a_r ^{(3)} }{q_3 ^{r/2} -q_3 ^{-r/2}} \right] =  \left(q_3 ^{\frac{d}{2}} \otimes 1 \right) V_3(X_1),
\end{split}
\end{align}
where we defined the vertex operator $V_3 (X_1) := \prod_{j=0} ^\infty \varphi_3 ^+ (q_3 ^{-j} X_1)   =\exp \left[ \sum_{r=1} ^\infty \frac{\k_r}{r} \frac{X_1 ^{-r}}{1-q_3 ^r} \otimes \frac{a_r ^{(3)} }{q_3 ^{r/2} -q_3 ^{-r/2}} \right]$. The rest is given by 
\begin{align} \label{eq:rrest3}
    \bar{R}^{(3)} = 1- e^{-a_0 ^{(3)} \frac{\log q_1 \log q_2}{\log q_3}} q_3 ^{-1} \xi_3 (X_1) q_3 ^{-D_{X_1}}
\end{align}
Note that $X_1 \in \text{M2}_{0,0,1}$ commutes with itself so that the vertex operators with the argument $X_1$ are well-defined.

\vspace{4mm}

\subsection{Non-transverse M2-M5 intersections} \label{subsec:nontrans}
We investigate the intersection between a single M2-brane and a single M5-brane that share one topological $\BR^2$-plane in their supports. We fix the support of the M2-brane as $\BR^2 _{\ve_3} \times \BR_t$. All other cases follow simply by the triality. The holomorphic part of the support of the M5-brane is chosen as $\BC^\times _X$. 

Since we are assuming the support of the M5-brane shares $\BR^2 _{\ve_3}$, the only two options are $\BR^2 _{\ve_{c+1}} \times \BR^2 _{\ve_{c-1}} \times \BR_t \times \BC^\times _X$ with $c \in \{1,2\}$. We also use the notation $\{\bar{c} \} = \{1,2\} \setminus \{c\}$.

\subsubsection{Intertwiners from intersections with ends} \label{subsubsec:intert}
Unlike the transverse case, the M2-brane can end on the M5-brane from either direction of the line $\BR_t$. We show such non-transverse M2-M5 intersections with an end are associated to intertwiners of the quantum toroidal algebra of $\fgl(1)$.

First, we consider the case of the M2-brane ending on the M5-brane from the negative direction of $\BR_t$. In other words, the M2-brane is stretched between the negative infinity and the intersection point in $\BR_t$. It is helpful to view the other semi-infinite line as being occupied by a trivial line defect, so that the local operator supported at the intersection is an element of\footnote{Compare with the transverse R-matrix $R^{(3)}$ \eqref{eq:transrmat}, where both semi-infinite lines above and below from the intersection point support the same line defect $\CalL_{0,0,1}$. There, we had $\text{M2}_{0,0,1} = \text{End}(\CalV_3 (u))$ in turn.}
\begin{align}
\begin{split}
    &\Phi_{3,c} \in \text{Hom}(\CalV_3(u), \BC ) \otimes \text{M5}_c ^{\mathbf{(1,0)}}, \qquad c \in\{1,2\}.
\end{split}
\end{align}

The intersection must be gauge-invariant. As argued in section \ref{subsec:intrmat} for the line defect passing through the surface defect, the classical gauge-invariance constraint for the local operator $\Phi_{3,c}$ reads
\begin{align}
    \Phi_{3,c}  \left(\r_{\text{M2}_{0,0,1}} (g) \otimes \text{id} + \text{id} \otimes \r_{\text{M5}_{c} ^{\mathbf{(1,0)}} }  (g) \right)   = \r_{\text{M5}_{c} ^{\mathbf{(1,0)}} }  (g) \Phi_{3,c}
\end{align}
for any $g \in U\left(\mathscr{O}_{q_2} (\BC^\times \times \BC^\times) \right)$.

Turning on $q_1 \neq 1$, the gauge-invariance condition for intersection gets quantum corrections, which are non-linear in the generators. By a similar argument for the transverse M2-M5 intersection, we expect that the full quantum gauge-invariance condition reads
\begin{align} \label{eq:intconst}
    &\Phi_{3,c}  \D_{\text{M2}_{0,0,1},\text{M5} _c ^{\mathbf{(1,0)}}} (g) =  \r_{\text{M5}^{\mathbf{(1,0)}} _c}(g) \Phi_{3,c} ,\qquad \text{for any } g\in \qta,
\end{align}
for each $c \in \{1,2\}$.\footnote{Using the counit $\epsilon = \r_{\text{M2}_{0,0,0}}$ associated to the trivial line defect (see footnote \ref{fn:counit}), the gauge transformation on right hand side of \eqref{eq:intconst} can be rephrased as $(\r_{\text{M2}_{0,0,0}} \otimes \r_{\text{M5}_{0,0,1} ^{\mathbf{(1,0)}}} ) \Delta^{\text{op}}(g) = (1 \otimes \r_{\text{M5}_{0,0,1} ^{\mathbf{(1,0)}}}) (\epsilon \otimes \text{id} ) \Delta^{\text{op}}(g) 
 =\r_{\text{M5}_{0,0,1} ^{\mathbf{(1,0)}}} (g)$.} The gauge-invariance condition for all the other cases $\Phi_{c,c'} \in \text{Hom}(\CalV_c, \BC ) \otimes \text{M5}_{c'} ^{\mathbf{(1,0)}}$, $c\neq c'$ simply follows by the triality. In this sense, the non-transverse M2-M5 intersection provides the intertwiner between the representations $\text{M2}_{0,0,1}$ and $\text{M5}_c ^{\mathbf{(1,0)}}$ of $\qta$. The constraint \eqref{eq:intconst} determines $\Phi_{3,c}$ up to an overall scaling, yielding \cite{Feigin2009ACA}
\begin{align} \label{eq:vechfint}
\begin{split}
\Phi_{3,c}  &=  e^ {  \frac{\log q_{\bar{c}}}{\log q_c} \left( {\log X_1} \otimes a_0 ^{(c)} \right)}  \left(1 \otimes q_{\bar{c}} ^{d^\perp}\right)\exp \left[ - \sum_{r=1} ^\infty \frac{\k_r}{r} \frac{q_3 ^{\frac{r}{2}}  }{ \left(q_c ^\frac{r}{2} - q_c ^{-\frac{r}{2}} \right)^2 \left( q_3 ^{\frac{r}{2}}  - q_3 ^{-\frac{r}{2}}\right) } X_1 ^r \otimes a_{-r} ^{(c)}  \right] \\
&\qquad \times \exp \left[ \sum_{r=1} ^\infty \frac{\k_r}{r} \frac{q_3 ^{-\frac{r}{2}} q_c ^{-\frac{r}{2}} }{ \left(q_c ^\frac{r}{2} - q_c ^{-\frac{r}{2}} \right)^2 \left( q_3 ^{\frac{r}{2}}  - q_3 ^{-\frac{r}{2}}\right) } X_1 ^{-r} \otimes a_{r} ^{(c)} \right] .
\end{split}
\end{align}
Note that $\log X_1$ makes sense as an element of $\text{Hom}(\CalV_3 (u),\BC)$ since $X_1$ acts on $\CalV_3 (u)$ as an element of $\BC^\times$. Noting that $\CalV_3 (u) ^{*} = \text{Hom}(\CalV_3 (u) ,\BC)$, we may also expand $\Phi_{3,c}$ in the dual basis $\{ [u]_i ^{(3)*} \; \vert \; i\in \BZ \}$ satisfying $[u]_i ^{(3)*} ([u]_j ^{(3)}) = \d_{i,j}$ as $\Phi_{3,c} = \sum_{i\in \BZ} (\Phi_{3,c})_i [u]^{(3)*} _i $, where the coefficients $(\Phi_{3,c})_i$ is obtained simply by replacing $X_1$ by $u q_3 ^i$ due to its action on $[u]_i ^{(3)} \in \CalV_{3} (u)$.\\

It should be stressed that $\Phi_{3,c}$ provides a reconstruction of a vertex operator in the $q$-boson quantum vertex algebra in the realm of the representation theory of the quantum toroidal algebra $\qta$. Moreover, in the degeneration limit of the $q$-boson to the free boson $\widehat{\fgl}(1)$ vertex algebra, it reduces to the usual degenerate vertex operator. The degenerate limit is $R\to 0$ where $q_a ^{R \ve_a}$, in which the $q$-boson commutation relation \eqref{eq:qbos} reduces to that of $\widehat{\fgl}(1)$, given by
\begin{align}
    [\tilde{a}_m ^{(c)} , \tilde{a}_n ^{(c)}] = - \frac{1}{\ve_{c+1}\ve_{c-1}} m \d_{m+n,0},\qquad m,n \in \BZ,
\end{align}
with the redefinition $\tilde{a}_n ^{(c)} = \frac{a_n ^{(c)}}{\log q_c}$. Then, we indeed have $\Phi_{3,c} (X_1) \to e^{ \ve_{\bar{c}}\phi(X_1)}$, where $\phi(X_1) = d^{\perp} + \tilde{a}_0 ^{(c)} \log X_1 - \sum_{n\in \BZ \setminus \{0\}} \frac{1}{n} \tilde{a}_{n} ^{(c)} X_1 ^{-n} $ is the free boson field.\\

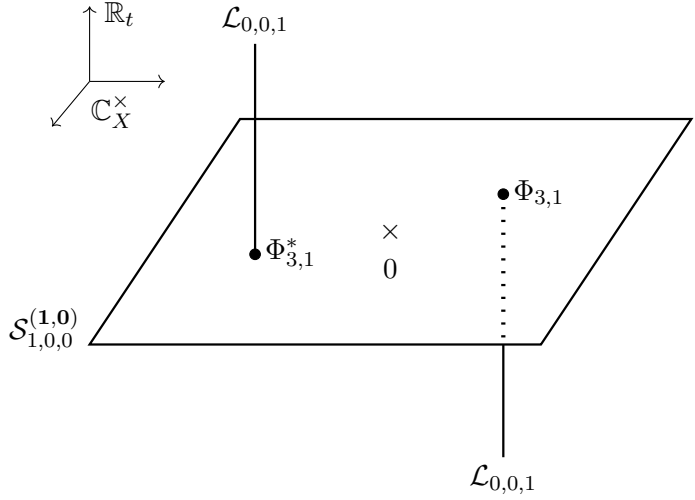
\begin{figure}[h!]\centering
\begin{tikzpicture} 
\draw[line width=0.3mm] (-1,1.5) -- (5,1.5) -- (3,-1.5) -- (-3,-1.5) 
      -- cycle;

\draw[line width=0.3mm, loosely dotted, ultra thick] (2.5,0.5) -- (2.5,-1.5);
\draw[line width=0.3mm] (2.5,-1.5) -- (2.5,-3);

\draw[line width=0.3mm] (-0.8,2.5) -- (-0.8,-0.3);

\filldraw[black] (-0.8,-0.3) circle (2pt) node[anchor=west]{$\Phi^* _{3,1}$};

\filldraw[black] (2.5,0.5) circle (2pt) node[anchor=west]{$\Phi_{3,1}$};

 \node at (1,0) {$\times$};
 \node at (1,-0.5) {$0$};

   \node at (-3.6,-1.3) {$\mathcal{S}_{1,0,0} ^{\mathbf{(1,0)}}$};
  \node at (2.5,-3.3) {$\mathcal{L}_{0,0,1} $};
    \node at (-0.8,2.8) {$\mathcal{L}_{0,0,1} $};

\draw[->] (-3,2) -- (-3,3);
\draw[->] (-3,2) -- (-2,2);
\draw[->] (-3,2) -- (-3.5,1.4);

\node at (-2.7,1.6) {$\BC^\times _X$};
\node at (-2.6,2.9) {$\BR_t$};
    
\end{tikzpicture} \caption{Non-transverse M2-M5 intersections} \label{fig:m2m5intnon} 
\end{figure}

Next, we turn to the M2-brane ending on the M5-brane from the positive $\BR_t$-direction. The M2-brane lies on the semi-infinite line between the intersection point and the positive infinity in $\BR_t$, and we treat the other semi-infinite line as supporting a trivial line defect. The local operator at the intersection thus lies in 
\begin{align}
    \Phi^* _{3,c} \in \text{Hom}(\BC,\CalV_3 (u))\otimes \text{M5}_c ^{\mathbf{(1,0)}}, \qquad c\in \{1,2\}.
\end{align}
By the argument similar to the previous case, the gauge-invariance condition at the quantum level $q_1 \neq 1$ is expected to be written as
\begin{align} \label{eq:dintconst}
    \Phi^* _{3,c} \r_{\text{M5}_c ^{\mathbf{(1,0)}}} (g) = \D^{\text{op}}_{\text{M2}_{0,0,1},\text{M5} _c ^{\mathbf{(1,0)}}} (g) \Phi^* _{3,c} ,\qquad c \in\{1,2\},
\end{align}
The constraint can be solved to determine $\Phi^* _{3,c}$ up to an overall scaling as
\begin{align} \label{eq:vechfdualint}
\begin{split}
     \Phi^* _{3,c}  &=  \left(1 \otimes q_{\bar{c}} ^{-d^\perp}\right) e^{ -\frac{\log q_{\bar{c}}}{\log q_c} \left( {\log X_1} \otimes a_0 ^{(c)} \right)}  \exp \left[  \sum_{r=1} ^\infty \frac{\k_r}{r} \frac{q_3 ^{\frac{r}{2}} q_c ^{\frac{r}{2}}  }{ \left(q_c ^\frac{r}{2} - q_c ^{-\frac{r}{2}} \right)^2 \left( q_3 ^{\frac{r}{2}}  - q_3 ^{-\frac{r}{2}}\right) } X_1 ^r \otimes a_{-r} ^{(c)}  \right] \\
    &\qquad \times \exp \left[ -\sum_{r=1} ^\infty \frac{\k_r}{r} \frac{q_3 ^{-\frac{r}{2}}   }{ \left(q_c ^\frac{r}{2} - q_c ^{-\frac{r}{2}} \right)^2 \left( q_3 ^{\frac{r}{2}}  - q_3 ^{-\frac{r}{2}}\right) } X_1 ^{-r} \otimes a_{r} ^{(c)} \right] .
\end{split}
\end{align}
Noting that $\CalV_3 (u) \simeq \text{Hom}(\BC,\CalV_3 (u))$, we may expand $\Phi^* _{3,c}$ in the basis elements as $\Phi^* _{3,c} = \sum_{i\in \BZ} (\Phi^* _{3,c})_i [u]_i ^{(3)}$. The coefficients $(\Phi^* _{3,c})_i$ are obtained simply by replacing $X_1$ by $u q_3 ^i$ due to the action of $[u]^{(3)*} _i$ on $X_1$.\\

Again, we observe that $\Phi^* _{3,c}$ gives a reconstruction of a vertex operator in the $q$-boson algebra as an intertwiner of representations of $\qta$. In the degeneration limit of the $q$-boson to the free boson $\widehat{\fgl}(1)$, it is straightforward to see it reduces to the degenerate vertex operator, $\Phi^* _{3,c} (X_1) \to e^{-\ve_{\bar{c}} \phi(X_1)}$.

\subsubsection{R-matrices for non-transverse intersections and fusion of intertwiners} \label{subsubsec:nontransr}
Having established the non-transverse intersection where M2-brane ends on the M5-brane, now we turn to the non-transverse intersection where M2-brane passes through the M5-brane. The local operator at the intersection is an element of
\begin{align}
    R^{(c)} \in \text{M2}_{0,0,1} \widehat{\otimes}\, \text{M5}_c ^{\mathbf{(1,0)}} ,\qquad c\in\{1,2\}.
\end{align}
The gauge-invariance condition reads
\begin{align} 
\begin{split}
    &R^{(c)} \D_{\text{M2}_{0,0,1},\text{M5} _c ^{\mathbf{(1,0)}}} (g) = \D^{\text{op}}_{\text{M2}_{0,0,1},\text{M5} _c ^{\mathbf{(1,0)}}} (g) R^{(c)},\qquad c\in\{1,2\}. 
\end{split}
\end{align}
The constraint can be solved exactly. As we demonstrate in appendix \ref{sec:solver}, the constraint determines $R^{(c)}$ up to an overall scaling, yielding
\begin{align} \label{eq:rmatexp}
    R^{(c)} = \bar{R}^{(c)} K^{(c)},\qquad c\in\{1,2\},
\end{align}
where the Cartan part is given by
\begin{align} \label{eq:cartanc}
\begin{split}
\begin{split}
    K ^{(c)} &= \left(q_c ^{\frac{d}{2}} \otimes 1 \right)  \exp \left[ \sum_{r=1} ^\infty \frac{\k_r}{r} \frac{X_1 ^{-r}}{1-q_3 ^r} \otimes \frac{a_r ^{(c)} }{q_c ^{r/2} -q_c ^{-r/2}} \right] \\
    &=  \left(q_c ^{\frac{d}{2}} \otimes 1 \right)  V_c (X_1) = V_c (q_c ^{\frac{1}{2}} X_1)  \left(q_c ^{\frac{d}{2}} \otimes 1 \right) ,
\end{split}
\end{split}
\end{align}
where we defined the vertex operator $V_c(X_1) :=  \prod_{j=0} ^\infty \varphi_c ^+ (q_3 ^{-j} X_1) = \exp \left[ \sum_{r=1} ^\infty \frac{\k_r}{r} \frac{X_1 ^{-r}}{1-q_3 ^r} \otimes \frac{a_r ^{(c)} }{q_c ^{r/2} -q_c ^{-r/2}} \right] $. The rest is expanded with respect to the $d^\perp$-grading as
\begin{align} \label{eq:rbar}
\begin{split}
    \bar{R}^{(c)} &= \sum_{m=0} ^\infty \k_1 ^m \sum_{k_1,k_2,\cdots k_m \in \BZ} \left( \r_{\text{M2}_{0,0,1}} \otimes \r_{\text{M5}_{c} ^{\mathbf{(1,0)}} } \right) \left( E_{k_1} E_{k_2}  \cdots E_{k_m} \otimes  F_{-k_1} F_{-k_2} \cdots F_{-k_m} \right) \\
    &= \sum_{m=0} ^\infty q_{\bar{c}} ^m e^{- m a_0 ^{(c)} \frac{\log q_{c+1} \log q_{c-1}}{\log q_c} }  \left(\prod_{j=0} ^{m-1} \frac{1-q_c q_3^{-j}}{1-q_3 ^{-j-1}} \right) : \prod_{j=0} ^ {m-1} \xi_c (q_3 ^{-j} X_1) : q_3 ^{-m D_{X_1}}.
\end{split}
\end{align}
Here, the second equality follows from the normal-ordering relation,
\begin{align} \label{eq:xinormal}
    \xi_c (X) \xi_c (X') = \frac{\left(1- \frac{X'}{X} \right)\left(1- \frac{q_c X'}{X} \right)}{\left(1- \frac{q_{c-1} ^{-1}  X'}{X} \right) \left(1- \frac{q_{c+1} ^{-1} X'}{X} \right) } : \xi_c (X) \xi_c (X'): , \qquad c\in \{1,2,3\}.
\end{align}
In fact, the above expressions are valid even at $c=3$ for the transverse M2-M5 intersection. It is obvious that the expression for the Cartan part $K^{(c)}$ \eqref{eq:cartanc} applies to $c=3$ to reproduce $K^{(3)}$ \eqref{eq:rcartan3}. For \eqref{eq:rbar}, if we set $c=3$ the term in the parenthesis in the second line vanishes when $m \geq 2$. The infinite summation thus truncates to the first two terms, recovering \eqref{eq:rrest3} that we already obtained for the transverse R-matrix.\\

Having obtained the exact expressions for the R-matrices $R^{(c)}$,  we can also reorganize them into a suggestive form. Let us consider the non-transverse cases $c\in \{1,2\}$. It is crucial to consider the following normal-ordered product of the intertwiners,
\begin{align}
    :  \Phi^* _{3,c} (X_1) \Phi_{3,c} (q_c ^{\frac{1}{2}} q_3 ^{-m} X_1): = q_{\bar{c}}^{\frac{1}{2}(1\otimes a_0 ^{(c)} )} e^{-m a_0 ^{(c)} \frac{\log q_{\bar{c}} \log q_3}{\log q_c}} : \prod_{j=0} ^{m-1} \xi_c (q_3 ^{-j} X_1): V_c( q_c ^{\frac{1}{2}} q_3 ^{-m} X_1 ).
\end{align}
Using this identity, the R-matrix \eqref{eq:rmatexp} turns into
\begin{align}\label{eq:Rsum}
\begin{split}
    {R}^{(c)} &= \sum_{m=0} ^\infty (-1)^m q_c ^{-m} q_3 ^{\frac{m(m-1)}{2}} q_{\bar{c}}^{-\frac{1}{2} (1\otimes a_0 ^{(c)} ) } \frac{(q_3;q_3)_{\n_c}}{(q_3;q_3)_m (q_3;q_3)_{\n_c -m}} \\ 
    &\qquad \times : \Phi^* _{3,c} (X_1) \Phi_{3,c} (q_c ^{\frac{1}{2}} q_3 ^{-m} X_1 )  : (Z_1 ^m \otimes 1) \left(q_c ^\frac{d}{2} \otimes 1 \right) ,
\end{split}
\end{align}
where $\n_c = \frac{\log q_c}{\log q_3}$ and $(z;q)_{\n} = \frac{(z;q)_{\infty}}{(q^\n z;q)_{\infty}}$ is the $q$-Pochhammer symbol. We can further convert this infinite summation into a contour integral as
\begin{align} \label{eq:rcontnormal}
\begin{split}
    R^{(c)} &= - (q_3;q_3)_{\n_c} q_{\bar{c}} ^{-\frac{1}{2} (1\otimes a_0 ^{(c)})} \oint \frac{dX}{2\pi i X} \frac{\left( q_{\bar{c}}^{-1}  \frac{X}{X_1} ;q_3 \right)_\infty}{\left( \frac{X}{X_1};q_3 \right)_\infty} e^{ -\frac{\log q_{\bar{c}}}{\log q_3}\log \frac{X}{X_1} } \\
    &\qquad\qquad \qquad\qquad \qquad \qquad  \times :\Phi ^*_{3,c} (X_1) \Phi_{3,c} (q_c ^{\frac{1}{2}} X): Z_1 ^{-\frac{\log \frac{X}{X_1}}{\log q_3}} \left(q_c ^\frac{d}{2} \otimes 1 \right) ,
\end{split}
\end{align}
where the contour is chosen in such a way that the integral picks up the residues from all the simple poles at $X = X_1 q_3 ^{-m}$, $m\in \BZ_{\geq 0}$. Finally, using the normal-ordering relation between the two intertwiners,
\begin{align} \label{eq:noint}
    \Phi^*_{3,c} (X) \Phi_{3,c}(q_c ^{\frac{1}{2}} X') = e^{- \frac{\log q_{\bar{c}}}{\log q_3} \left(\log \frac{X'}{X} + \frac{1}{2}\log q_c \right)} \frac{\left( q_{\bar{c}}^{-1} \frac{X'}{X} ;q_3 \right)_\infty}{\left( \frac{X'}{X};q_3 \right)_\infty} : \Phi^*_{3,c} (X) \Phi_{3,c}(q_c ^{\frac{1}{2}} X') :,
\end{align}
we can undo the normal-ordering in the integrand, yielding
\begin{align} \label{eq:nontransfin}
\begin{split}
    R^{(c)} &= - (q_3;q_3)_{\n_c} q_{\bar{c}} ^{\frac{1}{2} \left(\frac{\log q_c}{\log q_3}- 1\otimes a_0 ^{(c)} \right)} \left( \oint \frac{dX}{2\pi i X} \, \Phi ^* _{3,c} (X_1) \Phi_{3,c} (q_c ^{\frac{1}{2}} X) \, Z_1 ^{-\frac{\log \frac{X}{X_1}}{\log q_3}} \right) \left(q_c ^\frac{d}{2} \otimes 1 \right).
\end{split}
\end{align}
Here, we view the integral in the parenthesis as an element in $\text{M2}_{0,0,1} \widehat{\otimes} \, \text{M5}_c ^{\mathbf{(1,0)}}$ by first performing the integral in $X$ and then invoking $X_1,Z_1 \in \text{M2}_{0,0,1}$. 

\begin{figure}[h!]\centering
\resizebox{.4\textwidth}{!}{
\begin{tikzpicture}
\draw[line width=0.3mm] (-1,1.5) -- (5,1.5) -- (3,-1.5) -- (-3,-1.5) 
      -- cycle;

\draw[line width=0.3mm, loosely dotted, ultra thick] (2.5,0.5) -- (2.5,-1.5);
\draw[line width=0.3mm] (2.5,-1.5) -- (2.5,-3);

\draw[line width=0.3mm] (-0.8,2.5) -- (-0.8,-0.3);

\filldraw[black] (-0.8,-0.3) circle (2pt) node[anchor=west]{$\Phi^*_{3,1}$};

\filldraw[black] (2.5,0.5) circle (2pt) node[anchor=west]{$\Phi_{3,1}$};

 \node at (1,0) {$\times$};
 \node at (1,-0.5) {$0$};

   \node at (-3.6,-1.3) {$\mathcal{S}_{1,0,0} ^{\mathbf{(1,0)}}$};
  \node at (2.5,-3.3) {$\mathcal{L}_{0,0,1} $};
    \node at (-0.8,2.8) {$\mathcal{L}_{0,0,1} $};

\draw[->] (-3,2) -- (-3,3);
\draw[->] (-3,2) -- (-2,2);
\draw[->] (-3,2) -- (-3.5,1.4);

\node at (-2.7,1.6) {$\BC^\times _X$};
\node at (-2.6,2.9) {$\BR_t$};
    
\end{tikzpicture}} \hspace{3mm} \raisebox{12\height}{\begin{tikzpicture}
    \draw[->] (-0.5,0) -- (0.5,0);
\end{tikzpicture}} \hspace{0.3mm} \resizebox{.4\textwidth}{!}{
\begin{tikzpicture}
\draw[line width=0.3mm] (-1,1.5) -- (5,1.5) -- (3,-1.5) -- (-3,-1.5) 
      -- cycle;

\draw[line width=0.3mm, loosely dotted, ultra thick] (2.5,0.5) -- (2.5,-1.5);
\draw[line width=0.3mm] (2.5,-1.5) -- (2.5,-3);

\draw[line width=0.3mm] (2.5,2.5) -- (2.5,0.5);

\filldraw[black] (2.5,0.5) circle (2pt) node[anchor=west]{$R^{(1)}$};

 \node at (1,0) {$\times$};
 \node at (1,-0.5) {$0$};

   \node at (-3.6,-1.3) {$\mathcal{S}_{1,0,0} ^{\mathbf{(1,0)}}$};
  \node at (2.5,-3.3) {$\mathcal{L}_{0,0,1} $};
    \node at (2.5,2.8) {$\mathcal{L}_{0,0,1} $};   
\end{tikzpicture}}
\caption{Non-transverse R-matrix from fusion of intertwiners} \label{fig:intfusion} 
\end{figure}
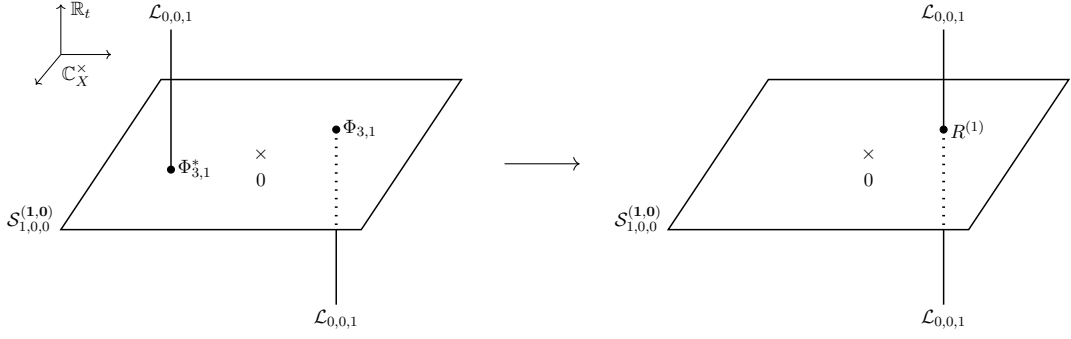

It is remarkable that the non-transverse R-matrices can be reconstructed by using the intertwiners as building blocks. We can visualize this as a fusion of intertwiners initiated by two non-transverse M2-M5 intersections approaching each other (see Figure \ref{fig:intfusion}). The two M2-branes, ending on the M5-brane from below and from above respectively, can join at a single point on $\BC^\times _X$ across the M5-brane. The outcome is a single M2-brane passing through the M5-brane, which gives rise to the non-transverse R-matrix $R^{(c)}$ discussed above.\\

The transverse R-matrix $R^{(3)}$ can also be expressed as a contour integral, even though extra care is needed: recall that in that case, there is no gauge-invariant intersection where the M2-brane ends on the M5-brane, since the two do not actually meet at all in the 11-dimensional spacetime. However, we will still define $\Phi_{3,3}$ and $\Phi^*_{3,3}$ as elements in $\text{M2}_{0,0,1} \otimes \text{M5}_{0,0,1} ^{\mathbf{(1,0)}}$, mimicking $\Phi_{3,c}$ and $\Phi^*_{3,c}$ for $c\in \{1,2\}$, as follows:
\begin{align}
\begin{split}
    &\Phi_{3,3} =  e^ {  \frac{\log q_1 \log q_2}{(\log q_3)^2} \left( {\log X_1} \otimes a_0 ^{(3)} \right)} e^{\frac{\log q_1 \log q_2}{\log q_3} (1 \otimes d^\perp)} \exp \left[ - \sum_{r=1} ^\infty \frac{\k_r}{r} \frac{q_3 ^{\frac{r}{2}}  }{  \left( q_3 ^{\frac{r}{2}}  - q_3 ^{-\frac{r}{2}}\right)^3 } X_1 ^r \otimes a_{-r} ^{(3)}  \right] \\
&\qquad \qquad\qquad\qquad\qquad \times \exp \left[ \sum_{r=1} ^\infty \frac{\k_r}{r} \frac{q_3 ^{-{r}}  }{  \left( q_3 ^{\frac{r}{2}}  - q_3 ^{-\frac{r}{2}}\right)^3 } X_1 ^{-r} \otimes a_{r} ^{(3)} \right],   \\
    &\Phi^*_{3,3} =   e^{-\frac{\log q_1 \log q_2}{\log q_3} (1 \otimes d^\perp)}   e^{ -\frac{\log q_{1} \log q_2}{(\log q_3)^2} \left( {\log X_1} \otimes a_0 ^{(3)} \right)}  \exp \left[  \sum_{r=1} ^\infty \frac{\k_r}{r} \frac{q_3 ^{r}   }{  \left( q_3 ^{\frac{r}{2}}  - q_3 ^{-\frac{r}{2}}\right)^3 } X_1 ^r \otimes a_{-r} ^{(3)}  \right] \\
    &\qquad \qquad\qquad\qquad\qquad \times \exp \left[ -\sum_{r=1} ^\infty \frac{\k_r}{r} \frac{q_3 ^{-\frac{r}{2}}   }{ \left( q_3 ^{\frac{r}{2}}  - q_3 ^{-\frac{r}{2}}\right)^3 } X_1 ^{-r} \otimes a_{r} ^{(3)} \right]  .
\end{split}
\end{align}
Let us stress again that individual $\Phi_{3,3}$ and $\Phi^*_{3,3}$ are \textit{not} gauge-invariant, and they do not provide intertwiners for the associated representations of the quantum toroidal algebra $\qta$. Nevertheless, they can be used together to express the gauge-invariant objects such as the transverse R-matrix $R^{(3)}$. Indeed, let us note that their normal-ordered product leads to
\begin{align}
\begin{split} 
        &:\Phi^*_{3,3} (X_1) \Phi_{3,3} (q_3 ^{\frac{1}{2}} X_1): = e^{\frac{1}{2} \frac{\log q_1 \log q_2}{\log q_3} (1\otimes a_0 ^{(3)}) }  V_3 (q_3 ^{\frac{1}{2}} X_1), \\
    &:\Phi^*_{3,3} (X_1) \Phi_{3,3} (q_3 ^{-\frac{1}{2}} X_1): = e^{-\frac{1}{2} \frac{\log q_1 \log q_2}{\log q_3} (1\otimes a_0 ^{(3)}) } \xi_3 (X_1) V_3 (q_3 ^{-\frac{1}{2}} X_1).
\end{split}
\end{align}
Thus, the transverse R-matrix $R^{(3)}$ is reorganized as
\begin{align} \label{eq:rcontnormal3}
\begin{split}
    R^{(3)} &= e^{-\frac{1}{2} \frac{\log q_1 \log q_2}{\log q_3} (1\otimes a_0 ^{(3)})}   : \Phi^*_{3,3} (X_1) \left( 1- q_3 ^{-1} q_3 ^{-D_{X_1}} \right)  \Phi_{3,3} (q_3 ^{\frac{1}{2}}  X_1 )  :  \left(q_3 ^{\frac{d}{2}} \otimes 1\right) \\
    & = (1-q_3^{-1}) e^{-\frac{1}{2} \frac{\log q_1 \log q_2}{\log q_3} (1\otimes a_0 ^{(3)})}  \\ 
 &\quad \times \left(\oint \frac{dX}{2\pi i X} \frac{1}{\left( 1- \frac{X_1}{X} \right) \left( 1- \frac{q_3 ^{-1} X_1}{X} \right)} : \Phi^*_{3,3} (X_1) \Phi_{3,3} (q_3 ^{\frac{1}{2} }  X )  :  Z_1 ^{- \frac{\log \frac{X}{X_1}}{\log q_3}} \right)  \left(q_3^{\frac{d}{2}} \otimes 1\right) ,
\end{split}
\end{align}
where the contour is chosen to enclose two simple poles at $X=X_1$ and $X= X_1 q_3 ^{-1}$. Thus, we observe that the transverse R-matrix $R^{(3)}$ can be re-expressed in terms of the vertex operators $\Phi_{3,3} $ and $\Phi^* _{3,3}$. 

Nevertheless, we cannot undo the normal-ordering of the two vertex operators to further simplify the expression in this case, unlike the non-transverse analogue \eqref{eq:nontransfin}. In this sense, the two vertex operators here are not truly isolated within itself. This is consistent with the fact that individual $\Phi_{3,3}$ or $\Phi^*_{3,3}$ is not gauge-invariant; that is, the gauge-invariant transverse M2-M5 intersection $R^{(3)}$ cannot be a composite of two independent gauge-invariant M2-M5 intersections. 

\paragraph{Remark}
We may define the inverse of the intertwiners simply by flipping the signs of the exponents. For instance,
\begin{align}
\begin{split}
  \left( \Phi^* _{3,c} \right)^{-1}  &:=  e^{ \frac{\log q_{\bar{c}}}{\log q_c} \left( {\log X_1} \otimes a_0 ^{(c)} \right)} \left(1 \otimes q_{\bar{c}} ^{d^\perp}\right)  \exp \left[ - \sum_{r=1} ^\infty \frac{\k_r}{r} \frac{q_3 ^{\frac{r}{2}} q_c ^{\frac{r}{2}}  }{ \left(q_c ^\frac{r}{2} - q_c ^{-\frac{r}{2}} \right)^2 \left( q_3 ^{\frac{r}{2}}  - q_3 ^{-\frac{r}{2}}\right) } X_1 ^r \otimes a_{-r} ^{(c)}  \right] \\
    &\qquad \times \exp \left[ \sum_{r=1} ^\infty \frac{\k_r}{r} \frac{q_3 ^{-\frac{r}{2}}   }{ \left(q_c ^\frac{r}{2} - q_c ^{-\frac{r}{2}} \right)^2 \left( q_3 ^{\frac{r}{2}}  - q_3 ^{-\frac{r}{2}}\right) } X_1 ^{-r} \otimes a_{r} ^{(c)} \right] .
\end{split}
\end{align}
Note that there is a non-trivial relation
\begin{align}
    \left( \Phi^* _{3,c} (X_1) \right)^{-1}  = q_{\bar{c}} ^{- \frac{1}{2} (1 \otimes a_0 ^{(c)} )} \Phi_{3,c} (q_c ^{\frac{1}{2}} X_1) V_c (q_c ^{\frac{1}{2}} X_1) ^{-1} .
\end{align}
Using this relation, the $\bar{R}^{(c)}$ part of the R-matrix can be reorganized as
\begin{align} \label{eq:rbar2}
\begin{split}
    \bar{R}^{(c)} &= \sum_{m=0} ^\infty (-1)^m q_c ^{-m} q_3 ^{\frac{m(m-1)}{2}}  \frac{(q_3;q_3)_{\n_c}}{(q_3;q_3)_m (q_3;q_3)_{\n_c -m}}  : \Phi^* _{3,c} (X_1) \, q_3 ^{-m D_{X_1}} \left(\Phi^* _{3,c} (X_1) \right)^{-1}  :  \\
    &= \sum_{m=0} ^\infty (-1)^m q_c ^{-m} q_3 ^{\frac{m(m-1)}{2}}  \frac{(q_3;q_3)_{\n_c}}{(q_3;q_3)_m (q_3;q_3)_{\n_c -m}}  : \left( \Phi^* _{3,c} (X_1) \, q_3 ^{- D_{X_1}} \left(\Phi^* _{3,c} (X_1) \right)^{-1}  \right)^m: \\
    &= :\left( q_c ^{-1} \Phi^* _{3,c} (X_1) \, q_3 ^{- D_{X_1}} \left(\Phi^* _{3,c} (X_1) \right)^{-1}  ;q_3\right)_{\n_c} : \\
    &= :\left( \L(X_1) q_3 ^{-D_{X_1}} ;q_3 \right)_{\n_c} :,
\end{split}
\end{align}
where the $m$-th power is taken out to the whole vertex operator with the help of the normal-ordering in the second line, and we used the $q$-binomial expansion,
\begin{align}
    (z;q)_\n = \sum_{m=0} ^\infty (-z)^m q^{\frac{m(m-1)}{2}} \frac{(q;q)_{\n}}{(q;q)_m (q;q)_{\n-m}},
\end{align}
for the third line. 

Note that \eqref{eq:rbar2} does not satisfy the gauge-invariance condition \eqref{eq:rmatexp} by itself, and therefore is not an R-matrix. In section \ref{subsubsec:miuram2m5}, we will refer to the full R-matrix $R^{(c)} = \bar{R}^{(c)} K^{(c)}$ as the Miura operator, establishing the Miura transformation for the $q$-deformed $W$- and $Y$-algebras using the R-matrix properties. We will show that the expression generalizing \eqref{eq:rbar2} can be obtained for the product of Miura operators, recovering the ones used in \cite{Harada:2021xnm}.

\vspace{4mm}

\subsection{Miura transformation from M2-M5 intersections}
We show that the R-matrices $R^{(c)}$ that we obtained from the M2-M5 intersections indeed follow from mapping the universal R-matrix of $\qta$ to the associated representations. We also derive the Miura transformations for the $q$-deformed $Y$-algebras using the R-matrix properties and their expressions in terms of the intertwiners.

\subsubsection{Universal R-matrix of quantum toroidal algebra of $\fgl(1)$} \label{subsubsec:univr}
The universal R-matrix of the quantum toroidal algebra $\qta$ is an element $\CalR \in \qta \, \widehat\otimes \, \qta$ satisfying
\begin{subequations} \label{eq:univr}
\begin{align}
    &\CalR \Delta (g) = \Delta^{\text{op}} (g) \CalR,\qquad \text{for any } g\in \qta , \label{eq:univr1} \\
        & (\text{id} \otimes \Delta) \CalR = \CalR_{13} \CalR_{12} \label{eq:univr3}\\
    &(\Delta \otimes \text{id} ) \CalR = \CalR_{13} \CalR_{23} \label{eq:univr2}.
\end{align}
\end{subequations}
A direct consequence of \eqref{eq:univr} is the universal Yang-Baxter equation,
\begin{align}\label{eq:univyb}
    \CalR_{12} \CalR_{13} \CalR_{23} = \CalR_{23} \CalR_{13} \CalR_{12} .
\end{align}

Using the Drinfeld double construction of the quantum toroidal algebra of $\fgl(1)$, the universal R-matrix of $\qta$ is obtained by the product of two parts \cite{Garbali:2021qko}:
\begin{align}
    \CalR = \bar{\CalR} \CalK ,
\end{align}
where the Cartan part is
\begin{align}
    \CalK=  e^{\log C \otimes d + d \otimes \log C +\log C^\perp \otimes d^\perp + d^\perp \otimes \log C^\perp} \exp \left[ \sum_{r=1} ^\infty \frac{\k_r}{r} H_{-r} \otimes H_{r} \right],
\end{align}
and the rest is expanded according to the $d^\perp$-grading as
\begin{align}
    \bar{\CalR} = 1 \otimes 1 + \k_1 \sum_{k \in \BZ} E_k \otimes F_{-k}+ \cdots.
\end{align}
The generic expressions for the higher terms are not available as of now, due to the lack of understanding of the Poincar\'{e}-Birkhoff-Witt (PBW) basis of the quantum toroidal algebra $\qta$.\\

However, we do not need the generic expression for the universal R-matrix for our purpose, but only the ones mapped to the representations associated to the M2- and M5-branes. In particular, we claim that the R-matrices $R^{(c)}$, \eqref{eq:transr} and \eqref{eq:rmatexp}, that we established from the M2-M5 intersections are recovered when the universal R-matrix is represented on $\text{M2}_{0,0,1} \otimes \text{M5}_{c} ^{\mathbf{(1,0)}}$, justifying the terminology. 

Namely, it turns out that
\begin{align} \label{eq:rrep}
    R^{(c)} = (\r_{\text{M2}_{0,0,1}} \otimes \r_{\text{M5}_{c}^{\mathbf{(1,0)}}} ) \CalR, \qquad c\in \{1,2,3\},
\end{align}
as we now show. For each $c\in \{1,2,3\}$, it is clear from \eqref{eq:univr1} and \eqref{eq:rrep} that the right hand side (which we just denote by $R^{(c)}$ for brevity) should satisfy
\begin{align} \label{eq:rmacond1}
    R^{(c)} \D_{\text{M2}_{0,0,1}, \text{M5}_c ^{\mathbf{(1,0)}}} (g) = \D^\text{op}_{\text{M2}_{0,0,1}, \text{M5}_c ^{\mathbf{(1,0)}}} (g) R^{(c)} ,\qquad \text{for any } g\in \qta.
\end{align}
Now it is straightforward to see, on one hand,
\begin{align} \label{eq:rmacond2}
    K^{(c)} = (\r_{\text{M2}_{0,0,1}} \otimes \r_{\text{M5}_{c}^{\mathbf{(1,0)}}} ) \CalK,
\end{align}
since $\r_{\text{M2}_{0,0,1}} (C,C^\perp) = (1,1)$ and $\r_{\text{M5}_{c} ^{\mathbf{(1,0)}} } (C,C^\perp) = (q_c ^{\frac{1}{2}},1)$, in particular. On the other hand, it is also obvious that
\begin{align}  \label{eq:rmacond3}
\begin{split}
     (\r_{\text{M2}_{0,0,1}} \otimes \r_{\text{M5}_{c}^{\mathbf{(1,0)}}} ) \bar{\CalR} &= 1 \otimes 1 + \k_1 \sum_{k \in \BZ}  (\r_{\text{M2}_{0,0,1}} \otimes \r_{\text{M5}_{c}^{\mathbf{(1,0)}}} )  (E_k \otimes F_{-k}) + \cdots \\
     &= 1 - q_c ^{-1} \xi_c (X_1) Z_1 + \cdots
\end{split}
\end{align}
precisely matches with the first two terms of the $d^\perp$-graded expansion of $\bar{R}^{(c)}$. In appendix \ref{sec:solver}, we show that there is unique solution, up to an overall scaling, for $R^{(c)} = \bar{R}^{(c)} K^{(c)}$ satisfying \eqref{eq:rmacond1}, \eqref{eq:rmacond2}, and \eqref{eq:rmacond3}. It is none other than our R-matrices $R^{(c)}$ obtained in \eqref{eq:transr} and \eqref{eq:rmatexp} associated to the M2-M5 intersections.

\subsubsection{Miura transformation from multiple M2-M5 intersections} \label{subsubsec:miuram2m5}
Having verified that the M2-M5 intersections correspond to the R-matrices of the quantum toroidal algebra, we are now ready to construct the Miura transformation for the $q$-deformed $Y$-algebras using the R-matrix properties. The crucial point is that the concatenation of the R-matrices under coproduct operations, derived from the axiom \eqref{eq:univr3}, translates into the Miura transformation being the product of Miura operators.

 Let us take $ L+M+N$ M5-branes which share the holomorphic support $C^{(1,0)} = \BC^\times _X$ but are assigned with arbitrary choices for their topological supports, dictated by $c: I \mapsto c_I \in \{1,2,3\}$, $I=1,2,\cdots, L+M+N$. Let us set $\vert c^{-1} (1) \vert = L$, $\vert c^{-1} (2) \vert = M$, and $\vert c^{-1} (3) \vert = N$. Let us consider the fused surface defect algebra $\text{M5}_{L,M,N} ^{\mathbf{(1,0)}}$ obtained by the surjective homomorphism
 \begin{align}
     \r_{\text{M5}_{L,M,N} ^{\mathbf{(1,0)}}} = \left( \r_{\text{M5}_{c_1} ^{\mathbf{(1,0)}}} \otimes \cdots \otimes \r_{\text{M5}_{c_{L+M+N}} ^{\mathbf{(1,0)}}} \right)\Delta^{L+M+N-1} : \qta \twoheadrightarrow \text{M5}_{L,M,N} ^{\mathbf{(1,0)}}.
 \end{align}
Although the construction appears to depend on the ordering of the M5-branes in the tensor product, we will demonstrate in section \ref{subsubsec:orderingyb} that changing the ordering results in an isomorphic algebra. Assuming this ordering-independence, we will simply denote this algebra as $\text{M5}_{L,M,N} ^{\mathbf{(1,0)}}$.
 
\begin{figure}[h!]\centering
\begin{tikzpicture}
\draw[line width=0.3mm] (-1,1.5) -- (5,1.5) -- (3,-0.5) -- (-3,-0.5) 
      -- cycle;

\draw[line width=0.3mm] (-1,0.8) -- (5,0.8) -- (3,-1.2) -- (-3,-1.2) 
      -- cycle;

\draw[line width=0.3mm] (-1,-1.5) -- (5,-1.5) -- (3,-3.5) -- (-3,-3.5) 
      -- cycle;

\draw[line width=0.3mm, loosely dotted, ultra thick] (2.5,0.5) -- (2.5,-2.5);

\draw[line width=0.3mm, loosely dotted, ultra thick] (2.5,-2.5) -- (2.5,-3.5);

\draw[line width=0.3mm] (2.5,2.5) -- (2.5,0.5);
\draw[line width=0.3mm] (2.5,-3.5) -- (2.5,-4.5);

\filldraw[black] (2.5,0.5) circle (2pt) node[anchor=west]{$R_{L+M+N} ^{(c_{L+M+N})}$};

\filldraw[black] (2.5,-0.2) circle (2pt) node[anchor=west]{$R_{L+M+N-1} ^{(c_{L+M+N-1})}$};

\filldraw[black] (2.5,-2.5) circle (2pt) node[anchor=west]{$R_1 ^{(c_1)}$};

 \node at (1,0.5) {$\times$};
 \node at (1,-2.9) {$0$};

  \node at (1,-0.2) {$\times$};
    \node at (1,-2.5) {$\times$};

   \node at (-3.8,-0.4) {$\mathcal{S}_{c_{L+M+N}} ^{\mathbf{(1,0)}}$};
      \node at (-3.9,-1.1) {$\mathcal{S}_{c_{L+M+N-1}} ^{\mathbf{(1,0)}}$};
      \node at (-3.6,-3.4) {$\mathcal{S}_{c_1} ^{\mathbf{(1,0)}}$};
\draw[line width=0.3mm, loosely dotted, ultra thick] (-3.6,-1.8) -- (-3.6,-2.7);
      
  \node at (2.5,-4.8) {$\mathcal{L}_{0,0,1} $};
    \node at (2.5,2.8) {$\mathcal{L}_{0,0,1} $};

\draw[->] (-3,2) -- (-3,3);
\draw[->] (-3,2) -- (-2,2);
\draw[->] (-3,2) -- (-3.5,1.4);

\node at (-2.7,1.6) {$\BC^\times _X$};
\node at (-2.6,2.9) {$\BR_t$};
    
\end{tikzpicture} 
\caption{Miura transformation for the $q$-deformed $Y$-algebra from the intersection between a single M2-brane and multiple M5-branes}\label{fig:miurat}
\end{figure}
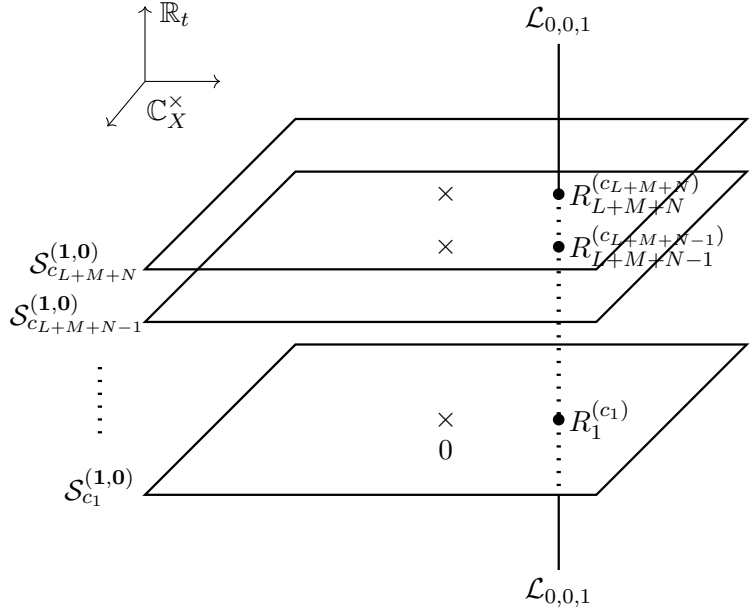

Then, we consider the R-matrix between the representations $\text{M2}_{0,0,1}$ and $\text{M5}_{L,M,N} ^{\mathbf{(1,0)}}$. By repetitively applying \eqref{eq:univr3}, it is given by the product of the basic R-matrices \eqref{eq:rmatexp},
\begin{align} \label{eq:rmatwhole}
    R_{\text{M2}_{0,0,1} , \text{M5}_{L,M,N} ^{\mathbf{(1,0)}}} = R_{L+M+N} ^{(c_{L+M+N})} \cdots R_2 ^{(c_2)} R_1 ^{(c_1)} \in \text{M2}_{0,0,1} \widehat{\otimes} \, \text{M5}_{L,M,N} ^{\mathbf{(1,0)}},
\end{align}
where the product is taken as elements in $\text{M2}_{0,0,1}$. Such a decomposition of the R-matrix into the product of basic R-matrices can be visualized as a single M2-brane passing through multiple M5-branes, where a basic R-matrix is assigned at each intersection point (see Figure \ref{fig:miurat}). 

Now, we may expand the product \eqref{eq:rmatwhole} as a series in the difference operator $Z_1 = q_3 ^{-D_{X_1}} \in \text{M2}_{0,0,1}$. Explicitly, the expansion gives
\begin{align} \label{eq:miuraexpand}
    R_{\text{M2}_{0,0,1 }, \text{M5}_{L,M,N}^{\mathbf{(1,0)}}}= \sum_{m=0} ^\infty (-1)^m T_m (X_1) q_3 ^{-m D_{X_1}}.
\end{align}
By construction, when each coefficient $T_m (X_1)$ of $Z_1 ^m$ is further expanded as a Laurent series in $X_1 \in \text{M2}_{0,0,1}$, the Laurent coefficients are valued in $\text{M5}_{L,M,N} ^{\mathbf{(1,0)}}$ since $ R_{\text{M2}_{0,0,1 }, \text{M5}_{L,M,N}^{\mathbf{(1,0)}}}$ is the R-matrix. We will show in section \ref{sssec:screeningsM2} that the coefficients in fact generate the whole algebra $\text{M5}_{L,M,N} ^{\mathbf{(1,0)}}$, verifying that the R-matrix $ R_{\text{M2}_{0,0,1 }, \text{M5}_{L,M,N}^{\mathbf{(1,0)}}}$ indeed provides the Miura transformation for the $q$-deformed $Y$-algebra $\text{M5}_{L,M,N} ^{\mathbf{(1,0)}}$. Moreover, we will also show that they are completely characterized as commutants of certain screening charges in the tensor product of the $q$-boson algebras, giving an alternative definition of the $q$-deformed $Y$-algebra.

Let us remark that in the case of parallel M5-branes with $L=M=0$, the expansion \eqref{eq:miuraexpand} of the R-matrix truncates, since each $R^{(3)}$ is only first-order in $Z_1 = q_3^{-D_{X_1}}$. As a result, there are only $N$ non-trivial generating currents $T_m(X_1)$, with $m=1,2,\cdots, N$. We will see this is precisely the specialization that yields the $q$-deformed $W$-algebra.\\

The R-matrix \eqref{eq:rmatwhole} admits contour integral presentations. First, let us derive the contour integral expression where the integrand is normal-ordered. Substituting the contour integral presentations \eqref{eq:rcontnormal} and \eqref{eq:rcontnormal3} for the basic R-matrices into \eqref{eq:rmatwhole}, we get 
\begin{align}\label{eq:generalRmatrix}
\begin{split}
    &R_{\text{M2}_{0,0,1} , \text{M5}_{L,M,N} ^{\mathbf{(1,0)}}} = (-1)^{L+M+N} q_3 ^{-N} \prod_{I=1} ^{L+M+N} (q_3;q_3)_{\n_{c_I}}  e^{-\sum_{I=1} ^{L+M+N} \frac{\log q_{c_I+1}  \log q_{c_I-1} }{\log q_3} a_0 ^{[I]}}  \\
    & \qquad \times \oint \prod_{I=1} ^{L+M+N} \frac{dX^{(I)}}{2\pi i X^{(I)}} 
\prod_{I \in c^{-1}\{1,2\} } \frac{\left( q_{\bar{c}_I} ^{-1}  \frac{X^{(I)}}{X^{(I+1)}} ;q_3 \right)_\infty }{\left(  \frac{X^{(I)}}{X^{(I+1)}} ;q_3 \right)_\infty} \prod_{I\in c^{-1}(3) } \frac{1}{\left( 1- \frac{ X^{(I+1)}}{X^{(I)}} \right)\left( 1- \frac{q_3 ^{-1}   X^{(I+1)} }{X^{(I)}} \right)} \\
&\qquad \times e^{-\sum_{I \in c^{-1} \{1,2\}} \frac{\log q_{\bar{c}_I}}{ \log q_3} \log \frac{X^{(I)}}{ X^{(I+1)}} } \prod_{I=1} ^{L+M+N} : \Phi^* _I \left( X^{(I+1)} \prod_{J=I+1} ^{L+M+N} q_{c_J} ^{\frac{1}{2}} \right)  \Phi_I \left( X^{(I)} \prod_{J=I} ^{L+M+N} q_{c_J} ^{\frac{1}{2}} \right) :  \\
&\qquad \times Z_1 ^{-\sum_{I=1} ^{L+M+N} \frac{\log \frac{X^{(I)}}{ X_1 }}{\log q_3} } \prod_{I=1} ^{L+M+N} q_{c_I} ^{\left.\frac{d}{2} \right\vert_{\text{M2}_{0,0,1}}}
\end{split}
\end{align}
where we introduced $L+M+N$ integration variables $X^{(I)}$, $I=1,2,\cdots, L+M+N$ (these should not be confused with $X_1$, which is a generator of $\text{M2}_{0,0,1}$). For brevity, we also used the notation $X^{(L+M+N+1)} \equiv X_1$ and denoted $\Phi_{3,c_I} \equiv \Phi_I$ and $\Phi_{3,c_I} ^* \equiv \Phi^* _I$, while using the abbreviation $a_r ^{[I]} \equiv \left(a_r ^{(c_I)}\right)^{[I]}$ for the $I$-th $q$-boson modes. There is no issue of ordering in the product of vertex operators, since each term only involves a single $q$-boson representation. The integration is performed starting from $X^{(1)}$ up to $X^{(L+M+N)}$, picking up the residues from all the simple poles at $X^{(I)} = X^{(I+1)} q_3 ^{-k}$, where $k\geq 0$ if $I \in c^{-1} \{1,2\}$ and $k= 0,1$ if $I \in c^{-1} (3)$.

We may partially undo the normal-ordering in the integrand, by using the normal-ordering relation \eqref{eq:noint} between the intertwiners when $I \in c^{-1} \{1,2\}$. As a result, we can get rid of the ratio of $q$-Pochhammer symbols and obtain
\begin{align}
\begin{split}
    &R_{\text{M2}_{0,0,1} , \text{M5}_{L,M,N} ^{\mathbf{(1,0)}}} = (-1)^{L+M+N} q_3 ^{-N} \prod_{I=1} ^{L+M+N} (q_3;q_3)_{\n_{c_I}}  e^{-\sum_{I=1} ^{L+M+N} \frac{\log q_{c_I+1} \log q_{c_I-1} }{\log q_3} a_0 ^{[I]}}  \\
    & \qquad \times \oint \prod_{I=1} ^{L+M+N} \frac{dX^{(I)}}{2\pi i X^{(I)}} 
\prod_{I\in c^{-1}(3) } \frac{: \Phi^* _I \left( X^{(I+1)} \prod_{J=I+1} ^{L+M+N} q_{c_J} ^{\frac{1}{2}} \right)  \Phi_I \left( X^{(I)} \prod_{J=I} ^{L+M+N} q_{c_J} ^{\frac{1}{2}} \right)  :}{\left( 1- \frac{ X^{(I+1)}}{X^{(I)}} \right)\left( 1- \frac{q_3 ^{-1}   X^{(I+1)} }{X^{(I)}} \right)}  \\
&\qquad \times e^{\frac{L+M}{2}\frac{\log q_1 \log q_{2}}{ \log q_3}  } \prod_{I\in c^{-1}\{1,2\}}  \Phi^* _I \left( X^{(I+1)} \prod_{J=I+1} ^{L+M+N} q_{c_J} ^{\frac{1}{2}} \right)  \Phi_I \left( X^{(I)} \prod_{J=I} ^{L+M+N} q_{c_J} ^{\frac{1}{2}} \right)   \\
&\qquad \times Z_1 ^{-\sum_{I=1} ^{L+M+N} \frac{\log \frac{X^{(I)}}{ X_1 }}{\log q_3} } \prod_{I=1} ^{L+M+N} q_{c_I} ^{\left.\frac{d}{2} \right\vert_{\text{M2}_{0,0,1}}}.
\end{split}
\end{align}
This contour integral presentation explicitly shows that the R-matrix can be reconstructed from multiple gauge-invariant M2-M5 intersections, each of which either splits into two with ends (for $I \in c^{-1} \{1,2\}$) or cannot be split (for $I \in c^{-1} (3)$) across the $I$-th M5-brane. \\

\paragraph{Remark} 
Recall that the basic R-matrices \eqref{eq:transr} and \eqref{eq:rmatexp} are decomposed into the product $R^{(c)} = \bar{R}^{(c)} K^{(c)}$. For the Miura transformation \eqref{eq:rmatwhole}, we can push all the Cartan parts to the right and express it as
\begin{align}
     R_{\text{M2}_{0,0,1} , \text{M5}_{L,M,N} ^{\mathbf{(1,0)}}} =  \bar{R}_{\text{M2}_{0,0,1} , \text{M5}_{L,M,N} ^{\mathbf{(1,0)}}} K_{\text{M2}_{0,0,1} , \text{M5}_{L,M,N} ^{\mathbf{(1,0)}}},
\end{align}
where 
\begin{align}
\begin{split}
    K_{\text{M2}_{0,0,1} , \text{M5}_{L,M,N} ^{\mathbf{(1,0)}}}&= K^{(c_{L+M+N})} _{L+M+N} \cdots K^{(c_1)} _1 \\
    & = \prod_{I=1} ^{L+M+N} V_{c_I} ^{[I]} \left( X_1 \prod_{J=I} ^{L+M+N} q_{c_J} ^{\frac{1}{2}} \right) \times \prod_{I=1} ^{L+M+N} q_{c_I} ^{\left.\frac{d}{2} \right\vert_{\text{M2}_{0,0,1}}}.
\end{split}
\end{align}
Here, we remind that $V_{c_I} ^{[I]} (X) = \prod_{j=0} ^\infty \varphi^{+,[I]} _{c_I} (q_3 ^{-j} X ) = \exp \left[ \sum_{r=1} ^\infty \frac{\k_r}{r} \frac{X^{-r}}{1-q_3 ^r} \otimes \frac{a_r ^{[I]}}{q_{c_I} ^{r/2} -q_{c_I} ^{-r/2}} \right]$. Then, we may separate out $\bar{R} _{\text{M2}_{0,0,1} , \text{M5}_{L,M,N} ^{\mathbf{(1,0)}}}$ from the whole R-matrix \eqref{eq:rmatwhole} as
\begin{align} \label{eq:partialmiura}
    \bar{R}_{\text{M2}_{0,0,1} , \text{M5}_{L,M,N} ^{\mathbf{(1,0)}}}= \bar{R}_{L+M+N} \cdots \bar{R}_1 ,
\end{align}
where each $\bar{R}_I$, $I=1,2,\cdots, L+M+N$, is computed to be
\begin{align} \label{eq:partialr}
\begin{split}
    \bar{R}_I &= \sum_{m=0} ^\infty (-1)^m q_{c_I} ^{-m} q_3 ^{\frac{m(m-1)}{2}} \frac{(q_3;q_3)_{\n_{c_I}}}{(q_3;q_3)_m (q_3;q_3)_{\n_{c_I} -m}}  : f_I (X_1) q_3^{-m D_{X_1}} f_I (X_1)^{-1} : \\
    &= \sum_{m=0} ^\infty (-1)^m q_{c_I} ^{-m} q_3 ^{\frac{m(m-1)}{2}} \frac{(q_3;q_3)_{\n_{c_I}}}{(q_3;q_3)_m (q_3;q_3)_{\n_{c_I} -m}}  : \left( f_I (X_1) q_3^{- D_{X_1}} f_I (X_1)^{-1} \right)^m : \\
    &= :\left( q_{c_I} ^{-1} f_I (X_1) q_3 ^{-D_{X_1}} f_I (X_1)^{-1} ;q_3 \right)_{\n_{c_I}}: \\
    & \equiv :\left(\L_I (X_1) q_3 ^{-D_{X_1}} ;q_3 \right)_{\n_{c_I}}:,
\end{split}
\end{align}
where we defined
\begin{align}
    f_I (X_1) :=  \Phi^* _I \left(X_1 \prod_{J=I+1} ^{L+M+N }q_{c_J} ^{\frac{1}{2}} \right) \prod_{J=I+1} ^{L+M+N} V_{c_J} ^{[J]} \left( X_1 \prod_{K=J} ^{L+M+N} q_{c_K} ^{\frac{1}{2}}  \right).
\end{align}
and
\begin{align}\label{eq:Lambdaops}
\begin{split}
    \L_I (X_1) &:= q_{c_I} ^{-1} :f_I (X_1) f_I (q_3^{-1} X_1)^{-1}: \\
    &= q_{c_I} ^{-1} e^{- \frac{\prod_{a=1}^3 \log q_a}{(\log q_{c_I})^2 } a_0 ^{[I]}} \exp \left[ \sum_{r=1} ^\infty \frac{\k_r}{r} \frac{\prod_{J=I} ^{L+M+N} q_{c_J} ^{r/2}}{(q_{c_I} ^{r/2} - q_{c_I} ^{-r/2})^2} a_{-r} ^{[I]} X_1 ^r \right] \\
    & \times \exp \left[ \sum_{r=1} ^\infty \frac{\k_r}{r} \frac{\prod_{J=I+1} ^{L+M+N} q_{c_J} ^{-r/2}}{(q_{c_I}^{r/2} - q_{c_I} ^{-r/2})^2} a_r ^{[I]} X_1 ^{-r} + \sum_{r=1} ^\infty \frac{\k_r}{r} \sum_{J=I+1} ^{L+M+N}  \frac{\prod_{K=J} ^{L+M+N} q_{c_K} ^{-r/2} }{q_{c_J} ^{r/2} - q_{c_J} ^{-r/2}} a_r ^{[J]} X_1 ^{-r} \right].
\end{split}
\end{align}
The operator \eqref{eq:partialr} is precisely what was suggested to be the Miura operator for the $q$-deformed $Y$-algebra in \cite{Harada:2021xnm}, after properly adjusting the convention. In this sense, the Miura transformation \eqref{eq:rmatwhole} that we established is related to the one there, realized by \eqref{eq:partialmiura}, by a multiplication of the vertex operators $V_{c_I} ^{[I]}$ to the generating currents $T_m$. Note, however, that neither \eqref{eq:partialmiura} nor individual \eqref{eq:partialr} is an R-matrix of the quantum toroidal algebra $\qta$, unlike our Miura transformation \eqref{eq:rmatwhole} and Miura operators \eqref{eq:rmatexp}. This is why \eqref{eq:partialr} involves not only the $I$-th $q$-boson but also the \textit{mixing} from all the $J$-th $q$-bosons with $J>I$.

\vspace{4mm}

\subsubsection{Screening charges from M2-branes between M5-branes}
\label{sssec:screeningsM2}

To fully characterize $\text{M5}_{L,M,N} ^{\mathbf{(1,0)}}$ as a subalgebra of the tensor product of $q$-bosons, we will identify $\text{M5}_{L,M,N} ^{\mathbf{(1,0)}}$ as commutants of a certain set of operators called screening charges. First, we will construct the screening charges by the M2-brane placed between each adjacent pair of M5-branes in the tensor product, and show that they commute with all the generators of $\text{M5}_{L,M,N} ^{\mathbf{(1,0)}}$. Then, we show that there is no more operator that commutes with all the generators produced by the product of Miura operators. We thereby prove that the product of Miura operators indeed gives the Miura transformation for $\text{M5}_{L,M,N} ^{\mathbf{(1,0)}}$, the $q$-deformed $Y$-algebra.

\begin{figure}[h!]\centering
\resizebox{.4\textwidth}{!}{
\begin{tikzpicture}
\draw[line width=0.3mm] (-1,1.5) -- (5,1.5) -- (3,-0.5) -- (-3,-0.5) 
      -- cycle;

\draw[line width=0.3mm] (-1,-0.1) -- (5,-0.1) -- (3,-2.1) -- (-3,-2.1) 
      -- cycle;

\draw[line width=0.3mm, loosely dotted, ultra thick] (2.5,0.5) -- (2.5,-0.5);
\draw[line width=0.3mm] (2.5,-0.5) -- (2.5,-1.1);

\filldraw[black] (2.5,0.5) circle (2pt) node[anchor=west]{$\Phi_{c\pm 1 ,c}$};

 \node at (2,-0.3) {$\CalL_{c\pm 1}$};

\filldraw[black] (2.5,-1.1) circle (2pt) node[anchor=west]{$\Phi ^*_{c\pm 1 , c}$};

 \node at (1,0.5) {$\times$};
 \node at (1,0.1) {$0$};

  \node at (1,-1.1) {$\times$};
 \node at (1,-1.5) {$0$};

   \node at (-3.6,-0.4) {$\mathcal{S}_{c} ^{\mathbf{(1,0)}}$};
      \node at (-3.6,-2) {$\mathcal{S}_{c} ^{\mathbf{(1,0)}}$};

\draw[->] (-3,2) -- (-3,3);
\draw[->] (-3,2) -- (-2,2);
\draw[->] (-3,2) -- (-3.5,1.4);

\node at (-2.7,1.6) {$\BC^\times _X$};
\node at (-2.6,2.9) {$\BR_t$};
    
\end{tikzpicture}} \hspace{10mm}  \resizebox{.4\textwidth}{!}{
\begin{tikzpicture}
\draw[line width=0.3mm] (-1,1.5) -- (5,1.5) -- (3,-0.5) -- (-3,-0.5) 
      -- cycle;

\draw[line width=0.3mm] (-1,-0.1) -- (5,-0.1) -- (3,-2.1) -- (-3,-2.1) 
      -- cycle;

\draw[line width=0.3mm, loosely dotted, ultra thick] (2.5,0.5) -- (2.5,-0.5);
\draw[line width=0.3mm] (2.5,-0.5) -- (2.5,-1.1);

 \node at (2,-0.3) {$\CalL_{c\mp  1}$};

\filldraw[black] (2.5,0.5) circle (2pt) node[anchor=west]{$\Phi_{c\mp 1 , c\pm 1}$};

\filldraw[black] (2.5,-1.1) circle (2pt) node[anchor=west]{$\Phi^*_{c\mp 1 , c}$};

 \node at (1,0.5) {$\times$};
 \node at (1,0.1) {$0$};

  \node at (1,-1.1) {$\times$};
 \node at (1,-1.5) {$0$};

   \node at (-3.6,-0.4) {$\mathcal{S}_{c\pm 1} ^{\mathbf{(1,0)}}$};
      \node at (-3.6,-2) {$\mathcal{S}_{c} ^{\mathbf{(1,0)}}$};

\end{tikzpicture}}
\caption{Screening charges from M2-branes stretched between M5-branes} \label{fig:screen} 
\end{figure}
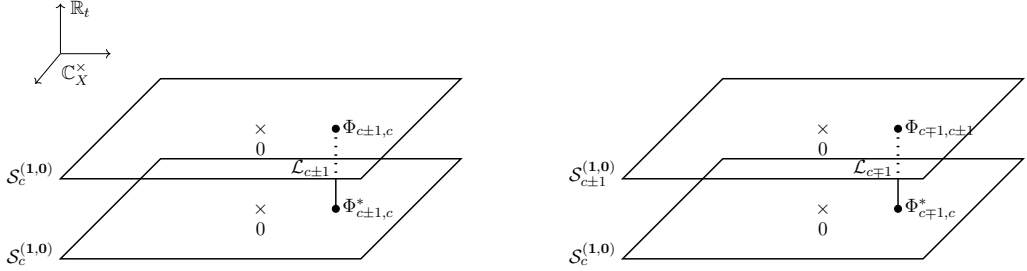

Take any consecutive pair; the $i$-th and the $(i+1)$-th M5-branes. The only two possibility for the topological part of their support are the following: 1) the two share two topological planes $\BR^2 _{\ve_{c+1}} \times \BR^2 _{\ve_{c-1}}$; 2) the $i$-th M5-brane is supported on $\BR^2 _{\ve_{c+1}} \times \BR^2 _{\ve_{c-1}}$ and the $(i+1)$-th M5-brane is supported on $\BR^2 _{\ve_{c}} \times \BR^2 _{\ve_{c\mp 1}}$, so that the two share only one topological plane $\BR^2 _{\ve_{c \mp 1}}$. In the first case, the M2-branes supported on $\BR^2 _{\ve_{c\pm 1}} \times \{\text{finite interval} \} $ can be placed between the two M5-branes (in the $\BR_t$-direction), ending on the two M5-branes from above and from below, respectively. In the second case, the M2-brane supported on $\BR^2 _{\ve_{c\mp 1}} \times \{\text{finite interval} \} $ can be placed between the M5-branes, ending on the two from above and from below, respectively. See Figure \ref{fig:screen}.

In section \ref{subsubsec:intert}, we have associated the intertwiners $\Phi$ and $\Phi^*$ of the quantum toroidal algebra of $\fgl(1)$ to the M2-M5 intersections with ends. We construct the screening charges by composing them along the finite support of the M2-brane. Specifically, for the first case where the two adjacent M5-branes share two topological planes $\BR^2 _{\ve_{c+1}} \times \BR^2 _{\ve_{c-1}}$, there are two screening charges obtained as
\begin{align} \label{eq:screen1}
\begin{split}
    &S_{c, c} ^{c\pm 1}  := \s \circ\left( \Phi_{c\pm 1,c}  \otimes \Phi ^*_{c\pm 1,c} \right) \in \text{M5}_{c_{i}=c} ^{\mathbf{(1,0)}}  \otimes \text{M5}^{\mathbf{(1,0)}}_{c_{i+1} =c}
\end{split}
\end{align}
The term in the parenthesis should be understood as the composition of $\Phi_{c\pm 1,c}$ and $\Phi^* _{c \pm 1 , c}$ as elements in $\text{Hom}(\CalV_{c\pm 1} (u) ,\BC)$ and $\text{Hom}(\BC,\CalV_{c\pm 1} (u))$, respectively, so that $\Phi_{c\pm 1,c}  \otimes \Phi ^*_{c\pm 1,c} \in \text{M5}^{\mathbf{(1,0)}}_{c_{i+1} =c} \otimes \text{M5}^{\mathbf{(1,0)}}_{c_{i} =c} $. Then, the exchange operator $\s$ rearranges the ordering of the two $q$-boson representations in the tensor product as above. 

Inserting the complete set of the orthonormal basis states, $\text{id} = \sum_{i\in \BZ} [u]_i ^{(c\pm 1)} \otimes [u]^{(c\pm 1)*} _i \in \text{End}(\CalV_{c\pm 1} (u))$ ($[u]_j ^{(c\pm 1)*} ([u]_i^{(c\pm 1)}) = \d_{i,j}$), we can express the above screening charges \eqref{eq:screen1} as
\begin{align}\label{eq:screeningdef}
    S^{c\pm 1} _{c,c} = \sum_{i\in \BZ} (\Phi^* _{c\pm 1,c} )_i \otimes \left(\Phi _{c\pm 1,c} \right)_i \equiv \sum_{i\in \BZ} \ES_{c,c} ^{c\pm 1} (u q_{c\pm 1} ^i).
\end{align}
The summand $\ES_{c,c} ^{c\pm 1} (X)$ is called the screening current. The summation over integers can be thought of as a trace over the Hilbert space $\CalV_{c\pm 1} (u) $ of the line defect worldline theory, which was to be expected since the M2-brane is stretched between the M5-branes. Inserting the components of the intertwiners, the two screening charges are explicitly given by
\begin{align}\label{eq:screeningdef2}
\begin{split}
    S_{c,c} ^{c\pm 1} &= q_{c\mp 1}^{-d^\perp \otimes 1 + 1\otimes d^\perp} \sum_{i\in \BZ} (u q_{c\pm 1} ^i)^{-\frac{\log q_{c\mp 1}}{\log q_c} (a_0 ^{(c)} \otimes 1 - 1 \otimes a_0 ^{(c)} )- \frac{\log q_{c\mp 1}}{\log q_{c\pm 1}}}  \\
    & \qquad\qquad\qquad \times \exp \left[ \sum_{r=1} ^\infty \frac{1}{r} \frac{q_{c \mp 1} ^{r/2} - q_{c \mp 1} ^{-r/2}}{q_c ^{r/2} - q_c ^{-r/2}} q_{c\mp 1} ^{-r/2} ( a_{-r} ^{(c)} \otimes 1 -   1 \otimes q_c ^{-r/2} a^{(c)} _{-r}  ) (u q_{c\pm 1} ^i)^r \right] \\
    &\qquad\qquad\qquad \times \exp \left[- \sum_{r=1} ^\infty \frac{1}{r} \frac{q_{c\mp 1} ^{r/2} - q_{c\mp 1} ^{-r/2}}{q_c ^{r/2} - q_c ^{-r/2}}q_{c\mp 1 }^{r/2} ( q_{c } ^{r/2} a_r ^{(c)} \otimes 1 - 1 \otimes  a_r ^{(c)}) (u q_{c\pm 1} ^i)^{-r}\right],
\end{split}
\end{align}
This precisely recovers the bosonic screening charges that was used to define the $q$-deformed $W$-algebras, up to the decoupling of a single $q$-boson \cite{Feigin:1995sf,2015arXiv151208779B}.

On the other hand, in the second case where the adjacent M5-branes share only one topological plane $\BR^2 _{\ve_{c\mp 1}}$ in their support, there is a single screening charge obtained as
\begin{align}\label{eq:screeningdef3}
        &S_{c, c \pm 1} ^{c\mp 1}  :=  \s \circ \left(\Phi_{c \mp 1,c\pm 1} \otimes \Phi ^* _{c \mp 1,c} \right)  \in \text{M5}_{c_i=c} ^{\mathbf{(1,0)}} \otimes \text{M5}^{\mathbf{(1,0)}}_{c_{i+1} = c\pm 1}.
\end{align}
Inserting the complete set of basis $\text{id} = \sum_{i\in \BZ} [u]_i ^{(c\mp 1)} \otimes [u]_i ^{(c\mp 1)*}$, we can express it into
\begin{align}
    S_{c, c \pm 1} ^{c\mp 1} = \sum_{i\in \BZ} (\Phi ^* _{c \mp 1,c}  )_i \otimes (\Phi_{c \mp 1,c\pm 1})_i \equiv \sum_{i\in \BZ} \ES_{c,c\pm 1}^{c\mp 1}(u q_{c\mp 1} ^i).
\end{align}
Explicitly, we have
\begin{align}\label{eq:screeningdef3}
\begin{split}
    S_{c, c \pm 1} ^{c\mp 1} &= q_{c\pm 1}^{-d^\perp \otimes 1} q_c^{1\otimes d^\perp }  \sum_{i\in \BZ} (u q_{c\mp 1} ^i )^{-\frac{\log q_{c\pm 1}}{\log q_c} (a_0 ^{(c)}\otimes 1) + \frac{\log q_c}{\log q_{c\pm 1}    } (1 \otimes a_0 ^{(c\pm 1)}  )   - \frac{\log q_c}{\log q_{c\mp 1 }} } \\
    & \times \exp \left[ \sum_{r=1} ^\infty \frac{1}{r} \left( \frac{q_{c\pm 1} ^{r/2} -q_{c\pm 1} ^{-r/2}}{q_{c} ^{r/2} -q_{c} ^{-r/2}} q_{c\pm 1} ^{-r/2} ( a_{-r } ^{(c)} \otimes 1) - \frac{ q_{c} ^{r/2} -q_{c} ^{-r/2} }{q_{c\pm 1} ^{r/2} -q_{c\pm 1} ^{-r/2}} q_{c\mp 1} ^{r/2}( 1\otimes  a_{-r} ^{(c\pm 1)} ) \right) (u q_{c\mp 1} ^i)^r \right] \\
    &\times \exp \left[- \sum_{r=1} ^\infty \frac{1}{r} \left(  \frac{q_{c\pm 1} ^{r/2} -q_{c\pm 1} ^{-r/2}}{q_{c} ^{r/2} -q_{c} ^{-r/2}} q_{c\mp 1} ^{-r/2} (a_{r } ^{(c)} \otimes 1 ) - \frac{ q_{c} ^{r/2} -q_{c} ^{-r/2} }{q_{c\pm 1} ^{r/2} -q_{c\pm 1} ^{-r/2}} q_{c} ^{r/2}( 1 \otimes a_{r} ^{(c\pm 1)} ) \right) (u q_{c\mp 1} ^i)^{-r} \right] .
\end{split}
\end{align}
Note that this precisely matches with the fermionic screening charge obtained in \cite{2015arXiv151208779B}.\\

The gauge-invariance conditions \eqref{eq:intconst} and \eqref{eq:dintconst} for the intertwiners directly imply that the screening charges commute with the generators of the fused surface defect algebra. Namely,
\begin{align} \label{eq:screencomm}
    \left[S_{c_i,c_{i+1}} ^{c_{\text{M2}}}  ,  \D_{\text{M5}_{c_i}^{\mathbf{(1,0)}}, \text{M5}_{c_{i+1}}^{\mathbf{(1,0)}}  }(g) \right] = 0, \qquad \text{for any } g\in \qta,
\end{align}
where $c_{\text{M2}} = c+ 1 \; \text{or} \; c-1$ if $(c_i,c_{i+1}) = (c,c)$ and $c_{\text{M2}} = c\mp 1$ if $(c_i,c_{i+1}) = (c,c\pm 1)$. Indeed, a straightforward computation gives
\begin{align}
\begin{split}
    &\D_{\text{M5}_{c_i}^{\mathbf{(1,0)}}, \text{M5}_{c_{i+1}}^{\mathbf{(1,0)}}  }(g) S_{c_i ,c_{i+1}} ^{c_{\text{M2}}} \\
    &=  \s  \left( \r_{\text{M5}_{c_{i+1}}^{\mathbf{(1,0)}}} (g_{(2)}) \otimes \r_{\text{M5}_{c_{i+1}}^{\mathbf{(1,0)}}} (g_{(1)})  \right) \left( \Phi_{c_{\text{M2}}, c_{i+1}} \otimes \Phi^* _{c_{\text{M2}}, c_i} \right) \\
    & =  \s \left[\Phi_{c_{\text{M2}},c_{i+1} } \left( \r_{\text{M2}} \otimes \r_{\text{M5}_{c_{i+1}}^{\mathbf{(1,0)}}} \otimes  \r_{\text{M5}_{c_{i}}^{\mathbf{(1,0)}}} \right)  ( g_{(2),(1)} \otimes g_{(2),(2)} \otimes g_{(1)}) \Phi^* _{c_{\text{M2}}, c_i} \right] \\
    & = \s \left[\Phi_{c_{\text{M2}},c_{i+1} } \left( \r_{\text{M2}} \otimes \r_{\text{M5}_{c_{i+1}}^{\mathbf{(1,0)}}} \otimes  \r_{\text{M5}_{c_{i}}^{\mathbf{(1,0)}}} \right) (\text{id}\otimes \s) (\s \otimes \text{id}) (g_{(1)} \otimes g_{(2),(1)} \otimes g_{(2),(2)}) \Phi^* _{c_{\text{M2}}, c_i} 
 \right] \\
    & = \s \left[\Phi_{c_{\text{M2}},c_{i+1} } \left( \r_{\text{M2}} \otimes \r_{\text{M5}_{c_{i+1}}^{\mathbf{(1,0)}}} \otimes  \r_{\text{M5}_{c_{i}}^{\mathbf{(1,0)}}} \right) (\text{id}\otimes \s) (\s \otimes \text{id}) (g_{(1),(1)} \otimes g_{(1),(2)} \otimes g_{(2)}) \Phi^* _{c_{\text{M2}}, c_i} 
 \right] \\
 &=\s \left[\Phi_{c_{\text{M2}},c_{i+1} } \left( \r_{\text{M2}} \otimes \r_{\text{M5}_{c_{i+1}}^{\mathbf{(1,0)}}} \otimes  \r_{\text{M5}_{c_{i}}^{\mathbf{(1,0)}}} \right)  ( g_{(1),(2)} \otimes g_{(2)} \otimes g_{(1),(1)}) \Phi^* _{c_{\text{M2}}, c_i} \right] \\
 & = \s  \left[\left(\Phi_{c_{\text{M2}},c_{i+1} } \otimes \Phi^* _{c_{\text{M2}}, c_i} \right)\left( \r_{\text{M5}_{c_{i+1}}^{\mathbf{(1,0)}}} (g_{(2)}) \otimes \r_{\text{M5}_{c_{i}}^{\mathbf{(1,0)}}} (g_{(1)})  \right) \right] \\
 & =  S_{c_i, c_{i+1}} ^{c_{\text{M2}}} \D_{\text{M5}_{c_i}^{\mathbf{(1,0)}}, \text{M5}_{c_{i+1}}^{\mathbf{(1,0)}}  }(g),
\end{split}
\end{align}
where we used the Sweedler's notation $\D(g) = g_{(1)} \otimes g_{(2)}$ omitting the summation. We used \eqref{eq:intconst} for the second equality; the coassociativity $(\Delta \otimes \text{id})\Delta = (\text{id}\otimes \Delta)\Delta$ for the fourth equality; and \eqref{eq:dintconst} for the sixth equality. The commutativity \eqref{eq:screencomm} was proven in \cite{2015arXiv151208779B,Harada:2021xnm} through brute-force computation. It is worth noting that realizing the screening charges in terms of the gauge-invariant M2-M5 intersections allows us to avoid the labor.\\

Further, we can show that the set of screening charges that we obtained is maximal; in other words, there is no more operator in the tensor product $\text{M5}_{c_1} ^{\mathbf{(1,0)}} \otimes \text{M5}_{c_2} ^{\mathbf{(1,0)}}$ of two $q$-bosons which commutes with all the generators produced by the product $R^{(c_2)} _2 R^{(c_1)} _1$ of two R-matrices.\footnote{We make a mild assumption on the form of the screening charge. See footnote \ref{fn:assump}.} We emphasize that, in deriving this, it is crucial to have the generating series factorized into the product of two R-matrices, providing a systematic way to solve the commutativity constraint. Details can be found in appendix \ref{sec:screeningmaximal}.\\

This completes the proof of the two equalities, as promised. First, the product \eqref{eq:rmatwhole} of R-matrices, when expanded in the generators of $\text{M2}_{0,0,1}$ as in \eqref{eq:miuraexpand}, produces the full set of generators of the $q$-deformed $Y$-algebra $\text{M5}_{L,M,N}^{\mathbf{(1,0)}}$. Thus, our R-matrix \eqref{eq:rmatwhole} can be properly called the Miura transformation for the $q$-deformed $Y$-algebra. Second, the $q$-deformed $Y$-algebra $\text{M5}_{L,M,N} ^{\mathbf{(1,0)}}$ is completely characterized as the commutants of the screening charges identified in \eqref{eq:screeningdef2} and \eqref{eq:screeningdef3}. We may regard this statement as an alternative definition of the $q$-deformed $Y$-algebra.

\subsubsection{Ordering-independence by Yang-Baxter equation} \label{subsubsec:orderingyb}
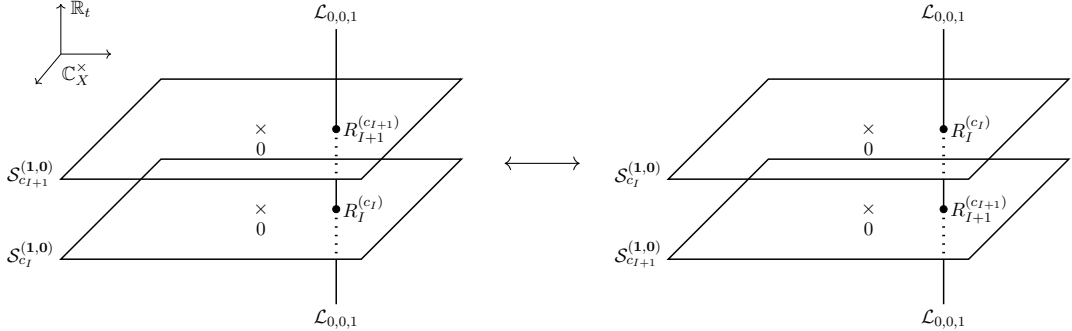
\begin{figure}[h!]\centering
\resizebox{.4\textwidth}{!}{
\begin{tikzpicture}
\draw[line width=0.3mm] (-1,1.5) -- (5,1.5) -- (3,-0.5) -- (-3,-0.5) 
      -- cycle;

\draw[line width=0.3mm] (-1,-0.1) -- (5,-0.1) -- (3,-2.1) -- (-3,-2.1) 
      -- cycle;

\draw[line width=0.3mm, loosely dotted, ultra thick] (2.5,0.5) -- (2.5,-0.5);
\draw[line width=0.3mm] (2.5,-0.5) -- (2.5,-1.1);
\draw[line width=0.3mm, loosely dotted, ultra thick] (2.5,-1.1) -- (2.5,-2.1);
\draw[line width=0.3mm] (2.5,-2.1) -- (2.5,-3);

\draw[line width=0.3mm] (2.5,2.5) -- (2.5,0.5);

\filldraw[black] (2.5,0.5) circle (2pt) node[anchor=west]{$R_{I+1} ^{(c_{I+1})}$};

\filldraw[black] (2.5,-1.1) circle (2pt) node[anchor=west]{$R_I ^{(c_I)}$};

 \node at (1,0.5) {$\times$};
 \node at (1,0.1) {$0$};

  \node at (1,-1.1) {$\times$};
 \node at (1,-1.5) {$0$};

   \node at (-3.6,-0.4) {$\mathcal{S}_{c_{I+1}} ^{\mathbf{(1,0)}}$};
      \node at (-3.6,-2) {$\mathcal{S}_{c_I} ^{\mathbf{(1,0)}}$};
  \node at (2.5,-3.3) {$\mathcal{L}_{0,0,1} $};
    \node at (2.5,2.8) {$\mathcal{L}_{0,0,1} $};

\draw[->] (-3,2) -- (-3,3);
\draw[->] (-3,2) -- (-2,2);
\draw[->] (-3,2) -- (-3.5,1.4);

\node at (-2.7,1.6) {$\BC^\times _X$};
\node at (-2.6,2.9) {$\BR_t$};
    
\end{tikzpicture}} \hspace{3mm} \raisebox{12\height}{\begin{tikzpicture}
    \draw[<->] (-0.5,0) -- (0.5,0);
\end{tikzpicture}} \hspace{0.3mm} \resizebox{.4\textwidth}{!}{
\begin{tikzpicture}
\draw[line width=0.3mm] (-1,1.5) -- (5,1.5) -- (3,-0.5) -- (-3,-0.5) 
      -- cycle;

\draw[line width=0.3mm] (-1,-0.1) -- (5,-0.1) -- (3,-2.1) -- (-3,-2.1) 
      -- cycle;

\draw[line width=0.3mm, loosely dotted, ultra thick] (2.5,0.5) -- (2.5,-0.5);
\draw[line width=0.3mm] (2.5,-0.5) -- (2.5,-1.1);
\draw[line width=0.3mm, loosely dotted, ultra thick] (2.5,-1.1) -- (2.5,-2.1);
\draw[line width=0.3mm] (2.5,-2.1) -- (2.5,-3);

\draw[line width=0.3mm] (2.5,2.5) -- (2.5,0.5);

\filldraw[black] (2.5,0.5) circle (2pt) node[anchor=west]{$R_I ^{(c_I)}$};

\filldraw[black] (2.5,-1.1) circle (2pt) node[anchor=west]{$R_{I+1} ^{(c_{I+1})}$};

 \node at (1,0.5) {$\times$};
 \node at (1,0.1) {$0$};

  \node at (1,-1.1) {$\times$};
 \node at (1,-1.5) {$0$};

   \node at (-3.6,-0.4) {$\mathcal{S}_{c_I} ^{\mathbf{(1,0)}}$};
      \node at (-3.6,-2) {$\mathcal{S}_{c_{I+1}} ^{\mathbf{(1,0)}}$};
  \node at (2.5,-3.3) {$\mathcal{L}_{0,0,1} $};
    \node at (2.5,2.8) {$\mathcal{L}_{0,0,1} $};  
\end{tikzpicture}}
\caption{Change of ordering in surface defect fusion, and product of R-matrices related by a conjugation of the $q$-deformed Maulik-Okounkov R-matrix.} \label{fig:mor} 
\end{figure}

The Miura transformation for the $q$-deformed $Y$-algebra should not depend on the ordering of the Miura operators. Since we identified the Miura operators as R-matrices of the quantum toroidal algebra, the change of ordering of the Miura operators is simply implemented by the universal Yang-Baxter equation (see Figure \ref{fig:mor}).

Specifically, let us consider the R-matrices between the $q$-boson representations by
\begin{align}
    R_{\text{M5}_{c} ^{\mathbf{(1,0)}}  , \text{M5}_{c'} ^{\mathbf{(1,0)}} } = \left(\r_{\text{M5}_{c} ^{\mathbf{(1,0)}} } \otimes \r_{\text{M5}_{c'} ^{\mathbf{(1,0)}} } \right)\CalR \in \text{M5}_{c} ^{\mathbf{(1,0)}} \widehat{\otimes} \, \text{M5}_{c'} ^{\mathbf{(1,0)}},\qquad c,c' \in \{1,2,3\}.
\end{align}
We may call these R-matrices the $q$-deformed Maulik-Okounkov R-matrices. For other studies on these R-matrices between $q$-boson representations, see \cite{Fukuda:2017qki,Garbali:2020sll,Negut:2020npc,Garbali:2021qko}.

Now, the universal Yang-Baxter equation \eqref{eq:univyb} represented on $\text{M2}_{0,0,1} \otimes \text{M5}_{c_{I}} ^{\mathbf{(1,0)}}  \otimes \text{M5}_{c_{I+1}} ^{\mathbf{(1,0)}} $ directly gives
\begin{align}
     R_{\text{M5}_{c_I} ^{\mathbf{(1,0)}}  , \text{M5}_{c_{I+1}} ^{\mathbf{(1,0)}} } R_{I+1} ^{(c_{I+1})} R_{I} ^{(c_{I})} = R_{I} ^{(c_{I})} R_{I+1} ^{(c_{I+1})} R_{\text{M5}_{c_I} ^{\mathbf{(1,0)}}  , \text{M5}_{c_{I+1}} ^{\mathbf{(1,0)}} } ,\qquad \text{for any } I.
\end{align}
As a result, the Miura transformations with different orderings of Miura operators produce isomorphic $q$-deformed $Y$-algebras, which thereby can be written as $\text{M5}_{L,M,N} ^{\mathbf{(1,0)}}$ without ambiguity.

\subsubsection{R-matrices with multiple M2-branes}
So far in this section, we have considered the intersections between a single M2-brane and multiple M5-branes, leading to the Miura transformation for the $q$-deformed $Y$-algebras. Here, we will consider the intersections between multiple M2-branes and a single M5-brane.

Let us fix the support of the M5-brane to be $\BR^2 _{\ve_1} \times \BR^2 _{\ve_2} \times \BC^\times _X$. The support of the $i$-th M2-brane is determined according to the function $c: i \mapsto c_i \in \{1,2,3\}$, to be $\BR^2 _{c_i}\times \BR_t$. In particular, we set $\vert c^{-1} (1) \vert = l$, $\vert c^{-1} (2) \vert = m$, and $\vert c^{-1} (3) \vert = n$. The algebra of local operators on the resulting line defect can be reconstructed by the composing the tensor product of the respective representations and the coproduct,
\begin{align}\label{eq:multipleM2}
    \r_{\text{M2}_{l,m,n}} = \left( \r_{\text{M2}_{c_{l+m+n}}} \otimes \cdots \otimes \r_{\text{M2}_{c_1}} \right) \D^{l+m+n-1} : \qta \twoheadrightarrow \text{M2}_{l,m,n}.
\end{align}
The resulting M2-brane algebra does not depend on the ordering of the M2-branes in the tensor product, as we will show shortly. Thus, we will denote this algebra as $\text{M2}_{l,m,n}$ without ambiguity.

\begin{figure}[h!]\centering
\begin{tikzpicture} 
\draw[line width=0.3mm] (-1,1.5) -- (5,1.5) -- (3,-1.5) -- (-3,-1.5) 
      -- cycle;

\draw[line width=0.3mm] (3.4,2.5) -- (3.4,0.3);
\draw[line width=0.3mm, loosely dotted, ultra thick] (3.4,0.3) -- (3.4,-1);
\draw[line width=0.3mm] (3.4,-0.935) -- (3.4,-2.4);

\draw[line width=0.3mm] (0.5,2.5) -- (0.5,0.1);
\draw[line width=0.3mm, loosely dotted, ultra thick] (0.5,0.1) -- (0.5,-1.5);
\draw[line width=0.3mm] (0.5,-1.5) -- (0.5,-3);

\draw[line width=0.3mm] (1.7,2.5) -- (1.7,0.5);
\draw[line width=0.3mm, loosely dotted, ultra thick] (1.7,0.5) -- (1.7,-1.5);
\draw[line width=0.3mm] (1.7,-1.5) -- (1.7,-3);

\filldraw[black] (2.2,2) circle (1pt);
\filldraw[black] (2.5,2) circle (1pt);
\filldraw[black] (2.8,2) circle (1pt);

\filldraw[black] (0.5,0.1) circle (2pt) node[anchor=east]{$R_1$};

\filldraw[black] (3.4,0.3) circle (2pt) node[anchor=west]{$R_{l+m+n}$};

\filldraw[black] (1.7,0.5) circle (2pt) node[anchor=west]{$R_2$};

 \node at (1,0) {$\times$};
 \node at (1,-0.5) {$0$};

   \node at (-3.6,-1.3) {$\mathcal{S}_{0,0,1} ^{\mathbf{(1,0)}}$};

    \node at (0.5,2.8) {$\mathcal{L}_{c_{1}} $};
        \node at (1.7,2.8) {$\mathcal{L}_{c_{2}} $};
            \node at (3.4,2.8) {$\mathcal{L}_{c_{l+m+n}} $};

\draw[->] (-3,2) -- (-3,3);
\draw[->] (-3,2) -- (-2,2);
\draw[->] (-3,2) -- (-3.5,1.4);

\node at (-2.7,1.6) {$\BC^\times _X$};
\node at (-2.6,2.9) {$\BR_t$};
    
\end{tikzpicture} \caption{Intersection between multiple M2-branes and a single M5-brane, and associated R-matrices. Here, we abbreviated $R_i \equiv R_{\text{M2}_{c_{i}} , \text{M5}_{0,0,1} ^{\mathbf{(1,0)}}}$.} \label{fig:multim2m5} 
\end{figure}
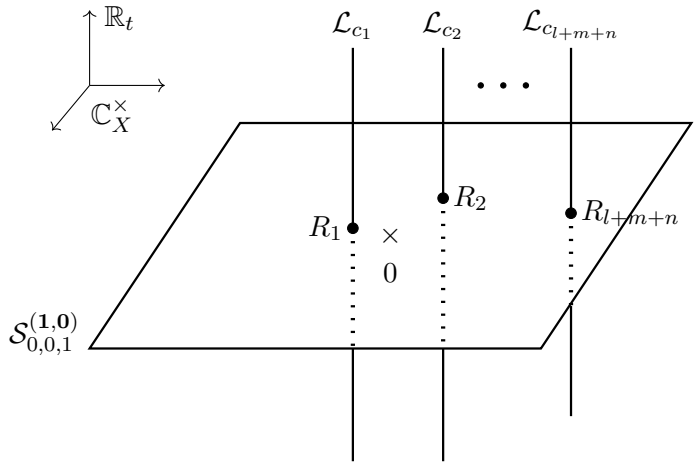

The intersection between the fused line defect and the surface defect is decomposed into multiple intersections between the individual line defects and the surface defect (see Figure \ref{fig:multim2m5}). Correspondingly, by repetitively applying \eqref{eq:univr2} the R-matrix between the representations $\text{M2}_{l,m,n}$ and $\text{M5}_{0,0,1} ^{\mathbf{(1,0)}}$ is obtained by taking the product of the basic R-matrices:
\begin{align}
    R_{\text{M2}_{l,m,n},\text{M5}_{0,0,1} ^{\mathbf{(1,0)}}} = R_{\text{M2}_{c_{l+m+n}} , \text{M5}_{0,0,1} ^{\mathbf{(1,0)}}} \cdots R_{\text{M2}_{c_{2}} , \text{M5}_{0,0,1} ^{\mathbf{(1,0)}}} R_{\text{M2}_{c_{1}} , \text{M5}_{0,0,1} ^{\mathbf{(1,0)}}} \in \text{M2}_{l,m,n} \, \widehat{\otimes} \, \text{M5}_{0,0,1} ^{\mathbf{(1,0)}}.
\end{align}
Thus, when the product on the right hand side is expanded, the terms recombine into the elements in $ \text{M2}_{l,m,n} \, \widehat{\otimes} \, \text{M5}_{0,0,1} ^{\mathbf{(1,0)}}$ by construction. In the case of $\CalC = \BC \times \BC^\times$, such a recombination of the product of Miura operators was observed in \cite{Gaiotto:2020dsq} at the example of two M2-branes passing through a single M5-brane. Our construction of the Miura operators as R-matrices provides a natural explanation for this recombination.\\

Moreover, let us consider the R-matrices between the quantum torus algebras,
\begin{align}
    R_{\text{M2}_c ,\text{M2}_{c'}} = \left( \r_{\text{M2}_c} \otimes \r_{\text{M2}_{c'}}\right) \CalR \in \text{M2}_c \, \widehat{\otimes}\, \text{M2}_{c'} ,\qquad c,c' \in \{1,2,3\}.
\end{align}
Then, the universal Yang-Baxter equation \eqref{eq:univyb} implies
\begin{align}
    R_{\text{M2}_{c_{i+1}} ,\text{M2}_{c_i}} R_{\text{M2}_{c_{i+1}} , \text{M5}_{0,0,1} ^{\mathbf{(1,0)}}} R_{\text{M2}_{c_{i}} , \text{M5}_{0,0,1} ^{\mathbf{(1,0)}}} = R_{\text{M2}_{c_{i}} , \text{M5}_{0,0,1} ^{\mathbf{(1,0)}} }R_{\text{M2}_{c_{i+1}} , \text{M5}_{0,0,1} ^{\mathbf{(1,0)}}}  R_{\text{M2}_{c_{i+1}} ,\text{M2}_{c_i}} ,
\end{align}
so that changing the order of any two adjacent basic R-matrices results in an equivalent R-matrix, up to a conjugation by $R_{\text{M2}_{c_{i+1}}, \text{M2}_{c_i}}$.

\vspace{4mm}

\section{Miura transformation from IIB string theory}
\label{sec:typeIIB}

So far, we have analyzed the modules of the quantum toroidal algebra in terms of branes in twisted M-theory. In this section, we will give an alternate derivation of our results in terms of type IIB string theory. It is well known that M-theory compactified on a 2-torus is the same theory as type IIB compactified on a circle. Here, we have two natural tori in M-theory on which we can perform the duality to type IIB: one possibility is to consider the torus of  phases in the internal Calabi-Yau manifold $\mathbb{R}^2_{\ve_1} \times \BR^2_{\ve_2} \times \BR^2_{\ve_3}$, whose action on local coordinates is
\beq
(z_1, z_2, z_3) \mapsto (e^{i\alpha_1}\, z_1, e^{i\alpha_2}\, z_2,e^{-i(\alpha_1+\alpha_2)}\, z_3) \; .
\eeq
The Calabi-Yau geometry used in the M-theory compactification is mapped to a $\mathbf{(p,q)}$ web of 5-branes in type IIB. In that picture, the M5-branes become D3-branes supported on 4-manifolds with codimension-2 ``corners'', where the chiral algebras $Y_{L,M,N}$ live \cite{Gaiotto:2017euk}, while the M2-branes become strings.\\

Alternatively, we can view $\mathbb{C}^\times_X\times\mathbb{C}^\times _Z$ in the M-theory background as a torus fibration:
\beq\label{fibration}
S^1_X(R_X)\times S^1_Z(R_Z)\rightarrow \mathbb{C}^\times_X\times\mathbb{C}^\times_Z \rightarrow \mathbb{R}_x\times\mathbb{R}_z \; .
\eeq 
In this picture, the complex structure $\tau=i\, R_X/R_Z$ of the M-theory torus is identified with the type IIB coupling (dilaton field\footnote{We are assuming throughout that the axion field of type IIB is turned off, or equivalently that the torus in M-theory is rectangular.}) $\tau=i\, g^{-1}_s$. The duality turns the M5-branes into $\mathbf{(p,q)}$ 5-branes, and the M2-branes into D3-branes, all wrapping a common circle $S^1(R)$ of radius
\beq
R=\frac{1}{m^3_{pl}\,R_x\,R_z} \; ,
\eeq
with $m_{pl}$ the Planck mass. The support of the branes is displayed in table \ref{table:2am}. This type IIB frame makes it possible to understand the underlying quantum toroidal symmetry in a very direct way. To our knowledge, the earliest example comes from the work \cite{Awata:2011ce}, which reinterpreted 5-brane junctions as intertwiners of Fock modules in the toroidal algebra, see also \cite{Bourgine:2016vsq,Bourgine:2017jsi,Mironov:2016yue}. Since then, the dictionary has been extended in a series of paper by Zenkevich \cite{Zenkevich:2018fzl,Zenkevich:2020ufs,2022arXiv221214808Z,Zenkevich:2023cza} and unpublished work by Aganagic-Zenkevich: the quantum toroidal generators have been identified as $\mathbf{(p,q)}$ strings, the vector representations as D3-branes, the Fock representations as $\mathbf{(p,q)}$ 5-branes, and the MacMahon representations as $\mathbf{(p,q)}$ 7-branes (though this last identification is still at the conjectural stage). See also \cite{Awata:2017lqa,Bourgine:2019phm} for extensions to the quantum toroidal algebras of higher-rank.\\

For instance, back in section \ref{subsubsec:singlem2}, we showed that the algebra of local operators on a single M2-brane in the twisted M-theory is given by a quantum torus algebra, furnishing a vector representation of $U_{q_1, q_2, q_3} (\widehat{\widehat{\fgl}}(1))$; we can now provide an alternative perspective for this statement in the dual IIB theory, by mapping the M2-brane to a D3-brane: the vector representation has central charges $(C,C^{\perp})=(1,1)$, so in particular is invariant under the Miki automorphism action $C^{\perp}\rightarrow C\rightarrow (C^{\perp})^{-1} \rightarrow C^{-1}  \rightarrow C^{\perp}$, but this is precisely the property of a D3-brane, the only $SL(2,\mathbb{Z})$-invariant object of type IIB (regardless of whether one turns on an $\Omega$-background or not). Furthermore, recall from \eqref{eq:vectorr} that this vector representation can be realized via the action of $q_c$-difference operators on states $[u]_i ^{(c)}$, for $u\in\mathbb{C}^\times$ and $i\in\BZ$, where each state belongs in the vector space $\CalV_c (u) = \bigoplus_{i\in \BZ} \BC [u]_i ^{(c)}$. If we label each state $[u]_i ^{(c)}$ as a function $f(X)$ of the $X$-coordinate, then note that the ``raising'' mode $E_0$ and the ``Cartan'' mode $H_1$  act as 
\beq\label{actionsvec}
E_0\, f(X)\sim  f(q_c \, X) \; ,\qquad\qquad H_1\, f(X)\rightarrow X\, f(X) \; ,
\eeq
according to \eqref{eq:vecrep}.
Since the grading of $E_0$ is $(d^{\perp},\,d)=(1,\,0)$ and the grading of $H_1$ is $(d^{\perp},\,d)=(0,\,1)$, and since $Z$ and $X$ are literally complex coordinates on the M-theory torus, these actions are naturally interpreted as adding 1 unit of Kaluza-Klein momentum along $S^1_Z$ and along $S^1_X$ on the torus, respectively.\footnote{And more generally, $Z^m\, X^n$ carries $m$ units of KK momentum along $S^1_Z$ and $n$ units of KK momentum along $S^1_X$ on the M-theory torus.} In type IIB, this diagonal action of $H_1$ is precisely that of a fundamental string wrapped around $S^1(R)$ and ending on a D3-brane, adding one unit of electric charge.\\ 

Although it is not immediately apparent, the action of $E_0$ is likewise that of a D-string wrapped around $S^1(R)$ ending on the D3-brane; this is because a D-string turns on one unit of vortex flux on a D3-brane. One would naively expect a soliton singularity to be present at the D-string endpoint location, but this is not so in an $\Omega$-background \cite{Nekrasov:2010ka}: remarkably,  turning on $k$ units of vortex flux there has the sole effect of shifting  $X \rightarrow q^k_c\, X$ . In more detail, consider a D3-brane supported on $\BR^2_{\ve_c}\times\mathbb{R}_t\times S^1(R)$. Its fluctuation in the $\mathbb{R}_x$ direction is a Coulomb modulus, complexified by the holonomy around $S^1(R)$, and denoted as $X$. In the $\Omega$-background, $X$ always appears as part of a product $X \,  e^{R\,\ve_c\,y_c\, D_{y_c}}$ in the low energy effective Lagrangian on the brane, with $D_{y_c}= \partial_{y_c}+ A_{y_c}$ the covariant derivative on $\BR^2_{\ve_c}$ of complex coordinate $y_c$. A vortex of charge $k$ induces a gauge field profile $A_{y_c}= k/y_c$, or equivalently a shift of the D3-brane modulus as $X \rightarrow q^k_c\, X$, as claimed; for $k=1$, this is exactly the $E_0$ action \eqref{actionsvec}.\\

It is far more involved to see directly that $\mathbf{(p,q)}$ 5-branes are Fock modules of the quantum toroidal algebra. The idea is that  $({\bf{p}},{\bf{q}})$ strings are now instantons in the 5-branes, meaning the quantum toroidal algebra generators are acting on the instanton moduli space of the 5-brane theory. More precisely, for 5-branes supported on $\BR^2_{\ve_{c+1}}\times\BR^2_{\ve_{c-1}}\times l^{({\bf{p}},{\bf{q}})} \times S^1(R)$, this is an action on the equivariant K-theoretic moduli space of instantons on $\BR^2_{\ve_{c+1}}\times\BR^2_{\ve_{c-1}}$. The equivariant action is the corresponding $U(1)\times U(1)$ on this part of the $\Omega$-background, with parameters $q_a$ and $q_b$. In the limit $R\rightarrow 0$, the Hilbert space of BPS states would be the cohomology of the moduli space, that is the Higgs branch of the instanton quantum mechanics on $l^{({\bf{p}},{\bf{q}})}$. For us, $S^1(R)$ has finite size, and the instantons are strings wrapping that circle. The resulting Kaluza-Klein tower of states implies that the Hilbert space of BPS states is the K-theory of the moduli space rather than its cohomology. 

It has been shown that this instanton moduli space admits a realization as a Fock space \cite{2009arXiv0904.1679F,2009arXiv0905.2555S}.\footnote{Note that for a Fock representation of central charges $(C,C^{\perp})=(q_c^{N/2},\,1)$, this realization implies, in the $R\rightarrow 0$ limit, a mathematical proof of the  Alday-Gaiotto-Tachikawa-Wyllard conjecture \cite{Alday:2009aq,Wyllard:2009hg}, for the algebra ${\cal W}(\text{sl}(N))\oplus u(1)$ \cite{2012arXiv1202.2756S,2012arXiv1211.1287M}.} It follows from their construction that a $({\bf{p}},{\bf{q}})$ 5-brane is a Fock representation with central charges $(C,C^{\perp})=(q^{\mathbf{p}/2}_c,q^{\mathbf{q}/2}_c)$.\\

An elementary Miura operator is therefore realized as the intersection of an NS5-brane and a D3-brane in the type IIB frame. We will now analyze how it can be recovered from a gauge theory supersymmetric index.

\vspace{4mm}

\subsection{Miura operators from supersymmetric index}
\label{ssec:Miurahalfindex}

A generic configuration of type IIB fivebranes and D3-branes in table \ref{table:2am} will preserve at least 2 supercharges, even in the presence of the $\Omega$-background; the supersymmetry in question is the reduction of 2d $\EN=(0,2)$ supersymmetry to 1d. The spectrum of protected BPS operators in such a brane system is captured by an index of the type IIB string theory on $M_{10}$ in the presence of the branes, which is the index of this supersymmetric quantum mechanics along the thermal circle $S^1(R)$.\\ 

The simplest configuration consists of a single NS5-brane, say a NS5$_{1,0,0}$ supported on $\BR^2_{\ve_2}\times\BR^2_{\ve_3}$, intersecting a single D3-brane in spacetime, say a D3$_{0,0,1}$ brane supported on $\BR^2_{\ve_3}$. In section \ref{subsec:nontrans}, we proposed that such an intersection stood for the insertion of a Miura operator $R^{(1)}(x)$ at position $x\in{\mathbb{R}}_x\times S^1(R)$, or equivalently an R-matrix valued in the tensor product of modules ${\mathcal V}_3 \otimes {\cal F}_1^{(1,0)}$, with spectral parameter $x$.\\ 

Because this brane configuration preserves a full 3-dimensional ${\EN} = 4$ supersymmetry, it is somewhat more direct to count BPS operators by defining a supersymmetric index directly in 3 dimensions on $\BR^2_{\ve_3}\times S^1(R)$, instead of via the quantum mechanics on $S^1(R)$. We can think of the NS5/D3 intersection as two D3-branes, both semi-infinite along ${\mathbb{R}}_u$, ending on opposite sides of the NS5-brane. The quantization of string zero modes at this locus yields charged matter,  a 3d  ${\EN} = 4$ hypermultiplet in the bifundamental representation $(1,-1)$ of $U(1)\times U(1)$, with both $U(1)$ groups understood as non-dynamical. When the two D3-branes break along the NS5-brane in the ${\mathbb{R}}_x$ direction, the relative distance between them is proportional to a real equivariant mass for this hypermultiplet.\\

When the two halves of the D3-brane coincide along ${\mathbb{R}}_x$ and combine into a single brane, the hypermultiplet becomes massless. In that case, the full D3-brane can instead be pulled away from the NS5-brane in the ${\mathbb{R}}_z$ direction without breaking supersymmetry, probing the Higgs phase.\\

We explicitly break supersymmetry to $\EN=2$ as follows by turning on a real equivariant mass $\ve_2$ for the $U(1)_{\ve_2}$ flavor symmetry rotating $\BR^2_{\ve_2}$. Both the 3d $\EN=2$ bifundamental chiral multiplets and the adjoint chiral multiplet sitting inside the $\EN=4$ vector multiplet\footnote{By adjoint representation, we really mean the charge 0 irrep of $U(1)$ here, since there is only one D3-brane present.} are charged under this symmetry. After turning on the mass, the resulting supersymmetry is often denoted as 3d $\EN=2^*$.\\ 

%The remaining relative distances between the D3 and NS5-branes do not play a role our discussion, so we set them to zero for convenience. 

All the above global symmetry parameters are complexified by the D3-brane $U(1)$ holonomy along the circle $S^1(R)$, with notation 
\begin{align}\label{holonomy}
q_c=e^{-R\left(\ve_c+i\,A^{\theta}_{\ve_c}\right)}\; , \;\;\;\qquad X_d=e^{-R\left(x_d+i\,A^{\theta}_{x_d}\right)} \; .
\end{align}
For a single D3-brane, the Coulomb branch moduli space is that of a free vector multiplet, with background gauge field $A^{\theta}$. In what follows, we do not reduce this moduli space by factoring out the center of mass position of D3-branes.\footnote{Because our 3d theories are all supported on an $\Omega$-background $S^1(R)\times \BR^2_{\ve_c}$, it is customary to treat each $U(1)$ modulus as a physical equivariant Coulomb parameter.}\\

The 3-dimensional index has been defined and studied both by physicists and mathematicians. 
In physics, the BPS count is the partition function of a 3d ${\EN} = 2$ SCFT on the 3-manifold $S^1(R)\times\BR^2$, which is well-defined only after regularizing the non-compactness of $\BR^2$. One such regularization prescription is to introduce a massive deformation of the background, denoted as $M_3 = S^1(R)\times \BR^2_{\ve_3}$, understood as a $\BR^2$ bundle over $S^1(R)$: let $(X_{S^1},X_3)$ be coordinates on $S^1(R)\times\BR^2$, and identify 
\begin{align}\label{omega3d}
(0, X_3) \sim  (2\pi R, q_3\, X_3) \; .
\end{align}
This background $M_3$ is known as a Melvin cigar, or 3-dimensional $\Omega$-background with equivariant parameter $q_3=e^{R\,\ve_3}$; the fugacity $\log(q_3)$ is conjugate to the Cartan generator $J_3$ of the $U(1)_{J_3}$ action rotating $\BR^2_{\ve_3}$. The BPS states counted by the partition function are fixed points of this equivariant action, so they are codimension-2 particles (vortices) localized at the origin of $\BR^2_{\ve_3}$. A partition function on $M_3$  is known as a holomorphic block \cite{Beem:2012mb}. Because boundary conditions have to be specified for the fields ``at infinity'' of $\BR^2_{\ve_3}$, holomorphic blocks are intrinsically defined in the IR.\footnote{Such boundary conditions label the choice of a 3d vacuum, and are often called Picard-Lefschetz thimbles in supersymmetric physics \cite{Cecotti:1991me,Cecotti:1992rm,Hori:2000ck,Gaiotto:2015zna,Gaiotto:2015aoa}.}\\

In mathematics, this index is called vertex function, or generating function of quasimaps $\mathbb{CP}^1\rightarrow X$ of all degrees in equivariant quantum K-theory, where $X$ is the Higgs branch of the 3d theory \cite{2009arXiv0908.4446C,2010arXiv1005.4125K,2011arXiv1106.3724C,Okounkov:2015spn}. Quasimaps are solutions to the vortex equations of the 3d theory,  subjected to certain regularity conditions on the gauge and matter fields, and K-theory enters here instead of cohomology because we are studying a 3-dimensional theory compactified on a circle instead of a purely 2-dimensional one.\\

It is also possible to define the index straight from a 3d ${\EN} = 2$ UV Lagrangian: one simply replaces $\BR^2_{\ve_3}$ with a finite disk/hemisphere $D^2$, and imposes a specific set of 1/2-BPS $\EN=(0,2)$ boundary conditions at finite distance on $T^2 = S^1(R)\times S^1_{D^2}$, which will flow to a superconformal point in the IR. One requires the $\EN=(0,2)$ boundary theory to have an unbroken $U(1)_R$ R-symmetry, so that an index can be defined in the first place. Often, this $U(1)_R$ R-symmetry is identified with  the one from the 3d $\EN=2$ bulk, but it does not need to be so in general.\footnote{The $U(1)_R$ charge is sometimes a linear combination of the bulk R-symmetry charge and other $U(1)$ charges in the maximal torus of the UV flavor symmetry group; for instance, this will be the case for the distinguished Dirichlet boundary conditions of section \ref{ssec:Dirichletcond}.}

\vspace{4mm}

\subsection{Index as an integral and Neumann boundary conditions}
 
This UV partition function, sometimes called half-index, is most easily computed via equivariant localization, as a trace over the Hilbert space of states on $D^2$ \cite{Yoshida:2014ssa}. Alternatively, the UV half-index has yet another definition, as a character over the vector space of local operators at the boundary of 3d $\EN=2$ Minkowski spacetime $\mathbb{R}^{1,1}\times\mathbb{R}_{\leq 0}$, with a $\EN=(0,2)$-preserving boundary supported on $\mathbb{R}^{1,1}\times\{0\}$. These two UV definitions of the half-index happen to agree up to 1-loop contributions of certain boundary mixed 't Hooft anomalies, which will be safely ignored for our purposes \cite{Bullimore:2020jdq}.

In the Minkowski formulation, one introduces coordinates $(x^0, x^1)$ on the boundary $\mathbb{R}^{1,1}\times\{0\}$ and a bulk coordinate $x^\bot\leq 0$ on the half-line $\mathbb{R}_{-}$, with boundary located at $x^\bot=0$. The 3d $\EN=2$ algebra has four real supercharges, or two complex spinors $Q_{\alpha}$ and $\overline{Q}_{\alpha}$, where $\alpha=+,-$. These supercharges $(Q_+, Q_-, \overline{Q}_+, \overline{Q}_-)$ have $U(1)_R$ R-charge $(-1,-1,1,1)$, and satisfy the relations
\begin{align}
&\{Q_{\pm},\overline{Q}_{\pm}\}=\mp 2 \, P_{\pm}\; , \nonumber \label{commutatorQ}\\
&\{Q_{\pm},\overline{Q}_{\mp}\}=\mp 2\, i \, (P_{\bot}\mp i \, Z)\; ,
\end{align}
where $P_\alpha$ denotes the momentum on $\mathbb{R}^{1,1}$, and $Z$ is a real central charge.\\

It directly follows from \eqref{commutatorQ} that a 1/2-BPS $\EN=(0,2)$ subalgebra  generated by the supercharges  $Q_+$ and $\overline{Q}_+$ is preserved on the boundary. Then, the half-index is defined in $\overline{Q}_+$-cohomology, as a character over the vector space of local operators at $x^\bot=0$:
\beq
\label{3dhalfindex}
{\mathcal Z}_{3d}  = {\rm Tr}\left[(-1)^F\, {q_1}^{J_1}\, {q_2}^{J_2} \, {q_3}^{J_3} \;  X^{\ft_H} \;  v^{\ft_C} \right]\;\; .
\eeq
The fermion number operator is denoted as $F=2 J_3$. The operator $\ft_H$ stands for the Cartan generator of the flavor symmetries $F_{H}$ preserved on the boundary; the conjugate fugacities are the hypermultiplet masses, collectively denoted as $\log(x)$. The operator $\ft_C$ stands for the Cartan generator of an additional global symmetry $U(1)_{\cal J}$, called topological symmetry. The conjugate fugacity is the Fayet-Iliopoulos (FI) parameter, denoted as $\log(v)$; we will have more to say about this symmetry momentarily.\\

From the point of view of the quantum mechanics on $S^1(R)$, the fugacities $\log(q_{c})$ for $c=1,2,3,$ are treated on an equal footing, and are all conjugate to R-symmetries.  From the 3-dimensional viewpoint, the symmetry between them is broken, according to which $\BR^2_{\ve_{c}}$ plane the D3-brane is supported on; in our setup, we have fixed a D3$_{0,0,1}$ brane, so the fugacity $\log(q_{3})$ is a geometric parameter for the $\Omega$-background, conjugate to the Cartan generator $J_3$ of the $U(1)$ action rotating the boundary plane (or the disk/hemisphere $D^2$ in the Euclidean formulation); only $\log(q_{1})$ and $\log(q_{2})$ represent R-symmetries. In detail, our D3-NS5 brane configuration preserves a full 3d $\EN=4$ supersymmetry, with R-symmetry $SU(2)_H\times SU(2)_C$. We denote the maximal torus as $U(1)_H\times U(1)_C$, or as $U(1)_V\times U(1)_A$ on the $\EN=(2,2)$ boundary. Then, $\log(q_2)$ is conjugate to the Cartan generator $J_2$ of the $U(1)_A$ action on vector multiplet scalars at the boundary; these scalars are the moduli describing the fluctuation of the D3$_{0,0,1}$ brane along the NS5$_{1,0,0}$ brane worldvolume on $\BR^2_{\ve_2}$. The fugacity $\log(q_1)$ is conjugate to the Cartan generator $J_1$ of the $U(1)_V$ action on hypermultiplet scalars at the boundary.

In the above conventions, $J_{1,2,3}$ all have half-integer eigenvalues, so in particular the index can be expressed as a Taylor series in ${q_3}^{1/2}$.\\

In 3d $\EN=2$ notation, we can invoke the constraint $q_1 = q_2^{-1}\, q_3^{-1}$ and change notations to $J_1=-V/2$ and $J_2=-A/2$, so that the index takes a more familiar form \cite{Gadde:2013wq,Gadde:2013sca,Dimofte:2017tpi}: 
\beq
\label{3dhalfindexmore2}
{\mathcal Z}_{3d}  = {\rm Tr}\left[(-1)^F\, {q_3}^{J_3+\frac{V}{2}}\; {q_2}^{\frac{V-A}{2}} \;  X^{\ft_H} \;  v^{\ft_C}  \right]\;\; .
\eeq 
That is, the generator $J_1$ of $U(1)_V$ is now identified with the generator of the surviving $U(1)_R$ R-symmetry. Moreover, $J_2-J_1=(V-A)/2$  generates the $U(1)_{\ve_2}$ flavor symmetry rotating $\BR^2_{\ve_2}$. In this convention, $V$ and $A$ both have integer eigenvalues, and $R=V$ is the $U(1)_R$  R-symmetry generator.

\vspace{4mm}

\subsubsection{Index for non-transverse $\text{D3}_{0,0,1}$-$\text{NS5}_{1,0,0}$ intersection}
\label{sssec:nontransverse1}

Let $x_1$ and $x_2$ be the positions of the two semi-infinite D3-branes on ${\mathbb{R}}_x$, respectively, on opposite sides of the NS5-brane. The 3d $\EN=4$ hypermultiplet decomposes into a pair of $\EN=2$ chiral multiplets $(\Phi^+,\Phi^-)$, which contribute 
\begin{align}
\label{index3d1}
\frac{\left(\sqrt{q_3/q_2} \; X_0/X_1\, ; q_3\right)_\infty}{\left(\sqrt{q_3 q_2} \; X_0/X_1\, ; q_3\right)_\infty} \; .
\end{align}
to the index. The denominator is the contribution of the $\EN=2$ bifundamental chiral multiplet $\Phi^+$ with Neumann (N) boundary conditions: boundary scalars (and their $n$-th derivatives) are counted in $\overline{Q}_+$-cohomology with charge
\beq\label{chiral1}
(\text{bif chiral}, \text{N}) : \;\;\; \left(1,-1,n,1,\frac{1}{2}\right)\;\; \text{under}\;\; U(1)_{X_0}\times U(1)_{X_1}\times U(1)_{J_3}\times U(1)_R\times U(1)_{\ve_2} .
\eeq 
The numerator is the contribution of the $\EN=2$ bifundamental chiral multiplet $\Phi^-$ with Dirichlet (D) boundary conditions: boundary fermions (and their $n$-th derivatives) are counted in $\overline{Q}_+$-cohomology with charge
\beq\label{chiral2}
(\overline{\text{bif}}\;\text{chiral}, \text{D}) : \;\;\; \left(1,-1,n+\frac{1}{2},0,\frac{-1}{2}\right)\;\; \text{under}\;\; U(1)_{X_0}\times U(1)_{X_1}\times U(1)_{J_3}\times U(1)_R\times U(1)_{\ve_2} .
\eeq

In writing \eqref{index3d1}, we assumed a certain holomorphic Lagrangian splitting of the $\EN=4$ bifundamental hypermultiplet matter, called a \emph{polarization}: in terms of the $\EN=2$ chiral multiplets $(\Phi^+,\Phi^-)$, a polarization is a choice of sign $\epsilon=\pm$, where $\epsilon=-$ labels (D,N) boundary conditions for $(\Phi^+,\Phi^-)$, while $\epsilon=+$ labels (N,D) boundary conditions for $(\Phi^+,\Phi^-)$. The index contribution \eqref{index3d1} therefore corresponds the polarization $(+)$.\footnote{It is also possible to flip boundary conditions to have both chiral multiplets with D, or both with N, at the expense of shifting the Chern-Simons level to a nonzero value at 1-loop.}\\

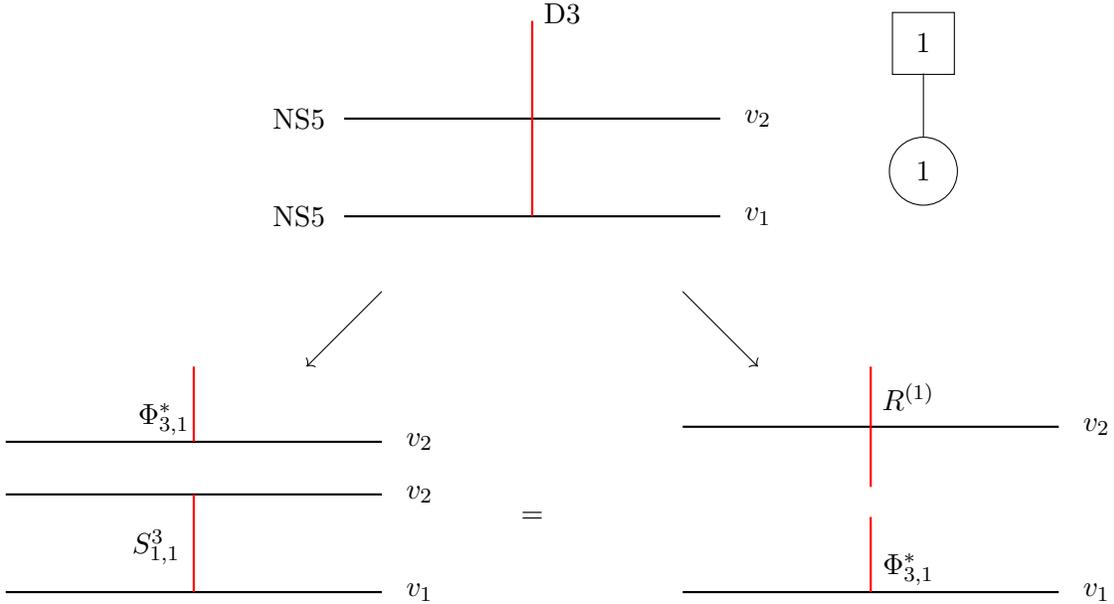
\begin{figure}[h!]\centering
\begin{tikzpicture}
\draw[thick] (-2.5,1.3) -- (2.5,1.3);
\draw[thick] (-2.5,0) -- (2.5,0);
\node at (3,1.3) {$v_2$};
\node at (3,0) {$v_1$};

\node at (-3.1,1.3) {NS5};
\node at (-3.1,0) {NS5};

\draw[thick, red] (0,0) -- (0,2.6);
\node at (0.4,2.7) {D3};

\node at (5.2,2.3) [circle] (flavor) {$1$};
\node [rectangle,draw,fit=(flavor),inner sep=2] {};
\node at (5.2,0.6) [circle,draw,inner sep=6] (gauge) {$1$};

\draw (gauge) -- (5.2,1.9);

\draw[thick] (-7,-3.7) -- (-2,-3.7);
\draw[thick] (-7,-5) -- (-2,-5);
\draw[thick] (-7,-3) -- (-2,-3);
\draw[thick,red] (-4.5,-5) -- (-4.5,-3.7);
\draw[thick,red] (-4.5,-3) -- (-4.5,-2);

\draw[thick] (7,-5) -- (2,-5);
\draw[thick] (7,-2.8) -- (2,-2.8);
\draw[thick,red] (4.5,-5) -- (4.5,-4);
\draw[thick,red] (4.5,-3.6) -- (4.5,-2);

\node at (0,-4) {$=$};
\draw[->] (-2,-1) -- (-3,-2);
\draw[->] (2,-1) -- (3,-2);

\node at (-1.5,-5) {$v_1$};
\node at (-1.5,-3.7) {$v_2$};
\node at (-1.5,-3) {$v_2$};

\node at (-5,-4.4) {$S_{1,1} ^3$};
\node at (-4.9,-2.7) {$\Phi^* _{3,1}$};

\node at (7.5,-2.8) {$v_2$};
\node at (7.5,-5) {$v_1$};

\node at (5,-4.7) {$\Phi^* _{3,1}$};
\node at (5,-2.4) {$R^{(1)}$};

\end{tikzpicture}
\caption{Top: Two NS5$_{1,0,0}$ branes and a semi-infinite D3$_{0,0,1}$ brane drawn in the $(\mathbb{R}_x,\mathbb{R}_t)$ plane, and the low energy effective 3d $\EN=4$ gauge theory: a $U(1)$ vector multiplet with one hypermultiplet. Bottom Left: The 3d half-index is manifestly equal to the vev of a screening charge composed with a dual intertwiner on the second Fock space of the quantum toroidal algebra. Bottom Right: Equivalently, the 3d half-index can be reinterpreted as the vev of a dual intertwiner on the first Fock space composed with an R-matrix.} \label{fig:electricvsmagnetic}
\end{figure}

So far, we have only analyzed the index of a bare D3-NS5 brane intersection; in order to make contact with the integral form of the Miura operator, we now argue that we need to further gauge the $U(1)$ symmetry on one of the two half-infinite D3-branes located on either side of the NS5-brane. In string theory, this is done by simply adding another NS5-brane for that D3 to end on, separated from the first NS5-brane along the $\mathbb{R}_t$-direction. Then, we consider two NS5$_{1,0,0}$ branes and one semi-infinite D3$_{0,0,1}$ brane, meaning $(L,M,N)=(2,0,0)$ and $(l,m,n)=(0,0,1)$ in our M-theory notation. It is convenient to think of this configuration as made up of two parts: a D3$_{0,0,1}$ brane segment, of finite extent between the two NS5$_{1,0,0}$ branes, and a semi-infinite D3$_{0,0,1}$ brane ending on one of the NS5$_{1,0,0}$ branes; see the top of figure \eqref{fig:electricvsmagnetic}.\\

In this Hanany-Witten setup \cite{Hanany:1996ie}, the low energy theory on the D3-branes is again a 3d $\EN=4$ $U(1)$ gauge theory on $\BR^2_{\ve_3}\times S^1(R)$, with a bifundamental hypermultiplet in representation $(1,-1)$ of $U(1)\times U(1)$; this time around, one of the two $U(1)$'s is gauged, say the first one. Correspondingly, the motion of the D3-brane segment  along ${\mathbb{R}}_x$ between the two NS5-branes is a real Coulomb modulus ${\mathfrak a}$ for this $U(1)$ gauge group. The position of the semi-infinite D3-brane  on ${\mathbb{R}}_x$ is a mass parameter, as before.

When the two parts of the D3-brane  combine along ${\mathbb{R}}_x$ into a single brane, the hypermultiplet becomes massless. In that case, a real  FI term can be turned on by pulling away the middle NS5-brane along the ${\mathbb{R}}_z$ direction. We denote the corresponding FI parameter as 
\beq\label{eq:FIreal}
\xi = \xi_1 - \xi_2 \; ,
\eeq 
where $\xi_i$ denotes the position of the $i$-th NS5-brane along ${\mathbb{R}}_z$.
In the 3d gauge theory, the FI term arises in the action from coupling the $U(1)$ gauge field on the D3-brane to a global symmetry group $U(1)_{\cal J}$ usually called topological symmetry. This symmetry arises from the conserved current ${\cal J}=\frac{1}{2\pi}* F$, where $*F$ is the Hodge dual of the $U(1)$ field strength. The charge associated to this topological symmetry is called vortex number $k$, and the equivariant mass is the FI parameter $\xi$. In what follows, we work with $\xi$ positive and generic. The Coulomb modulus and FI parameter are once again complexified by the holonomy along $S^1(R)$, with notation 
\begin{align}\label{holonomy2}
X=e^{-R\left({\mathfrak a}+i\,A^{\theta}_{\mathfrak a}\right)}\; , \;\;\;\qquad v=e^{-R\left(\xi+i\,A^{\theta}_{\xi}\right)} \; .
\end{align}
The 3d index is evaluated as before,  with the caveat that additional boundary conditions need to be specified for the dynamical vector multiplet. Imposing Neumann  boundary conditions ${\bf N}$, the gauge symmetry is preserved at the boundary, and the index trace becomes a finite-dimensional contour integral over the gauge moduli, projecting to gauge-invariant operators. For a $U(1)$ gauge group, there is a single such modulus $X$, which is integrated over, with contribution
\beq\label{veccon}
\frac{\left(q_3\, ; q_3\right)_\infty}{\left(q^{-1}_2\, ; q_3\right)_\infty} \oint \frac{dX}{2\pi i\,X} \; .
\eeq 
Indeed, the $\EN=4$ vector multiplet decomposes as a $\EN=2$ vector multiplet and a $\EN=2$ adjoint chiral multiplet.
The numerator is the contribution of the vector multiplet with boundary conditions ${\bf N}$: boundary gauginos (and their $n$-th derivatives) are counted in $\overline{Q}_+$-cohomology  with charge
\beq\label{vec}
(\text{vector}, {\bf N})  : \;\;\; \left(\text{adj},n+\dfrac{1}{2},1,0\right)\;\; \text{under}\;\; U(1)\times U(1)_{J_3}\times U(1)_R\times U(1)_{\ve_2}\; .
\eeq 
The denominator is the contribution of the adjoint chiral multiplet with boundary conditions N: boundary scalars (and their $n$-th derivatives) are counted in $\overline{Q}_+$-cohomology with charge
\beq\label{chiraladjoint}
(\text{adj chiral}, \text{N}) : \qquad \left(\text{adj},n,0,-1\right)\;\; \text{under}\;\; U(1)\times U(1)_{J_3}\times U(1)_R\times U(1)_{\ve_2}\; ,
\eeq 
Note that the above charge assignment is consistent with the existence of a $\EN=4$ superpotential: this is the familiar superpotential term coupling the $\EN=2$ bifundamental chiral multiplets \eqref{chiral1} and \eqref{chiral2} to the adjoint chiral  multiplet \eqref{chiraladjoint}.\\

In type IIB, the contribution \eqref{veccon} is obtained from quantizing the zero modes of strings starting and ending on the D3-brane segment, between the two NS5-branes. Meanwhile, the contribution \eqref{index3d1} is obtained from quantizing the zero modes of strings starting on the D3-brane segment, and ending on the semi-infinite part of the D3-brane. Note the fugacity $X_0$ there (or $X$ here)  is now integrated over.\\

There are further contributions of boundary 't Hooft anomalies at 1-loop. Crucially, we must ensure there is no gauge-gauge anomaly, since the gauge symmetry is preserved by the boundary condition ${\bf N}$ on the gauge multiplet. Here, both the $\EN=4$ vector multiplet and the $\EN=4$  hypermultiplet are separately gauge anomaly-free: 
\begin{itemize}
    \item For the $\EN=4$ vector multiplet, the anomaly polynomial is 
    \beq
+\frac{1}{2}f^2_R - \frac{1}{2}f^2_R = 0 \; ,
    \eeq
where $f_R$ denotes the curvature of the $U(1)_R$ R-symmetry. The first term is the contribution of 
 the  $\EN=2$ gauginos, which cancels the second term, the contribution of the fermions in the $\EN=2$ adjoint chiral multiplet.
\item
For the $\EN=4$ bifundamental chiral multiplet, the anomaly polynomial is 
\beq
-\frac{1}{2}(f-f_{X_1}+f_{\ve_2})^2 + \frac{1}{2}(f-f_{X_1}-f_{\ve_2})^2 = -2\,f\,f_{\ve_2} + 2\,f_{X_1}\,f_{\ve_2}\; ,
\eeq
where $f, f_{X_1}, f_{\ve_2}$, are the curvatures of the $U(1)$ gauge symmetry, $U(1)_{X_1}$ and  $U(1)_{\ve_2}$ flavor symmetries, respectively. The $\EN=2$ chiral multiplet with boundary condition N contributes the first term on the left-hand side, while the $\EN=2$ chiral multiplet with boundary condition D contributes the second term. In particular, the gauge anomaly is canceled, but there are remaining \emph{mixed} 't Hooft anomalies, which indicate $U(1)_{X_1}$ and  $U(1)_{\ve_2}$ are both broken at the boundary.\footnote{The analysis also applies to the non-abelian case: for $K$ D3-brane segments of the same type instead of just 1, the gauge group will be $U(K)$, in which case the gauginos contribute additional terms $+K\text{Tr}(f^2)-(\text{Tr}f)^2$ to the anomaly polynomial, but these are again exactly canceled by the fermions in the adjoint chiral multiplet with boundary condition N. Similarly, the gauge anomaly for the hypermultiplets is canceled as in the abelian case.}
\item There is a mixed topological-gauge anomaly, with polynomial 
\beq
-2 \,(f- f_{X_1})\, f_v \; ,
\eeq
where $f_v$ is the curvature of the  $U(1)_{{\cal J}}$ topological symmetry. 
\end{itemize}
Correspondingly, inside the index integrand, the mixed anomalies manifest themselves as the additional factors
\beq\label{anomalyFI}
e^{-\frac{\ln(q_2) \ln(X/X_1)}{\ln(q_3)}}\, e^{-\frac{\ln(v) \ln(X/X_1)}{\ln(q_3)}} \; .
\eeq
For our purposes, we only really need to keep track of the dynamical contributions to the index, which are the ones \emph{inside} the integrand. ``Constant'' $q$-Pochhammer symbols and boundary 't Hooft anomaly contributions of global symmetries factor out of the integral, and our index computations will only be correct up to overall normalization by such factors.\\

Note that none of our expressions depend on moduli in the $\mathbb{R}_t$ direction. This is by design, since the $\mathbb{R}_t$ direction is topological in our background.\\

All in all, the index comes out to be
\beq\label{index3d2}
{\mathcal Z}^{({\bf N},+)}_{\text{D3}_{0,0,1}-\text{NS5}_{1,0,0}}=\frac{\left(q_3\, ; q_3\right)_\infty}{\left(q^{-1}_2\, ; q_3\right)_\infty}\oint \frac{dX}{2\pi i\,X}\,e^{-\frac{\ln(q_2) \ln(X/X_1)}{\ln(q_3)}}\, e^{-\frac{\ln(v) \ln(X/X_1)}{\ln(q_3)}}  \frac{\left(\sqrt{q_3/q_2} \; X/X_1\, ; q_3\right)_\infty}{\left(\sqrt{q_3 q_2} \; X/X_1\, ; q_3\right)_\infty}  \; .
\eeq
We recognize the vev of a screening current \eqref{eq:screeningdef} and a dual intertwiner,
\beq\label{eq:2pointintertwiner2}
{\mathcal Z}^{({\bf N},+)}_{\text{D3}_{0,0,1}-\text{NS5}_{1,0,0}}=\left\langle\oint \frac{dX}{2\pi i\,X} \left(1\otimes \Phi_{3,1}^*(X_1)\right)\,{\ES}^{3}_{1,1}(X) \,  \right\rangle \; ,
\eeq
where the bracket notation labels the vacuum
\beq\label{eq:vacuum1}
\left\langle  \ldots  \right\rangle = {[u]^{(3)*}_{0}}\otimes\langle v_1\, q_2^{-1},\varnothing|\otimes\langle v_2,\varnothing| \;\;\ldots\;\; |v_1,\varnothing\rangle\otimes|v_2,\varnothing\rangle \; .
\eeq
In particular, the complexified FI parameter is $v=v_1/v_2$, and appears in the integrand via the zero modes of the operators acting on the Fock vacua, see \eqref{eq:zeromode}. The integrand should be understood as the composition of two homomorphisms: the screening current 
\beq
{\ES}^{3}_{1,1} \; : \;\; \CalF_1 ^{\mathbf{(1,0)}}(v_1)\otimes \CalF_1 ^{\mathbf{(1,0)}}(v_2) \rightarrow \CalF_1 ^{\mathbf{(1,0)}}(v_1\, q^{-1}_2)\otimes  \CalF_1 ^{\mathbf{(1,0)}}(v_2\, q_2) \; ,
\eeq
and the dual intertwiner
\beq
\Phi_{3,1}^*\; :\;\; \CalF_1 ^{\mathbf{(1,0)}}(v_2\, q_2)\rightarrow \CalV_3 (u)\otimes \CalF_1 ^{\mathbf{(1,0)}}(v_2) \; .
\eeq 
%\beq
%S^{3}_{1,1}(X) \, \left(1\otimes \Phi_{3,1}^*(X_1)\right)\; : \;\; \CalF_1 ^{\mathbf{(1,0)}}(v_1)\otimes \CalF_1 ^{\mathbf{(1,0)}}(v_2) \rightarrow \CalF_1 ^{\mathbf{(1,0)}}(v_1\, q^{-1}_2)\otimes\CalV_3 (u)\otimes \CalF_1 ^{\mathbf{(1,0)}}(v_2) \; ,
%\eeq
Above, $\CalF_1 ^{\mathbf{(1,0)}}(v_1)$ labels the Fock space of the first $\text{NS5}_{1,0,0}$ brane (the one the D3-brane ends on), and $\CalF_1 ^{\mathbf{(1,0)}}(v_2)$ labels the Fock space of the second intersecting $\text{NS5}_{1,0,0}$ brane (the notation was introduced above equation \eqref{eq:zeromode}). Moreover, the state in $\CalV_3 (u)$ is annihilated by the dual state $[u]^*_{0}$ from the vacuum: if we expand $\Phi^* _{3,1}$  in terms of basis elements as
$\Phi^* _{3,1} = \sum_{i\in \BZ} (\Phi^* _{3,1})_i [u]_i ^{(3)}$, then the insertion of the vacuum state ${[u]^{(3)*}_{0}}$ results in $\sum_{i\in \BZ} (\Phi^* _{3,1})_i \,{[u]^{(3)*}_{0}}\, [u]_i ^{(3)}=(\Phi^* _{3,1})_0$, by orthonormality of the basis.\\

Equivalently, the half-index is also the vev of an R-matrix, as pictured on the bottom of Figure \ref{fig:electricvsmagnetic}. Namely, the integrand can also be understood as the
%if we write the screening current in terms of intertwiners as we did in \eqref{eq:screeningdef},
%\begin{align}\label{eq:screeningdefagain}
%    S^{3} _{1,1} = \sum_{i\in \BZ} (\Phi^* _{3,1} )_i \otimes \left(\Phi _{3,1} \right)_i \; ,
%\end{align}
composition of two different homomorphisms: the dual intertwiner
\beq
\Phi_{3,1}^*\; :\;\; \CalF_1 ^{\mathbf{(1,0)}}(v_1)\rightarrow \CalV_3 (u)\otimes \CalF_1 ^{\mathbf{(1,0)}}(v_1\, q^{-1}_2) \; .
\eeq 
and the non-transverse R-matrix 
\beq
R^{(1)} \; : \;\; \CalV_3 (u)\otimes\CalF_2 ^{\mathbf{(1,0)}}(v_2) \rightarrow \CalV_3 (u)\otimes\CalF_2 ^{\mathbf{(1,0)}}(v_2)  \; .
\eeq
Explicitly, define an integrand
\beq
R^{(1)}(X_1) = \oint \frac{dX}{2\pi i\,X}\,{\ER}^{(1)}(X,X_1) \; ,
\eeq
for the integral presentation \eqref{eq:rcontnormal} of the R-matrix $R^{(1)}$; then the half-index is the vev
\beq\label{eq:index3d2transverse2}
{\mathcal Z}^{({\bf N},+)}_{\text{D3}_{0,0,1}-\text{NS5}_{1,0,0}} = \left\langle \oint \frac{dX}{2\pi i\,X}\,{\ER}^{(1)}(X,X_1)\,  (\Phi_{3,1}^*(X)\otimes 1)\right\rangle \; .
\eeq

\vspace{4mm}

\subsubsection{Index for non-transverse $\text{D3}_{0,0,1}$-$\text{NS5}_{0,1,0}$ intersection}
\label{sssec:nontransverse2}

We start again with a semi-infinite D3$_{0,0,1}$ brane ending on an NS5$_{1,0,0}$ brane. This time, we make the D3-brane intersect a new  NS5$_{0,1,0}$ brane, located some distance away from NS5$_{1,0,0}$ along $\mathbb{R}_t$. We would like to understand the half-index of the resulting configuration.\\

Quantizing the zero modes of strings at the  D3$_{0,0,1}$-NS5$_{0,1,0}$ intersection yields a 3d $\EN=4$ hypermultiplet as before. In the index, this is the contribution \eqref{index3d1}, with a swap of fugacities $q_2 \leftrightarrow q_1$:
\begin{align}
\label{index3new}
\frac{\left(\sqrt{q_3/q_1} \; X/X_1\, ; q_3\right)_\infty}{\left(\sqrt{q_3 q_1} \; X/X_1\, ; q_3\right)_\infty} \; .
\end{align}

Quantizing the zero modes of strings starting and ending on the D3-brane segment, between the NS5$_{1,0,0}$ and NS5$_{0,1,0}$ branes, the supersymmetry is broken from 8 to 4 supercharges, and the surviving degrees of freedom are those of a 3d $\EN=2$ vector multiplet.
Indeed, the NS5$_{0,1,0}$ brane is rotated in the $\BR^2_{\ve_2}$ plane compared to the NS5$_{1,0,0}$ brane; this is a complex mass deformation which explicitly breaks 3d $\EN=4$ supersymmetry to $\EN=2$ \cite{Kitao:1998mf,Bergman:1999na}. The $x^3$ and $x^4$ moduli for the position of the D3 brane on $\BR^2_{\ve_2}$ are lifted. After the mass deformation, the $\EN=2$ adjoint chiral multiplet inside the $\EN=4$ vector multiplet no longer contributes to the index, and only the $\EN=2$ vector multiplet remains. 

Imposing boundary conditions ${\bf N}$ on this multiplet, the index contribution is
\beq\label{vecconnew}
\left(q_3 ; q_3\right)_\infty \oint \frac{dX}{2\pi i \,X} \; .
\eeq 
Note the absence of the factor $\left(q^{-1}_2\, ; q_3\right)^{-1}_\infty$,  which stood for the contribution of the adjoint chiral multiplet in \eqref{veccon}.\\

In our abelian setup, the 3d gauge theory has no gauge anomaly, but it has a $U(1)_R$ R-symmetry anomaly: the $\EN=2$ gaugino contribution to the boundary anomaly polynomial is $+\frac{1}{2}f^2_R$, with $f_R$ the curvature of $U(1)_R$.\footnote{For $K$ D3-branes and a $U(K)$ gauge group, there would now be a gauge anomaly $+K\text{Tr}(f^2)-(\text{Tr}f)^2$ to cancel: if we do not introduce additional matter (as new 3d $\EN=2$ chiral multiplets and 2d $\EN=(0,2)$  chiral multiplets at the boundary), one possibility is to turn on bare Chern-Simons levels in the bulk. In fact, for a $U(K)$ gauge group, we are allowed to turn on two levels; for the $SU(K)$ part of the gauge group, a Chern-Simons level $k_1$ contributes $k_1\text{Tr}(f^2)$ to the anomaly polynomial. For the $U(1)$ part, an effective Chern-Simons level $k_1+k_2$ contributes $\frac{k_2}{K}(\text{Tr}f)^2$ to the anomaly polynomial. We would therefore choose $k_1=-K$ and $k_2=K$.} We will ignore this contribution as it factors out of the index integral. The remaining mixed anomalies are as before, again after swapping $q_2 \leftrightarrow q_1$.
It follows that the index evaluates to
\beq\label{eq:index3new}
{\mathcal Z}^{({\bf N},+)}_{\text{D3}_{0,0,1}\text{-}\text{NS5}_{0,1,0}}=\left(q_3\, ; q_3\right)_\infty\oint \frac{dX}{2\pi i\,X}\,e^{-\frac{\ln(q_1) \ln(X/X_1)}{\ln(q_3)}}\, e^{-\frac{\ln(v) \ln(X/X_1)}{\ln(q_3)}} \frac{\left(\sqrt{q_3/q_1} \; X/X_1\, ; q_3\right)_\infty}{\left(\sqrt{q_3 q_1} \; X/X_1\, ; q_3\right)_\infty} \; .
\eeq
This is the vev of a different screening current \eqref{eq:screeningdef3} and a dual intertwiner,
\beq\label{eq:2pointintertwinernew}
{\mathcal Z}^{({\bf N},+)}_{\text{D3}_{0,0,1}\text{-}\text{NS5}_{0,1,0}}=\left\langle\oint \frac{dX}{2\pi i\,X} \left(1\otimes \Phi_{3,1}^*(X_1)\right)\,{\ES}^{3}_{1,2}(X) \,  \right\rangle \; ,
\eeq
where the bracket notation labels the vacuum
\beq
\left\langle  \ldots  \right\rangle = {[u]^{(3)*}_{0}}\otimes\langle v_1\, q_2^{-1},\varnothing|\otimes\langle v_2,\varnothing| \;\;\ldots\;\; |v_1,\varnothing\rangle\otimes|v_2,\varnothing\rangle \; .
\eeq
%\beq
%S^{3}_{1,1}(X) \, \left(1\otimes \Phi_{3,1}^*(X_1)\right)\; : \;\; \CalF_1 ^{\mathbf{(1,0)}}(v_1)\otimes \CalF_1 ^{\mathbf{(1,0)}}(v_2) \rightarrow \CalF_1 ^{\mathbf{(1,0)}}(v_1\, q^{-1}_2)\otimes\CalV_3 (u)\otimes \CalF_1 ^{\mathbf{(1,0)}}(v_2) \; ,
%\eeq
Equivalently, this is the vev of an R-matrix: if we define
\beq
R^{(2)}(X_1) = \oint \frac{dX}{2\pi i\,X}\,{\ER}^{(2)}(X,X_1) \; ,
\eeq
where $R^{(2)}$ is the R-matrix integral \eqref{eq:rcontnormal}, we obtain
\beq\label{eq:index3d2transversenew}
{\mathcal Z}^{({\bf N},+)}_{\text{D3}_{0,0,1}-\text{NS5}_{0,1,0}} = \left\langle \oint \frac{dX}{2\pi i\,X}\,{\ER}^{(2)}(X,X_1)\,  (\Phi_{3,2}^*(X)\otimes 1)\right\rangle \; .
\eeq

\vspace{4mm}

\subsubsection{Index for transverse $\text{D3}_{0,0,1}$-$\text{NS5}_{0,0,1}$ intersection}
\label{sssec:transverse}

The index is further able to capture the contribution of open strings at a fully transverse D3-NS5-brane intersection. For concreteness, we consider a D3$_{0,0,1}$ brane and a NS5$_{0,0,1}$ brane; this configuration is the type IIB frame dual to the M-theory setup of section \ref{subsec:transr}. As before, it will be convenient to make the D3$_{0,0,1}$ end on one end on a NS5$_{1,0,0}$ brane, separated from the NS5$_{0,0,1}$ brane in the $\mathbb{R}_t$-direction.\\

When $q_c=1$ for $c=1,2,3$, the brane configuration only preserves a 1d $\EN=4$ quantum mechanics on the circle $S^1(R)$, by which we mean a reduction of 2d $\EN=(0,4)$ supersymmetry to one dimension. The supersymmetry is not broken further by turning on the $\Omega$-background, $q_c\neq 1$.\footnote{Explicit expressions for the preserved supercharges can be found in \cite{Hwang:2014uwa}. A detailed string theory derivation of the preserved supersymmetry is beyond the scope of this paper, but can be derived following the analysis of \cite{NaveenNikita}.}\\

The BPS states counted by the index are those of this $\EN=4$ quantum mechanics. If we view the expression \eqref{3dhalfindex} as the Witten index of this quantum mechanics, it is straightforward to count the degrees of freedom directly from the relevant 1-dimensional BPS multiplets. In fact, we could have proceeded that way already for the previous 3d $\EN=4$ brane intersections, after decomposing the 3d multiplets in terms of 1d multiplets, and identifying  the 3d supercharges  $Q_+$ and $\overline{Q}_+$ as the ones contributing to the quantum mechanical index. In the type IIB picture, this is done by explicitly introducing dynamical D1-branes, and quantizing the zero modes of strings stretching between them and the remaining heavier D-branes.\\

Instead, we chose to interpret \eqref{3dhalfindex} from a 3-dimensional point of view.\footnote{For further details on the equivalence between the indices of the 3d theory and its vortex quantum mechanics, see for instance \cite{Hwang:2017kmk,Haouzi:2019jzk,Haouzi:2020bso,Crew:2020jyf}.} In this spirit, we will count the 1d $\EN=4$ multiplets of the D3-NS5 intersection as arising from a coupled 3d/1d  half-index, by which we mean the half-index of a 3d bulk theory supported on $\mathbb{R}^2_{\ve_3}\times S^1(R)$, coupled to a codimension-2 line defect along $S^1(R)$.

From this perspective, the D3$_{0,0,1}$-NS5$_{1,0,0}$ and D3$_{0,0,1}$-NS5$_{0,1,0}$ intersections from section \ref{sssec:nontransverse1} and \ref{sssec:nontransverse2} are both captured by a 3d/1d index whose sole contributions come from the 3d bulk, while the defect is trivial and contributes the identity. By contrast, the present D3$_{0,0,1}$-NS5$_{0,0,1}$ intersection will be captured instead by a 3d/1d index whose  1d contributions are nontrivial, while the 3d bulk ones are essentially trivial (up to charge 0 $U(1)$ contributions and boundary mixed 't Hooft anomalies).\\

We denote by $X$ and $X_1$ the (complexified) positions of the D3$_{0,0,1}$ brane and the NS5$_{0,0,1}$ brane along $\mathbb{R}_x$, respectively. These are both flavor symmetry masses  from the point of view of the 1d quantum mechanics. We couple this 1d theory to the bulk by gauging these flavor symmetries with 3d background vector multiplets. Correspondingly, in the 3d/1d index, the 1d masses are turned into the scalars of the corresponding 3d vector multiplets. In what follows, we choose the scalar $X$ to be a modulus, meaning the associated vector multiplet is dynamical. Meanwhile, we choose the scalar $X_1$ to remain fixed as a mass even in 3d, meaning the associated vector multiplet is non-dynamical.\\
 
The fastest way to identify the contribution of the D3$_{0,0,1}$-NS5$_{0,0,1}$ strings is to note that because the intersection is fully transverse, the modulus $X$ should really be \emph{frozen} to the locus $X=X_1$. If we choose Neumann boundary conditions on the 3d $U(1)$ gauge field, this is easily achieved by considering the integral
\begin{equation}
\oint {dX}\; \frac{1}{1-X_1/X}\,\;\ldots
\end{equation}
That is, we force $X=X_1$ via the contribution of a 1d $\EN=2$ chiral multiplet.   
Because the preserved supersymmetry is $\EN=4$, this cannot be the full story:  the 1d $\EN=2$ chiral multiplet must pair up with another 1d $\EN=2$ chiral multiplet to restore $\EN=4$ supersymmetry.

With the support of the D3$_{0,0,1}$ brane along $\mathbb{R}^2_{\ve_3}$, this second chiral multiplet will have  $U(1)_{J_3}$ charge -1 (for a derivation, see for instance \cite{Hwang:2014uwa}). Then, after including the contribution of the bulk mixed topological-gauge anomaly, we obtain the index of the D3$_{0,0,1}$-NS5$_{0,0,1}$ intersection:
\begin{equation}\label{index3d2transverse}
{\mathcal Z}^{({\bf N})}_{\text{D3}_{0,0,1}-\text{NS5}_{0,0,1}}=\oint {dX}\,e^{-\frac{\ln(v) \ln(X/X_1)}{\ln(q_3)}} \frac{1}{\left(1-X_1/X\right) \left(1-q^{-1}_3\, X_1/X\right)}  \; .
\end{equation}
Correspondingly, the contour is chosen to enclose not only the simple pole at $X=X_1$, as we argued classically, but also the second simple pole at $X=X_1\,q^{-1}_3$, understood as a quantum contribution. For another derivation of the contribution \eqref{index3d2transverse}, see the work \cite{Nieri:2017ntx}.\\

Up to overall normalization, we recognize the vev of the totally transverse R-matrix: if we define ${\ER}^{(3)}(X,X_1)$ as the integrand,
\beq
R^{(3)}(X_1) = \oint \frac{dX}{2\pi i\,X}\,{\ER}^{(3)}(X,X_1) \; ,
\eeq
of the integral expression \eqref{eq:rcontnormal3} for the R-matrix $R^{(3)}$, we obtain\footnote{We stress that $\Phi_{3,3}^*(X)$ is \emph{not} an intertwiner, as explained in section \ref{subsubsec:nontransr}.}
\beq\label{eq:index3d2transversenewnew}
{\mathcal Z}^{({\bf N})}_{\text{D3}_{0,0,1}\text{-}\text{NS5}_{0,0,1}} = \left\langle \oint \frac{dX}{2\pi i\,X}\,{\ER}^{(3)}(X,X_1)\,  (\Phi_{3,3}^*(X)\otimes 1)\right\rangle \; .
\eeq
The bracket notation labels the vacuum
\beq
\left\langle  \ldots  \right\rangle = [u]_{0}^{(3)*}\otimes\langle v_1\, q_2^{-1},\varnothing|\otimes\langle v_2,\varnothing| \;\;\ldots\;\; |v_1,\varnothing\rangle\otimes|v_2,\varnothing\rangle \; .
\eeq

\vspace{4mm}

\paragraph{Remark}
In an S-dual frame, our D3$_{0,0,1}$-NS5$_{0,0,1}$ configuration turns into  a D3$_{0,0,1}$-D5$_{0,0,1}$ configuration, with 8 Dirichlet-Neumann directions. This is T-dual to the well-studied D0-D8 system with (0,8) supersymmetry, where the massless open string modes are known to be chiral fermions \cite{Banks:1997zs,Bachas:1997kn,Hung:2006nn,Dijkgraaf:2007sw,Tong:2014yna,NaveenNikita}. The index for this intersection would therefore count degrees of freedom found in $\EN=2$ Fermi multiplets, instead of our chiral multiplets. It would be interesting to generalize this observation to more general brane configurations, and systematically analyze the behavior of the 3d/1d half-index under S-duality, in the spirit of \cite{Okazaki:2019bok}.

\vspace{4mm}

\subsection{Index as a sum and Dirichlet boundary conditions}
\label{ssec:Dirichletcond}

In the above, we imposed Neumann boundary conditions ${\bf N}$ on the 3d $\EN=2$ gauge multiplet, and showed that the resulting half-index was a vev of the Miura R-matrix (or a product of such R-matrices in general), presented as a contour integral. In the present section, we will instead impose Dirichlet boundary conditions ${\bf D}$ on the gauge multiplet, and show that the half-index is again the same vev, but presented as a sum over monopole charges.\\

First, recall that in the Neumann case, the gauge symmetry was unbroken at the boundary, which is why the half-index took the form of an integral, projecting to gauge-invariant operators on the boundary of the 3-manifold. 
In contrast, the Dirichlet boundary condition sets $A_{\mu|\partial}=0$ for the gauge field, which implies the local gauge symmetry on the boundary is broken to a global symmetry $G_{|\partial}=U(1)_{\partial}$. We introduce a fugacity $u\in\mathbb{C}^\times$ and a generator $\ft_\partial$ for this $U(1)_{\partial}$ symmetry. Explicitly, the index now reads
\beq
\label{3dhalfindexmore3}
{\mathcal Z}^{({\bf D})}  = {\rm Tr}\left[(-1)^F\, {q_3}^{J_3+\frac{V}{2}}\; {q_2}^{\frac{V-A}{2}} \; u^{\ft_\partial} \;  X^{\ft_H} \;  v^{\ft_C} \right]\;\; .
\eeq 
In fact, we will seek to break $U(1)_{\partial}$ entirely, as follows: we will generate a flow  along the boundary to a new boundary condition by giving one of the 3d chiral fields a vev $\mathfrak{c}\neq 0$. 
At the level of the index, the weight of the corresponding field must be set to 1 to preserve supersymmetry. In the terminology of \cite{Dimofte:2017tpi}, we say that the chiral field is given deformed Dirichlet boundary condition $\text{D}_{\mathfrak{c}}$.\\

To see how this works in practice, consider again the non-transverse brane configuration of section \ref{sssec:nontransverse1}, with two NS5$_{1,0,0}$ branes of the same type and a semi-infinite D3$_{0,0,1}$ brane ending on one of them while intersecting the other. Recall that the supersymmetry preserved by this configuration is 3d $\EN=4$, and that the low energy effective theory is a $U(1)$ gauge theory with a single hypermultiplet. Imposing the Dirichlet boundary conditions on the $\EN=4$ vector multiplet, the index contribution is
 \beq\label{pertDirichlet}
\frac{\left(q_3\,q_2\, ; q_3\right)_\infty}{\left(q_3\, ; q_3\right)_\infty} \; .
\eeq
The denominator is the contribution of the $\EN=2$ vector multiplet with boundary conditions ${\bf D}$. The gauge symmetry is broken along the boundary, but there is a residual gauge symmetry orthogonal to it, and the gauge-invariant operators there (and their $n$-th derivatives) generate the $\overline{Q}_+$-cohomology.\footnote{In detail, the 3d gauge-invariant linear multiplet decomposes on the boundary as a $(0,2)$ gauge multiplet, whose top component is the gaugino, and a $(0,2)$ chiral multiplet, whose top component is $\sigma+i\,A_{\perp}$, where $\sigma$ is a real scalar and $A_{\perp}$ is the gauge field component perpendicular to the boundary. The index with boundary condition ${\bf N}$ counts the contributions of the gauge-invariant operators built out of the gaugino, while the index with boundary condition ${\bf D}$ counts the contributions of gauge-invariant operators built out of $\sigma+i\,A_{\perp}$.\label{fnlabel}} They contribute to the index with charge
\beq\label{vecdir}
(\text{vector}, {\bf D})  : \;\;\; \left(\text{adj},n+1,0,0\right)\;\; \text{under}\;\; U(1)_{\partial}\times U(1)_{J_3}\times U(1)_R\times U(1)_{\ve_2}\; .
\eeq  
The numerator is the contribution of the $\EN=2$ adjoint chiral multiplet with boundary conditions D: the gauginos (and their $n$-th derivatives) are counted in $\overline{Q}_+$-cohomology with charge  
\beq\label{adjointdir}
(\text{adj chiral}, {\bf D})  : \;\;\; \left(\text{adj},n+\dfrac{1}{2},1,1\right)\;\; \text{under}\;\; U(1)_{\partial}\times U(1)_{J_3}\times U(1)_R\times U(1)_{\ve_2}\; .
\eeq 
The $\EN=4$ hypermultiplet decomposes into a pair of $\EN=2$ chiral multiplets $(\Phi^+,\Phi^-)$. Here, we decide to impose Dirichlet boundary conditions for both $\Phi^+$ and $\Phi^-$, with index contribution 
\begin{align}
\label{index3dirrr}
\left(\sqrt{q_3/q_2}\; X_1/u\, ; q_3\right)_\infty \, \left(\sqrt{q_3/q_2} \;u/X_1\, ; q_3\right)_\infty \; .
\end{align} 
We will justify this choice below. The first factor is the contribution of $\Phi^+$ with boundary conditions D: boundary fermions (and their $n$-th derivatives) are counted in $\overline{Q}_+$-cohomology with charge
\beq\label{chiraldir}
(\text{bif chiral}, \text{D}) : \;\;\; \left(-1,1,n+\frac{1}{2},0,\frac{-1}{2}\right)\;\; \text{under}\;\; U(1)_{\partial}\times U(1)_{X_1}\times U(1)_{J_3}\times U(1)_R\times U(1)_{\ve_2}.
\eeq 
The second factor is the contribution of $\Phi^-$ with boundary conditions D: boundary fermions (and their $n$-th derivatives) are counted in $\overline{Q}_+$-cohomology with charge 
\beq\label{chiraldir2}
(\overline{\text{bif}}\;\text{chiral}, \text{D}) : \;\;\; \left(1,-1,n+\frac{1}{2},0,\frac{-1}{2}\right)\;\; \text{under}\;\; U(1)_{\partial}\times U(1)_{X_1}\times U(1)_{J_3}\times U(1)_R\times U(1)_{\ve_2} .
\eeq 

Crucially, the contribution \eqref{index3dirrr} receives non-perturbative corrections at the boundary coming from BPS monopole operators. Given a 3d $\EN=2$ theory, a monopole operator in the bulk is a disorder operator whose profile is a singular solution to
\beq
F = \star D \sigma\, , \qquad\; D \star\sigma = 0 \; ,
\eeq
where  $\sigma$ is the real scalar of footnote \ref{fnlabel}, and $F$ is the field strength of the gauge field. A monopole operator at the origin will be surrounded by a hemisphere $D^2$, with boundary $\partial(D^2)=S^1_{D^2}$. Because the scalar $\sigma$ and field strength component $F_{z\bot}$ and are left unconstrained at the boundary ($z$ is a holomorphic coordinate on $D^2$), this monopole is compatible with the Dirichlet boundary condition  ${\bf D}$, which we impose to trivialize the $U(1)$-bundle at the boundary $S^1_{D^2}$. After this trivialization, the Dirac quantization condition takes the form:
 \beq
 \frac{1}{2\pi}\oint_{D^2}F +\frac{1}{2\pi}\oint_{S^1_{D^2}} A = m \in\mathbb{Z} \; .
 \eeq
 But the boundary condition $A_{\mu|\partial}=0$ implies $\oint_{S^1_{D^2}} A=0$, so the monopole charge $m$ is nothing but the quantized flux through the hemisphere:\footnote{More generally, for a non-abelian gauge group $G$, we embed the Dirac monopole solution into $G$, meaning $m$ is an element of the cocharacter lattice  $m\in\Lambda_{cochar}=\text{Hom}(U(1),T_{G})$, the space of maps to the maximal torus of $G$. In our abelian setup, $\text{Hom}(U(1),U(1))=\mathbb{Z}$ and $m$ is just an integer.}
\beq
\frac{1}{2\pi}\oint_{D^2}F = m\in\mathbb{Z} \; .
\eeq

In the end, the half-index takes the form of a sum over all magnetic fluxes $m$ through the hemisphere $D^2$. Moreover, all electrically charged states acquire spin, meaning the fugacity $u$ for the boundary  symmetry $U(1)_{\partial}$ is shifted everywhere as $u\rightarrow q_3^m\, u$ \cite{Dimofte:2017tpi}:
\beq
\label{3dhalfindexdir}
{\mathcal Z}^{({\bf D},\text{D},\text{D})}_{\text{D3}_{0,0,1}-\text{NS5}_{1,0,0}} = \frac{\left(q_3\,q_2\, ; q_3\right)_\infty}{\left(q_3\, ; q_3\right)_\infty}\sum_{m\in\mathbb{Z}} \; \left(q_3^{-m}\sqrt{q_3/q_2}\; X_1/u\, ; q_3\right)_\infty \, \left(q_3^{m}\sqrt{q_3/q_2} \;u/X_1\, ; q_3\right)_\infty \; \ldots \; .
\eeq
The $``\ldots"$ stands for the 1-loop contributions of boundary 't Hooft anomalies, which we will come back to shortly.\\

In order to implement the deformed Dirichlet boundary condition $\text{D}_{\mathfrak{c}}$, we generate an RG flow along the boundary to a new boundary condition, where a chiral field acquires a vev $\mathfrak{c}\neq 0$ in the supersymmetric vacuum. If this chiral field has charge $(1,-1,1)$ under $U(1)_V\times U(1)_{\partial}\times U(1)_{X_1}$, then the linear combinations $V+\ft_\partial\equiv V'$ and $\ft_{X_1} + \,\ft_\partial\equiv\ft'_{X_1}$ are preserved along this flow; the generators of the new vector R-symmetry and flavor symmetry on the boundary are denoted by $V'$ and $\ft'_{X_1}$ in the index \eqref{3dhalfindexmore3}, respectively. Concretely, the weight of the corresponding chiral field is set to 1, which translates to a constraint $u^{-1}\, q_3^{1/2}\, q_2^{1/2}\, X_1\,  =1$ on $u$. Put differently, we simply substitute $u = q_3^{1/2}\, q_2^{1/2}\, X_1$ in the entire summand \eqref{3dhalfindexdir}. We obtain
\beq
\label{3dhalfindexdirdef}
{\mathcal Z}^{({\bf D},\text{D},\text{D}_{\mathfrak{c}})}_{\text{D3}_{0,0,1}-\text{NS5}_{1,0,0}} = \frac{\left(q_3\,q_2\, ; q_3\right)_\infty}{\left(q_3\, ; q_3\right)_\infty}\sum^{\infty}_{m=0} \; \left(q_3^{-m}q^{-1}_2\; \, ; q_3\right)_\infty \, \left(q_3^{1+m}\, ; q_3\right)_\infty \; \ldots \; ,
\eeq
Note the summation index $m$ only runs over the positive integers now, because of the factor $\left(q_3^{1+m}\, ; q_3\right)_\infty$.\\

\paragraph{Remark} 
An alternative way to motivate the deformed Dirichlet boundary condition is as follows: consider again the hypermultiplet contribution to the index \eqref{index3dirrr}, where we rescale $u$ by the fugacity $q^{1/2}_3$ out of convenience:
\beq\label{newhyper}
\left(q_3^{-m}\sqrt{1/q_2}\; X_1/u\, ; q_3\right)_\infty \, \left(q_3^{1+m}\sqrt{1/q_2} \;u/X_1\, ; q_3\right)_\infty \; .
\eeq
In the $U(1)$ 3d gauge theory, the flavor symmetries acts by Hamiltonian isometries on the complex scalars of the chiral multiplets $(\Phi^+, \Phi^-)$ as
\beq\label{eq:flavorhamiltonians}
U(1)_{X_1}:\;(\phi^+, \phi^-)\; \rightarrow (e^{+i\theta}\phi^+, e^{-i\theta}\phi^-)\; , \qquad U(1)_{\ve_2}:\;(\phi^+, \phi^-)\; \rightarrow (e^{+i\theta/2}\phi^+, e^{+i\theta/2}\phi^-)\; ,
\eeq
and the gauge symmetry acts as 
\beq\label{eq:gaugehamiltonians}
U(1):\;(\phi^+, \phi^-)\; \rightarrow (e^{-i\theta}\phi^+, e^{+i\theta}\phi^-) \; .
\eeq
Correspondingly, the complex and real superpotentials are
\begin{align}
\label{superpot}
&\widetilde{W}=\varphi\,\phi^+\,\phi^-\nonumber\\
&W=\left(-\sigma + x_1 +\frac{\ve_2}{2}\right)|\phi^+|^2 + \left(\sigma - x_1 +\frac{\ve_2}{2}\right)|\phi^-|^2 + \xi\, \sigma \; ,
\end{align}
where $\sigma$ and $\phi$ are the real and complex scalars of the $\EN=4$ vector multiplet, while $\xi$ and $m$ are the real FI parameter and hypermultiplet mass, respectively. The (classical) vacua are critical points of these superpotentials, which are solutions of the moment map equations
\begin{align}\label{momentmap}
\left(|\phi^+|^2 - |\phi^-|^2\right) &= \xi\; , &\phi^+ \, \phi^- &=0\; ,\\
(-\sigma+x_1+\frac{\ve_2}{2})\,\phi^+ &= 0\; , &\varphi\, \phi^+ &= 0\; ,\\
(+\sigma-x_1+\frac{\ve_2}{2})\,\phi^- &= 0\; , &\varphi\, \phi^- &= 0\; .
\end{align} 
We are interested in the Higgs vacuum where either $\phi^+$ or $\phi^-$ acquires a nonzero vev.  When $\xi >0$, the vev is $|\phi^+|^2 = \xi$ (with $\phi^-=0=\varphi$), and $-\sigma + x_1 +\frac{\ve_2}{2} = 0$. When $\xi <0$, the vev is $|\phi^-|^2 = -\xi$, with $\sigma - x_1 +\frac{\ve_2}{2} = 0$.
We choose to work with $\xi >0$ throughout\footnote{With this convention, the index has a positive radius of convergence as a power series in the complexified FI parameter $v=e^{-R\left(\xi+i\,A^{\theta}_{\xi}\right)}$ around 0.}, and therefore set $\sigma = x_1 +\frac{\ve_2}{2}$.\\ 

Because the 3d theory is compactified on $S^1(R)$, we likewise impose $A^{\theta}_{\sigma} = A^{\theta}_{x_1} +\frac{A^{\theta}_{\ve_2}}{2}$ for the holonomies.
In the end, substituting  $u=q^{1/2}_2\, X_1$ for the complexified variables in \eqref{newhyper}, we recover the summand of the index  \eqref{3dhalfindexdirdef}.\\

Imposing Dirichlet boundary conditions on both $\EN=2$ chiral multiplets of the hypermultiplet has introduced a boundary anomaly at 1-loop, meaning there is now an \emph{effective} Chern-Simons coupling at level 1. Correspondingly, a bare monopole operator of charge $m$ will induce a nontrivial electric charge $m$ and spin $\frac{1}{2}m^2$ in the theory, meaning an insertion $u^m\, q^{\frac{m^2}{2}}_3$ inside the summand of the index. After performing the deformed Dirichlet substitution $u = q_3^{1/2}\, q_2^{1/2}\, X_1$, and redefining the topological fugacity $v \rightarrow -v\,X^{-1}_1$, we find:
 \beq
 \label{3dhalfindexdirdef}
 {\mathcal Z}^{({\bf D},\text{D},\text{D}_{\mathfrak{c}})}_{\text{D3}_{0,0,1}-\text{NS5}_{1,0,0}} = \frac{\left(q_3\,q_2\, ; q_3\right)_\infty}{\left(q_3\, ; q_3\right)_\infty}\sum^{\infty}_{m= 0} (-1)^m \,q^m_2\, q^{\frac{m(m+1)}{2}}_3\;v^m \; \left(q_3^{-m}q^{-1}_2\; \, ; q_3\right)_\infty \, \left(q_3^{1+m}\, ; q_3\right)_\infty \; .
 \eeq
Up to normalization by perturbative contributions outside the summand, we recognize the vev of the  R-matrix \eqref{eq:Rsum} (for $c=1$), composed with a dual intertwiner:
\beq\label{eq:index3d2transverse2}
{\mathcal Z}^{({\bf D},\text{D},\text{D}_{\mathfrak{c}})}_{\text{D3}_{0,0,1}-\text{NS5}_{1,0,0}}  = \left\langle {R}^{(1)}\,  (\Phi_{3,1}^*\otimes 1)\right\rangle \; .
\eeq
Above,  $\Phi^* _{3,1} = \sum_{i\in \BZ} \Phi^* _{3,1}(u\,q^i_3)\, [u]_i ^{(3)}$ is understood as an expansion in the basis elements $[u]_i ^{(3)}$. The bracket once again denotes the vacuum \eqref{eq:vacuum1}, so the sum over $i\in\BZ$ reduces to a single term, by orthonormality of the basis.\\

In fact, up to the perturbative normalization, the above Dirichlet index coincides with the residue sum of the Neumann index \eqref{eq:2pointintertwiner2}, where the contour encloses ``$\infty$'' and the infinite number of poles at
\beq
X=q^{-m}_3\, X_1 \; , \qquad \;\;\; m=0,1,2,\ldots
\eeq
For instance, the residues of the mixed topological-gauge anomaly contribution are
\beq
e^{-\frac{\ln(v) \ln(q^{-m}_3\, X_1/X_1)}{\ln(q_3)}}= v^m \; ,
\eeq
which we recognize as the correct factor in the summand of the Dirichlet index. See also section 3.4.2 of \cite{Dimofte:2017tpi}.

%For our purposes, we will slightly simplify the analysis and only keep track of the dynamical contributions to the index, which are the ones \emph{inside} the integrand. As such, we henceforth simply ignore the $q$-Pochhammer "constant" symbols which factor out of the integral, as well as the boundary 't Hooft anomaly contributions at 1-loop involving the global symmetries.\\

\vspace{4mm}

\subsection{Arbitrary D3-NS5 intersections}

In summary, for all configurations considered so far, a D3$_{0,0,1}$ brane ends on a NS5$_{1,0,0}$ brane on one end, and intersects another NS5-brane in a supersymmetric way:

\begin{itemize}
    \item A non-transverse  D3$_{0,0,1}$-NS5$_{1,0,0}$ intersection, preserving 3d $\EN=4$ supersymmetry.
    \item A non-transverse D3$_{0,0,1}$-NS5$_{0,1,0}$ intersection, preserving 3d $\EN=2$ supersymmetry.
    \item A  transverse D3$_{0,0,1}$-NS5$_{0,0,1}$ intersection, preserving 1d $\EN=4$ supersymmetry.
\end{itemize}
These building blocks can be combined to study more general configurations of D3- and NS5-branes. Namely, consider again a semi-infinite D3$_{0,0,1}$ brane ending on a NS5$_{1,0,0}$ brane on one end, and intersecting $L$ NS5$_{1,0,0}$, $M$ NS5$_{0,1,0}$, and $N$ NS5$_{0,0,1}$ branes on the other end. The low energy effective theory is now an abelian quiver gauge theory \cite{Hanany:1996ie}, of 3d gauge group $U(1)^{L+M+N}$, with $L+M$ 3d hypermultiplets and $N$ 1d hypermultiplets (coupled via line defects).\\

It follows that if we give boundary conditions ${\bf N}$ for the vector multiplets, and polarization ``$+$'' to all $L+M$ hypermultiplets, the 3d/1d half index of this configuration is the integral 
\begin{align}\label{eq:generalRmatrixIIB}
\begin{split}
    &{\mathcal Z}^{({\bf N},+,\ldots,+)}_{\text{D3}_{0,0,1} , \text{NS5}_{L,M,N}} = \oint \prod_{I=1} ^{L+M+N} \frac{dX^{(I)}}{2\pi i X^{(I)}}  e^{-\sum_{I \in c^{-1} \{1,2\}} \frac{\log q_{\bar{c}_I}}{ \log q_3} \log \frac{X^{(I)}}{ X^{(I+1)}} }\; e^{-\sum_{I=1} ^{L+M+N}\frac{\log X^{(I)}}{ \log q_3} \log \frac{v_{I}}{v_{I+1}}}\\
    & \qquad \qquad\qquad\qquad\times
\prod_{I \in c^{-1}\{1,2\} } \frac{\left( q_{\bar{c}_I} ^{-1}  \frac{X^{(I)}}{X^{(I+1)}} ;q_3 \right)_\infty }{\left(  \frac{X^{(I)}}{X^{(I+1)}} ;q_3 \right)_\infty} \prod_{I\in c^{-1}(3) } \frac{1}{\left( 1- \frac{ X^{(I+1)}}{X^{(I)}} \right)\left( 1- \frac{q_3 ^{-1} X^{(I+1)} }{X^{(I)}} \right)}\; ,
\end{split}
\end{align}
up to overall normalization by perturbative contributions outside the integral. We used the notation $(\vert c^{-1} (1) \vert, \vert c^{-1} (2) \vert, \vert c^{-1} (3) \vert) = (L,M,N)$, and declare $X^{(L+M+N+1)}\equiv X_1$, as well as $v_{L+M+N+1}=v_{X_1}$.\\

We recognize the vev of a product of R-matrices \eqref{eq:generalRmatrix}, or equivalently the vev of the Miura transformation:
\beq\label{eq:index3dMiura}
{\mathcal Z}^{({\bf N},+,\ldots,+)}_{\text{D3}_{0,0,1} , \text{NS5}_{L,M,N}} = \left\langle \oint \prod_{I=1} ^{L+M+N} \frac{dX^{(I)}}{2\pi i X^{(I)}} \,{\ER}^{(c_{L+M+N})}_{L+M+N}\ldots\,{\ER}^{(c_{2})}_{2}\,{\ER}^{(c_{1})}_{1}\,  (\Phi_{3,1}^*(X^{(0)})\otimes 1)\right\rangle \; .
\eeq
In the bracket $\left\langle  \ldots  \right\rangle$ for the vev, the in-state on the right stands for 
\beq
|v_0,\varnothing\rangle\otimes|v_1,\varnothing\rangle\otimes|v_2,\varnothing\rangle\otimes\ldots\otimes|v_{L+M+N},\varnothing\rangle \; .
\eeq
and the out-state on the left stands for
\beq
{[u]^{(3)*}_{0}}\otimes\langle v_0\, q_2^{-1},\varnothing|\otimes\langle v_1,\varnothing|\otimes\langle v_2,\varnothing|\otimes\ldots\otimes\langle v_{L+M+N},\varnothing| \; .
\eeq

\vspace{4mm}

To close this section, recall that the brane configurations we studied all had in common a semi-infinite D3$_{0,0,1}$ brane ending on a NS5$_{1,0,0}$ brane. Choosing different orientations for this D3-brane and NS5-brane is easily achieved by simply permuting the parameters $q_1$, $q_2$ and $q_3$ in the index we have already computed. We can also relax the semi-infinite requirement for the D3-brane: making the brane infinite simply increases the number of intersections with NS5-branes by 1, which results in one additional hypermultiplet in the gauge theory, and one additional R-matrix insertion in the index. In that way, any arbitrary configuration of D3-branes intersecting NS5-branes in $\Omega$-deformed type IIB can be studied via our R-matrix formalism.

\vspace{4mm}

\section{$qq$-characters from M2-branes on web of M5-branes} \label{sec:qqchar}
So far, we have restricted our discussion to the M5-branes with the holomorphic part of the support being $C^{\mathbf{(1,0)}} = \BC^\times _X$. In this section, we introduce the M5-branes supported on more general holomorphic curves $C^{\mathbf{(p,q)}}$. We explain the gauge-invariant non-transverse intersections of M5-branes provide intertwiners between the $q$-boson representations. We elucidate the M2-branes, which transversally intersect the M5-brane web, give rise to the $qq$-characters of the 5d $\EN=1$ effective gauge theory on the worldvolume of the M5-brane web.

\vspace{4mm}

\subsection{Non-transverse intersections of M5-branes}
We discuss the M5-branes whose holomorphic part of the support is generic holomorphic curve $C^{\mathbf{(p,q)}}$. We find $q$-boson representations associated to them, and study their intersections in terms of the intertwiners of the $q$-boson representations.

\subsubsection{M5-branes on holomorphic curves and $SL(2,\BZ)$-dual $q$-bosons}\label{sssec:pqdualbranes}
In section \ref{subsubsec:singlem5}, we have argued that the worldvolume theory on single M5-brane, supported on $\BR^2 _{\ve_{c+1}} \times \BR^2 _{\ve_{c-1}} \times \BC^\times _X$, gets localized to a free $q$-boson on $\BC^\times _X$. For the single M5-brane supported on $\BR^2 _{\ve_{c+1}} \times \BR^2 _{\ve_{c-1}} \times C^{\mathbf{(p,q)}}$, the only modification is the holomorphic support, so that the algebra of local operators should again be the $q$-boson algebra supported on the holomorphic curve $C^{\mathbf{(p,q)}}$.

In fact, there exists a symmetry that exchanges the M5-branes supported on different holomorphic curves $C^{\mathbf{(p,q)}}$ with one another: the $SL(2,\BZ)$ symmetry of reparametrization of the holomorphic surface $\BC^\times _X \times \BC^\times _Z$. Specifically, let us reparametrize this complex two-dimensional space by $X'  = X^a Z^b$ and $Z' = X^c Z^d$. If $ad-bc =1$, the 3-form background of the twisted M-theory is invariant under the reparametrization since $\frac{d \bar{X}'}{\bar{X}'} \wedge \frac{d \bar{Z}'}{\bar{Z}'} = \frac{d \bar{X}}{\bar{X}} \wedge \frac{d \bar{Z}}{\bar{Z}}$. This $SL(2,\BZ)$ symmetry maps the holomorphic curve $C^{\mathbf{(p,q)}}$ to $C^{\mathbf{(p',q')}}$, where
\begin{align}
    \begin{pmatrix}
        \mathbf{p'} \\ \mathbf{q'} 
    \end{pmatrix} = g \begin{pmatrix}
         \mathbf{p} \\ \mathbf{q}
    \end{pmatrix}, \qquad g = \begin{pmatrix}
        a & b \\ c & d 
    \end{pmatrix} \in SL(2,\BZ).
\end{align}
Thus, the $SL(2,\BZ)$ maps the M5-branes supported on those holomorphic curves accordingly. After the dimensional reduction and T-dualization, this precisely matches with the $SL(2,\BZ)$ symmetry of the IIB theory, which maps the $\mathbf{(p,q)}$-fivebrane to the $\mathbf{(p',q')}$-fivebrane.

The $SL(2,\BZ)$ symmetry of the twisted M-theory descends to the 5d non-commutative $\fgl(1)$ Chern-Simons theory on $\BR_t \times \BC^\times _X \times \BC^\times _Z$. Thus, the $SL(2,\BZ)$ symmetry should be reflected on the universal associative algebra as its automorphism group. Indeed, the quantum toroidal algebra $\qta$ is known to possess an $SL(2,\BZ)$ automorphism group \cite{Miki:2007}. Then, the $q$-boson representations associated to the M5-branes supported on different holomorphic curves should be mapped to one another under the action of this $SL(2,\BZ)$ automorphism group.

The $SL(2,\BZ)$ group is generated by two elements,
\begin{align}
    T = \begin{pmatrix}
        1 & 0 \\ 1 & 1
    \end{pmatrix} ,\qquad S = \begin{pmatrix}
        0 & -1 \\ 1 & 0
    \end{pmatrix}.
\end{align}
Let us first consider the subgroup generated by $T$. When acted on $C^{\mathbf{(1,0)}} = \BC^\times _X$, the subgroup generates all the holomorphic curves of the form $C^{\mathbf{(1,n)}}$, $\mathbf{n}\in \BZ$. The $q$-boson representations associated to the M5-brane on these curves can thus be obtained by applying the $T$-transformations to the $q$-boson representation $\r_{\text{M5}_c ^{\mathbf{(1,0)}}}$ \eqref{eq:singlem5},
\begin{align} \label{eq:qbos1n}
    \r_{\text{M5}_c ^{\mathbf{(1,n)}} } :\qta \twoheadrightarrow \text{M5}_c ^{\mathbf{(1,n)}} \hookrightarrow \text{End}(\CalF^{\mathbf{(1,n)}} _c (v)).
\end{align}
This turns out to be a slight modification of the $q$-boson representation $\r_{\text{M5}_{c} ^{\mathbf{(1,0)}}}$ \eqref{eq:singlem5}, given by
\begin{align} \label{eq:hrep}
\begin{split}
    &E(X) \mapsto -e^{a_0 ^{(c)} \frac{\log q_{c+1 } \log q_{c-1}}{\log q_c}} \frac{1-q_c }{\k_1} \left(\frac{q_c^{1/2}}{X}\right)^{\mathbf{n}} \eta_c (X),\\
    &F(X)\mapsto  e^{-a_0 ^{(c)} \frac{\log q_{c+1 } \log q_{c-1}}{\log q_c}}  \frac{1-q_c  ^{-1} }{\k_1} \left(\frac{X}{q_c^{1/2}}\right)^{\mathbf{n}}  \xi_c (X), \\  
    &K^\pm (X) \mapsto q_c ^{\mp \mathbf{n}/2} \varphi^\pm _c (X)  ,
\end{split}
\end{align}
Note $(C,C^\perp) \mapsto \left( q_c ^{1/2},q_c ^{\mathbf{n}/2} \right)$, in particular.\\

Meanwhile, the $q$-boson representation associated to the M5-brane wrapping $C^{\mathbf{(0,1)}} = \BC^\times _Z$ can be understood in a different way. Recall that one way of obtaining the $q$-boson representation is reducing the twisted M-theory along the circle of $\BC^\times _Z \simeq \BR_z \times S^1$, after which this M5-brane becomes a D4-brane. The $q$-boson representation is viewed as the algebra of local operators on the line $\BR_z$ acting on the equivariant K-theory of the moduli space of $U(1)$ instantons. Thus, the $q$-boson representation arising in this case is the one where the representation constructed in \cite{2009arXiv0904.1679F,2009arXiv0905.2555S} directly applies. Namely, we have
\begin{align} \label{eq:qbos01}
    \r_{\text{M5}_c ^{\mathbf{(0,1)}} } :\qta \twoheadrightarrow \text{M5}_c ^{\mathbf{(0,1)}} \hookrightarrow \text{End}(\CalF^{\mathbf{(0,1)}} _c (a) ).
\end{align}
Let us recall the basis states of the Fock representation $\CalF_c^{\mathbf{(0,1)}} (a) = \bigoplus_{\{\l\}} \BC \vert a , \l \rangle ^{(c)}$ are enumerated by a single Young diagram $\l$. The action of the generating currents on the state $\vert a,\l\rangle ^{(c)}$ is given by
\begin{align} 
\begin{split}
    &E(X) \vert a ,\l \rangle^{(c)} = \frac{1-q_c}{\k_1} \sum_{\Box \in \p_+ \l} \d(\chi_\Box /X) \underset{X = \chi_\Box}{\text{Res}} {X^{-1} \EY_\l ^{(c)} (X,a) ^{-1}} \vert a,\l+\Box\rangle ^{(c)}, \\
    &F(X) \vert a,\l \rangle ^{(c)} =  - \frac{1-q_c ^{-1}}{\k_1} q_c ^{-1/2} \sum_{\Box \in \p_- \l} \d (\chi_\Box/X) \underset{X= \chi_\Box}{\text{Res}} X^{-1} \EY _\l ^{(c)} (X q_c ^{-1} ,a) \vert a,\l-\Box\rangle ^{(c)} , \\
    &K^\pm (X) \vert a,\l \rangle ^{(c)} = q_c ^{-1/2} \left(\frac{\EY_\l ^{(c)} (X q_c ^{-1},a) }{ \EY_\l ^{(c)} (X,a) } \right)_\pm \vert a,\l \rangle ^{(c)}, \qquad C\vert a,\l \rangle ^{(c)} = \vert a,\l \rangle ^{(c)},
\end{split}
\end{align}
where we defined $\chi_{\Box_{(i,j)}} = a q_{c+1} ^{i-1} q_{c-1} ^{j-1}$ and
\begin{align}
    \EY^{(c)} _\l (X,a) = \left( 1- \frac{a}{X} \right) \prod_{\Box \in \l} \frac{\left(1- \frac{ q_{c+1} \chi_\Box}{X} \right) \left(1- \frac{ q_{c-1} \chi_\Box}{X} \right)}{\left(1- \frac{  \chi_\Box}{X} \right) \left(1- \frac{ q_{c} ^{-1} \chi_\Box}{X} \right)} = \frac{\prod_{\Box \in \p_+ \l} \left(1- \frac{\chi_\Box}{X}  \right) }{\prod_{\Box \in \p_- \l} \left( 1- \frac{q_c ^{-1} \chi_\Box}{X} \right) }.
\end{align}
Here, $\p_+ \l$ (resp. $\p_- \l$) is the set of boxes that can be added to (resp. removed from) the Young diagram $\l$. Also, $f(X) _+$ (resp. $f(X)_-$) is the expansion of $f(X)$ in the domain $\vert X \vert >\!\!> 1$ (resp. $\vert X \vert <\!\!<1$). Note that $(C,C^\perp) \mapsto (1, q_c^{1/2})$, in particular.\footnote{In some literature, \eqref{eq:qbos1n} is referred to as the \textit{horizontal} Fock representation, while \eqref{eq:qbos01} is referred to as the \textit{vertical} Fock representation. Since the terminology may cause confusion when compared to the $\mathbf{(p,q)}$-fivebrane web, we will avoid using this terminology.}

It was indeed shown that the $q$-boson representations $\r_{\text{M5}_c ^{\mathbf{(1,0)}}}$ and $\r_{\text{M5}_c ^{\mathbf{(0,1)}}}$ are exchanged with each other under the action of the $S$-transformation, also known as the Miki automorphism \cite{Miki:2007}, of the quantum toroidal algebra $\qta$ \cite{Bourgine:2018fjy}. This is consistent with the corresponding reparametrization swapping the $\BC^\times _X$-plane and the $\BC^\times _Z$-plane.

\subsubsection{$q$-boson intertwiners for non-transverse M5-brane intersections}
The M5-branes supported on different holomorphic curves $C^{\mathbf{(p,q)}}$ can make transverse or non-transverse intersections. Since it is difficult to visualize this web of M5-branes directly, it is helpful to pass to the IIB dual frame where the M5-brane supported on the curve $C^{\mathbf{(p,q)}}$ becomes the $\mathbf{(p,q)}$-fivebrane (see section \ref{subsec:duality}). Recall that $\mathbf{(p,q)}$-fivebranes are supported on $\BR^2 _{\ve_{c+1}} \times \BR^2 _{\ve_{c-1}} \times  l^{\mathbf{(p,q)}} \times S^1$, where $l^{\mathbf{(p,q)}} = \{ \mathbf{q} x -\mathbf{p} z  = \text{const} \}$ is a line on the two-dimensional plane $\BR_x \times \BR_z$. 

When the two fivebranes only share one of the three topological planes in their supports (namely, $c_1 \neq c_2$), the two fivebranes cannot end on each other as they pass through along the mutually transverse directions. By a similar argument for the transverse M2-M5 intersection, the local operator at such a \textit{transverse} intersection of M5-branes would provide the R-matrix between the associated $q$-boson representations; explicitly, $R_{\text{M5}_{c_1} ^{\mathbf{(p_1,q_1)}},\text{M5}_{c_2} ^{\mathbf{(p_2,q_2)}} } = (\r_{\text{M5}_{c_1} ^{\mathbf{(p_1,q_1)}}} \otimes \r_{\text{M5}_{c_2} ^{\mathbf{(p_2,q_2)}}} ) \CalR$. We will not discuss this transverse M5-brane intersection further in this work.

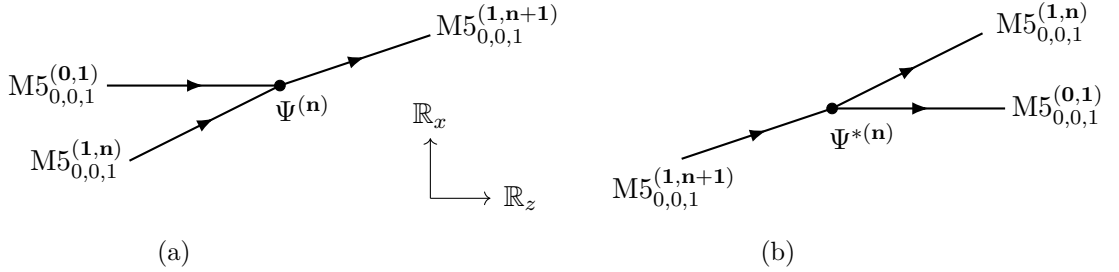
\begin{figure}[h!]\centering
 \begin{subfigure}[b]{0.3\textwidth} \centering
\begin{tikzpicture}[
    mid arrow/.style={
        postaction={decorate,decoration={
            markings,
            mark=at position .5 with {\arrow{Latex[length=6.4pt, sep=-3.2pt -1]}}
    }}
  },
]

\draw[mid arrow, thick] (-2,0) -- (0,1);
\draw[mid arrow, thick] (-2.3,1) -- (0,1);
\draw[mid arrow, thick] (0,1) -- (2,1.67);

\node at (2.9,1.8) {$\text{M5}_{0,0,1} ^{\mathbf{(1,n+1)}}$};

\node at (-3,1) {$\text{M5}_{0,0,1} ^{\mathbf{(0,1)}}$};

\node at (-2.7,0) {$\text{M5}_{0,0,1} ^{\mathbf{(1,n)}}$};

\filldraw[black] (0,1) circle (2pt) ;
\node at (0.3,0.7) {$\Psi^{\mathbf{(n)}}$};

\draw[->] (2,-0.5) -- (2,0.3);
\draw[->] (2,-0.5) -- (2.8,-0.5);

\node at (2,0.6) {$\BR_x$};
\node at (3.2,-0.5) {$\BR_z$};

\end{tikzpicture} \caption{} 
\end{subfigure} \hspace{0.2\textwidth}
\begin{subfigure}[b]{0.3\textwidth} \centering
\begin{tikzpicture}[
    mid arrow/.style={
        postaction={decorate,decoration={
            markings,
            mark=at position .5 with {\arrow{Latex[length=6.4pt, sep=-3.2pt -1]}}
    }}
  },
]

\draw[mid arrow, thick] (3,1.67) -- (5,2.67);
\draw[mid arrow, thick] (3,1.67) -- (5.3,1.67);
\draw[mid arrow, thick] (1,1) -- (3,1.67);

\node at (0.9,0.6) {$\text{M5}_{0,0,1} ^{\mathbf{(1,n+1)}}$};

\node at (5.8,2.8) {$\text{M5}_{0,0,1} ^{\mathbf{(1,n)}}$};

\node at (6,1.7) {$\text{M5}_{0,0,1} ^{\mathbf{(0,1)}}$};

\filldraw[black] (3,1.67) circle (2pt) ;
\node at (3.4,1.3) {$\Psi^{*\mathbf{(n)}}$};

 \end{tikzpicture} \caption{}
 \end{subfigure} 
\caption{Basic non-transverse intersections of fivebranes and associated intertwiners of $q$-boson representations. We call (a) the first kind; (b) the second kind.} \label{fig:trivalent}
\end{figure}

When the two fivebranes share two of the three topological planes (namely, $c_1 = c_2$), they can make an intersection at a point on $\BR_x \times \BR_z$ if the $\mathbf{(p,q)}$ charge is conserved \cite{Aharony:1997ju}. From now on, we will focus on non-transverse M5-brane intersections. Let us fix $c=3$ so that all the fivebranes have $\BR^2 _{\ve_1} \times \BR^2 _{\ve_2}$ as the topological part of their supports. By consecutively joining their intersections, we can form $\mathbf{(p,q)}$-web of fivebranes. The basic building blocks for the web are the trivalent intersections depicted in Figure \ref{fig:trivalent}.

The local operators on the fivebranes are ordered in the increasing $\BR_x$-direction, which is equivalent to the radial-ordering on $\BC^\times _X$ in the M-theory setting. These local operators can act on the Hilbert space attached to each fivebrane, formed as the Fock space $\CalF^{\mathbf{(p,q)}} _3 (v)$ for the associated $q$-boson representation $\text{M5}_{0,0,1} ^{\mathbf{(p,q)}}$. In this sense, the local operators supported at the first kind of the trivalent intersection provides a map,
\begin{align}
    \Psi^{\mathbf{(n)}} (v,v') : \CalF^{\mathbf{(0,1)}} _{3} (v') \otimes  \CalF^{\mathbf{(1,n)}} _{3} (v) \to  \CalF^{\mathbf{(1,n+1)}} _{3} (-v v'), 
\end{align}
between the Fock spaces attached to the three fivebranes. By an argument similar to that of the non-transverse M2-M5 intersection (see section \ref{subsubsec:nontransr}), we expect that the gauge-invariance condition for this intersection at full quantum level reads 
\begin{align} \label{eq:fockint}
    \r_{\text{M5}_{0,0,1} ^{\mathbf{(1,n+1)}}} (g) \Psi^{\mathbf{(n)}} = \Psi^{\mathbf{(n)}} \Delta_{\text{M5}_{0,0,1} ^{\mathbf{(0,1)}} , \text{M5}_{0,0,1} ^{\mathbf{(1,n)}}} (g).
\end{align}
In other words, the local operator at the trivalent intersection provides an intertwiner between the $q$-boson representations of the quantum toroidal algebra $\qta$. In fact, this constraint can be solved to uniquely determine the intertwiner, up to an overall scaling \cite{Awata:2011ce}. We will not review the derivation of this intertwiner here, but only present its expression in appendix \ref{sec:fockint}.

Similarly, let us consider the local operator supported at the second kind of the trivalent intersection. The local operator gives a map,
\begin{align}
    \Psi^{*\mathbf{(n)}} (v,v') : \CalF^{\mathbf{(1,n+1)}} _{3} (-v v')  \to \CalF^{\mathbf{(0,1)}} _{3} (v') \otimes  \CalF^{\mathbf{(1,n)}} _{3} (v), 
\end{align}
between the Fock spaces attached to the three M5-branes. The gauge-invariance condition at the quantum level is expected to be
\begin{align} \label{eq:dualfockint}
     \Delta^{\text{op}}_{\text{M5}_{0,0,1} ^{\mathbf{(0,1)}} , \text{M5}_{0,0,1} ^{\mathbf{(1,n)}}} (g) \Psi^{*\mathbf{(n)}} = \Psi^{*\mathbf{(n)}} \r_{\text{M5}_{0,0,1} ^{\mathbf{(1,n+1)}}} (g)
\end{align}
The constraint exactly determines the intertwiner up to an overall scaling \cite{Awata:2011ce}. We give the expression for this intertwiner in appendix \ref{sec:fockint}.

\vspace{4mm}

\subsection{$qq$-characters from Miura operators on web of $q$-bosons}
Having introduced the $q$-boson representations associated to M5-branes supported on holomorphic curves $C^{\mathbf{(p,q)}}$ and the intertwiners associated to their non-transverse intersections, now we can compose these M5-brane intersections to form a brane web engineering 5-dimensional $\EN=1$ gauge theories compactified on a circle. Moreover, we can introduce an M2-brane which is transverse to the M5-branes. We have studied such a transverse M2-M5 intersection giving rise to the R-matrix between the $q$-bosons and the quantum torus algebra (namely, a product of Miura operators). Here, we will show that the insertion of the R-matrix in the web of $q$-bosons establishes the \textit{qq-characters} \cite{Nikita:I}.

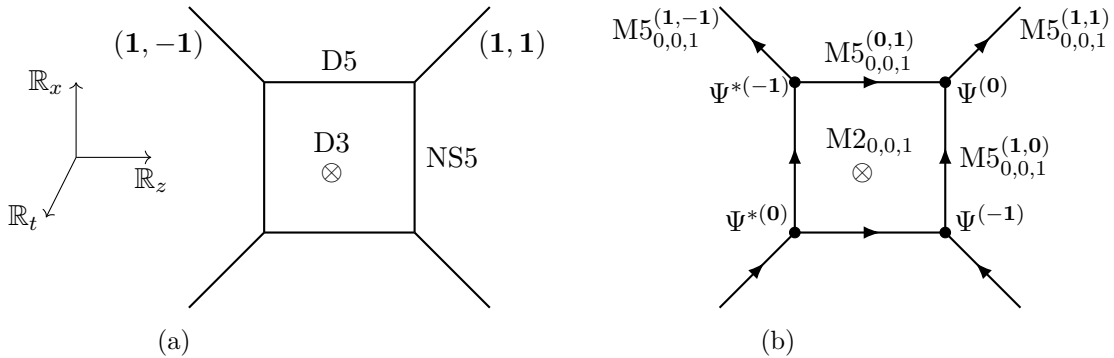
\begin{figure}[h!]\centering
 \begin{subfigure}[b]{0.3\textwidth} \centering
\begin{tikzpicture}[
    mid arrow/.style={
        postaction={decorate,decoration={
            markings,
            mark=at position .5 with {\arrow{Latex[length=6.4pt, sep=-3.2pt -1]}}
    }}
  },
]

\draw[thick] (-1,1) -- (-2,2) ;
\draw[thick] (1,1) -- (2,2);
\draw[thick] (-1,1) -- (1,1) ;
\draw[thick] (-1,-1) -- (-1,1);
\draw[thick] (1,-1) -- (1,1);
\draw[thick] (-1,-1) -- (1,-1);
\draw[thick]  (-2,-2) -- (-1,-1);
\draw[thick] (2,-2) -- (1,-1);
\node at (-0.1,-0.2) {$\otimes$};

\node at (1.5,0) {NS5};
\node at (0,1.3) {D5};
\node at (-0.1,0.2) {D3};
\node at (2.3,1.5) {$\mathbf{(1,1)}$};
\node at (-2.4,1.5) {$\mathbf{(1,-1)}$};

\draw[->] (-3.5,0) -- (-3.5,1);
\draw[->] (-3.5,0) -- (-2.5,0);
\draw[->] (-3.5,0) -- (-3.9,-0.8);

\node at (-3.9,1) {$\BR_x$};
\node at (-2.5,-0.3) {$\BR_z$};
\node at (-4.2,-0.8) {$\BR_t$};

\end{tikzpicture} \caption{} 
\end{subfigure} \hspace{0.2\textwidth}
 \begin{subfigure}[b]{0.3\textwidth} \centering
 \begin{tikzpicture}[
    mid arrow/.style={
        postaction={decorate,decoration={
            markings,
            mark=at position .5 with {\arrow{Latex[length=6.4pt, sep=-3.2pt -1]}}
    }}
  },
]
 
\draw[mid arrow, thick] (-1,1) -- (-2,2) ;
\draw[mid arrow, thick] (1,1) -- (2,2);
\draw[mid arrow, thick] (-1,1) -- (1,1) ;
\draw[mid arrow, thick] (-1,-1) -- (-1,1);
\draw[mid arrow, thick] (1,-1) -- (1,1);
\draw[mid arrow, thick] (-1,-1) -- (1,-1);
\draw[mid arrow, thick]  (-2,-2) -- (-1,-1);
\draw[mid arrow, thick] (2,-2) -- (1,-1);
\node at (-0.1,-0.2) {$\otimes$};

\node at (1.8,0) {$\text{M5}_{0,0,1} ^{\mathbf{(1,0)}}$};
\node at (0,1.4) {$\text{M5}_{0,0,1} ^{\mathbf{(0,1)}}$};
\node at (0,0.2) {$\text{M2}_{0,0,1}$};

\node at (2.6,1.7) {$\text{M5}_{0,0,1} ^{\mathbf{(1,1)}}$};
\node at (-2.7,1.7) {$\text{M5}_{0,0,1} ^{\mathbf{(1,-1)}}$};

\node at (1.5,0.9) {$\Psi ^{\mathbf{(0)}}$};
\node at (-1.6,0.9) {$\Psi ^{*\mathbf{(-1)}}$};
\node at (-1.5,-0.8) {$\Psi ^{*\mathbf{(0)}}$};
\node at (1.6,-0.8) {$\Psi ^{\mathbf{(-1)}}$};

\filldraw[black] (-1,1) circle (2pt) ;
\filldraw[black] (1,1) circle (2pt) ;
\filldraw[black] (1,-1) circle (2pt) ;
\filldraw[black] (-1,-1) circle (2pt) ;

 \end{tikzpicture} \caption{}
 \end{subfigure} 
\caption{(a) $\mathbf{(p,q)}$-fivebrane web of IIB theory engineering the 5d $\EN=1$ pure $SU(2)$ gauge theory. The $\mathbf{(p,q)}$-fivebranes are supported on $\BR^2 _{\ve_1} \times \BR^2 _{\ve_2} \times \{\text{line on the figure}\} \times S^1$. The $qq$-character as a codimension-4 defect of the 5d $\EN=1$ gauge theory is introduced by a transverse D3-brane supported on $\BR^2 _{\ve_3} \times \BR_t \times S^1$; (b) Web of $q$-bosons and intertwiners. At the transverse intersection with the D3-brane, the R-matrix between the $q$-bosons and the quantum torus algebra $\text{M2}_{0,0,1}$ is assigned.} \label{fig:puresu2}
\end{figure}

We will only exemplify the simplest case: the 5-dimensional $\EN=1$ pure $SU(2)$ gauge theory. In the IIB dual frame, the associated brane web is given as in Figure \ref{fig:puresu2}. Note, however, that the principle directly generalizes to arbitrary M5-brane web. The $qq$-characters of the 5d $\EN=1$ effective gauge theory on the worldvolume of the M5-brane web would arise by the additional insertion of transverse M2-branes. We emphasize that the correspondence between the transverse M2-M5 intersections and the $qq$-characters is anticipated due to a string duality. Specifically, the original crossed D-brane configuration for the $qq$-characters \cite{Nikita:I,Nikita:III} transforms into the M2-branes transversally intersecting the M5-brane web through T-duality and uplift to M-theory (see section \ref{subsec:duality}). The example we establish below is confirmation of this expected correspondence.

In our example, there are four trivalent intersections of the fivebranes, each of which is assigned with an intertwiner of the associated $q$-boson representations. Let us break the $\mathbf{(p,q)}$-fivebrane web into the lower half and the upper half, splitting the intertwiners into two groups. For the lower half of the web, we can compose the two intertwiners through the space $\CalF^{\mathbf{(0,1)}} _3 $ of states of the intermediate M5-brane as
\begin{align}
\begin{split}
    &\s\left(\Psi^{\mathbf{(-1)}} (v_2 , a_1) \otimes \Psi^{* \mathbf{(0)}} (v_1 a_2,a_1)\right)\s  \\
 &\qquad: \CalF^{\mathbf{(1,1)}} _3 (-v_1 a_1 a_2) \otimes \CalF_3 ^{\mathbf{(1,-1)}} (v_2 )   \to \CalF_3 ^{\mathbf{(1,0)}} (v_1 a_2) \otimes \CalF_3 ^{\mathbf{(1,0)}} (-v_2 a_1) ,
\end{split}
\end{align}
where we swapped the ordering of the Fock spaces in the tensor product using the exchange operators. Similarly, for the upper half of the web we have
\begin{align}
\begin{split}
    &\s\left(\Psi^{\mathbf{(0)}} ( -v_2 a_1,a_2) \otimes \Psi^{* \mathbf{(-1)}} (-v_1  , a_2) \right)\s \\
    &\qquad : \CalF_3 ^{\mathbf{(1,0)}} (v_1 a_2) \otimes \CalF_3 ^{\mathbf{(1,0)}} (-v_2 a_1) \to  \CalF_3 ^{\mathbf{(1,-1)}}(-v_1 )  \otimes \CalF^{\mathbf{(1,1)}} _3 (v_2  a_1 a_2)  ,
\end{split}
\end{align}
Then, the web as a whole provides a \textit{correlation function} of intertwiners,
\begin{align} \label{eq:measure}
\begin{split}
    &\CalT:= \s\left(\Psi^{\mathbf{(0)}} ( -v_2 a_1,a_2) \otimes \Psi^{* \mathbf{(-1)}} (-v_1  , a_2)\right) \circ \left(\Psi^{\mathbf{(-1)}} (v_2 , a_1) \otimes \Psi^{* \mathbf{(0)}} (v_1 a_2,a_1) \right)\s \\
    &\qquad: \CalF^{\mathbf{(1,1)}} _3 (-v_1 a_1 a_2) \otimes \CalF_3 ^{\mathbf{(1,-1)}} (v_2)   \to 
\CalF_3 ^{\mathbf{(1,-1)}}(-v_1)\otimes \CalF^{\mathbf{(1,1)}} _3 (v_2 a_1 a_2) .
\end{split}
\end{align}
We compute its vacuum expectation value in the $q$-boson representations by evaluating it at the respective vacua,
\begin{align}
    \CalZ(a;\qe) := (\langle \varnothing \vert \otimes \langle \varnothing \vert ) \CalT ( \vert \varnothing\rangle \otimes \vert \varnothing \rangle ).
\end{align}
By normal-ordering all the vertex operators in \eqref{eq:measure}, we can explicitly compute this vacuum expectation value. The outcome is (see appendix \ref{sec:fockint} for notations)
\begin{align}
\begin{split}
    &\CalZ(a;\qe) = \CalZ^{\text{1-loop}} (a) \sum_{\{\l_1,\l_2 \}}  \qe^{\vert \l_1 \vert + \vert \l_2 \vert} \frac{1}{\prod_{\a,\b=1,2} N_{\l_\a \l_\b} (a_\a/a_\b;q_1,q_2)  } \\
    &\CalZ^{\text{1-loop}} (a) = \exp \left[ - \sum_{r=1} ^\infty \frac{(a_1/a_2)^r}{r} \frac{1+q_3 ^{-r}}{(1-q_1 ^r)(1-q_2 ^r)} \right]
\end{split},
\end{align}
where we introduced $\qe = q_3 ^{-1}\frac{v_2}{v_1}$. This exactly reproduces the partition function of the 5d $\EN=1$ pure $SU(2)$ gauge theory on $\BR^2 _{\ve_1} \times \BR^2 _{\ve_2} \times S^1$, with the complexified gauge coupling $\qe$ and the Coulomb parameter $a=a_1/a_2$. Therefore, we establish the correspondence between the 5d $\EN=1$ gauge theory partition function and the vacuum expectation value of the \textit{web} of holomorphic surface defects in the 5d Chern-Simons theory, as anticipated by string duality (see section \ref{subsec:duality}). We stress that this reconstruction is not computationally new, but provides a physical explanation of the algebraic formulation obtained in \cite{Awata:2011ce}.\\

Now, let us introduce an M2-brane supported on $\BR^2 _{\ve_3} \times \BR_t$ on top of the M5-brane web. The resulting transverse M2-M5 intersection assigns the R-matrix $R_{\text{M2}_{0,0,1},\text{M5}_{0,0,2} ^{\mathbf{(1,0)}}} \in \text{M2}_{0,0,1} \otimes \text{M5}_{0,0,2} ^{\mathbf{(1,0)}} $. Recalling that ${\text{M5}_{0,0,2} ^{\mathbf{(1,0)}}} \hookrightarrow {\text{M5}_{0,0,1} ^{\mathbf{(1,0)}}} \otimes {\text{M5}_{0,0,1} ^{\mathbf{(1,0)}}} \simeq \text{End}(\CalF_3 ^{\mathbf{(1,0)}}) \otimes \text{End}(\CalF_3 ^{\mathbf{(1,0)}})$, this R-matrix is inserted in the middle between the lower and the upper halves of the web, producing
\begin{align}
\begin{split}
   \EX &:= \left(\s( \Psi^{\mathbf{(0)}} \otimes \Psi^{* \mathbf{(-1)}} )\s \right)  R_{\text{M2}_{0,0,1},\text{M5}_{0,0,2} ^{\mathbf{(1,0)}}} \left( \s ( \Psi^{\mathbf{(-1)}} \otimes \Psi^{* \mathbf{(0)}}) \s \right) \\ 
   & \in \text{M2}_{0,0,1}  \otimes \text{Hom} \left(\CalF^{\mathbf{(1,1)}} _3 (-v_1 a_1 a_2) \otimes \CalF_3 ^{\mathbf{(1,-1)}} (v_2)  , \CalF_3 ^{\mathbf{(1,-1)}}(-v_1)\otimes \CalF^{\mathbf{(1,1)}} _3 (v_2, a_1 a_2)  \right).
\end{split}
\end{align}
We can also define its vacuum expectation value by evaluating it at the vacua of the $q$-bosons and the basis elements of $\CalV_3 (u)$,
\begin{align} \label{eq:vevqq}
    \left\langle \EX(u) \right\rangle_k:= \left([u q_3^{-1}]_{i+k}^* \otimes \langle \varnothing \vert \otimes \langle \varnothing \vert \right) \EX \left( [u]_i \otimes \vert \varnothing \rangle \otimes \vert \varnothing \rangle  \right). 
\end{align}
Even though $i\in \BZ$ enumerating the states $[u]_i \in \CalV_3 (u)$ is left arbitrary, the non-trivial dependence is only on the \textit{distance} $k \in\BZ$ between the vector $[u]_i$ and the dual vector $[uq_3 ^{-1}]_{i+k} ^*$, not on the specific choice of $i $. Let us call the vacuum expectation value $ \left\langle \EX \right\rangle_k$ the $k$-th fundamental \textit{qq-character}.\\

We can explicitly evaluate the vacuum expectation values. As we studied in section \ref{subsubsec:miuram2m5}, the R-matrix is given by the product of Miura operators as
\begin{align} \label{eq:miuraqqex}
\begin{split}
    &R_{\text{M2}_{0,0,1},\text{M5}_{0,0,2} ^{\mathbf{(1,0)} }} = R_2 ^{(3)} R_1 ^{(3)} \\
    &= \left[ 1- q_3 ^{-1}\left( e^{-a_0 ^{[2]} \frac{\log q_1 \log q_2}{\log q_3}} \xi^{[2]} (X_1) + e^{-a_0 ^{[1]} \frac{\log q_1 \log q_2}{\log q_3}} \xi^{[1]} (q_3 ^\frac{1}{2} X_1) \varphi^{+,[2]} (q_3 ^{\frac{1}{2}} X_1)   \right)Z_1 \right. \\
    &\quad \left. + q_3 ^{-2} e^{-(a_0 ^{[1]} + a_0 ^{[2]} ) \frac{\log q_1 \log q_2}{\log q_3} } \xi^{[1]} (q_3 ^{-\frac{1}{2}} X_1) \xi^{[2]} (X_1) \varphi^{+,[2]} (q_3 ^{-\frac{1}{2}} X_1)  Z_1 ^2 \right] q_3 ^{d\vert_{\text{M2}_{0,0,1}}}V^{[1]} (X_1) V^{[2]} (q_3 ^{-\frac{1}{2}} X_1).
\end{split}
\end{align}
Here the square bracket indicates which of the two $q$-bosons the operators are valued in. Plugging this expression into \eqref{eq:vevqq}, it is apparent that $\langle \EX \rangle_k \neq 0$ only if $k=0,1,2$ since the R-matrix is a degree-two polynomial in $Z_1$. Also note that the module parameter $u \in \BC^\times$ for $\CalV_3 (u)$ gets shifted by $q_3 ^{-1}$ due to $q_3 ^{d\vert_{\text{M2}_{0,0,1}}}$, as appearing in \eqref{eq:vevqq}.\\

The $0$-th fundamental $qq$-character turns out to be exactly identical to the partition function $\CalZ$ itself, $\left\langle \EX \right\rangle_0 = \CalZ$, since the terms emergent from normal-ordering $V^{[1]}(X_1)$ and $V^{[2]} (q_3 ^{-\frac{1}{2}} X_1)$ to the right cancel with each other. The second fundamental $qq$-character, though seemingly more complicated, also yields more or less trivial expression since the terms appearing from normal-ordering $\xi^{[1]} (q_3 ^{-\frac{1}{2}} X_1)$ and $ \xi^{[2]} (X_1) \varphi^{+,[2]} (q_3 ^{-\frac{1}{2}} X_1)$ also cancel with each other: $\left\langle \EX \right\rangle_2 = -q_3 ^{-2} v_1 ^{-1} v_2 ^{-1} a_1 ^{-1} a_2 ^{-1} \CalZ$.

Finally, let us compute the first fundamental $qq$-character, which is the only one that gives a non-trivial expression in our example.\footnote{To have more than one non-trivial fundamental $qq$-characters, we need to consider the M5-brane web which engineers quiver gauge theories. If the quiver has $n$ gauge nodes, the transverse M2-brane assigns the R-matrix which is a product of $n+1$ Miura operators. It is a degree-$(n+1)$ polynomial in $Z_1$, yielding $\langle \EX \rangle_k \neq 0$ only if $k=0,1,\cdots, n+1$. Moreover, $k=0$ and $k=n+1$ give trivial expressions, so that there are $n$ non-trivial fundamental $qq$-characters. This exactly matches with the number of fundamental $qq$-characters for the linear quiver gauge theory with $n$ gauge nodes (see section \ref{subsec:duality}, where $n=l-2$) \cite{Nikita:I,Nikita:III}.} A straightforward computation gives 
\begin{align}
\begin{split}
     \left\langle \EX(X_1) \right\rangle_1 &= \CalZ^{\text{1-loop}} (a)  \sum_{\{\l_1,\l_2\}} \qe^{\vert \l_1 \vert + \vert \l_2 \vert} \frac{1}{\prod_{\a,\b=1,2} N_{\l_\a \l_\b} (a_\a/a_\b;q_1,q_2)} \\ 
     &\qquad \qquad \qquad \quad \times \frac{q_3 ^{-1/2}}{v_2} \left( \frac{q_3^{-1}}{a_1 a_2}  X_1 \EY (X_1 q_3 ^{-1/2} ) + \qe \frac{1}{ X_1 \EY ( X_1 q_3 ^{1/2})} \right),
\end{split}
\end{align}
where $\EY (X) := \prod_{\a=1,2} \EY_{\l_\a} (X,a_\a)$. We abused the notation and denoted $uq_3 ^i$ simply by $X_1$ (recall the former is the eigenvalue of the latter on the state $[uq_3 ^{-1} ]_{i+1}$). Up to an overall constant, this precisely recovers the fundamental $qq$-character as an observable of the 5d $\EN=1$ pure $SU(2)$ gauge theory \cite{Nikita:I}. Thus, we establish the simplest example of our general prescription: the Miura operators inserted on the web of $q$-bosons, realized by the transverse M2-branes passing through the web of M5-branes, give rise to the $qq$-characters as codimension-4 defect observables of the 5d $\EN=1$ effective gauge theory on the worldvolume of the M5-brane web.\\

\paragraph{Remark}
Note that the degree-one term in the R-matrix \eqref{eq:miuraqqex}, which is responsible for the non-trivial $qq$-character, is proportional to $\Delta_{\text{M5}_{0,0,1} ^{\mathbf{(1,0)}},\text{M5}_{0,0,1} ^{\mathbf{(1,0)}}}(F(X_1))$. It was previously observed in \cite{Bourgine:2016vsq} that the $qq$-characters are realized by inserting such generators into the web of $q$-bosons. In this sense, at the computational level, our findings do not introduce new information. The emphasis, however, is on the origin of the generator insertion. In our prescription, the generators are not simply inserted by hand, but appear as a part of the R-matrix (i.e., product of Miura operators) which is present due to the M2-brane transversally intersecting the M5-brane web. In the perspective of the 5d Chern-Simons theory, the $qq$-characters are re-interpreted as the vacuum expectation values of \textit{network} of line and surface defects. As we emphasized earlier, our construction of the $qq$-characters by the M2-M5 intersection is compatible with the original crossed D-brane configuration in the gauge origami \cite{Nikita:I,Nikita:III}, related to each other by a string duality (see section \ref{subsec:duality}).

\vspace{4mm}

\section{Discussion} \label{sec:discussion}
There are many interesting directions that deserve further studies.

\paragraph{MacMahon representations} We may replace $\CalC=\mathbb{C}^\times_X \times\mathbb{C}^\times_Z$ in the twisted M-theory background by a (multiplicative) $A_P$ singularity, that is to say a $\mathbb{C}^\times_Z$ fibration over $\mathbb{C}^\times_X$, where the fiber degenerates over $P$ points on the base. In the type IIB web, this is dual to adding $P$ D7 branes supported on $\BR^2_{\ve_1}\times\BR^2_{\ve_2}\times\BR^2 _{\ve_3}\times S^1(R)\times \mathbb{R}_t$ at points on $\mathbb{R}_x$. It has been suggested that such branes are natural candidates to be MacMahon representations of the quantum toroidal algebra \cite{Zenkevich:2020ufs}. It would be important to make this point precise, and ultimately further our understanding of 7-brane physics via the representation theory of quantum toroidal algebras. See \cite{Gaiotto:2019wcc} for the case of $\CalC = (\BC\times \BC)/\BZ_{P+1}$.\\

\paragraph{Miki automorphism and 5-branes} In section \ref{sec:typeIIB}, we saw that the Miura R-matrices were captured by crossings of D3 and NS5-branes in type IIB. In an S-dual frame, we would study crossings of  D3 and D5-branes instead, related to the previous setup by action of the Miki automorphism, see section \ref{sssec:pqdualbranes}. At such crossings, we would construct S-dual Miura operators at such crossings, where quantum toroidal vector representations are taken as before, while the Fock representations are no longer ``horizontal'' $q$-boson vertex representations, but instead the ``vertical'' representations  \eqref{eq:qbos01}. In a fully generic S-dual frame, the Miura operator will be realized via crossing of a D3-brane with a $\mathbf{(p,q)}$ 5-brane. We leave the derivation of such S-dual operators to future work.\\

\paragraph{3d non-abelian gauge theory} We only considered only a single D3-brane in the type IIB web; for an arbitrary number of NS5-branes, the low energy 3d theory was a quiver whose gauge content was a product of abelian groups. We can increase the total number of D3-branes, which increases the rank of the gauge group(s) in the quiver. This non-abelian case brings about new subtleties with regards to the cancellation of the gauge anomaly, which generically requires turning on bare Chern-Simons terms, but the computation of the half-index remains straightforward: in the simple case of $K$ semi-infinite D3-branes intersecting a single NS5-brane (and ending on another NS5-brane), the half-index is a product of $K$ R-matrices of the form  \eqref{eq:multipleM2}; it follows from section \ref{sssec:parallelM2} that if we expand this product as a series in $q$-shift operators, the coefficients can all be expressed in terms of generators of spherical DAHA. It would be nice to work out nontrivial examples directly from the type IIB frame.\\

\paragraph{Geometric Langlands Program} Let $\fg$ be a simple Lie algebra. The quantum $q$-Langlands correspondence is the statement on the Riemann surface $\mathbb{C}^{\times}$, there exists an isomorphism of spaces  of $q$-conformal blocks for two different algebras :
\beq
{\cal W}_{q_1,q_2}(\fg) \; \leftrightarrow \; U_{\hbar}(\widehat{^L\fg}_{{}^L k}) \; ,
\eeq
where the parameters $q_1$ and $q_2$ on the left-hand side are related to the quantization parameter $\hbar$ and the affine Kac-Moody level $^L k$ on the right-hand side, all taken complex and not equal to roots of unity.
At the present time, the correspondence has been shown to hold only for blocks valued in tensor products of fundamental representations  \cite{Aganagic:2017smx} and Verma representations \cite{Haouzi:2023doo,Tamagni:2023wan}. It is a tantalizing possibility that the quantum $q$-Langlands correspondence could be embedded into the quantum toroidal algebra, at least for $\fg=\fsl(r)$. Indeed, in that case, the algebra on the left-hand side is the one governing the Fock representations of $U_{q_1,q_2,q_3} (\widehat{\widehat{\fgl}} (1))$ at level $r$, and each $q$-conformal block likewise has an elegant interpretation as the vev of specific operators in $U_{q_1,q_2,q_3} (\widehat{\widehat{\fgl}} (1))$, following section \ref{ssec:Miurahalfindex}. The puzzle lies in the appearance of the algebra $U_{\hbar}(\widehat{^L\fg}_{^L k})$ on the right-hand side; in particular,  the interpretation of its $q$-conformal blocks will require a representation theoretic understanding of K-theoretic stable envelopes \cite{Okounkov:2015spn}. Taking the degeneration limit, it would also be desirable to relate our study of M-brane intersections and R-matrices to the 4d $\EN=2$ gauge theory formulation of the ($\hbar$-)Langlands correspondence  \cite{Jeong:2018qpc,Jeong:2023qdr,Jeong:2024hwf}. This is work in progress.

\appendix

\section{Quantum toroidal algebra of $\fgl(1)$} \label{app:qta}
\subsection{Definition}
The quantum toroidal algebra of $\fgl(1)$, denoted by $U_{q_1,q_2,q_3} (\widehat{\widehat{\fgl}} (1))$, is an associative algebra with the generators $E_k, F_k, K_{\pm r} ^\pm$ $(k \in \BZ,\, r \in \BZ_{\geq 0})$ and $C$. We introduce the generating currents as
\begin{align}
    E(X) = \sum_{k \in \BZ} E_k X^{-k} ,\quad F(X) = \sum_{k \in \BZ} F_k X^{-k},\quad  K ^\pm (X) = \sum_{r =0} ^\infty K_{\pm r} ^\pm X^{\mp r}.
\end{align}
In terms of these generating currents, the defining relations are
\begin{subequations}   
\begin{align}
& C \text{ is central} \\
& 1= K^+ _0 K^- _0 = K^- _0 K^+ _0 \\
& K^\pm (X) K^\pm (X') = K^\pm (X') K^\pm (X) \label{eq:eq:cartan2} \\
&\frac{g(C^{-1} X,X')}{g(CX,X')} K^- (X)K^+ (X') = \frac{g(X',C^{-1} X)}{g(X',C X)} K^+ (X') K^- (X) \label{eq:cartans} \\
&g(X,X') E(X)E(X') + g(X',X) E(X')E(X)=0 \label{eq:EE}\\
&g(X',X) F(X)F(X') + g(X,X') F(X')F(X)=0 \label{eq:FF} \\
&g(X,X') K^\pm (C^{(1\mp 1 )/2} X)E(X')+ g(X',X) E(X') K^\pm (C^{(1\mp 1)/2} X) = 0 \label{eq:KE}\\
&g(X',X) K^\pm (C^{(1\pm 1 )/2} X)F(X')+ g(X,X') F(X') K^\pm (C^{(1\pm 1)/2} X) = 0  \label{eq:KF}\\
& [E(X),F(X')] = \tilde{g} \left\{ \d \left(C \frac{X'}{X} \right) K^+(X) - \d \left( C \frac{X}{X'} \right) K^- (X') \right\} \label{eq:EF} ,
\end{align}
\end{subequations}
augmented by the Serre relations
\begin{subequations}
\begin{align}
    &0 = \text{Sym}_{i_1,i_2,i_3} [E_{i_1} , [E_{i_2+1}, E_{i_3-1}]] \\
    &0 = \text{Sym}_{i_1,i_2,i_3} [F_{i_1} , [F_{i_2+1}, F_{i_3-1}]] .
\end{align}
\end{subequations}
Here, $g(X,X')$ is the structure function defined by
\begin{align} \label{eq:structure}
    g(X,X') := \prod_{a=1} ^3 (X- q_a X').
\end{align}
Note that $\tilde{g}$ only affects the relative normalization of the currents $E(X)$, $F(X)$, and $K^\pm (X)$. We will always choose it to be $\tilde{g} =\k_1 ^{-1}$, where we defined
\begin{align} \label{eq:kappa}
    \k_n = \prod_{i=1} ^3 (q_i ^{\frac{n}{2}} - q_i ^{-\frac{n}{2}}) =\prod_{i=1} ^3 (1- q_i ^{-n}) = - \k_{-n}.
\end{align}
The quantum toroidal algebra $\qta$ is parametrized by two complex numbers, $(q_1,q_2,q_3) \in \left(\BC^\times \right)^{\times 3}$ constrained by $q_1 q_2 q_3 = 1$. They enter in the definition through \eqref{eq:structure} and \eqref{eq:kappa}. Note that there is an explicit triality of permuting $q_1$, $q_2$, and $q_3$.

We reorganize $\left( K^\pm _{\pm r} \right)_{r=0} ^\infty$ in terms of $C^\perp$ and $\left( H_{\pm r} \right)_{r=0} ^\infty$ by
\begin{align}
        K^\pm (X) = (C^\perp) ^{\mp 1} \exp \left( \pm \sum_{r=1} ^\infty \frac{\k_r}{r} H_{\pm r} X^{\mp r} \right),
\end{align}
where $K^- _0 = (K^+ _0)^{-1} = C^\perp$ is a central element. 

It is helpful to recast the defining relations in terms of the modes. It is straightforward to see \eqref{eq:eq:cartan2} and \eqref{eq:cartans} are equivalent to the $q$-boson commutation relation satisfied by $H_r$,
\begin{align}
    [H_r,H_s] = \d_{r+s,0} \frac{r}{\k_r} (C^r - C^{-r}).
\end{align}
Also, the relations \eqref{eq:EE}, \eqref{eq:FF}, and \eqref{eq:EF} read
\begin{subequations}
\begin{align}
    & [E_{k+2}, E_{l-1} ] - (q_1 + q_2 + q_3) E_{k+1} E_l + (q_1 ^{-1} + q_2 ^{-1} + q_3 ^{-1} ) E_l E_{k+1} + (k \leftrightarrow l) =0 \\
    & [F_{k+2}, F_{l-1} ] - (q_1 ^{-1} + q_2 ^{-1} + q_3 ^{-1}) F_{k+1} F_l + (q_1  + q_2   + q_3  ) F_l F_{k+1} + (k \leftrightarrow l) =0 \\
        & [E_k,F_l] = \tilde{g}  \left( C^{-l} K^+ _{k+l} \th_{k+l \geq 0}  - C^{-k} K^- _{k+l} \th_{k+l \leq 0} \right),
 \end{align}
\end{subequations}
where in the last relation we used the notation
\begin{align}
    \th_{\{\text{condition}\}} = \begin{cases}
     1 \qquad \text{if the condition is true} \\
     0 \qquad \text{if the condition is false}
    \end{cases}.
\end{align}
Finally, the relations \eqref{eq:KE} and \eqref{eq:KF} are equivalent to
\begin{subequations}
\begin{align}
    &[H_r, E_k] =  C^{(r-\vert r\vert)/2} E_{k+r} \\
    &[H_r , F_k] = -C^{(r+\vert r \vert)/2} F_{k+r}.
\end{align}
\end{subequations}
Again, note the explicit invariance under the triality of exchanging $(q_1,q_2,q_3)$.

\subsection{Coproduct}
The quantum toroidal algebra of $\fgl(1)$ is a Hopf algebra. In particular, it has a (formal) coproduct $\D : U_{q_1, q_2, q_3} (\widehat{\widehat{\fgl}}(1)) \to \qta \, \widehat{\otimes} \, \qta$ defined by
\begin{subequations}\label{eq:cprod}
    \begin{align}
    &\D (E(X)) = E(X) \otimes 1 + K^- (C_1 X) \otimes E(C_1 X) \\
    &\D (F(X)) = 1 \otimes F(X) + F(C_2 X) \otimes K^+ (C_2 X) \\
    & \D (K^+ (X)) = K^+ (X) \otimes K^+ (C_1 ^{-1} X) \\
    & \D(K^- (X)) = K^- (C_2 ^{-1} X) \otimes K^- (X) \\
    & \D (C) = C\otimes C,
\end{align}
\end{subequations}
where we used the notation $C_{1} = C \otimes 1 $ and $C_2  = 1 \otimes C$. The image of the formal coproduct contains infinite summations of elements, and thus has to be understood as lying in the completed tensor product. The infinite summation is expected to truncate to a finite summation when represented on appropriate modules over $\qta$.

\subsection{Grading operators} \label{subsec:grading}
The quantum toroidal algebra of $\fgl(1)$ is equipped with a $\BZ^2$-grading with two grading operators $d$ and $d^\perp$, which act by
\begin{align} \label{eq:grading}
\begin{split}
  &[d, E(X)] = -X\p_X E(X) ,\quad [d, F(X)] = -X \p_X F(X)  ,\quad [d, K^\pm (X)]= -X \p_X K^\pm (X),\\
    &[d^\perp,  E(X)] = E( X),\quad [d^\perp ,F(X)] = -F( X)  ,\quad [d^\perp ,K^\pm (X)] = 0, \\  
    &[d,C]= [d^\perp , C] = 0.
\end{split}
\end{align}
We define $\text{deg} = (d^\perp , d): \qta \to \BZ^2$, so that we have
\begin{align}
    \deg E_k = (1,k),\quad \deg F_k = (-1,k),\quad \deg H _{r} = (0, r) ,\quad \deg C = \deg C^\perp = (0,0).
\end{align}
In particular, for any parameter $q \in \BC^\times$ it follows from \eqref{eq:grading} that
\begin{align} \label{eq:gradingact}
    q^{d} E(X) = E(q^{-1 }X) q^d,\quad q^{d} F(X) = F(q^{-1 }X) q^d ,\quad q^{d} K^\pm (X) = K^\pm (q^{-1 }X) q^d.
\end{align}

\section{Spherical double affine Hecke algebras} \label{app:sdaha}

\subsection{Double affine Hecke algebras}
The double affine Hecke algebra (DAHA) of $GL(m)$, denoted by $\mathbf{\ddot{H}} ^{(m)} _{q,t^{-1}}$, is an associative algebra generated by $T_i ^{\pm 1}$ ($i=1,2,\cdots, m-1$), $X_i ^{\pm 1}$, and $Y_i ^{\pm 1}$ ($i=1,2,\cdots, m$) subject to the following relations:
\begin{subequations} \label{eq:defdaha}
\begin{align}
&(T_i - t^{\frac{1}{2}} )(T_i+ t^{-\frac{1}{2}}) = 0 ,\qquad T_i T_{i+1} T_i = T_{i+1} T_i T_{i+1} \\
&T_i T_j = T_j T_i \qquad \text{if}\;\; \vert i-j \vert >1 \\
&X_i X_j = X_j X_i,\qquad Y_i Y_j = Y_j Y_i \\
&T_i X_i T_i = X_{i+1},\qquad T_i ^{-1}Y_i T_i ^{-1} = Y_{i+1} \\
&T_i X_j = X_j T_i ,\qquad T_i Y_j = Y_j T_i \qquad \text{if}\;\;j \neq i, i+1 \\
&Y_1 X_1\cdots X_m = q X_1 \cdots X_m Y_1 \\
&X_1 ^{-1} Y_2 = Y_2 X_1 ^{-1} T_1 ^{-2}.
\end{align}
\end{subequations}
Here, the subscripts are understood modulo $m$. The DAHA is parameterized by two complex numbers $q,t^{-1} \in \BC^\times$ (we intentionally use $t^{-1}$ instead of $t$, since it is $q$ and $t^{-1}$ which get identified with two out of the triple $(q_1,q_2,q_3)$, parametrizing the quantum toroidal algebra of $\fgl(1)$ under the algebra homomorphism).

The DAHA is endowed with the automorphism group given by the braid group on three strands,
\begin{align}
    B_3 = \left\langle\left. \t_\pm \; \right\vert \; \t_+ \t_- ^{-1} \t_+ = \t_- ^{-1} \t_+ \t_- ^{-1} \right\rangle.
\end{align}
The action on the generators is given by
\begin{align}
    \t_+ : \begin{cases}
         T_i \mapsto T_i \\
        X_i \mapsto X_i Y_i (T_{i-1} \cdots T_1)(T_1 \cdots T_{i-1}) \\
        Y_i \mapsto Y_i
    \end{cases},\quad \t_- : \begin{cases}
        T_i \mapsto T_i \\
        X_i \mapsto X_i \\
        Y_i \mapsto Y_i X_i (T_{i-1} ^{-1} \cdots T_1 ^{-1})(T_1 ^{-1}\cdots T_{i-1} ^{-1})
    \end{cases}.
\end{align}
Note that
\begin{align} \label{eq:sigma}
    \s := \t_+ \t_- ^{-1} \t_+ = \t_- ^{-1} \t_+ \t_- ^{-1} : \begin{cases}
        T_i \mapsto T_i \\
        X_i \mapsto Y_i ^{-1}\\ 
        Y_i \mapsto Y_i X_i Y_i ^{-1}
    \end{cases}.
\end{align}
The braid group $B_3$ is an extension of $SL(2,\BZ)$ by $\BZ$,
\begin{align}
    0 \to \BZ \to B_3 \to SL(2,\BZ) \to 1,
\end{align}
where the kernel of the projection $B_3 \to SL(2,\BZ)$ is generated by $(\t_+ \t_- ^{-1} \t_+)^4$. Moreover, we have
\begin{align}
     \t_+ &\mapsto \begin{pmatrix}
         1 & 1 \\ 0 & 1
     \end{pmatrix},\qquad \t_-  \mapsto \begin{pmatrix}
         1 & 0 \\ 1 & 1 
     \end{pmatrix},\qquad \s \mapsto \begin{pmatrix}
         0 & 1 \\ -1 & 0 
     \end{pmatrix}.
\end{align}

There is a faithful representation $\varphi_m$ of $\ddot{\mathbf{H}}_{q,t^{-1} } ^{(m)} $ on the space $\BC(q,t^{\frac{1}{2}})[X_1 ^{\pm 1} ,\cdots, X_m ^{\pm 1}]$ defined by \cite{cherednik2005}
\begin{align} \label{eq:polyrep}
\begin{split}
    &\varphi_m (X_i ) = X_i, \\
    &\varphi_m (T_i )= t^{\frac{1}{2}} s_i + \frac{t^{\frac{1}{2}} - t^{-\frac{1}{2}}}{X_i / X_{i+1} -1} (s_i -1), \\
    &\varphi_m (Y_i ) = \varphi_m (T_i)\cdots \varphi_m (T_{m-1}) \o \varphi_m (T_1 ^{-1}) \cdots \varphi_m (T_{i-1} ^{-1}),
\end{split}
\end{align}
where $s_i$ stands for the transposition $X_i \leftrightarrow X_{i+1}$; $\o = s_{m-1} \cdots s_1 q^{D_{X_1}}$; and $q^{D_{X_i}}$ is the difference operator $q^{D_{X_i}} f(X_j) = f(q^{\d_{i,j}} X_j)$.

\subsection{Spherical double affine Hecke algebras}
Let $s_i \in S_m$ be the transposition $s_i = (i,i+1)$, and let $l:S_m \to \BZ_{\geq 0}$ be the length function. For a reduced decomposition $w = s_{i_1} \cdots s_{i_r}$ of $w \in S_m$, we define $T_w = T_{i_1} \cdots T_{i_r}$. It is easy to show that
\begin{align}
    S= \left(\prod_{i=1} ^m \frac{1-t}{1-t^i} \right)\sum_{w\in S_m} t^{\frac{l(w)}{2}} T_w
\end{align}
is an idempotent element in $\mathbf{\ddot{H}} ^{(m)} _{q,t^{-1}}$, i.e., $S^2 = S$. Also, for any $i$ we have $T_i S = S T_i = t^{\frac{1}{2}} S$. The spherical DAHA is defined to be the subalgebra
\begin{align}
    \sh^{(m)} _{q,t^{-1}} = S \mathbf{\ddot{H}}^{(m)} _{q,t^{-1}} S.
\end{align}

The spherical DAHA is generated by
\begin{align}
\begin{split}
    &P^{(m)} _{0,r} = S \sum_{i=1} ^m Y_i ^r S,\qquad P^{(m)} _{0,-r} = q^r S \sum_{i=1} ^m Y_i ^{-r} S,\\
    &P^{(m)} _{r,0} = q^r S \sum_{i=1} ^m X_i ^r S,\quad  P^{(m)} _{-r,0} = S \sum_{i=1} ^m X_i ^{-r} S,
    \end{split} \qquad r \in \BZ_{>0},
\end{align}
with the relations following from \eqref{eq:defdaha}. The action of the braid group $B_3$ on $\mathbf{\ddot{H}}^{(m)} _{q,t^{-1}}$ preserves the subalgebra $\sh^{(m)} _{q,t^{-1}}$. Moreover, the corresponding action factors through the projection $B_3 \to SL(2,\BZ)$; namely, $(\t_+ \t_- ^{-1} \t_+)^4$ acts trivially on $\sh^{(m)} _{q,t^{-1}}$. The remaining $SL(2,\BZ)$-action on $\mathbf{S\ddot{H}}^{(m)} _{q,t^{-1}}$ is given by 
\begin{align}
    P^{(m)} _{a,b} \mapsto P^{(m)} _{g(a,b)},\qquad g\in SL(2,\BZ).
\end{align}
Some useful formulas for the other generators obtained by the action of $SL(2,\BZ)$ are
\begin{align}
    \begin{split}
    &P^{(m)} _{1,k} = q \frac{1- t^m}{1-t} S X_1 Y_1 ^k S,\quad P^{(m)} _{-1,k} = \frac{1-t^{-m}}{1-t^{-1}} S Y_1 ^k X_1 ^{-1} S, \\
    &P^{(m)} _ {k,1} = \frac{1-t^{-m}}{1-t^{-1}} S Y_1 X_1 ^k S,\quad P^{(m)} _ {k,-1} = q \frac{1-t^m}{1-t} S X_1 ^k Y_1 ^{-1} S
\end{split}\qquad k\in \BZ.
\end{align}

The polynomial representation \eqref{eq:polyrep} of DAHA, $\mathbf{H}_{q,t^{-1}} ^{(m)}$, restricts to a faithful representation of the spherical DAHA, $\sh_{q,t^{-1}} ^{(m)}$, on the space of symmetric polynomials $\BC(q,t)[X_1 ^{\pm 1} ,\cdots, X_m ^{\pm 1}]^{S_m}$, given by \cite{MacDonald1979SymmetricFA} (after redefining the difference operators by a simple function of $(X_i)_{i=1}^m$)
\begin{align} \label{eq:polsh}
\begin{split}
    &\varphi_m (S e_r (X) S) = e_r (X), \\
    &\varphi_m (S e_r (Y) S) = \hat{H}_r := t^{\frac{r(r-1)}{2}} \sum_{\substack{I \subset \{1,2,\cdots, m\} \\ \vert I \vert = r}} \left( \prod_{\substack{i \in I \\ j \notin I}} \frac{t X_i -  X_j }{X_i - X_j} \right) \prod_{i \in I} q^{D_{X_i}},
\end{split}
\end{align}
where $e_r (X) = \sum_{1 \leq i_1 <\cdots < i_r \leq m} X_{i_1} \cdots X_{i_r}$ is the $r$-th elementary symmetric polynomial of $(X_i)_{i=1} ^m$, and similarly for $e_r(Y)$. In particular, the difference operators $(\hat{H}_r)_{r=1}^m$ are precisely the Macdonald operators, which provide the quantum Hamiltonians of the trigonometric Ruijsenaars-Schneider model of type $GL(m)$.

In section \ref{sssec:parallelM2}, we used the polynomial representation $\varphi_m \circ \s$ obtained by composing \eqref{eq:polsh} with the automorphism $\s$ \eqref{eq:sigma}. This representation can be written explicitly as
\begin{align}
\begin{split}
    &P_{0,r} ^{(m)} \mapsto q^r \sum_{i=1} ^m X_i ^r,\qquad P_{0,-r} ^{(m)} \mapsto  \sum_{i=1} ^m X_i ^{-r},\qquad r \in \BZ_{>0} \\
    \begin{split}
    &P_{1,k} ^{(m)} \mapsto q \sum_{i=1} ^m \left( \prod_{j\neq i} \frac{ t^{-1} X_i -  X_j }{X_i -X_j} \right) X_i ^k q^{-D_{X_i}} \\
    &P_{-1,k} ^{(m)} \mapsto \sum_{i=1} ^m \left( \prod_{j\neq i} \frac{ t X_i -  X_j }{X_i -X_j} \right) q^{D_{X_i}} X_i ^k
    \end{split} ,\qquad k \in \BZ.
\end{split}
\end{align}

\subsection{Spherical DAHA as representation of quantum toroidal algebra} \label{subsec:daharep}
At each $m\in \BZ_{>0}$, there is a surjective algebra homomorphism $\qta \twoheadrightarrow \sh^{(m)} _{q_2,q_1}$ given by \cite{Schiffmann_Vasserot_2011}
\begin{align}
\begin{split}
    &E_k \mapsto \frac{1}{q_2 (1-q_2)} P^{(m)} _{1,k},\quad F_k \mapsto \frac{1}{1-q_2 ^{-1}} P^{(m)} _{-1,k},\quad H_{\pm r} \mapsto \mp \frac{1}{1-q_2 ^r} P^{(m)} _{0,\pm r} \\
    &C\mapsto 1,\quad C^\perp \mapsto 1,
\end{split}
\end{align}
with $k\in \BZ$ and $r \in \BZ_{>0}$.

\section{Solving the R-matrix}\label{sec:solver}
The constraint for the R-matrix $R^{(c)} = (\r_{\text{M2}_{0,0,1}} \otimes \r_{\text{M5}_{c} ^{\mathbf{(1,0)}}} ) \CalR$, $c\in \{1,2,3\}$, reads
\begin{align}
        R^{(c)} \D_{\text{M2}_{0,0,1}, \text{M5}_c ^{\mathbf{(1,0)}}} (g) = \D^\text{op}_{\text{M2}_{0,0,1}, \text{M5}_c ^{\mathbf{(1,0)}}} (g) R^{(c)} ,\qquad \text{for any } g\in \qta.
\end{align}
We solve the constraint by making an ansatz,
\begin{align}
    R^{(c)} = \bar{R}^{(c)} K^{(c)} \in \text{M2}_{0,0,1} \, \widehat{\otimes} \, \text{M5}_c ^{\mathbf{(1,0)}},
\end{align}
where
\begin{align}
\begin{split}
    &K^{(c)} = \left(q_3 ^{\frac{d}{2}} \otimes 1 \right)  \exp \left[ \sum_{r=1} ^\infty \frac{\k_r}{r} \frac{X_1 ^{-r}}{1-q_3 ^r} \otimes \frac{a_r ^{(c)} }{q_c ^{r/2} -q_c ^{-r/2}} \right] \\
    &\bar{R}^{(c)} = \sum_{m=0} ^\infty \a_m \bar{R} ^{(c) }_m.
\end{split}
\end{align}
Here, we set
\begin{align} \label{eq:ansa}
    \bar{R}_m ^{(c)} = \sum_{k_1,k_2,\cdots, k_m \in \BZ} \left(\r_{\text{M2}_{0,0,1}} \otimes \r_{\text{M5}_c ^{\mathbf{(1,0)}}} \right) \left( E_{k_1}E_{k_2}\cdots E_{k_m} \otimes F_{-k_1}F_{-k_2} \cdots F_{-k_m} \right)
\end{align}
so that $\bar{R} ^{(c)}$ is expanded with respect to the $d^\perp$-grading, where $\a_m \in \BC$ are coefficients to be determined.

Let us denote $\Delta_{\text{M2}_{0,0,1} , \text{M5}_c ^{\mathbf{(1,0)}}} = (\r_{\text{M2}_{0,0,1}} \otimes \r_{\text{M5}_c ^{\mathbf{(1,0)}}}) \Delta  $ simply by $\Delta$ for brevity. Let us also omit explicit representations, so that $g_1 \otimes g_2$ is to be read as $ \left(\r_{\text{M2}_{0,0,1}} \otimes \r_{\text{M5}_c ^{\mathbf{(1,0)}}}  \right) (g_1 \otimes g_2)$. Moreover, we write the zero-mode term in the $q$-boson representation as $v \equiv e^{ a_0 ^{(c)} \frac{\log q_{c+1 } \log q_{c-1}}{\log q_c}}$. 

First, we show that, for each $m \in \BZ_{\geq 0} $,
\begin{align}
    \bar{R}^{(c)}_m K^{(c)} \Delta (K^+ (X)) = \Delta^{\text{op}} (K^+ (X))  \bar{R}^{(c)}_m K^{(c)} .
\end{align}
A straightforward computation gives
\begin{align}
\begin{split}
    &\bar{R}^{(c)}_m K^{(c)} \Delta (K^+ (X)) = \bar{R}^{(c)}_m K^{(c)} \left( K^+ (X) \otimes \varphi^+ _c (X) \right) \\
    &= \frac{v^{-m}  (1-q_c ^{-1})^m}{\k_1 ^m (1-q_3)^m} \left( X_1 ^{k_1} (q_3 ^{-1} X_1)^{k_2} \cdots (q_3 ^{-m+1} X_1)^{k_m} \otimes \left(\xi_c\right)_{k_1} \cdots \left(\xi_c \right) _{k_m} \right)  \\
     & \qquad\qquad\qquad\qquad \times (Z_1 ^m  \otimes 1) \left( K^+ (q_c ^{-\frac{1}{2}} X) \otimes \varphi^+ _c (X) \right) K^{(c)} \\
     &= \frac{v^{-m}  (1-q_c ^{-1})^m}{\k_1 ^m (1-q_3)^m} \xi_c (X_1) \xi_c (q_3 ^{-1}X_1) \cdots \xi_c (q_3 ^{-m+1} X_1) \left( K^+ (q_c ^{-\frac{1}{2}+m} X) \otimes \varphi^+ _c (X) \right) Z_1^m K^{(c)} \\
     &= \exp\left[ \sum_{r=1} ^\infty \frac{\k_r}{r} \frac{1- q_3 ^{-mr}}{1-q_3 ^{-r}} q_c ^{\frac{r}{2}} \left(\frac{X_1}{X}\right)^r  \right] \left( K^+ (q_c ^{-\frac{1}{2}+m} X) \otimes \varphi^+ _c (X) \right) \bar{R}^{(c)}_m K^{(c)} \\
     & = \left( K(q_c ^{-\frac{1}{2}} X) \otimes \varphi^+ _c (X) \right) \bar{R}^{(c)}_m K^{(c)} \\
     &= \Delta^{\text{op}} (K^+ (X)) \bar{R}^{(c)}_m K^{(c)},
\end{split}
\end{align}
where we used \eqref{eq:gradingact} for the second equality and the normal-ordering relation 
\begin{align} \label{eq:normalpx}
\begin{split}
    \varphi^+ _c ( X) \xi(X') &= \exp \left[  - \sum_{r=1} ^\infty \frac{\k_r}{r} \left( 
\frac{q_c ^{\frac{1}{2}} X'}{X} \right)^r \right] :  \varphi^+ _c ( X) \xi(X'): \\ 
    &= \prod_{a=1} ^3 \frac{1- q_a q_c ^{\frac{1}{2}} \frac{ X'}{X}}{1- q_a^{-1} q_c ^{\frac{1}{2}}  \frac{ X'}{X}} :  \varphi^+ _c ( X) \xi(X'):,
\end{split}
\end{align}
in the fourth equality.

In a similar way, we can show that the following relation holds for each $m\in \BZ _{\geq 0}$:
\begin{align}
    \bar{R}^{(c)}_m K^{(c)} \Delta (K^- (X)) = \Delta^{\text{op}} (K^- (X))  \bar{R}^{(c)}_m K^{(c)}. 
\end{align}
A straightforward computation gives
\begin{align}
\begin{split}
   &\bar{R}^{(c)}_m K^{(c)} \Delta (K^- (X)) =    \bar{R}^{(c)}_m K^{(c)} \left( K^- (q_c ^{-\frac{1}{2}} X) \otimes \varphi^- _c (X) \right) \\
   &= \frac{v^{-m} (1-q_c ^{-1})^m}{\k_1 ^m (1-q_3)^m} \xi_c (X_1) \cdots \xi_c (q_3 ^{-m+1} X_1) (Z_1 ^m \otimes 1) \left( K^- (X) \otimes \varphi^- _c (X) \right) K^{(c)} \\
   &= \exp \left[ - \sum_{r=1} ^\infty \frac{\k_r}{r} \frac{1-q_3 ^{mr}}{1-q_3 ^r} \left( \frac{X}{X_1} \right)^r \right] \left( K^- (q_3^{m} X) \otimes \varphi^- _c (X) \right) \bar{R}^{(c)}_m K^{(c)} \\
   &=  \left( K^- ( X) \otimes \varphi^- _c (X) \right) \bar{R}^{(c)}_m K^{(c)}  \\
   &= \Delta^{\text{op}} (K^- (X))  \bar{R}^{(c)}_m K^{(c)},
\end{split}
\end{align}
where we used \eqref{eq:gradingact} and the normal-ordering relation,
\begin{align}
\begin{split}
    V_c (X_1) \varphi^- (X) &= \exp \left[- \sum_{r=1} ^\infty \frac{\k_r}{r} \frac{1-q_c ^{-r}}{1-q_3 ^r} \left(\frac{q_c ^{\frac{1}{2}} X}{X_1} \right)^r \right]:V_c (X_1) \varphi^- (X):\\
    &=K^- (q_c ^{\frac{1}{2}} X) K^- (q_c ^{-\frac{1}{2}} X) ^{-1} :V_c (X_1) \varphi^- (X):,
\end{split}
\end{align}
for the second equality; and the normal-ordering relation
\begin{align} \label{eq:normalxpm}
\begin{split}
\xi_c (X) \varphi^- _c (X ') &= \exp \left[ - \sum_{r=1} ^\infty \frac{\k_r}{r} \left( \frac{X'}{X} \right)^r \right] :\xi_c (X) \varphi^- _c ( X ')   : \\
&= \prod_{a=1} ^3 \frac{1- \frac{q_c  q_a X}{X'}}{1- \frac{q_c  q_a ^{-1} X}{X'} } :\xi_c (X) \varphi^- _c ( X ')   :,
\end{split}
\end{align}
for the third equality.

Next, we turn to the relation
\begin{align}
R^{(c)} \Delta (E(X)) = \Delta^{\text{op}} (E(X)) R^{(c)}.
\end{align}
Recall that $R^{(c)} $ is expanded with respect to the $d^\perp$-grading, so that the relation has to be satisfied for each degree of the grading. The degree-$m$ relation is found to be
\begin{align}\label{eq:erel}
\begin{split}
    &\bar{R}^{(c)}_m K^{(c)} \left( E(X) \otimes 1 \right) + \bar{R}^{(c)}_{m+1}K^{(c)} \left( K^{-} (X) \otimes \left(-v \frac{1-q_c}{\k_1} \right) \eta_c (X) \right) \\
     & \qquad = \left(1 \otimes \left(-v \frac{1-q_c}{\k_1} \right) \eta_c (X) \right) \bar{R}^{(c)}_{m+1} K^{(c)}  + \left(E(q_c ^{\frac{1}{2}} X) \otimes \varphi^- _c (q_c ^{\frac{1}{2}} X) \right)\bar{R}^{(c)}_{m} K^{(c)}. 
\end{split}
\end{align}
Note that there is a recursive relation
\begin{align}
    \bar{R}^{(c)}_{m} = \frac{\a_m}{\a_{m-1}}  \frac{v^{-1} (1-q_c ^{-1})}{\k_1 (1-q_3)} \xi_c (X_1) Z_1 \bar{R}^{(c)}_{m-1}.
\end{align}
Let us assume $\g \equiv \frac{\a_m}{\a_{m-1}}$, and look for its value in which \eqref{eq:erel} holds. We proceed by induction. Assume \eqref{eq:erel} holds for $0 \leq m \leq m'-1$. Then, for $m=m'$, the left hand side of \eqref{eq:erel} gives
\begin{align}
\begin{split}
& \g\frac{v^{-1} (1-q_c ^{-1})}{\k_1 (1-q_3)} \xi_c (X_1) Z_1 \\
& \times \left[ \left(1\otimes \left(-v\frac{1-q_c}{\k_1} \right) \eta_c (X) \right) \bar{R}^{(c)}_{m'} K^{(c)} +  \left(E(q_c ^{\frac{1}{2}} X ) \otimes \varphi^- _c (q_c ^{\frac{1}{2}} X) \right) \bar{R}^{(c)}_{m'-1} K^{(c)}  \right].
\end{split}
\end{align}
The second term in the square bracket vanishes, due to the normal-ordering relation \eqref{eq:normalxpm} and the delta function in $\r_{\text{M2}_{0,0,1}} (E (q_c^{\frac{1}{2}} q_3 X) )$. The remaining term is
\begin{align} \label{eq:emid}
\begin{split}
    &- \g \frac{(1-q_c ^{-1})(1-q_c)}{\k_1 ^2 (1-q_3)} \xi_c (X_1) \eta_c (X) Z_1  \bar{R}^{(c)}_{m'} K^{(c)}\\ 
    &=- \g  \frac{(1-q_c ^{-1})(1-q_c)}{\k_1 ^2 (1-q_3)} \left[\frac{(1-q_{c+1})(1-q_{c-1})}{1-q_c ^{-1}} \left\{ \d\left(\frac{q_c ^{\frac{1}{2}} X}{X_1} \right) \varphi^- _c (X_1) - \d\left( \frac{q_c ^{\frac{1}{2}} X_1}{X} \right) \varphi^+ _c (X) \right\}  \right.\\
    &\qquad\qquad\qquad \qquad\qquad\qquad \qquad \qquad+ \eta_c (X) \xi_c (X_1)   \Bigg]Z_1 \bar{R}^{(c)}_{m'} K^{(c)} \\
    &= \left(1\otimes \left(-v \frac{1-q_c}{\k_1} \right) \eta_c (X) \right)\bar{R}^{(c)}_{m'+1} K^{(c)}\\
    &\qquad+ \frac{\g}{\k_1}\left(E(q_c ^{\frac{1}{2}} X)\otimes \varphi^- (q_c ^{\frac{1}{2}}X) - E(q_c ^{-\frac{1}{2}}X) \otimes \varphi_c ^+ (X)    \right) \bar{R}^{(c)}_{m'} K^{(c)},
\end{split}
\end{align}
where we used the commutation relation
\begin{align}
    [\xi_c (X_1), \eta_c (X)]= \frac{(1-q_{c+1})(1-q_{c-1})}{1-q_c ^{-1}} \left\{ \d\left( \frac{q_c ^{\frac{1}{2}} X}{X_1} \right)\varphi^- _c (X_1) - \d\left( \frac{q_c ^{\frac{1}{2}} X_1}{X} \right)\varphi^+ _c (X) \right\}
\end{align}
for the first equality. The second term in the last line of \eqref{eq:emid} vanishes, due to the normal-ordering relation \eqref{eq:normalpx} and $\r_{\text{M2}_{0,0,1}}(E(q_c ^{-\frac{1}{2}} X)) = \frac{1}{1-q_3} \d(q_c ^{\frac{1}{2}} X_1 /X )Z_1$. Once we pass $Z_1$ to the right, the numerator of \eqref{eq:normalpx} vanishes due to the delta function. All in all, we have
\begin{align}
\begin{split}
    &\bar{R}^{(c)}_{m'} K^{(c)} \left( E(X) \otimes 1 \right) + \bar{R}^{(c)}_{m'+1}K^{(c)} \left( K^{-} (X) \otimes \left(-v \frac{1-q_c}{\k_1} \right) \eta_c (X) \right) \\
     & = \left(1\otimes \left(-v \frac{1-q_c}{\k_1} \right) \eta_c (X) \right)\bar{R}^{(c)}_{m'+1} K^{(c)}+ \frac{\g}{\k_1} \left( E(q_c ^{\frac{1}{2}} X)\otimes \varphi^- (q_c ^{\frac{1}{2}}X) \right) \bar{R}^{(c)}_{m'} K^{(c)}.
\end{split}
\end{align}
 The desired relation \eqref{eq:erel} is exactly recovered if and only if $\g  = \k_1$.

Finally, we check the relation
\begin{align}
R^{(c)} \Delta (F(X)) = \Delta^{\text{op}} (F(X)) R^{(c)}.
\end{align}
The degree-$m$ relation reads
\begin{align} \label{eq:relf}
\begin{split}
    &\bar{R}^{(c)}_{m} K^{(c)} \left(1\otimes v^{-1} \frac{1-q_c ^{-1}}{\k_1} \xi_c (X) \right) + \bar{R}^{(c)}_{m+1} K^{(c)} \left(F(q_c ^{\frac{1}{2}} X) \otimes \varphi ^+ _c (q_c ^{\frac{1}{2}} X) \right) \\
    &\quad = (F(X)\otimes 1) \bar{R}^{(c)}_{m+1} K^{(c)}+ \left( K^+ (X) \otimes v^{-1} \frac{1-q_c ^{-1}}{\k_1} \xi_c (X) \right)\bar{R}^{(c)}_{m} K^{(c)}.
\end{split}
\end{align}
By the induction hypothesis, the left hand side gives
\begin{align}
\begin{split}
 &\g\frac{v^{-1} (1-q_c^{-1}) }{\k_1 (1-q_3)} \xi_c (X_1) Z_1 \\
 &\quad\times\left[ (F(X) \otimes 1) \bar{R}^{(c)}_{m} K^{(c)} +  \left( K^+ (X) \otimes v^{-1} \frac{1-q_c ^{-1}}{\k_1} \xi_c (X) \right) \bar{R}^{(c)}_{m-1} K^{(c)} \right].
\end{split}
\end{align}
The first term in the square bracket gives zero, due to the normal-ordering relation \eqref{eq:xinormal}. The remaining term is
\begin{align}
\begin{split}
    &\g \frac{v^{-2} (1-q_c ^{-1})^2}{\k_1^2 (1-q_3)} \xi_c (X_1)\left( K^+ (q_3 X) \otimes  \xi_c (X) \right)Z_1  \bar{R}^{(c)}_{m-1} K^{(c)} \\
    &= \frac{\left(1- \frac{X}{X_1} \right)\left(1- \frac{q_c X}{X_1} \right)}{\left(1- \frac{q_{c+1}^{-1} X}{X_1} \right) \left(1- \frac{q_{c-1} ^{-1} X}{X_1} \right)} \frac{\left(1- \frac{q_{c+1}^{-1} X_1}{X} \right)\left(1- \frac{q_{c-1} ^{-1} X_1}{X} \right)}{\left(1- \frac{ X_1}{X} \right) \left(1- \frac{q_{c}  X_1}{X} \right)} (K^+ (q_3 X) \otimes F(X)) \bar{R}^{(c)}_{m} K^{(c)} \\
    &= (K^+(X) \otimes F(X))\bar{R}^{(c)}_{m} K^{(c)} \\
    &\qquad - \left[ \frac{(1-q_{c+1} ^{-1})(1-q_{c-1} ) }{1- q_{c-1} q_{c+1}^{-1} } \d \left(  \frac{q_{c+1} X_1 }{X} \right) + \frac{(1-q_{c-1}^{-1}) (1-q_{c+1})}{1- q_{c+1}q_{c-1}^{-1}} \d \left( \frac{q_{c-1} X_1 }{X} \right) \right] \\
    &\qquad\qquad  \times \frac{\left(1- \frac{q_{c+1}^{-1} X_1}{X} \right)\left(1- \frac{q_{c-1} ^{-1} X_1}{X} \right)}{\left(1- \frac{ X_1}{X} \right) \left(1- \frac{q_{c}  X_1}{X} \right)}  \frac{\left( 1- \frac{q_1 X_1}{X} \right)\left( 1- \frac{q_2 X_1}{X} \right)}{\left( 1- \frac{X_1}{X} \right) \left( 1- \frac{q_3 ^{-1} X_1}{X} \right)} (1\otimes F(X)) \bar{R}^{(c)}_{m} K^{(c)} ,
\end{split}
\end{align}
where we used the normal-ordering relation \eqref{eq:xinormal} for the first equality and the relation
\begin{align}
    f(X)_+ - f(X)_- = \sum_{\a} \underset{X= X^{(\a)}}{\text{Res}} \left(\frac{f(X)}{X} \right) \d\left( \frac{X^{(\a)}}{X} \right)
\end{align}
for the second equality. Here, $f(X)$ is a rational function with simple poles at $X= X^{(\a)}$, and $f(X)_+$ (resp. $f(X)_-$) is the expansion of $f(X)$ in the domain $\vert X \vert >\!\!> 1$ (resp. $\vert X \vert <\!\!< 1$). Due to the delta functions, only one of the two terms in the square bracket survives, yielding
\begin{align}
\begin{split}
&\frac{1}{1-q_3 ^{-1} } \d\left( \frac{q_3 X_1}{X} \right)  \frac{v^{-1}(1-q_c ^{-1})}{1-q_3} \xi_c (X)\bar{R}^{(c)}_{m} K^{(c)} \\
 &\qquad = \frac{1}{1-q_3 ^{-1}} Z_1 ^{-1} \d\left(\frac{X_1}{X} \right) \frac{v^{-1} (1-q_c ^{-1})}{1-q_3} \xi_c (X_1) Z_1 \bar{R}_m ^{(c)} K^{(c)} \\
    &\qquad = \frac{\k_1}{\g} (F(X)\otimes 1) \bar{R}_{m+1} ^{(c)} K^{(c)}.
\end{split}
\end{align}
Therefore, we arrive at
\begin{align}
\begin{split}
    &\bar{R}^{(c)}_{m} K^{(c)} \left(1\otimes v^{-1} \frac{1-q_c ^{-1}}{\k_1} \xi_c (X) \right) + \bar{R}^{(c)}_{m+1} K^{(c)} \left(F(q_c ^{\frac{1}{2}} X) \otimes \varphi ^+ _c (q_c ^{\frac{1}{2}} X) \right) \\
    &\quad =\frac{\k_1}{\g} (F(X)\otimes 1) \bar{R}^{(c)}_{m+1} K^{(c)}+ \left( K^+ (X) \otimes v^{-1} \frac{1-q_c ^{-1}}{\k_1} \xi_c (X) \right)\bar{R}^{(c)}_{m} K^{(c)}.
\end{split}
\end{align}
Again, the desired relation \eqref{eq:relf} is precisely recovered if and only if $\g=\k_1$. By choosing the overall normalization by $\a_0 = 1$, we therefore obtain the solution $\a_m = \k_1 ^m$, $m\in \BZ_{\geq 0}$. This completes the derivation of the R-matrix between $\text{M2}_{0,0,1}$ and $\text{M5}_c ^{\mathbf{(1,0)}}$.

If $c=3$, the derivation could have been simplified in fact, since the ansatz \eqref{eq:ansa} truncates to the first two terms as a result of the normal ordering. In any case, the derivation we presented applies to all the cases $c\in\{1,2,3\}$.

\vspace{4mm}

\section{Maximality of set of screening charges} 
\label{sec:screeningmaximal}

In this appendix, we sketch a proof of the following proposition: the only operators in the tensor product of two $q$-bosons which commute with the generators produced by the Miura transformation are the screening operators of section \ref{sssec:screeningsM2}.\\

Let us first consider a single $\text{M2}_{0,0,1}$ brane, with support on $\BR^2_{\ve_3}\times\mathbb{R}_t$.
By definition, the Miura transformation is a product of Miura operators, which in the light of our results is a product of R-matrices:
\beq\label{eq:Rmatrixdefagain}
{R}^{(c_{L+M+N})}_{L+M+N}\, \cdots \, {R}^{(c_{2})}_2\,{R}^{(c_{1})}_1 \; ,\qquad\qquad\qquad \;  R^{(c_I)}_I = \bar{R}^{(c_I)}_I K^{(c_I)}_I,\;\; c\in\{1,2,3\} 
\eeq
To make this appendix self-contained, we rewrite the operator definitions here. The Cartan part reads
\begin{align} 
    K ^{(c_I)}_I =\left(q_{c_I} ^{\frac{d}{2}} \otimes 1 \right)  V^{(c_I)}_I (X) = \left(q_{c_I} ^{\frac{d}{2}} \otimes 1 \right)  \exp \left[ \sum_{r=1} ^\infty \frac{\k_r}{r} \frac{X ^{-r}}{1-q_3 ^r} \otimes \frac{a_r ^{[I]} }{q_{c_I} ^{r/2} -q_c ^{-r/2}} \right] 
\end{align}
with $V^{(c_I)}_I(X) := \exp \left[ \sum_{r=1} ^\infty \frac{\k_r}{r} \frac{X ^{-r}}{1-q_3 ^r} \otimes \frac{a_r ^{[I]} }{q_{c_I} ^{r/2} -q_{c_I} ^{-r/2}} \right]$, while the main part of the R-matrix reads
\begin{align}
    \bar{R}^{(c_I)}_I = \sum_{m=0} ^\infty q_{\bar{c_I}} ^m e^{- m a_0 ^{[I]} \frac{\log q_{c_I+1} \log q_{c_I-1}}{\log q_{c_I}} }  \left(\prod_{j=0} ^{m-1} \frac{1-q_{c_I} q_3^{-j}}{1-q_3 ^{-j-1}} \right) : \prod_{j=0} ^ {m-1} \xi^{(c_I)}_I (q_3 ^{-j} X) : q_3 ^{-m D_{X}}.
\end{align}
with vertex operator
\begin{align}
\xi^{(c_I)}_I (X) =  \exp \left( \sum_{r=1} ^\infty \frac{\k_r}{r} \frac{ q_c ^{r/2} }{ (q_{c_I} ^{r/2} - q_{c_I} ^{-r/2 } )^2 }a_{-r}^{[I]} X^r \right) \exp\left(  \sum_{r=1} ^\infty \frac{\k_r}{r} \frac{1}{ (q_{c_I} ^{r/2} - q_{c_I} ^{-r/2} )^2 } a_r ^{[I]} X^{-r} \right)\; .
\end{align}
The main property of the Miura transformation is that it is a generating function of the current generators for the $q$-boson algebra. The expansion parameter of this generating function is the shift operator $q_3^{-D_{X}}$. The screening charges of the algebra are defined as the commutant of these currents.\\

Then, let us consider a general ansatz\footnote{It is a priori possible to write down a more general ansatz than \eqref{eq:generalscreening}. Our currents coincide with the ansatz proposed in \cite{2015arXiv151208779B,Harada:2021xnm}, and reduce to familiar conformal screening currents \cite{Litvinov:2016mgi,Gaiotto:2020dsq} in the limit where the $q$-deformation is turned off. While we do not prove it in this work, we conjecture that if the commutator of any endomorphism ${\ES}'$ (not necessarily of the form \eqref{eq:generalscreening}) with our R-matrix is zero, then  ${\ES}'$ is necessarily a linear combination of our screening currents ansatz.\label{fn:assump}} for an endomorphism of the tensor product of two Fock spaces $\text{End}\left(\CalF_{c_1} ^{\mathbf{(1,0)}}\otimes \CalF_{c_2} ^{\mathbf{(1,0)}}\right)$:
\begin{align}\label{eq:generalscreening}
\begin{split}
{\ES}^{(3)}(Y)&=e^{{\widetilde{\gamma}}^{[1]}_0\, d^\perp \otimes 1+{\widetilde{\gamma}}^{[2]}_0\, 1 \otimes d^\perp}\; {Y}^{\gamma^{[1]}_0\, a^{[1]}_{0} \otimes 1 + \gamma^{[2]}_0\, 1 \otimes a^{[2]}_{0}+ \gamma^{[0]}_0 1\otimes 1}\\
&\qquad \times \exp\left(\sum_{r>0}\frac{Y^r}{r}(\gamma^{[1]}_{-r}\, a^{[1]}_{-r}\otimes 1 + \gamma^{[2]}_{-r}\, 1\otimes a^{[2]}_{-r} )\right)\\
&\qquad \times\exp\left(-\sum_{r>0}\frac{Y^{-r}}{r}(\gamma^{[1]}_{+r}\, a^{[1]}_{+r}\otimes 1 + \gamma^{[2]}_{+r}\, 1\otimes a^{[2]}_{+r} \right) \; ,
\end{split}
\end{align}
with five zero mode unknowns $\gamma^{[0]}_{0}, \gamma^{[1]}_{0}, \gamma^{[2]}_{0}$, and $\widetilde{\gamma}^{[1]}_{0}, \widetilde{\gamma}^{[2]}_{0}$, as well as an infinite number of nonzero mode unknowns: $\gamma^{[1]}_{\pm r}, \gamma^{[2]}_{\pm r}$ for $r>0$.\\

Our strategy will be to investigate  $[{R}^{(c_{2})}_2\,{R}^{(c_{1})}_1 \;,\;{\ES}^{(3)}(Y)]$ at each order $m$ in the shift operator $q_3^{-m D_{X}}$, and constrain the unknowns in the ansatz current ${\ES}^{(3)}(Y)$ for the commutator to be a $q$-derivative. Note that it suffices to consider a Miura transformation for the product of two R-matrices ${R}^{(c_{2})}_2\,{R}^{(c_{1})}_1$, since the ansatz is defined on the tensor product of precisely two Fock spaces.\\

At order $m=0$, the commutator takes the form
\beq
[V^{(c_2)}_2(X\, q^{1/2}_{c_2}) V^{(c_1)}_1(X\,q^{1/2}_{c_1} q^{1/2}_{c_2})\;, {\ES}^{(3)}(Y)]
\eeq
Imposing its vanishing (up to $q$-derivatives), we find constraints:
\begin{align}\label{eq:constraint1}
    \begin{split}
&\bullet \text{For} \; c_1=c_2, \;\; \gamma^{[1]}_{-r} = -q^{r/2}_{c_1}\,\gamma^{[2]}_{-r} \; ,\\[3mm]
&\bullet \text{For} \; c_1\neq c_2, \;\; \gamma^{[1]}_{-r} = -\left(\frac{q^{r/2}_{c_2} - q^{-r/2}_{c_2}}{q^{r/2}_{c_1} - q^{-r/2}_{c_1}}\right)^2\, q^{r/2}_{c_1}\,\gamma^{[2]}_{-r} \; .
\end{split}
\end{align}
Note that at this point, $c_I\in\{1,2,3\}$. At order $m=1$, the commutator is more involved:
\begin{align}
\begin{split}
&\left[q_{\bar{c_2}}\frac{1-q_{c_2}}{1-q^{-1}_3}\,\xi^{(c_2)}_2 (X) V^{(c_2)}_2(X\, q^{-1}_{3}q^{1/2}_{c_2}) V^{(c_1)}_1(X\,q^{-1}_{3}q^{1/2}_{c_1} q^{1/2}_{c_2})\,e^{a^{[2]}_0\frac{\log q_3\, \log q_{\bar{c_2}}}{\log q_{c_2}}}\right.\\
&\;\;\;+\left.q_{\bar{c_1}}\frac{1-q_{c_1}}{1-q^{-1}_3}\,V^{(c_2)}_2(X\, q^{1/2}_{c_2})\xi^{(c_1)}_1 (X\, q^{1/2}_{c_1})  V^{(c_1)}_1(X\,q^{1/2}_{c_1} q^{1/2}_{c_2})\, e^{a^{[1]}_0\frac{\log q_3\, \log q_{\bar{c_1}}}{\log q_{c_1}}}\; , \;{\ES}^{(3)}(Y) \right]
\end{split}
\end{align}
After normal-ordering the currents and making use of the constraints \eqref{eq:constraint1} at order $m=0$, we find that all the unknowns are determined: a necessary condition for the commutator to be null or proportional to a $q$-derivative is that $c_I\neq 3$, so $c_I\in\{1,2\}$. Another necessary condition is that ${\ES}^{(3)}$ must be one of the four screening currents ${\ES}^{(3)}_{1,1}$, ${\ES}^{(3)}_{2,2}$, ${\ES}^{(3)}_{1,2}$, or ${\ES}^{(3)}_{2,1}$ written in \eqref{eq:screeningdef2} and \eqref{eq:screeningdef3}. The first two currents are the ``bosonic'' screening endomorphisms, on the tensor product of two identical Fock spaces, while the latter two currents are the ``fermionic'' screening endomorphisms, on the tensor product of two distinct Fock spaces. In more detail, we find that for these four solutions, the commutator is proportional to a $q_3$-derivative, 
\beq
{\cal D}^Y_{q_3}\left[Y\,\delta\left(q^{j_{c_1}}_1\,q^{j_{c_2}}_2 Y/X\right)\; :\ldots :\;\right] \; ,
\eeq
where the derivative operator is defined as ${\cal D}^Y_{q_c}(f(Y))=\dfrac{f(Y)-f(q_c\, Y)}{Y}$ for any function $f$. The expression $:\ldots :$ stands for the normal ordering of the product of all operators, and $j_{c_I}$ is a half-integer which is uniquely fixed by the computation, for each  $c_I\in\{1,2\}$.\\

There are again at most four solutions if the initial brane is $\text{M2}_{1,0,0}$, supported on $\BR^2_{\ve_1}\times\mathbb{R}_t$, and four solutions if the initial brane is  $\text{M2}_{0,1,0}$, supported on $\BR^2_{\ve_2}\times\mathbb{R}_t$. These solutions are readily obtained by simply permuting the parameters $(q_1, q_2, q_3)$, thanks to the triality of the quantum toroidal algebra. This is precisely the exhaustive list of screenings written in \eqref{eq:screeningdef2} and \eqref{eq:screeningdef3}, for any $c\in\{1,2,3\}$. This completes the proof of the proposition.\\

\paragraph{Remark} 
The work \cite{Harada:2021xnm} defines the Miura transformation  slightly differently, as 
\beq
\bar{R}^{(c_{L+M+N})}_{L+M+N}(X)\, \cdots \, \bar{R}^{(c_{2})}_2(X)\,\bar{R}^{(c_{1})}_1(X) = \sum_{m=0}^{\infty} (-1)^m\, T_m(X)\, q^{-m D_X}_3 \; ,
\eeq
where $T_m(X)$ is the $m$-th generating current of the $qY_{L,M,N}$ algebra, expressed as a finite series in the operators \eqref{eq:Lambdaops}.
In particular, the authors find that the screening charges of the $qY_{L,M,N}$ algebra are in the commutant of the operators $T_m(X)$. In fact, it is equivalent to show that the screening charges commute with the product of two operators, say $\bar{R}^{(c_{2})}_2\,\bar{R}^{(c_{1})}_1$. In our work, we found that the \emph{same} screening charges are also in the commutant of the Miura operators ${R}^{(c_{2})}_2\,{R}^{(c_{1})}_1$, with  $R^{(c_I)}_I = \bar{R}^{(c_I)}_I K^{(c_I)}_I$. The advantage of using the Miura operators $R^{(c_I)}_I$ is that they have a direct interpretation as R-matrices of the quantum toroidal algebra, unlike the operators $\bar{R}^{(c_I)}_I$.

%\begin{align}
%\gamma^{[I]}_{\pm r}=\frac{q^{r/2}_{j_{\pm,I,1}} - q^{-r/2}_{\pm,j_{\pm,I,1}}}{q^{r/2}_{j_{\pm,I,2}} - q^{-r/2}_{j_{\pm,I,2}}\,q^{j'_{\pm,I,1} r}_{1}\,q^{j'_{\pm,I,2} r}_{2} 
%\end{align}

%For $I=1,2$, $j_{\pm,I,1},j_{\pm,I,2}\in \{1,2,3\},j_{\pm,I,1},j_{\pm,I,2}\in \frac{1}{2}\mathbb{Z}
%{\widetilde{\gamma}}^{[1]}_0$ and ${\widetilde{\gamma}}^{[2]}_0$ 

\vspace{4mm}

\section{$q$-boson intertwiners for non-transverse M5-brane intersections} \label{sec:fockint}
The intertwiner $\Psi^{(\mathbf{n})} (v,v')$ satisfying \eqref{eq:fockint} can be expanded in the dual basis of $\CalF^{\mathbf{(0,1)}} _3$ as,
\begin{align}
    \Psi^{(\mathbf{n})} (v,v') = \sum_{\{\l\}} \Xi_\l  \Psi^{\mathbf{(n)}} _\l (v,v') \langle v',\l \vert
\end{align}
where the coefficients are given by
\begin{align}
    \Psi^{\mathbf{(n)}} _\l (v,v')= t_\mathbf{n} (\l,v,v') : \Psi_\varnothing (v') \prod_{\Box \in \l} \eta(\chi_\Box):,
\end{align}
with
\begin{align}
\begin{split}
& t_{\mathbf{n}} (\l,v,v') = (-v v')^{\vert \l \vert} \prod_{\Box \in \l} \left(q_3^{1/2} / \chi_\Box \right)^{\mathbf{n}+1},\qquad \chi_{\Box_{(i,j)}} = v' q_1 ^{i-1} q_2 ^{j-1}, \\
 &   \Psi_\varnothing (v') =\exp \left[ \sum_{r=1} ^\infty \frac{1}{r} \frac{1} {1-q_3 ^{-r}} a_{-r} ^{(3)} v'^r \right]\exp \left[ -\sum_{r=1} ^\infty \frac{1}{r} \frac{q_3 ^{-r/2}} {1-q_3 ^{-r}} a_{r} ^{(3)} v'^{-r} \right].
\end{split}
\end{align}
Here, we used the normalization of the dual states such that (our case is $c=3$)
\begin{align}
    {}^{(c)}\langle u,\l \vert u,\m \rangle^{(c)} = \left(\Xi_\l  ^{(c)} \right)^{-1} \d_{\l,\m},
\end{align}
where we defined
\begin{align}
    \Xi_\l ^{(c)} = \frac{(u q_c ^{1/2})^{- \vert \l \vert} \prod_{\Box \in \l} \chi_\Box }{N^{(c)} _{\l\l} (1;q_1,q_2)},
\end{align}
with
\begin{align}
    N^{(c)} _{\l \m} (u_1/u_2;q_1;q_2) = \prod_{\Box \in \l} \left( 1- \frac{\chi_\Box}{q_c u_2} \right) \prod_{\blacksquare \in \m} \left( 1- \frac{u_1}{\chi_\blacksquare} \right) \prod_{\substack{\Box \in \l \\ \blacksquare \in \m}} \frac{\left( 1- \frac{q_{c+1} \chi_\square}{\chi_\blacksquare} \right) \left( 1- \frac{q_{c-1} \chi_\square}{\chi_\blacksquare} \right)}{\left( 1-  \frac{\chi_\square}{\chi_\blacksquare} \right) \left( 1- \frac{q_{c} ^{-1} \chi_\square}{\chi_\blacksquare} \right) }.
\end{align}

The intertwiner $\Psi^{*\mathbf{(n)}} (v,v')$ satisfying \eqref{eq:dualfockint} can be expanded in the basis of $\CalF^{\mathbf{(0,1)}} _3$ as
\begin{align}
    \Psi^{*\mathbf{(n)}} (v,v') = \sum_{\{\l\}} \Xi_\l \vert a,\l \rangle \Psi^{* \mathbf{(n)}} _\l (v,v')
\end{align}
where
\begin{align}
    \Psi^{* \mathbf{(n)}} _\l (v,v') = t_\mathbf{n} ^* (\l,v,v') : \Psi^*_\varnothing  (v') \prod_{\Box \in \l} \xi (\chi_\Box) :,
\end{align}
with
\begin{align}
\begin{split}
    &t_\mathbf{n}  ^* (\l,v,v') = (v q_3^{1/2})^{-\vert \l \vert} \prod_{\Box \in \l} (q_3 ^{-1/2} \chi_\Box)^{\mathbf{n}}, \\    
    &\Psi^*_\varnothing (v') = \exp \left[ -\sum_{r=1} ^\infty \frac{1}{r} \frac{q_3 ^{r/2}}{1-q_3 ^{-r}} a_{-r} ^{(3)} v'^r \right] \exp\left[ \sum_{r=1} ^\infty \frac{1}{r} \frac{1}{1-q_3 ^r} a_r ^{(3)} v'^{-r} \right].
\end{split}
\end{align}

\bibliographystyle{utphys}
\bibliography{reference}

\end{document}